\def\isarxiv{1} 

\ifdefined\isarxiv
\documentclass[11pt]{article}

\usepackage[numbers]{natbib}

\else
\documentclass{article}
\usepackage[main,final]{neurips_2026}
\fi

\usepackage{amsmath}
\usepackage{amssymb}
\usepackage{footnote}
\usepackage{algorithm}
\usepackage{algpseudocode}
\usepackage{graphicx}
\usepackage{grffile}
\usepackage{wrapfig}
\usepackage{url}
\usepackage{xcolor}
\usepackage{epstopdf}

\makesavenoteenv{tabular}
\makesavenoteenv{table}

\usepackage{bbm}
\usepackage{dsfont}

\allowdisplaybreaks

\ifdefined\isarxiv

\usepackage{microtype}
\usepackage{array}
\usepackage{tikz}
\usepackage{hyperref}  
\hypersetup{colorlinks=true,citecolor=blue,linkcolor=blue} 
\usetikzlibrary{arrows}
\usepackage[margin=1in]{geometry}

\else

\usepackage{microtype}
\usepackage{hyperref}

\fi
\usepackage{subfig}
\usepackage{amsthm}
\usepackage{epsfig}

\newcommand{\paperTitle}{Sublinear Time Quantum Sensitivity Sampling}

\ifdefined\isarxiv
\newcommand{\paperAuthor}{Zhao Song\thanks{\texttt{magic.linuxkde@gmail.com}. Independent Researcher.}
\and
David P. Woodruff\thanks{\texttt{dwoodruf@cs.cmu.edu}. Carnegie Mellon University and Google Research}
\and
Lichen Zhang\thanks{\texttt{lichenz@mit.edu}. Massachusetts Institute of Technology.}
}
\else
\newcommand{\paperAuthor}{%
  Zhao Song\\
  Independent Researcher\\
  \texttt{magic.linuxkde@gmail.com}
  \And
  David P. Woodruff\\
  Carnegie Mellon University and Google Research\\
  \texttt{dwoodruf@cs.cmu.edu}
  \And
  Lichen Zhang\\
  Massachusetts Institute of Technology\\
  \texttt{lichenz@mit.edu}
}
\fi

\newcommand{\wh}{\widehat}
\newcommand{\wt}{\widetilde}
\newcommand{\ov}{\overline}

\newcommand{\R}{\mathbb{R}}

\renewcommand{\d}{\mathrm{d}}

\renewcommand{\hat}{\wh}

\DeclareMathOperator*{\E}{{\mathbb{E}}}
\DeclareMathOperator*{\var}{\mathrm{Var}}

\DeclareMathOperator{\OPT}{OPT}

\DeclareMathOperator{\poly}{poly}

\DeclareMathOperator{\nnz}{nnz}

\DeclareMathOperator{\rank}{rank}
\DeclareMathOperator{\dist}{dist}
\DeclareMathOperator{\cost}{cost}

\DeclareMathOperator{\tr}{tr}

\newenvironment{CompactItemize}{
\begin{list}{\tiny$\bullet$}{%
\setlength{\leftmargin}{10pt}
\setlength{\itemindent}{0pt}
\setlength{\topsep}{-1pt}
\setlength{\itemsep}{0pt}
}}
{\end{list}}

\makeatletter
\newcommand*{\RN}[1]{\expandafter\@slowromancap\romannumeral #1@}
\makeatother

\newtheorem{theorem}{Theorem}[section]
\newtheorem{lemma}[theorem]{Lemma}
\newtheorem{definition}[theorem]{Definition}

\newtheorem{corollary}[theorem]{Corollary}

\newtheorem{assumption}[theorem]{Assumption}

\newtheorem{remark}[theorem]{Remark}
\newtheorem{claim}[theorem]{Claim}

\usepackage{lineno}

\ifdefined\isarxiv
\renewcommand{\citep}{\cite}
\renewcommand{\citet}{\cite}
\date{}
\fi

\ifdefined\isarxiv
\newsavebox{\tablesavebox}
\newcommand{\tablebox}[1]{\sbox{\tablesavebox}{#1}\ifdim\wd\tablesavebox>\textwidth\resizebox{\textwidth}{!}{\usebox{\tablesavebox}}\else\usebox{\tablesavebox}\fi}
\else
\newcommand{\tablebox}[1]{\resizebox{\textwidth}{!}{#1}}
\fi

\title{\paperTitle}
\author{\paperAuthor}

\begin{document}

\ifdefined\isarxiv
\begin{titlepage}
  \maketitle
  \begin{abstract}
We present a unified framework for quantum sensitivity sampling, extending the advantages of quantum computing to a broad class of classical approximation problems. Our unified framework provides a streamlined approach for constructing coresets and offers significant runtime improvements in applications such as clustering, regression, and low-rank approximation. Throughout, we assume that $n$ and $d$ are polynomially larger than $k$ and $1/\epsilon$, and for regression that $n$ is polynomially larger than $d$. Running times report the leading term. Our contributions include:

\begin{itemize}
    \item \textbf{$k$-median and $k$-means clustering:} For $n$ points in $d$-dimensional Euclidean space, we give an algorithm that constructs an $\epsilon$-coreset consisting only of points in the dataset, of size $\widetilde O(k^{4/3}\epsilon^{-2})$ for $k$-median and $\widetilde O(k^{3/2}\epsilon^{-2})$ for $k$-means, in time $\widetilde O(\epsilon^{-1}n^{0.5}dk^{1.67})$ for $k$-median and $\widetilde O(\epsilon^{-1}n^{0.5}dk^{1.75})$ for $k$-means. Our approach achieves a better dependence on $d$, with the running time depending only linearly on $1/\epsilon$, improving upon the main result of [Xue, Chen, Li and Jiang, ICML'23] in the high-dimensional regime $d\geq \Omega(k^{7/3})$ for $k$-median and $d\geq \Omega(k^{2.5})$ for $k$-means.
    
    \item \textbf{$\ell_p$ regression:} For $\min_{x\in \mathbb{R}^d}\|Ax-b\|_p$ where $A\in \mathbb{R}^{n\times d}$ and $b\in \mathbb{R}^n$, we construct an $\epsilon$-coreset of size $\widetilde O_p(d^{\max\{1, p/2\}}\epsilon^{-2})$ in time $\widetilde O_p(n^{0.5}d^{1.5}\epsilon^{-1})$ for $p\leq 2$ and $\widetilde O_p(n^{0.5}d^{p/4+1}(\epsilon^{-1}+d^{0.5}))$ for $p>2$, matching [Liu and Ji, ICML'26] for $p\leq 2$ and improving upon the prior best quantum sampling approach of [Apers and Gribling, SICOMP'26] for every $p>2$.
    
    \item \textbf{Low-rank approximation with Frobenius norm error:} We introduce the first quantum sublinear-time algorithm for low-rank approximation that approximates the best rank-$k$ solution to a matrix $A\in \mathbb{R}^{d\times n}$ and does not rely on data-dependent parameters, and, returning the factors in factored form, runs in $\widetilde O(n^{0.5}dk^{0.5}\epsilon^{-0.5})$ time. Materializing the factors explicitly costs an additional $\widetilde O(nk^2\epsilon^{-2})$ time. Additionally, we present quantum sublinear algorithms for kernel low-rank approximation and tensor low-rank approximation, broadening the range of achievable sublinear time algorithms in randomized numerical linear algebra.

    \item \textbf{Lower bounds:} We prove that any quantum algorithm for generic sensitivity sampling with total sensitivity $S$ that outputs an $\epsilon$-coreset of size $o(\sqrt{nS})$ must make $\Omega(\sqrt{nS})$ queries to the input, so the query complexity of our framework is optimal up to $\mathrm{poly}(\epsilon^{-1},\log n)$ and VC-dimension factors.
\end{itemize}
  \end{abstract}
  \thispagestyle{empty}
\end{titlepage}
\newpage
\else
\maketitle
\begin{abstract}

\end{abstract}
\fi


\section{Introduction}

Given a set of points $A = \{a_1, \ldots, a_n\} \subset \mathbb{R}^d$, a universe $X$, and a cost function ${\rm cost}: \mathbb{R}^d \times X \rightarrow \mathbb{R}_{\geq 0}$, we study the problem of constructing a \emph{coreset} of $A$: a weighted subset $B$ of points along with a nonnegative weight vector $w \in \mathbb{R}_{\geq 0}^{|B|}$ such that $\sum_{b \in B} w_b \cdot {\rm cost}(b, x) = (1 \pm \epsilon) \cdot {\rm cost}(A, x)$ for all $x \in X$, where ${\rm cost}(A, x) = \sum_{i=1}^n {\rm cost}(a_i, x)$. A coreset is particularly useful because it enables applying any existing approximation (or exact) algorithm on the smaller summary, yielding a good approximation to the original problem. Applications of coresets span clustering~\citep{c09,ls10,fl11,vx12,bflsz22,hv20,bjkw21,css21,clss22,clsss22,hlw24}, graph sparsification~\citep{bk96,st04,ss11,bss12}, hypergraph sparsification~\citep{bst19,kkty21,jls23,l23}, $\ell_p$ regression~\citep{dmm06,c05,ddh+09,cp15,wy23}, submodular optimization~\citep{ry22,jlls23_norm}, generalized linear models~\citep{mmr21,mop22,mmwy22,jlls23_linear}, and subspace approximation~\citep{cem+15,cmm17,wy24}.

Coresets can be constructed via \emph{sensitivity sampling}: the sensitivity of the $i$-th point is defined as $s_i = \max_{x \in X} \frac{{\rm cost}(a_i, x)}{{\rm cost}(A, x)}$. Sensitivity sampling draws point $i$ with probability proportional to $s_i$, reweighting each sampled point by the inverse of its sampling probability to ensure the estimator is unbiased. The seminal work of~\cite{ls10} shows that this yields a coreset with a simple and elegant proof:  for any \emph{fixed} $x\in X$, sampling proportional to sensitivity ensures low variance, and Bernstein’s inequality implies that $O(\epsilon^{-2}S)$ samples suffice, where $S = \sum_{i=1}^n s_i$ is the \emph{total sensitivity}. Recall that our goal is to approximate the cost for \emph{all} $x\in X$. To do so, we union bound over a discretization of $X$ with size $\exp({\rm dim}(X))$, where ${\rm dim}(X)$ is a notion akin to VC dimension~\citep{fl11}. Thus, a coreset can be constructed with $O(\epsilon^{-2} \cdot {\rm dim}(X) \cdot S \log(1/\delta))$ samples, where $\delta$ is the failure probability of the algorithm.


Algorithmically, a challenge arises: it is necessary to either compute or efficiently approximate $s_i$ or an upper bound on it. Most of the work in sensitivity sampling focuses on this task, and for many problems it can be achieved in nearly-linear time in $nd$~\citep{hv20,ss11,cp15,cmm17,wy24}. Since scanning the entire dataset already takes $\Omega(nd)$ time, achieving nearly-linear time is close to optimal.

Typically, coresets are constructed for downstream \emph{optimization problems}. For instance, coresets for $\ell_p$ regression help solve the original regression problem~\citep{jls22,akps24}, and coresets for subspace approximation yield column subset selection for low-rank approximation~\citep{cmm17}. In certain structured settings, some of these optimization problems admit \emph{sublinear time} algorithms. For example, if the input matrix $A$ is positive semidefinite (PSD)~\citep{mw17,bcw20} or Toeplitz~\citep{ms24}, one can obtain a rank-$k$ approximation in $n \cdot \mathrm{poly}(k/\epsilon)$ time, despite $A$ having size $n \times n$.

Can sensitivity sampling—and subsequently solving downstream optimization problems—be accomplished in sublinear time, even without structural assumptions? In this work, we explore this question through the lens of \emph{quantum computing}, analyzing the time complexity of sensitivity sampling under quantum algorithms. Notably, tasks like linear regression, low-rank approximation, and clustering have quantum algorithms running in $o(nd)$ time~\citep{kp16,kllp19,gst22,sj25}, though these often rely on special input representations that support efficient weighted sampling. Moreover, their runtimes often depend on data-specific parameters such as $\|A\|_F$, condition number $\kappa(A) = \sigma_{\max}(A)/\sigma_{\min}(A)$, or dataset radius. In contrast, we seek quantum algorithms that (1) operate in sublinear time, (2) are \emph{independent of input representation}, and (3) have \emph{runtime independent of data-specific parameters}.

In this work, we provide a generic quantum algorithm applicable to \emph{sensitivity sampling} in general.  Throughout the paper, we use $s$ to denote the size of the coreset and let ${\cal T}_{\rm sensitivity}(s, X)$ represent the time to approximate one sensitivity over a set of $s$ points and universe $X$. Our algorithm runs in time $\widetilde{O}(\sqrt{ns}) \cdot {\cal T}_{\rm sensitivity}(s, X)$  and makes $\widetilde O(\sqrt{ns})$ queries to the points in $A$, which implies that as long as $s^{0.5} \cdot {\cal T}_{\rm sensitivity}(s, X) = o(n^{0.5}d)$, we achieve sublinear runtime. Moreover, our algorithm avoids dependence on data-specific parameters: the sample size and sensitivity approximation time depend only on $n$, $d$, $1/\epsilon$, $1/\delta$, and other problem-related parameters (e.g., $k$ in clustering and low-rank approximation, or $p$ in $\ell_p$ regression). We summarize the main result in the following theorem.

\begin{theorem}[Informal version of Theorem~\ref{thm:general}]\label{thm:informal_main}
Let $A\in \R^{n\times d}$ and let $X$ be a universe. Then, there exists a quantum algorithm that constructs an $\epsilon$-coreset $C$ of expected size $s:=\wt O(\epsilon^{-2}\cdot \dim(X)\cdot S\log(1/\delta))$ with probability at least $1-\delta$, where $\dim(X)$ is the dimension of the range space induced by $X$ (Definition~\ref{def:dim}, a shatter-function exponent), $S$ is the total sensitivity and $\epsilon,\delta \in (0,1)$. Moreover, if there exists a classical oracle that can output a constant-factor two-sided approximation\footnote{By a constant-factor two-sided approximation, we mean an estimate $\wt s_i$ satisfying $c\cdot s_i\leq \wt s_i\leq C\cdot s_i$ for absolute constants $c, C>0$. The oracle receives a dataset of cardinality $s$ together with the universe $X$, and outputs such an estimate for a single sensitivity.} to one sensitivity over a set of $s$ points and universe $X$ in ${\cal T}_{\rm sensitivity}(s, X)$ time, then the quantum algorithm can be implemented in time $\wt O(\sqrt{ns}\cdot {\cal T}_{\rm sensitivity}(s, X))$, up to the cost of preparing the oracle once per recursion level (Theorem~\ref{thm:general}),  and $\wt O(\sqrt{ns})$ queries to the points in $A$.
\end{theorem}

We emphasize that the runtime statement is conditional on a fast sensitivity oracle: it is sublinear in the input size only when $s^{0.5}\cdot {\cal T}_{\rm sensitivity}(s, X)=o(n^{0.5}d)$. For the applications in this paper, the weights in question (leverage scores, ridge leverage scores, and Lewis weights) reduce to quadratic forms and admit $\wt O_p(d)$-time oracles via Johnson--Lindenstrauss sketching, so this condition holds, while $(k,p)$-clustering, whose sensitivities only admit one-sided approximations, is handled by a separate algorithm.

The query complexity of our framework is essentially optimal. In Section~\ref{sec:lb}, we prove that any quantum algorithm performing generic sensitivity sampling -- outputting a coreset of size $o(\sqrt{nS})$ for a cost function with total sensitivity $S$ -- must make $\Omega(\sqrt{nS})$ quantum queries (Theorem~\ref{thm:lb_generic}), via a reduction from the strong direct product theorem for quantum search~\citep{ksw07}, adapting Theorem 8.1 of~\cite{ag23}. Since the coreset size of Theorem~\ref{thm:informal_main} satisfies $s=\wt O(\epsilon^{-2}\dim(X)\cdot S)$, the $\wt O(\sqrt{ns})$ query count of our framework matches this lower bound up to $\poly(\epsilon^{-1}, \log n)$ and VC-dimension factors.

Our approach is simple and general: it constructs the sample by uniformly subsampling half of the points, recursively computing approximate sensitivities on this subset, and then resampling based on these estimates. This scheme was first used for leverage score sampling~\citep{clm+15} and in recent quantum linear programming algorithms~\citep{ag23}. We extend this strategy to sensitivity sampling.

As a key application, we adapt our framework to solve the low-rank approximation problem. Given a matrix $A \in \mathbb{R}^{d \times n}$, the goal is to find matrices $U, V$ of rank $k$ such that $\|A - UV^\top\|_F^2 \leq (1 + \epsilon) \cdot \|A - A_k\|_F^2$, where $A_k$ is the best rank-$k$ approximation to $A$. We provide a quantum algorithm that constructs a column coreset of $A$, resulting in a low-rank approximation algorithm that runs in time $\widetilde{O}(n^{0.5}dk^{0.5}\epsilon^{-0.5})$ (Theorem~\ref{thm:quantum_lra}). We note a recent result by~\cite{cgw24}, which provides a quantum algorithm for approximating the top-$k$ eigenvectors of a Hermitian matrix. Their method computes an orthonormal basis $W \in \mathbb{R}^{n \times k}$ such that $\|WW^\top - \sum_{i=1}^k v_i v_i^\top\| \leq \epsilon$, where $v_i$ is the $i$-th eigenvector of the Hermitian input matrix, in time $\widetilde{O}(n^{1.5}k/(\epsilon \gamma))$, where $\gamma$ is the spectral gap between $\lambda_k$ and $\lambda_{k+1}$. While powerful, their method targets spectral norm error and depends on $\gamma^{-1}$. In contrast, our algorithm selects and reweights subsets of rows and columns and approximates the Frobenius-norm optimal rank-$k$ solution, with no dependence on $\gamma$. This makes our method suitable for downstream applications for small $\gamma$. Another related work~\citep{cchlw22} considers quantum-inspired classical algorithms for low-rank approximation, where they assume data structure access to $A$ that can sample its rows and entries proportional to their $\ell_2^2$ norms in $O(\log(nd))$ time. We note that their result also has data-dependent parameters such as condition number and the ratio $\frac{\|A\|_F^2}{\sigma_k(A)^2}$ in its running times, and our algorithm does not rely on such data structures at all. More generally, our recursion requires samples from sensitivity, Lewis-weight, or ridge-leverage distributions that are defined \emph{adaptively} by the coreset constructed at the previous recursive level. Standard sample-and-query access provides samples from a fixed distribution, typically proportional to squared row or entry norms, and does not directly supply these adaptive target distributions. Constructing and maintaining them is precisely the difficult step. Two remarks make the comparison precise. First, sample-and-query access is built from the raw input in $O(\nnz(A))$ classical time~\citep{cglltw20}, and the exact row norms it provides already require reading every nonzero entry. Hence, from raw input every sample-and-query algorithm costs $\Omega(\nnz(A))=\Omega(nd)$ on dense $A$, whereas our low-rank algorithms (Corollary~\ref{thm:quantum_pcp}, Theorem~\ref{thm:quantum_lra}) make $\wt O(n^{0.5}\cdot \poly(d, k, 1/\epsilon))$ column and row queries (Table~\ref{tab:query_complexity}). Our query lower bounds (Section~\ref{sec:lb}) are stated in the row query model, in which sample-and-query access is not provided, so they are unaffected. Second, given sample-and-query access, leverage score sampling (Theorem~\ref{thm:ls}) is reachable classically after oversampling by $\kappa^2(A)$~\citep{cchlw22}, and the column coreset of Corollary~\ref{thm:quantum_pcp} after oversampling by $\lambda/2$ with $\lambda=\|A\|_F^2/\|A-A_k\|_F^2$~\citep{cglltw20}. In these cases the gain of our algorithms is the removal of these data-dependent factors and of the $\nnz(A)$ pass. We therefore do not expect a direct dequantization of our framework through this access model, although we do not claim a separation against all possible sample-and-query algorithms.

Beyond the recursive framework, we obtain new results for $(k, p)$-clustering, whose sensitivities admit only one-sided approximations and hence fall outside the framework. We instead give a quantum one-shot implementation of the coreset construction of~\cite{hlw24} seeded by a discrete constant-factor solution computed in $\wt O(n^{0.5}kd)$ time: the resulting algorithm outputs an $\epsilon$-coreset of input points only, of size $\widetilde O_p(k^{\frac{2p+2}{p+2}}\epsilon^{-2})$ as in~\cite{hlw24} up to polylogarithmic factors, in time $\widetilde O_p(\epsilon^{-1}n^{0.5}dk^{\frac{2p+3}{p+2}})$ (Theorem~\ref{thm:one_shot_main}). Compared with the recent quantum algorithm of~\cite{xclj23}, whose main bound (their Theorem 3.1) is $\widetilde O_p(n^{0.5}d^{1.5}k^{0.5}\epsilon^{-(p/2\lor 1)})$ time with coresets of size linear in $d$ (their Johnson--Lindenstrauss variant, Remark 3.2 there, trades $d^{1.5}$ for $d$ at the price of $\max\{\epsilon^{-4},\epsilon^{-p/2-3}\}$), our algorithm has a better dependence on $d$ than their main bound and no worse dependence on $1/\epsilon$, and it is faster than that bound, up to polylogarithmic factors, whenever $d\geq \Omega_p(k^{\frac{3p+4}{p+2}})$, i.e., $d\geq \Omega(k^{7/3})$ for $k$-median and $d\geq \Omega(k^{2.5})$ for $k$-means.

Our clustering algorithm has further implications for the \emph{data selection pipeline} used in training foundation models. As discussed in~\cite{ach+24}, if the loss function $\ell$ and all $k$-center solutions satisfy the $(p, \Lambda)$-well-behaved property, then a subset of $s = O(\epsilon^{-2})$ points suffices for training or fine-tuning: one first computes $k$ centers $x = (x_1, \ldots, x_k)$ using a clustering algorithm, then samples $s$ points according to the loss values $\ell(x_i)$. In our pipeline, the centers are computed from our coreset, and the second round of sampling requires at most $k$ queries to the loss function. Both steps run in $\widetilde{O}(n^{0.5} d \cdot \mathrm{poly}(k/\epsilon))$ time, whereas classical algorithms for this pipeline would require $\Omega(n)$ time. Our method is hence the first sublinear-time quantum algorithm for data selection pipelines.

We note that an independent and concurrent work of~\cite{lj25} studies quantum speedups for sparsifying generalized linear models (in turn they obtain improved quantum algorithms for various regression problems), via leverage score sampling over multiple scales~\citep{jlls23_linear}. While one could interpret the resulting sparsified model obtained by them as a coreset for generalized linear models, their approach significantly differs from ours as we directly operate with approximate sensitivities or proxies to sensitivities, and our generic framework applies to generating any coreset. For $\ell_p$ regression with $p\in (0, 2]$,~\cite{lj25} obtain $\wt O(rn^{0.5}d^{0.5}\epsilon^{-1})+\poly(d,1/\epsilon)$ time for row sparsity $r$, and Theorems~\ref{thm:lewis} ($p<2$) and~\ref{thm:ls} ($p=2$) match this on dense rows with $\wt O_p(n^{0.5}d^{1.5}\epsilon^{-1}+d^\omega)$ time. On $r$-sparse rows, our oracles cost only $\wt O_p(r)$ per query, so we also match their sparse bound (Remark~\ref{rem:row_sparsity}). We summarize our results in the following tables. Table~\ref{tab:coreset} compares our coreset construction runtimes with prior work, and Table~\ref{tab:optimization} compares runtimes for solving the corresponding optimization problems.

\begin{table}[!ht]
\caption{Comparison of running times for quantum algorithms to construct an $\epsilon$-coreset for the respective problems. Here $n$ points lie in $\R^d$, $k$ is the number of centers, $p$ the norm exponent and $\epsilon$ the accuracy. We assume $n, d\gg \poly(k,1/\epsilon)$ (and $n\gg \poly(d)$ for the regression rows) and ignore all dependencies on functions that only depend on $p$ for simplicity of presentation. The entry marked with $\dagger$ compares against the prior quantum Lewis-weight sampling approach. Our bound improves upon it for every $p>2$. The entry marked with $\star$ compares against the time to compute the sparsifier of the concurrent work~\cite{lj25}. For simplicity both entries assume dense points, and for points with at most $r$ nonzero entries both become $n^{0.5}d^{0.5}r\epsilon^{-1}$ (Remark~\ref{rem:row_sparsity}). The clustering bounds are established in Theorem~\ref{thm:one_shot_main} and the regression bounds in Theorems~\ref{thm:lewis} and~\ref{thm:ls}.}
\label{tab:coreset}
\begin{center}
\tablebox{%
\begin{tabular}{|l|l|l|l| } \hline
 & Reference  &  Previous & Ours\\ \hline
{$k$-Median Clustering} & \cite{xclj23}  & $n^{0.5}d^{1.5}k^{0.5}\epsilon^{-1}$ 
 & $n^{0.5}dk^{1.67}\epsilon^{-1}$ \\ \hline
 $k$-Means Clustering & \cite{xclj23}  & $n^{0.5}d^{1.5}k^{0.5}\epsilon^{-1}$ 
 & $n^{0.5}dk^{1.75}\epsilon^{-1}$ \\ \hline 
 $(k, p)$-Clustering & \cite{xclj23}  & $n^{0.5}d^{1.5}k^{0.5}\epsilon^{-(p/2\lor 1)}$ 
 & $n^{0.5}dk^{\frac{2p+3}{p+2}}\epsilon^{-1}$ \\ \hline
 $\ell_{p> 2}$ Regression~$\dagger$& \cite{ag23}  & $n^{0.5}(d^{7}+d^{p/4+3.5}\epsilon^{-1})$  
 & $n^{0.5}d^{p/4+1}(\epsilon^{-1}+d^{0.5})$ \\ \hline
 $\ell_{p\leq 2}$ Regression~$\star$& \cite{lj25}  & $n^{0.5}d^{1.5}\epsilon^{-1}$
 & $n^{0.5}d^{1.5}\epsilon^{-1}$ \\ \hline
\end{tabular}}
\end{center}
\end{table}
\begin{table}[!ht]
\caption{Comparison of running times for low-rank and subspace approximation problems. We compare with the prior best classical algorithms, as quantum algorithms for these problems either have not been studied, assume a data-structure input, or have data-dependent parameters in the running time. We assume all matrices/tensors are dense and $n, d\gg \poly(k,1/\epsilon)$ (and $n\gg \poly(d)$ for the Previous subspace entries), and ignore lower order terms. Low-rank factors are returned in factored form (Remark~\ref{rem:factored_output}). We use ${\cal T}_{\sf K}$ for the time to evaluate the kernel on two data points and $\omega$ for the matrix multiplication exponent. The tensor rows are $(2+\epsilon)$-approximations, versus $(1+\epsilon)$ in~\cite{swz19}. The row marked with $\ddagger$ omits, in both columns, a common factor $n^{O(\delta)}$ and, in ours, the hypotheses of Theorem~\ref{thm:quantum_fpt_two}. The row marked with $\S$ assumes the flatness precondition of Theorem~\ref{thm:quantum_lp}.}
\label{tab:optimization}
\begin{center}
\tablebox{%
\begin{tabular}{|l|l|l|l|} \hline
 & Reference  & Previous  
 & Ours \\ \hline
 Low-Rank & \cite{cmm17}  &  $nd$ 
 & $n^{0.5}d k^{0.5}\epsilon^{-0.5}$ \\ \hline
 PSD Low-Rank & \cite{bcw20}  & $nk^{\omega-1}\epsilon^{-(\omega-1)}$  
 & $n^{0.75} k^{2.25}\epsilon^{-2.25}$ \\ \hline
 Kernel Low-Rank  & \cite{bcw20}&  $nk\epsilon^{-1}({\cal T}_{\sf K}+k^{\omega-2}\epsilon^{2-\omega})$ 
 & $n^{0.75}k^{1.25}\epsilon^{-1.25}({\cal T}_{\sf K}+k\epsilon^{-1})$ \\ \hline
 Tensor, Rank-$k$~$\ddagger$ & \cite{swz19} &  $n^3 + 2^{O(k^2\epsilon^{-1})}$  
 & $n^2 k^{2}\epsilon^{-2}+ 2^{O(k^2\epsilon^{-1})}$  \\ \hline
 Tensor, Bicriteria  & \cite{swz19} & $n^3$ 
 & $n^2k^{0.5}\epsilon^{-1}$ \\ \hline
 Subspace, $1\leq p<2$~$\S$ & \cite{wy24}  & $nd$ 
& $n^{1-p/4}dk^{p/4}\epsilon^{-1}$ \\ \hline
Subspace, $p\geq 2$ & \cite{wy24}  & $nd$ & $n^{1-1/p}dk^{0.5}\epsilon^{-2/p}$ \\ \hline
\end{tabular}}
\end{center}
\end{table}

\ifdefined\isarxiv
\paragraph{Our contributions}
\else
\textbf{Our contributions.}
\fi 
We summarize our main contributions below:
\begin{CompactItemize}
\item We introduce a general quantum weighted sampling framework. Given weights $w\in \R^n_{\geq 0}$ satisfying mild conditions and access to a classical oracle that approximates the weight of a point over a small set, the framework constructs a coreset using $\widetilde{O}(\sqrt{ns})$ queries to the input, where $s$ is the size of the output coreset, and for an $\epsilon$-coreset, $s=\widetilde{O}(\epsilon^{-2}\cdot {\rm sum}(w))$ up to a dimension-dependent factor, where ${\rm sum}(w)\geq \sum_{i=1}^n w_i$ is an upper bound on the total weight. We show that sensitivity, leverage scores, and Lewis weights all meet these conditions, implying that coreset construction with these weights can be accelerated within our framework.
\item We design the first sublinear time quantum algorithms for several fundamental low-rank approximation tasks: Frobenius-norm low-rank approximation, PSD and kernel low-rank approximation, and tensor low-rank approximation. Our algorithms are purely sampling-based, and avoid dependence on data-dependent parameters.
\item We develop improved quantum algorithms for $(k,p)$-clustering in the high-dimensional regime $d\geq \Omega_p(k^{\frac{3p+4}{p+2}})$: we construct an $\epsilon$-coreset of size $\widetilde O_p(k^{\frac{2p+2}{p+2}}\epsilon^{-2})$, consisting of input points only, in time $\widetilde O_p(\epsilon^{-1}n^{0.5}dk^{\frac{2p+3}{p+2}})$ (Theorem~\ref{thm:one_shot_main}), and further demonstrate how our framework can accelerate data selection pipelines for training foundation models.
\item We complement our algorithms with an $\Omega(\sqrt{nS})$ query lower bound for generic sensitivity sampling with total sensitivity $S$ for coresets of size $o(\sqrt{nS})$ (Theorem~\ref{thm:lb_generic}), showing that our framework is query-optimal up to $\poly(\epsilon^{-1},\log n)$ and VC-dimension factors.
\end{CompactItemize}

We also clarify the relationship to prior work: Grover-based quantum sampling and the recursive halving scheme for leverage scores originate in prior work (see Sections~\ref{sec:tech} and~\ref{sec:tech:more:instantiations}). Our contribution lies in the sufficient conditions (Assumption~\ref{assumption:recurse} and Theorem~\ref{thm:general}) that make recursive quantum weighted sampling applicable in general, the proof that generic sensitivities satisfy these conditions, the stronger approximator formulations required for Lewis weights and ridge leverage scores, and the problem-specific classical reductions for the low-rank, kernel, and tensor variants.

\ifdefined\isarxiv
\paragraph{Roadmap.}
\else
\textbf{Roadmap.}
\fi
Section~\ref{sec:tech} gives a technical overview of the framework, its instantiations, the low-rank and clustering algorithms, and the lower bounds, and Section~\ref{sec:conclusion} discusses open problems. The appendix contains the full proofs. Appendix~\ref{sec:tech:more} continues the overview, Appendix~\ref{sec:preli} collects preliminaries, and Appendix~\ref{sec:general} proves the framework and its instantiations with leverage scores, Lewis weights, and ridge leverage scores. Appendices~\ref{sec:cmm}, \ref{sec:kernel}, \ref{sec:subspace_approx}, \ref{sec:tensor}, and~\ref{sec:clustering} treat low-rank approximation, kernel low-rank approximation, subspace approximation, tensor low-rank approximation, and clustering. Appendix~\ref{sec:lb} proves the lower bounds, Appendix~\ref{sec:QRAM} explains how the classical oracles are run in the quantum model, and Appendix~\ref{sec:query} lists the query complexities.

\ifdefined\isarxiv
\paragraph{Quantum computation model.}
\else
\textbf{Quantum computation model.}
\fi

We adopt the standard quantum computation model used in, e.g.,~\cite{aw22,ag23}. This model supports quantum subroutines operating on $O(\log n)$ qubits, allows quantum queries to the input, and grants access to a quantum-read/classical-write RAM (QRAM) of size $\mathrm{poly}(n)$ bits. Each quantum read or classical write to QRAM incurs unit cost. We measure \emph{time complexity} by the number of QRAM operations, and \emph{query complexity} by the number of input queries made by the algorithm. QRAM is a common model studied in quantum algorithms. However, it is known that practically realizing the QRAM architecture is challenging. Throughout the paper, we often refer to randomized, classical routines as classical oracles. Converting these routines into quantum circuits would require one to analyze the complexity of cleaning up the garbage states. We circumvent the issue by adopting a derandomization approach due to~\cite{aw22} that ``simulates'' access to a random oracle via a $k$-wise independent hash function. Specifically, we use the following result of~\cite{aw22}:
\begin{claim}[Claim 1 in~\cite{aw22} and Lemma 3.12 of~\cite{ag23}]
Consider any quantum algorithm with runtime $q$ that makes queries to a uniformly random string. We can simulate this algorithm with a quantum algorithm with runtime $\wt O(q)$ without random string, using an additional QRAM of $\wt O(q)$ bits.
\end{claim}
Given this claim, we can safely simulate any classical procedure without randomness. We refer to Section~\ref{sec:QRAM} for a more detailed discussion. We also note that the query complexities we obtained in this work are \emph{independent} of QRAM, and they remain sublinear in $n$. We defer the details to Section~\ref{sec:query}.

\section{Technical Overview}
\label{sec:tech}
Our entire framework can be stated in one sentence: \emph{recursively halve the dataset, use a coreset of the half to estimate every point's sampling weight up to a constant factor, Grover-sample against these estimates with $\wt O(\sqrt{ns})$ queries, and recurse to depth $\log n$.} Concretely, the pipeline is: (i) uniformly subsample half of the points, $A'\subset_{1/2} A$, (ii) recursively construct a coreset (or approximator) $C'$ of $A'$, (iii) use $C'$ to implement a cheap classical oracle for approximate weights of the points of $A$, and (iv) Grover-sample against these estimates using $\wt O(\sqrt{ns})$ queries (Lemma~\ref{lem:q_sampling_1_d}) and reweight, obtaining a coreset of $A$. The subsections below and Appendix~\ref{sec:tech:more} expand each ingredient and the applications. Table~\ref{tab:weights_overview} summarizes the weight notions we instantiate the framework with, the approximation guarantee each requires, and the per-query oracle cost that makes step (iii) fast.

\begin{table}[!ht]
\ifdefined\isarxiv\else\small\fi
\caption{The weight notions used in this paper, the approximation guarantee each requires for the recursion, and the per-query oracle cost. Clustering sensitivities admit only one-sided approximations, so they are handled by a separate, non-recursive algorithm (Section~\ref{sec:tech:clustering}).}
\label{tab:weights_overview}
\begin{center}
\ifdefined\isarxiv\else\footnotesize\fi\begin{tabular}{|l|l|l|l|}\hline
Weight & Approximation notion & Oracle time & Result \\ \hline
generic sensitivity & $\epsilon$-coreset & ${\cal T}_{\rm sensitivity}(s, X)$ & Theorem~\ref{thm:general} \\ \hline
leverage score & $\epsilon$-coreset (spectral) & $\wt O(d)$ & recovers~\cite{ag23} \\ \hline
$\ell_p$ Lewis weight & $\epsilon$-approximator (weights) & $\wt O_p(d)$ & Theorem~\ref{thm:lewis} \\ \hline
ridge leverage score & $\epsilon$-approximator (add.-mult.) & $\wt O(d)$ & Corollary~\ref{thm:quantum_pcp} \\ \hline
clustering sensitivity & one-sided only & $O(kd)$ & Theorem~\ref{thm:one_shot_main} \\ \hline
\end{tabular}
\end{center}
\end{table}


\subsection{A Generic Recursive Sampling Framework}
\label{sec:tech:grover}
Our framework rests on a single quantum primitive, a sampling procedure based on Grover search~\citep{g96}: associate each item with a value $q_i\in [0,1]$, and the goal is to sample a subset where the $i$-th item is included with probability $q_i$. Given $q_i$'s, we can sample the subset in $\wt O(\sqrt{n\sum_{i=1}^n q_i})$ time, and if $\sum_{i=1}^n q_i = o(n)$, this is sublinear in $n$. The procedure requires an oracle that returns the value of $q_i$ upon query (see Lemma~\ref{lem:q_sampling_1_d} for a formal statement). Grover-based speedups of this kind have previously been obtained for edit distance, Laplacian solving, and linear programming~\citep{beghs21,gjkt23,aw22,ag23}.


The difficulty is that the weights we wish to sample from depend on the entire dataset, so the oracle itself is expensive to implement directly. Our framework resolves this with a recursive scheme governed by three conditions. Let $w(A, B)\in \R_{\geq 0}^{|A|}$ denote generalized weights of $A$ measured with respect to a reference set $B$, and say $B$ is an \emph{$\epsilon$-approximator} of $A$ if for any $C$ and any point $c_i$ of $C$, we have $w_i(C, B)=(1\pm\epsilon) w_i(C, A)$ (for sensitivities, an $\epsilon$-coreset plays this role). Three conditions suffice:
\begin{CompactItemize}
    \item {\bf (C1) Consistent total weights:} For any subset $S\subseteq [n]$, $\sum_{i\in S} w_i(A_S, A_S) \leq {\rm sum}(w)$, where ${\rm sum}(w)$ is a finite upper bound on the total weight of any subset measured with respect to itself (for sensitivities, the total sensitivity). \emph{Role:} (C1) bounds the total weight of every recursive subproblem, in particular of the uniform half and of the sets the oracle is built from, and hence the expected number of samples at every level. A level that draws $s$ samples in expectation costs $\wt O(\sqrt{ns})$ Grover queries.
    \item {\bf (C2) Uniform sampling bound:} Let $A'\subset_{1/2} A$ be the subset that contains each point of $A$ independently with probability $1/2$. For a weighted set $B$, define the augmented weights $w'_i(A, A'; B)$ as $w_i(A, B)$ if $a_i\in A'$ and $w_i(A, B\cup \{a_i\})$ otherwise, and write $w'_i(A, A')=w'_i(A, A'; A')$. Then $w'_i(A, A')\geq w_i(A, A)$ for all $i\in [n]$. \emph{Role:} (C2) guarantees that weights computed from the halved dataset never underestimate. The oracle of the recursion evaluates $w'_i(A, A'; C')$ for an approximator $C'$ of $A'$, which is within $1\pm\epsilon$ of $w'_i(A, A')$ (for $a_i\in A'$ because $C'$ approximates $A'$, and for $a_i\notin A'$ because $C'\cup\{a_i\}$ approximates $A'\cup\{a_i\}$), so the recursion remains sound.
    \item {\bf (C3) Importance sampling bound:} Let $u_i\geq w_i(A, A)$ be overestimates of the weights. Suppose we sample according to $q_i=\min\{1, \alpha u_i\log(1/\delta)\}$ for a large enough $\alpha$, and reweight the sample by $1/q_i$, then with probability at least $1-\delta$, the weighted sample is an $\epsilon$-approximator of $A$ of size $O(\alpha\cdot \|u\|_1\log(1/\delta))$, which is $O(\alpha\cdot {\rm sum}(w)\log(1/\delta))$ in expectation for the overestimates the recursion produces, by (C1). \emph{Role:} (C3) shows that resampling against overestimates yields a valid approximator, closing the recursion.
\end{CompactItemize}
Let $s=O(\alpha\cdot {\rm sum}(w)\log(1/\delta))$. Together, (C1)--(C3) drive Algorithm~\ref{alg:quantum_recurse_main}: line 5 halves the dataset, line 7 recurses, line 8 implements the weight oracle $w'_i(A, A'; C')$ from the child coreset $C'$, whose estimates are valid overestimates (up to the factor $1-\epsilon$, absorbed by the constant $1000$) by (C2), and line 9 Grover-samples with $\wt O(\sqrt{ns})$ expected queries by (C1), since the expected number of samples is $O(\alpha\cdot {\rm sum}(w)\log(1/\delta))=O(s)$. Correctness of the resample follows from (C3). We obtain an algorithm that computes an $\epsilon$-coreset in the desired $\wt O(\sqrt{ns})\cdot {\cal T}_{\rm sensitivity}(s, X)$ time (Theorem~\ref{thm:informal_main}, formally Theorem~\ref{thm:general}). We summarize an informal version of the algorithm below, where we use $\textsc{QSample}$ to denote the Grover search procedure for sampling from a vector $q$ of probabilities (Lemma~\ref{lem:q_sampling_1_d}).
\begin{algorithm}[!ht]
\caption{Quantum recursive sampling for coreset.}
\label{alg:quantum_recurse_main}
\begin{algorithmic}[1]
\Procedure{QRecurseSample}{$A\in \R^{n\times d}, \alpha, \delta$}
\If{$n\leq \alpha\cdot{\rm sum}(w)\log(1/\delta)$}
\State \Return $(A, I_n)$
\EndIf
\State $A'\subset_{1/2} A$ \Comment{Sample uniformly half of the rows from $A$}
\State $s\gets \alpha\cdot{\rm sum}(w)\log(1/\delta)$
\State $(C', D')\gets \textsc{QRecurseSample}(A', \alpha,\delta)$
\State Implement a classical oracle for $w'_i(A, A'; C')$ 
\State $D\gets \textsc{QSample}(q)$ \Comment{$q_i=\min\{1, 1000\alpha\, w'_i(A, A'; C')\log(1/\delta)\}$, failure parameter as in Algorithm~\ref{alg:quantum_iterate}}
\State $C\gets D^\top A$
\State \Return $(C, D)$
\EndProcedure
\end{algorithmic}
\end{algorithm}

\subsection{Instantiations: Sensitivity and Lewis Weight Sampling}
\label{sec:tech:approximator}

\paragraph{Generic sensitivities.} We first instantiate the framework with generic sensitivities, taking $w_i(A, B)=s_i(A, B)$. The interesting condition to verify is the uniform sampling bound (C2). Consider the following simplified algorithm:  form $A'$ by sampling each point of $A$ with probability $1/2$ uniformly, and define the generalized sensitivity as $ s_i(A, A') = \max_{x \in X,\, {\rm cost}(A', x) \neq 0} \frac{{\rm cost}(a_i, x)}{{\rm cost}(A', x)}$, i.e., we change the denominator to ${\rm cost}(A', x)$. Note that this is \emph{not} necessarily an overestimate of $s_i$: if the maximizer $x^*$ for $s_i$ has ${\rm cost}(A', x^*) = 0$, then $s_i(A, A')$ may fall below $s_i$. Including the point itself repairs this: $s_i(A, A' \cup \{a_i\})\geq s_i$ always holds, since the denominator either stays positive and only shrinks relative to ${\rm cost}(A, x^*)$, or degenerates to ${\rm cost}(a_i, x^*)$, in which case the ratio equals $1$, an upper bound on any sensitivity (a short case analysis in the proof of Theorem~\ref{thm:sensitivity} verifies this). We note that $s_i(A, A' \cup \{a_i\})$ is in fact the overestimate used by~\cite{clm+15} to obtain their initial uniform sampling bound.

The recursive framework follows directly: uniformly sample half of the points $A'$, then compute a coreset of $A'$, called $C$. To compute the overestimates of $s_i$, we use $s_i(A, C)$ for $a_i\in A'$ and $s_i(A, C \cup \{a_i\})$ for $a_i\notin A'$ (the augmented weights $w'_i(A, A'; C)$ of Section~\ref{sec:general} with $w=s$), which is efficient since $C$ is a small-size coreset. Note that these are valid approximations of $s_i(A, A')$ and of $s_i(A, A' \cup \{a_i\})$, respectively -- as $C$ is a coreset of $A'$, it approximates the cost of $A'$ with respect to all $x \in X$, and they have the same kernel (given a set of points $A$ and a cost function, the kernel of $A$ is ${\rm ker}(A) = \{x \in X: {\rm cost}(A, x) = 0 \}$). Moreover, for $a_i\notin A'$, it is not hard to see that $C \cup \{a_i\}$ is also a coreset of $A' \cup \{a_i\}$ (Lemma~\ref{lem:coreset_add}), and thus the sensitivity is preserved in both cases. To summarize, in each round of recursion, we are given a size-$s$ coreset, and assuming we can approximate each $s_i(A, C)$ or $s_i(A, C \cup \{a_i\})$ in ${\cal T}_{\rm sensitivity}(s, X)$ time, then the overall runtime is $\wt O(\sqrt{ns}) \cdot {\cal T}_{\rm sensitivity}(s, X)$, with recursion depth at most $\log n$ as we halve the points at each step, giving the desired runtime for sensitivity sampling.

\paragraph{Lewis weights.} While the framework above handles \emph{all} sensitivity sampling, in many applications, the exact sensitivities can be difficult to compute, and thus proxies are often sought as efficient alternatives. Take the $\ell_p$ regression problem as an example, where the sensitivity is $s_i = \max_{x\in \mathbb{R}^d, Ax\neq 0}\frac{|a_i^\top x|^p}{\|Ax\|_p^p}$. For $p=2$, this corresponds to the leverage score, which can be quickly approximated. However, for general $p$, this is more complex, and algorithms for $\ell_p$ sensitivities tend to be less efficient than those for leverage scores~\citep{pwz23}. Instead, constructing a coreset for $\ell_p$ regression is typically done \emph{not} via sensitivity sampling, but through \emph{Lewis weights sampling}~\citep{blm89,lt91,t95,sz01,cp15,wy23}. These weights are defined as the fixed-point solution for the following equation: $w_i^{2/p} = a_i^\top (A^\top W^{1-2/p} A)^{-1} a_i$, where $w_i$ represents the $i$-th leverage score of the matrix $W^{1/2-1/p}A$. Lewis weights have several desirable properties, such as $\sum_{i=1}^n w_i = d$, and they upper bound the $\ell_p$ sensitivities, exactly for $p\leq 2$ and up to a factor $d^{p/2-1}$ for $p>2$. Moreover, Lewis weights can be approximated in nearly-linear time~\citep{cp15,l16,jls22,flps22,ags24,gsz26}.

We defer the instantiation with ridge leverage scores, and the adaptation of our framework to Lewis weights, to Section~\ref{sec:tech:more:instantiations}.

\subsection{Optimality of the Framework: Query Lower Bounds}
\label{sec:tech:lb}

The $\wt O(\sqrt{ns})$ query count of the framework cannot be improved in its dependence on $n$ and $S$, and the obstruction is generic sensitivity sampling itself rather than any one application. Consider $n$ points in $\R^2$ split into $S$ blocks, where each point carries a hidden bit that one query reveals, and a point is marked if its bit is $1$. The cost of a point under the $l$-th element of the universe is $1$ if the point is a marked point of block $l$ and $0$ otherwise. A marked point of a block has sensitivity equal to the inverse of the number of marked points in its block, and an unmarked point has sensitivity $0$, so the total sensitivity is at most $S$ and the dimension ${\rm dim}(X)$ of the range space (Definition~\ref{def:dim}) is $O(1)$. An $\epsilon$-coreset $B$ with $\epsilon<1/2$ must contain a marked point of every block that contains a marked point, so reading the bits of its points, with at most $|B|$ further queries, decides for every block whether it contains a marked point. Computing these $S$ disjunctions at once costs $\Omega(S\sqrt{n/S})=\Omega(\sqrt{nS})$ bounded-error quantum queries by the strong direct product theorem for ${\rm OR}$~\citep{ksw07}. Hence, any quantum algorithm that outputs an $\epsilon$-coreset of size $o(\sqrt{nS})$ with probability at least $2/3$ makes $\Omega(\sqrt{nS})$ queries to the points (Theorem~\ref{thm:lb_generic}). On this family the sensitivity oracle is trivially implementable, and Theorem~\ref{thm:general} produces a coreset of size $\wt O(\epsilon^{-2}S)$, which is $o(\sqrt{nS})$ whenever $S=\wt o(\epsilon^{4} n)$, with $\wt O(\epsilon^{-1}\sqrt{nS})$ queries. Thus, the framework is query-optimal in $n$ and $S$ up to $\poly(\epsilon^{-1},\log n)$ factors.

The dependence on $\epsilon$ is settled on the same family, under the hypotheses of Theorems~\ref{thm:lb_eps} and~\ref{thm:lb_eps_S}. For $S=1$ the weights of the coreset, together with the bits of its points, determine the number of marked points within a factor $1\pm\epsilon$, which is approximate counting. Hence, the counting lower bound of~\cite{nw99} shows that any quantum algorithm that outputs an $\epsilon$-coreset of size $o(\sqrt{n/\epsilon})$ with its weights makes $\Omega(\sqrt{n/\epsilon})$ queries. The counting algorithms and Grover search of~\cite{bhmt02} output an $\epsilon$-coreset of size $1$ with $O(\sqrt{n/\epsilon})$ queries, so the bound is tight on this family (Theorem~\ref{thm:lb_eps}). For general $S$, each block carries the partial function of~\cite{nw99} that distinguishes two consecutive counts of marked points, whose query complexity is $\Omega(\sqrt{n/(S\epsilon)})$. The direct product theorem of~\cite{lr13} for partial functions then gives $\Omega(\sqrt{nS/\epsilon})$ queries (cf.~\cite{xclj23}) for any algorithm that outputs an $\epsilon$-coreset of size $o(\sqrt{nS/\epsilon})$ with its weights. Running the $S=1$ algorithm on each block gives $O(\sqrt{nS/\epsilon}\log S)$ queries (Theorem~\ref{thm:lb_eps_S}). Thus, the answer on the block family is $\Theta(\sqrt{nS/\epsilon})$ up to a $\log S$ factor, against the $\wt O(\epsilon^{-1}\sqrt{nS})$ of the framework, a gap of $\epsilon^{-1/2}$ up to logarithmic factors. For leverage scores ($S=d$), Lemma 8.2 of~\cite{ag23} gives $\wt\Omega(\epsilon^{-1}\sqrt{nd})$ row queries, attained by Theorem~\ref{thm:ls}, when $n=O(d^2)$ and $\epsilon=\Omega(\sqrt{d/n})$, which force $S=\Omega(\epsilon^{-2})$. Whether the framework's $\epsilon^{-1}$ factor is necessary when $S=o(\epsilon^{-2})$ remains open. The bounds count queries to the points only.

\subsection{Pure-Sampling Framework For Low-Rank Approximation}
\label{sec:tech:sample}
Given $A\in \mathbb{R}^{d\times n}$, the rank-$k$ low-rank approximation problem seeks to find a pair of matrices $U\in \mathbb{R}^{d\times k}$, $V\in \mathbb{R}^{n\times k}$ such that $\|A-UV^\top\|_F^2 \leq (1+\epsilon) \|A-A_k\|_F^2$, where $A_k$ is the best rank-$k$ approximation of $A$. Low-rank approximation is closely related to the $(k, 2)$-subspace approximation coreset: let ${\cal F}_k$ be the set of all $k$-dimensional subspaces in $\mathbb{R}^d$, and define ${\rm cost}(a_i, x)=\|(I-P_x)a_i\|_2^2$ where $P_x$ is the orthogonal projection onto $x\in {\cal F}_k$. If we obtain a coreset $C$ for $A$, then we have for any $k$-dimensional orthogonal projection $P_x$, $\|(I-P_x)C\|_F^2 = (1\pm\epsilon) \|(I-P_x)A\|_F^2$, which is sufficient to show that choosing $P_x$ as the projection onto $C_k$ will give the desired low-rank approximation~\citep{cmm17}. Moreover, instead of $(k, 2)$-subspace sensitivities, one could sample according to the \emph{ridge leverage scores}, which can be computed quickly. To adapt our weighted sampling framework, we need to identify the right notion of an $\epsilon$-approximator for ridge leverage scores: a coreset $C$ of $A$ that additionally satisfies $(1-\epsilon) CC^\top-\epsilon \lambda_{A_k} I \preceq AA^\top \preceq (1+\epsilon) CC^\top+ \epsilon \lambda_{A_k} I$, where $\lambda_{A_k}=\|A-A_k\|_F^2/k$. Thus, our framework gives a column coreset for the $k$-subspace cost in $\wt O(n^{0.5}dk^{0.5}\epsilon^{-1})$ time (Corollary~\ref{thm:quantum_pcp}). For the low-rank approximation itself, the $\epsilon^{-1}$-oversampled column subset of Lemma~\ref{lem:cmm_thm7} brings the time to $\wt O(n^{0.5}dk^{0.5}\epsilon^{-0.5})$ (Theorem~\ref{thm:quantum_lra}, where the first term leads under the standing assumption $d\gg\poly(k,1/\epsilon)$).

We defer the recovery of the solution through the subsampled columns to Section~\ref{sec:tech:sample:more}, and the extension to tensor low-rank approximation via approximate regression by sampling responses to Section~\ref{sec:tech:response}.

\subsection{Improved Coreset for Clustering with Applications}
\label{sec:tech:clustering}

We also design an improved quantum algorithm for constructing an $\epsilon$-coreset of $(k,p)$-clustering. In contrast to the recursive sampling framework we developed in the preceding discussions, our algorithm is a quantum one-shot implementation of~\cite{hlw24}, where the idea is to first compute a set of approximate $k$-centers, then perform one round of importance sampling on top of it. Why could our recursive sampling framework not be applied here? This is because the sensitivities of $(k,p)$-clustering can only be \emph{overestimated}, and these overestimates in general do not satisfy the uniform sampling bound. In fact, a closer examination of our analysis shows that during the intermediate stages in the recursive sampling, we would need the sensitivities to be approximated in a \emph{two-sided} fashion, i.e., let $s_i$ be the exact sensitivities. We require the approximate sensitivities $\wt s_i$ to satisfy $(1-\epsilon)s_i\leq \wt s_i\leq (1+\epsilon)s_i$. Nevertheless, seeding this one-shot implementation with a discrete constant-factor solution computed from bicriteria centers and cell counts in $\wt O(n^{0.5}kd)$ time (Lemma~\ref{lem:discrete_constant}), we obtain an $\epsilon$-coreset of size $\wt O_p(k^{\frac{2p+2}{p+2}}\epsilon^{-2})$, consisting of input points only, in time $\wt O_p(\epsilon^{-1}n^{0.5}dk^{\frac{2p+3}{p+2}})$ (Theorem~\ref{thm:one_shot_main}). Compared to the previous work of~\cite{xclj23} where they obtain a coreset in $\wt O_p(n^{0.5}d^{1.5}k^{0.5}\epsilon^{-(p/2\lor 1)})$ time,
\begin{CompactItemize}
    \item Up to polylogarithmic factors, our algorithm is faster in the large dimension regime where $d\geq \Omega_p(k^{\frac{3p+4}{p+2}})$, i.e., $d\geq \Omega(k^{7/3})$ for $k$-median and $d\geq \Omega(k^{2.5})$ for $k$-means. The comparison is against their main bound, Theorem 3.1 of~\cite{xclj23}. In particular, the regime is independent of $\epsilon$.
    \item Our algorithm directly outputs a coreset of size $\wt O_p(k^{\frac{2p+2}{p+2}}\epsilon^{-2})$, while their main bound gives a coreset of size $\wt O_p(dk\epsilon^{-(2\lor p)})$. Obtaining our coreset size from their output by running the algorithm of~\cite{hlw24} on top would require an additional $\wt O_p(d^2 \poly(k,\epsilon^{-p}))$ time.
\end{CompactItemize}

As an application, we demonstrate that $(k,p)$-clustering can be used to bootstrap the construction of the data selection pipeline~\citep{ach+24}, as it enables the computation of approximate $k$-centers in sublinear time. Furthermore, we show that the quantum techniques developed for $(k,p)$-clustering can also be leveraged to obtain a sublinear-time quantum algorithm for data selection. We defer a more detailed discussion to Section~\ref{sec:clustering}.

\section{Conclusion}\label{sec:conclusion}

We present a quantum, sublinear-time algorithm for weighted sampling that yields a broad range of results in coreset construction. These include $(k, p)$-clustering, $\ell_p$ regression, $(k, p)$-subspace approximation, and low-rank approximation. For the low-rank approximation problem, we design specialized algorithms for multiple settings, including Frobenius norm error minimization, PSD low-rank approximation, kernel-based low-rank approximation, and tensor low-rank approximation. For $(k,p)$-clustering, we develop an improved quantum coreset construction that has a better $d$-dependence than the main bound of~\cite{xclj23} and outputs a coreset consisting of input points only (Theorem~\ref{thm:one_shot_main}), and we generalize this framework to address the data selection problem for foundation models. We highlight three major open problems arising from our work:

\begin{CompactItemize}
    \item \textbf{Two-sided approximation for clustering sensitivities.} Unlike regression and low-rank approximation, where coresets can be constructed efficiently via leverage scores or Lewis weights, the approximate sensitivities used in clustering are only known to be \emph{upper bounds}. This asymmetry significantly limits the applicability of the recursive sampling framework to clustering. It remains an open question whether one can design algorithms that compute \emph{two-sided} approximations to clustering sensitivities, unifying clustering with our weighted sampling framework.

    \item \textbf{Quantum algorithms for Frobenius norm tensor low-rank approximation.} While we achieve a $(1+\epsilon)$-approximation for matrix low-rank approximation in sublinear time, the scenario is more complex for tensors. For 3rd-order tensors, our bicriteria algorithm achieves a $(2+\epsilon)$-approximation with output rank $k^2$ (Theorem~\ref{thm:quantum_bicriteria_two}), and our fixed-parameter algorithm achieves a $(2+\epsilon)$-approximation with exact output rank $k$ (Theorem~\ref{thm:quantum_fpt_two}). A compelling open question is whether one can design a sublinear-time quantum algorithm that achieves a $(1+\epsilon)$-multiplicative approximation for tensor low-rank approximation. Remark~\ref{rem:towards_one_eps} gives an instance showing that the route of Theorem~\ref{thm:quantum_bicriteria_two}, which fixes a $k$-dimensional span for the first factors in its first stage, cannot reach $(1+\epsilon)$.

    \item \textbf{Query lower bounds for coreset construction.} In Section~\ref{sec:lb}, we establish an $\Omega(\sqrt{nS})$ query lower bound for generic sensitivity sampling (Theorem~\ref{thm:lb_generic}) that matches our framework up to $\poly(\epsilon^{-1},\log n)$ and VC-dimension factors. On the block family the $\epsilon$-dependence is settled at $\Theta(\sqrt{nS/\epsilon})$ up to a $\log S$ factor (Theorems~\ref{thm:lb_eps} and~\ref{thm:lb_eps_S}), a factor $\epsilon^{-1/2}$ (up to logarithmic factors) below the framework's $\wt O(\epsilon^{-1}\sqrt{nS})$ on that family. An intriguing direction for future research is to capture the $\epsilon$-dependence when $S=o(\epsilon^{-2})$, since for larger $S$ the framework's $\epsilon^{-1}$ factor can be necessary (Section~\ref{sec:tech:lb}), and to extend these bounds to the remaining applications.
\end{CompactItemize}


\ifdefined\isarxiv
\section*{Acknowledgments and Disclosures}
David P. Woodruff would like to acknowledge support from a Simons Investigator Award and Office of Naval Research (ONR) award number  N000142112647. Lichen Zhang is supported by a Simons Dissertation Fellowship in Mathematics. The authors declare no competing interests. Most of this paper was written by the authors and polished with the help of large language models (LLMs). Some of the results, including the lower bounds, the $\ell_p$ regression results, and the tighter bounds for $(k,p)$-clustering and $(k,p)$-subspace approximation, were improved with the help of LLMs (Claude Opus 5.5 in Claude Code) after the acceptance of the paper, and the authors verified all proofs.

\bibliographystyle{alpha}
\bibliography{ref}
\else
\newpage
\begin{ack}
David P. Woodruff would like to acknowledge support from a Simons Investigator Award and Office of Naval Research (ONR) award number  N000142112647. Lichen Zhang is supported by a Simons Dissertation Fellowship in Mathematics. The authors declare no competing interests. Most of this paper was written by the authors and polished with the help of large language models (LLMs). Some of the results, including the lower bounds, the $\ell_p$ regression results, and the tighter bounds for $(k,p)$-clustering and $(k,p)$-subspace approximation, were improved with the help of LLMs (Claude Opus 5.5 in Claude Code) after the acceptance of the paper, and the authors verified all proofs.

\end{ack}
\bibliographystyle{plainnat}
\bibliography{ref}
\fi

\appendix

\section{Additional Technical Overview}
\label{sec:tech:more}

This section contains the parts of the technical overview of Section~\ref{sec:tech} deferred from the main text.

\subsection{Ridge Leverage Score and Lewis Weight Sampling}
\label{sec:tech:more:instantiations}

\paragraph{Ridge leverage scores.} For $\ell_2$ regression the weights are the ordinary leverage scores, and instantiating the framework with them recovers the Grover-based sampler of~\cite{ag23}, which combines the recursive halving scheme of~\cite{clm+15} with a Johnson--Lindenstrauss sketch of the pseudoinverse~\citep{jl84,ss11}. We do not elaborate on this known case. The instantiation that is new to this work is the \emph{ridge leverage score} of Definition~\ref{def:ridge_lg}, which governs low-rank approximation. The $k$-subspace sensitivity is at most $(1+1/k)$ times the ridge leverage score, as shown in Section~\ref{sec:ridge}, and~\cite{cmm17} showed that sampling by ridge leverage scores yields a projection-cost preserving sample. Compared with ordinary leverage scores, the new difficulty is that the regularization $\lambda_{A_k}$ is itself a data-dependent quantity, so a small reference set must simultaneously approximate the Gram matrix \emph{and} the ridge term. Therefore, we define an $\epsilon$-approximator as an $\epsilon$-coreset of $A$ that also satisfies the additive-multiplicative spectral approximation guarantee of Section~\ref{sec:ridge}. By Lemma~\ref{lem:cmm_lem12}, such an approximator preserves every ridge leverage score up to a factor $1\pm 4\epsilon$, which validates the oracle implemented from the child approximator in the recursion. Moreover, uniform halving followed by resampling against the weights with respect to the uniform half yields an $\epsilon$-approximator of size $O(\epsilon^{-2}k\log(k/\delta))$ by Lemma~\ref{thm:ridge_via_unif}, supplying conditions (C2) and (C3). The oracle itself is cheap: given a size-$s$ approximator $C$, Algorithm~\ref{alg:classical_ridge_ls} stores two Gaussian sketches with $O(\log n)$ rows computed from the thin SVD of $C$, after which each query takes $\wt O(d)$ time. Plugging these ingredients into the framework yields an $\epsilon$-approximator of size $O(k\epsilon^{-2}\log k)$ in $\wt O(n^{0.5}dk^{0.5}\epsilon^{-1}+dk^{\omega-1})$ time by Corollary~\ref{thm:quantum_pcp}, which we use for our low-rank approximation results in Section~\ref{sec:tech:sample}.

To adapt our sensitivity sampling framework for Lewis weights sampling, we encounter a notable challenge: given a coreset $B$ of $A$, it is guaranteed that any vector in the subspace of $A$ has its $\ell_p$ norm preserved by $B$, but the Lewis weights instead measure the $\ell_2$ norm of the subspace after a density transformation induced by $W^{1/2-1/p}$. Consequently, the Lewis weights of $A$ might not be preserved by $B$. We instead use the notion of an \emph{$\epsilon$-approximator} of $A$ in Definition~\ref{def:approximator}, which preserves the Lewis quadratic form. Note that this is a different approximation notion than that of a coreset, as the cost becomes \emph{global} rather than local: for generic sensitivity-based arguments, one relies on the fact that if $B$ is a coreset of $A$, then $B \cup \{a\}$ is also a coreset of $A \cup \{a\}$, but this is not true for an approximator of $A$, as adding a single row to both $A$ and $B$ would potentially affect the weights to all existing rows. In~\cite{cp15}, they provide a classical recursive sampling algorithm by noting that, if we sample according to the generalized Lewis weights with respect to an approximator, scaled up by $d^{p/2}$ when $p\geq 2$, then the resulting weighted sample is \emph{also} an approximator, at constant accuracy (Lemma~\ref{lem:lewis_level}). This is the approximator property our framework requires, and the argument of~\cite{cp15} plays the role of the conditions (C2) and (C3). Theorem~\ref{thm:lewis} makes this formal. Thus, by using weighted sampling with Lewis weights, we generate a coreset for $\ell_p$ regression in $\wt O_p(\epsilon^{-1}n^{0.5}d^{1.5})$ time for $p<2$ (as Theorem~\ref{thm:ls} gives for $p=2$) and in $\wt O_p(n^{0.5}d^{p/4+1}(\epsilon^{-1}+d^{0.5}))$ time for $p\geq 2$. For $p>2$, this improves upon the prior quantum algorithm for Lewis weights sampling that is based on \emph{iterating leverage scores}~\citep{ag23}, whose iteration is the classical scheme of~\citet{ags24,gsz26}. For sampling, it suffices to approximate the Lewis weights up to a constant factor (Lemma~\ref{lem:wy23_thm13}), which Theorem 4.2 of~\cite{ag23} (arXiv v3) does after $\wt O_p(n^{0.5}d^7)$ preprocessing time for a dense matrix and $p\geq 2$. To construct a coreset, one then samples the $m=\wt O_p(d^{p/2}\epsilon^{-2})$ rows of Lemma~\ref{lem:wy23_thm13} with the Grover-based sampler of Lemma~\ref{lem:q_sampling_1_d}, using $\wt O(\sqrt{nm})$ weight queries of $O(p^3d^{7/2})$ time each, so the prior runtime is $\wt O_p(n^{0.5}(d^7+d^{p/4+3.5}\epsilon^{-1}))$. Our algorithm provides a speedup for any $p>2$, as both terms of our runtime $\wt O_p(n^{0.5}d^{p/4+1}(\epsilon^{-1}+d^{0.5}))$ are at least a factor $d^2$ smaller than the second term of the prior runtime, and the additive $\wt O_p(d^{p/2+\omega})$ term of Theorem~\ref{thm:lewis} is lower order under $n\gg\poly(d)$. For $p<2$ (which includes the popular $\ell_1$ regression), Theorem 4.2 of~\cite{ag23} does not apply, and we compare with the concurrent work of~\cite{lj25} in the introduction and in Remark~\ref{rem:row_sparsity}. It is worth noting that the main purpose of the work of~\cite{ag23} is to estimate Lewis weights up to $p=O(\log n)$ as they use it as a subroutine for solving linear programs, so their algorithm has no $p$ dependence on $d$, while our sampling algorithm is particularly suitable for small $p$.

\subsection{Recovering the Solution through the Subsampled Columns}
\label{sec:tech:sample:more}

Let $C\in \mathbb{R}^{d\times s}$ denote the weighted subset of columns of $A$ sampled by ridge leverage scores for $s=O(\epsilon^{-1}k\log k)$ (Lemma~\ref{lem:cmm_thm7_bernoulli}). Then it is guaranteed that $\min_{X: \rank(X)\leq k} \|CX-A\|_F^2 \leq (1+\epsilon) \|A_k-A\|_F^2$. Constructing an optimal $X$ would require computing $P_k(C^\dagger A)$ where $P_k$ is the projection onto the top-$k$ principal components. Directly computing $C^\dagger A$ is too expensive, and standard approaches mostly involve using an \emph{oblivious subspace embedding} (OSE) matrix, a random matrix that approximates the cost of all regression problems. Matrices such as CountSketch~\citep{ccfc02,cw13} could be applied in time $\nnz(A)$, but this is already too slow for our purpose. We address this with a pure-sampling framework for low-rank approximation: we demonstrate that it is possible to recover (or approximate) the solution $X$ via leverage score sampling.

In particular, let $S$ sample rows according to the leverage score distribution of $C$. A good rank-$k$ solution $CX^\star$ has column span inside that of $C$, so its row leverage scores are dominated by those of $C$ and $S$ oversamples them. The standard leverage score guarantee for unconstrained regression (Lemma~\ref{lem:leverage_score_optimal_solution}), applied to the design $CX^\star$, shows that $CX^\star(SCX^\star)^\dagger SA$ is a $(1+\epsilon)$-approximate rank-$k$ solution. Hence a good rank-$k$ solution exists in the row span of $SA$, and it is enough to solve the regression problem $\min_{Y: \rank(Y)\leq k} \|A-CYSA\|_F^2$. We further speed up the algorithm by employing two leverage score sampling matrices $T_1$ and $T_2$ on the left and right, respectively. For the new subsampled regression problem $\min_{Y: \rank(Y)\leq k} \|T_1AT_2-T_1CYSAT_2\|_F^2$, we can compute the subsampled $A$ in \emph{sublinear in $n, d$} time, because $T_1AT_2$ and $SAT_2$ both amount to selecting a $\poly(k/\epsilon)$ subset of entries of $A$, which, assuming random access to the entries of $A$, can be done in the same order of time.

\subsection{Approximate Regression via Sampling Responses}
\label{sec:tech:response}

For matrix low-rank approximation and its variants, ridge leverage score sampling is the crucial tool to compute a good approximate solution. Can we extend the framework to solve \emph{tensor} low-rank approximation? Unfortunately, even for a 3rd order tensor $A\in \mathbb{R}^{n\times n\times n}$, it is not always the case that it admits a low-rank approximation, due to the so-called border rank issue~\citep{dsl08}. Even when the low-rank approximation exists, under the Exponential Time Hypothesis (ETH) no $2^{o(k^{1-o(1)})}$-time algorithm approximates the rank of $A$ within a constant factor, and no $2^{o(\epsilon^{-1/4})}$-time algorithm computes a $(1+\epsilon)$-approximate rank-$1$ approximation~\citep[Theorems 1.6 and 1.7]{swz19}. If one relaxes the problem by allowing the output to be a higher-rank solution (bicriteria solution) or a running time that depends exponentially on $k$ and $1/\epsilon$ (fixed-parameter tractable, i.e., FPT), then~\cite{swz19} provides algorithms with leading running time term being $\nnz(A)$. Their core algorithm is as follows: for tensor $A\in \mathbb{R}^{n\times n\times n}$, let $A_1, A_2, A_3\in \mathbb{R}^{n\times n^2}$ be matrices such that the 1st, 2nd, and 3rd dimensions of the array are preserved, while the other 2 dimensions are collapsed and flattened into a dimension of size $n^2$. They then apply OSEs $S_1, S_2, S_3$ with only $\poly(k/\epsilon)$ columns to form $A_1S_1, A_2S_2$ and $A_3S_3$. 

Although one might attempt to replace the OSEs $S_1, S_2$, and $S_3$ with leverage score matrices for $A_1, A_2$, and $A_3$, \emph{this approach, unfortunately, does not work}. The argument of~\cite{swz19} is as follows: suppose the optimal rank-$k$ approximation $A_k$ exists, then $A_k=\sum_{i=1}^k U^*_i\otimes V^*_i\otimes W^*_i$. To reduce the problem dimension, the goal is to demonstrate that a good approximate solution exists in the column span of $A_1S_1$ and $A_2S_2$. In particular, suppose we have access to $V^*$ and $W^*$, and let $Z_1\in \mathbb{R}^{k\times n^2}$ have $V_i^*\otimes W_i^*$ as its $i$-th row. Then the optimal $U^*$ could be recovered by solving $\min_{U\in \mathbb{R}^{n\times k}} \| UZ_1-A_1\|_F^2$, as $\| UZ_1-A_1\|_F^2=\|\sum_{i=1}^k U_i\otimes V_i^*\otimes W_i^*-A\|_F^2$. This multiple response regression problem can be accelerated by applying an OSE on the right and instead solving $\min_U \|UZ_1S_1-A_1S_1\|_F^2$, and its optimal solution $\wh U=A_1S_1(Z_1S_1)^\dagger$ is in the column span of $A_1S_1$.

For sampling, this setup is more challenging. If we were to replace $S_1$ with the leverage score sampling matrix, we would require the \emph{leverage score matrix of $Z_1$} in order to preserve the cost of the optimal solution, and it is unclear how to sample according to the leverage scores of the unknown matrix $Z_1$. On the other hand, we do have access to the response matrix $A_1$. However, a simple counterexample demonstrates that sampling directly from $A_1$ fails: suppose $A_1$ is a single column equal to $e_n$, and the design matrix $Z_1$ is $e_i+e_n$ for $i$ randomly chosen from $1$ to $n-1$. Any sampling scheme based on $A_1$ will likely sample the $n$-th entry but miss the $i$-th entry with high probability. This would lead to a solution to the original problem that has twice the optimal cost.

Surprisingly, we show that this factor-$2$ loss is essentially the worst that can happen: if one instead samples from the ridge leverage score distribution of $A_1$, then there exists a solution $\wh U$ in the column span of $A_1S_1$ ($S_1$ is the ridge leverage score sampling matrix of $A_1$) such that $\|\wh UZ_1-A_1\|_F^2 \leq (2+\epsilon)\cdot\min_{U} \|UZ_1-A_1\|_F^2$. This result is particularly surprising as one might expect an adversarial choice of $A_1$ that would disrupt ridge leverage score sampling. However, ridge leverage scores provide the so-called projection-cost preserving guarantee: for any rank-$k$ projection $P$, we have that $(1-\epsilon)\|(I-P)A_1\|_F^2 \leq \|(I-P)A_1S_1\|_F^2 \leq (1+\epsilon) \|(I-P)A_1\|_F^2$, where setting $P_k$ as the projection onto the top-$k$ principal components of $A_1S_1$ minimizes $\|(I-P_k)A_1S_1\|_F^2$. Additionally, the optimal cost of the regression can be bounded by $\|[A_1]_k-A_1\|_F^2$. Setting $\wh U=P_kA_1Z_1^\dagger$, we get
\begin{align*}
   \|\wh UZ_1-A_1\|_F^2= & \|P_kA_1Z_1^\dagger Z_1-A_1\|_F^2 \\
    = & \|(P_kA_1-A_1)(Z_1^\dagger Z_1)-A_1(I-Z_1^\dagger Z_1)\|_F^2 \\
    = & \| (P_kA_1-A_1)(Z_1^\dagger Z_1)\|_F^2+\| A_1(I-Z_1^\dagger Z_1)\|_F^2 \\
    \leq & \|(I-P_k)A_1\|_F^2 + \| A_1(I-Z_1^\dagger Z_1)\|_F^2 \\
    \leq & (2+\epsilon)\OPT,
\end{align*}
where $\OPT:=\min_{U} \|UZ_1-A_1\|_F^2$, and we use the Pythagorean theorem in the proof, along with the fact that $\|A_1-A_1Z_1^\dagger Z_1\|_F^2$ is the optimal cost. To see $\wh U$ is in the column span of $A_1S_1$, it is enough to observe that $P_k$ is the projection onto the top-$k$ principal components of $A_1S_1$, and hence $\wh U$ is in the column span of $P_k$, a subset of the column span of $A_1S_1$. Our bicriteria result pays the response-sampling loss only once: after the first stage fixes a $k$-dimensional span for the first factors, we treat the second and third modes \emph{jointly} as a single matrix low-rank approximation problem on an $n\times kn$ matrix. This gives a $(2+\epsilon)$ ratio with output rank $k^2$ by Theorem~\ref{thm:quantum_bicriteria_two}, and the same first stage followed by the classical $(1+\epsilon)$ fixed-parameter algorithm of~\cite{swz19} gives exact output rank $k$ with ratio $(2+\epsilon)$ by Theorem~\ref{thm:quantum_fpt_two}, whose hypotheses include that an optimal rank-$k$ solution exists.

\section{Preliminaries}
\label{sec:preli}

\subsection{Notation}
For any $n \in \mathbb{N}$, let $[n]$ denote the set $\{1, 2, \ldots, n\}$. We use $\wt O(\cdot)$ to hide polylogarithmic factors in $n$, $d$, $1/\epsilon$, $1/\delta$, and other problem-related parameters, such as $k$ and $p$. We use $O_p(\cdot)$ and $\wt O_p(\cdot)$ when the hidden constant may also depend on $p$, and in our clustering bounds this dependence is exponential in $p$. Throughout the paper, we assume that $n$ and $d$ are polynomially larger than $k$ and $1/\epsilon$, and we write $n, d\gg \poly(k,1/\epsilon)$. For the regression results, we also assume that $n$ is polynomially larger than $d$. The theorems state their full bounds, while the abstract, introduction, and tables report the leading term under these assumptions. For instance, the first term of Theorem~\ref{thm:quantum_lra} is the leading term once $d\geq \epsilon^{-3.5}k$. For two numbers $a$ and $b$, we use $a\lor b$ as a shorthand for $\max\{a, b\}$. We use $a=(1\pm \epsilon)b$ to denote $a\in [(1-\epsilon)b, (1+\epsilon)b]$. We use $\omega$ to denote the matrix multiplication exponent.

For a matrix or a tensor $A$, we use $\|A\|_2$ or simply $\|A\|$ to denote the spectral norm of $A$, which for a tensor is
$$\|A\| = \sup_{x,y,z \neq 0} \frac{|A(x, y, z)|}{\|x\| \cdot \|y\| \cdot \|z\|}.$$

Let $A\in \R^{n\times d}$ and $k\leq \min\{n, d\}$. We will use $A_k$ or $[A]_k$ to denote its best rank-$k$ approximation. Let $\|A\|_F$ denote the Frobenius norm of a matrix/tensor $A$, the square root of the sum of squares of its entries. For $1 \leq p < 2$, we use $\|A\|_p$ to denote the entry-wise $\ell_p$-norm of a matrix/tensor $A$, the $p$-th root of the sum of $p$-th powers of the absolute values of its entries.

Let $\text{nnz}(A)$ denote the number of nonzero entries of $A$, $A^\top$ its transpose and $A^\dagger$ its Moore-Penrose pseudoinverse. For a square matrix $A$, let $\det(A)$ denote its determinant, and let $A^{-1}$ denote its inverse when $A$ has full rank.

For a 3rd order tensor $A \in \R^{n \times n \times n}$, we use $A_{i,j,l}$ to denote its $(i, j, l)$-th element. We call the three kinds of fibers $A_{*,j,l}$ (mode-$1$), $A_{i,*,l}$ (mode-$2$), and $A_{i,j,*}$ (mode-$3$) its columns, rows, and tubes, respectively. Note that the columns of the unfolding $A_1$ are the columns of $A$ in this sense.

A tensor $A$ is symmetric if and only if for any $i, j, k$, $A_{i,j,k} = A_{i,k,j} = A_{j,i,k} = A_{j,k,i} = A_{k,i,j}$.

For a tensor $A \in \R^{n_1 \times n_2 \times n_3}$, we use $\top$ to denote rotation (3-dimensional transpose) so that $A^\top \in \R^{n_3 \times n_1 \times n_2}$. For a tensor $A \in \R^{n_1 \times n_2 \times n_3}$ and matrix $B \in \R^{n_3 \times k}$, we define the tensor-matrix dot product to be $A \cdot B \in \R^{n_1 \times n_2 \times k}$.

\subsection{Sensitivity and Coreset}

Throughout this paper, we will extensively work with sensitivities and coresets. Let $X$ be some universe of elements. Our main focus is the cost function: ${\rm cost}: \R^d\times X\rightarrow \R_{\geq 0}$, which measures the cost of an element $x\in X$ with respect to the first argument. We then define the notion of strong and weak coresets.

\begin{definition}[(Strong) Coreset]
\label{def:strong_coreset}
Let $B\subseteq A$ and $\epsilon\in (0, 1)$. We say that $B$ is an \emph{$\epsilon$-strong coreset} or $\epsilon$-coreset of $A$ if there exists a nonnegative weight vector $w\in \R^{|B|}_{\geq 0}$ such that for all $x\in X$,
\begin{align*}
    \sum_{b\in B} w_b \cdot {\rm cost}(b, x) = & ~ (1\pm\epsilon)\cdot {\rm cost}(A, x).
\end{align*}
\end{definition}

A strong coreset preserves the cost over all possible $x\in X$, but sometimes we only need the optimal cost preserved. We also introduce the notion of weak coreset.

\begin{definition}[Weak Coreset]
\label{def:weak_coreset}
Let $B\subseteq A$ and $\epsilon\in (0, 1)$. We say that $B$ is an \emph{$\epsilon$-weak coreset} if there exists a nonnegative weight vector $w\in \R^{|B|}_{\geq 0}$ such that
\begin{align*}
    \min_{x\in X}\sum_{b\in B} w_b \cdot {\rm cost}(b, x) = & ~ (1\pm\epsilon)\cdot {\rm OPT},
\end{align*}
where ${\rm OPT}=\min_{x\in X} {\rm cost}(A, x)$.
\end{definition}

\begin{remark}
Oftentimes, given a weighted subset $(B, w)$, we will use ${\rm cost}(B, x)$ as an abbreviation for $\sum_{b\in B} w_b\cdot {\rm cost}(b, x)$, as our analysis and algorithms on the subset of points work in both unweighted and weighted settings. Hence, when the weight is clear from context, we will abuse notation and use ${\rm cost}(b, x)$ to denote $w_b\cdot {\rm cost}(b, x)$.
\end{remark}

\begin{definition}[Sensitivity and Generalized Sensitivity]
Let $A=\{a_1,\ldots, a_n\}\subset \R^d$. We define the \emph{sensitivity of $a_i$} as
\begin{align*}
    s_i(A, A) = & ~ \max_{x\in X} \frac{{\rm cost}(a_i, x)}{{\rm cost}(A, x)}.
\end{align*}
Let $B\subset \R^d$. We define the \emph{sensitivity of $a_i$ with respect to $B$} as
\begin{align*}
    s_i(A, B) = & ~ \max_{x\in X, {\rm cost}(B, x)\neq 0} \frac{{\rm cost}(a_i, x)}{{\rm cost}(B, x)}.
\end{align*}
\end{definition}

\subsection{Leverage Score, Ridge Leverage Score, and Lewis Weights}

\begin{definition}[Statistical Dimension]\label{def:statistical_dimension}
For real value $\lambda \geq 0$ and a rank-$d$ matrix $A \in \R^{n \times d}$ with singular values $\sigma_i(A)$, the quantity $s_d^\lambda(A) := \sum_{i=1}^d \frac{1}{1 + \lambda/\sigma_i^2(A)}$ is the statistical dimension of the ridge regression problem with regularizing weight $\lambda$.
\end{definition}

\begin{definition}[Leverage Score]
Given matrix $A \in \R^{n \times d}$, leverage score can be defined as follows:
\begin{align*}
\tau_i(A) := a_i^\top (A^\top A)^{\dagger} a_i,    
\end{align*}
where $a_i^\top$ is the $i$-th row of $A$ for all $i \in [n]$.
\end{definition}

\begin{definition}[Ridge Leverage Score]\label{def:ridge_lg}
Given matrix $A \in \R^{n \times d}$, we denote the $i$-th ridge leverage score, for $i \in [n]$, as follows:
\begin{align*}
    \ov{\tau_i}(A, \lambda_{A_k}) := a_i^\top (A^\top A + \lambda_{A_k} I)^{\dagger} a_i,
\end{align*}
where $\lambda_{A_k} = \|A - A_k\|_F^2/k$ and $I \in \R^{d \times d}$ is the identity matrix (the pseudo-inverse coincides with the inverse unless $\rank(A)\leq k$, in which case $\lambda_{A_k}=0$ and the score is the leverage score). When the rank $k$ is clear from context, we may abbreviate $\ov \tau_i(A)$ as $\ov\tau_i(A,\lambda_{A_k})$.
\end{definition}

\begin{definition}[Generalized Ridge Leverage Score]\label{def:generilized_ridge_lg}
Let $A \in \R^{n \times d}$, $C \in \R^{n \times d'}$, and $i \in [d]$. We define the $i$-th generalized ridge leverage score of $A \in \R^{n \times d}$ with respect to $C \in \R^{n \times d'}$ as follows:
\begin{align*}
\ov{\tau}_i(A, C, \lambda_{C_k}) =
\begin{cases} 
a_i^\top  ( C C^\top  + \lambda_{C_k}  I_n  )^\dagger  a_i , & \mathrm{if~} a_i \in \mathrm{span} ( C C^\top + \lambda_{C_k} I_n ) , \\
\infty, & \mathrm{otherwise},
\end{cases}
\end{align*}
where $a_i\in \R^n$ denotes the $i$-th column of $A$ and $\lambda_{C_k}=\|C-C_k\|_F^2/k$ as in Definition~\ref{def:ridge_lg}. When the rank $k$ is clear from context, we may use $\ov \tau_i(A, C)$ as shorthand for $\ov \tau_i(A, C, \lambda_{C_k})$.
\end{definition}

\begin{definition}[Lewis Weights]
Let $p\in (0, \infty)$ and $A\in \R^{n\times d}$. We define the $\ell_p$ Lewis weights of $A$, denoted by $w_A$, as
\begin{align*}
    w_{A, i} = & ~ \tau_i(W_A^{1/2-1/p} A),
\end{align*}
or equivalently,
\begin{align*}
    w_{A, i}^{2/p} = & ~ a_i^\top (A^\top W_A^{1-2/p}A)^{-1}a_i.
\end{align*}
\end{definition}

\subsection{Matrix Approximations}

\begin{definition}[Subspace Embedding in~\cite{s06}]\label{def:subspace_embedding}
Let $\epsilon$, $\delta \in (0,1)$ and $n > d$. Given a matrix $U \in \R^{n \times d}$ which is orthonormal (i.e., $U^\top U = I_d$), we say $S \in \R^{m \times n}$ is an $\mathsf{SE}(\epsilon, \delta, n, d)$ subspace embedding for fixed $U$ if, for all $x\in \R^d$,
\begin{align*}
    (1 -\epsilon)\|Ux\|^2_2\leq \|SUx\|^2_2 \leq (1 + \epsilon)\|Ux\|^2_2
\end{align*}
holds with probability $1 - \delta$. This is equivalent to 
\begin{align*}
    \| U^\top S^\top S U - U^\top U \| \leq \epsilon.
\end{align*}
\end{definition}

\begin{definition}[Weak $\epsilon$-Affine Embedding, Theorem~39 in~\cite{cw13}]\label{def:weak_affine_embbedding}
Let matrices $A \in \R^{n \times r}$ and $B \in \R^{n \times d}$. Given matrix $S \in \R^{t \times n}$, we say $S$ is a weak $\epsilon$-affine embedding if the following holds: let $\hat{X} = \arg \min_{X} \|A X - B \|_F^2$ and $\hat{B} = A \hat{X} - B$. Then, for all $X\in \R^{r\times d}$, 
\begin{align*}
    \|S(AX -B)\|_F^2 - \| S \hat{B}\|_F^2 = (1  \pm \epsilon) \|A X - B \|_F^2 - \| \hat{B} \|_F^2.
\end{align*}
\end{definition}

\subsection{Properties of Leverage Score}

Sampling according to leverage score distribution yields a weak affine embedding property. Additionally, solving the subsampled problem results in an optimal solution whose cost is close to the original optimal cost.
\begin{lemma}[Theorem~42 in~\cite{cw13}]\label{lem:leverage_score_weak_affine}
Let matrix $A\in \R^{n \times r}$ with rank at most $k$, and let $B\in \R^{n \times d}$. If $S\in \R^{n \times n}$ is a sampling and rescaling diagonal matrix according to the leverage scores of $A$, let $m =O( \epsilon^{-2}k\log k)$ denote the number of nonzero entries on the diagonal of $S$. Then, with probability at least $0.99$, for all $X \in \R^{r \times d}$, we have:
\begin{itemize}
    \item $S$ is a weak $\epsilon$-affine embedding (see Definition~\ref{def:weak_affine_embbedding}).
    \item Equivalently, if $\hat{X} = \arg\min_{X}\|AX -B\|_F^2$, $\hat{B} = A \hat{X} - B$, and $C := \| S \hat{B}\|^2_F - \|\hat{B}\|^2_F$, then 
    \begin{align*}
        (1 - \epsilon )\cdot \| A X - B\|^2_F + C \leq \| S(AX - B) \|_F^2 \leq (1 + \epsilon) \cdot \|AX - B\|_F^2 + C.
    \end{align*}
\end{itemize}
\end{lemma}

\begin{lemma}[Leverage Score Preserves Optimal Cost, Lemma C.31 of~\cite{swz19}]\label{lem:leverage_score_optimal_solution}
Let $A\in \R^{n\times r}$ be a matrix with rank at most $k$, and let $B\in \R^{n\times d}$. If we sample $O(k\log k+k/\epsilon)$ rows of $A$ and $B$ proportional to the leverage scores of $A$ to obtain a sampling matrix $S$, then with probability at least $0.99$,
    \begin{align*}
        \|AY_*-B\|_F^2 \leq & ~ (1+\epsilon)\cdot \min_{Y} \|AY-B\|_F^2,
    \end{align*}
where $Y_*=\arg\min_Y \|SAY-SB\|_F^2$.
\end{lemma}

\begin{remark}\label{rem:delta_dependence}
Lemmas~\ref{lem:leverage_score_weak_affine} and~\ref{lem:leverage_score_optimal_solution} are proved in~\cite{swz19} for a constant failure probability, with $0.99$ in their Theorem C.50 and $0.999$ in their Lemma C.31. We read the $\delta$-dependence off their proofs, which combine two events. The first event is a subspace embedding by a matrix Chernoff bound, see their Lemma C.22, and the second event is an approximate matrix product for the residual by a second-moment bound and Markov's inequality, see their Lemma C.29. Hence, a failure probability $\delta$ enters only the sample size, as $\log(1/\delta)$ through the first event and as $1/\delta$ through the second event. Every invocation in this paper that runs either lemma with a failure probability $\delta$ uses a specific constant $\delta$, such as $0.04$, $0.004$, $0.002$, or $0.001$. This changes the sample size by a constant factor, and this factor is absorbed by the $O(\cdot)$.
\end{remark}

\subsection{Quantum Primitives}

Our core quantum primitive is a sampling algorithm based on Grover search.

\begin{lemma}[Claim 3 in~\cite{aw22}]
\label{lem:q_sampling_1_d}
Let $n$ be a positive integer, let $q_1,\ldots,q_n\in [0, 1]$, each specified with $O(\log n)$ bits of precision, and let $\delta\in (0,1)$. There is a quantum algorithm, $\textsc{QSample}(q, \delta)$, that with probability at least $1-\delta$ outputs a set $S\subseteq [n]$ containing each index $i$ independently with probability $q_i$ (precisely: its output coincides, except on an event of probability at most $\delta$, with such a sample), and runs in expected time $\wt O(\sqrt{n\max\{1, \sum_{i=1}^n q_i\}}\log(1/\delta))\cdot {\cal T}$ over its internal randomness, where ${\cal T}$ is the time to compute $q_i$. The precision hypothesis is without loss of generality throughout this paper: rounding each $q_i$ up to a multiple of $n^{-2}$ preserves the overestimate property of our sampling probabilities and increases $\sum_{i=1}^n q_i$ by at most $1/n$.
\end{lemma}

We note that this runtime bound could also be achieved via quantum rejection sampling~\citep{orr12}. Let $P=\sum_{i=1}^n q_i$, then $q_i/P$ for all $i\in [n]$ induces a probability distribution, which we denote by $\sigma$. Recall that rejection sampling aims to generate one sample from the target distribution $\sigma$ (where $\sigma_i=q_i/P$) using a uniform proposal distribution $\pi$ (where $\pi_i=1/n$). The query complexity is $\wt O(\max_{i\in [n]}\sqrt{\sigma_i/\pi_i})\cdot {\cal T}$. Since each $q_i\leq 1$, the ratio can be upper bounded by $\max_i (q_i/P)/(1/n)\leq n/P$. Thus, the complexity to generate \emph{one sample} is $\wt O(\sqrt{n/P})\cdot {\cal T}$. As $\sum_{i=1}^n q_i=P$, if we choose each index $i$ with probability $q_i$ independently, then the expected size is $P$, hence the total expected complexity is $\wt O(P \sqrt{n/P})\cdot {\cal T}=\wt O(\sqrt{nP})\cdot {\cal T}$, as desired. Claim 3 of~\cite{aw22} is stated in expected time without a failure probability, but it relies on the bounded-error search of their Claim 2, so we obtain the success probability $1-\delta$ of Lemma~\ref{lem:q_sampling_1_d} as follows. The number $K$ of marked indices is a sum of independent Bernoulli$(q_i)$ variables with mean $P$. For $j=1,2,\ldots$, the $j$-th Grover search is the search of Theorem 3 of~\cite{bhmt02} for a marked index among those not yet found, run with the cutoff $c\sqrt n\log(4j^2/\delta)$ for a suitable constant $c$, and the algorithm stops at the first search that reaches its cutoff without finding a marked index. When $t\geq 1$ marked indices remain, the current search finds one in expected time $C\sqrt{n/t}\cdot{\cal T}$ for an absolute constant $C$, and we run it in independent windows of $2C\sqrt n\cdot{\cal T}$ steps, restarting after each window. By Markov's inequality, a window succeeds with probability at least $1/2$, as $t\geq 1$, so with $c:=2C/\log 2$, the cutoff contains $\log_2(4j^2/\delta)$ windows and is reached with probability at most $\delta/(4j^2)$. Moreover, the expected time is still $O(\sqrt{n/t})\cdot{\cal T}$, since, by Markov's inequality, the first window is left with probability at most $1/(2\sqrt t)$, and the restarts then cost at most $2\cdot 2C\sqrt n\cdot{\cal T}$ in expectation. When no marked index remains, the search reaches the cutoff and the algorithm stops correctly. The total failure probability is at most $\sum_{j\geq 1}\delta/(4j^2)<\delta/2$. Conditioned on $K$, the expected running time is $\sum_{j=1}^{K}O(\sqrt{n/(K-j+1)})\cdot{\cal T}+O(\sqrt n\log(4(K+1)^2/\delta))\cdot{\cal T}=O(\sqrt{nK}+\sqrt n\,(\log(K+1)+\log(1/\delta)))\cdot{\cal T}$. Since $\E\sqrt K\leq\sqrt P$ and $K\leq n$, this is $\wt O(\sqrt{n\max\{1,P\}}\log(1/\delta))\cdot{\cal T}$. Since a candidate is checked with one query, a failed search can only miss marked indices and never reports an unmarked one. Hence, the output is a subset of the exact sample and coincides with it on the success event.

\emph{Convention.} Running-time and query bounds for algorithms that call $\textsc{QSample}$ are expected values over the internal randomness of the sampler calls. Theorems~\ref{thm:general} and~\ref{thm:ls} state this explicitly, with the expectation also taken over the random halves of the recursion and conditioned on the success event. A theorem that states a running time without the word ``expected'' should be read in this sense. Each proof below that invokes $\textsc{QSample}$ splits an explicit share of its failure budget equally among these calls, and the factor $O(\log(K/\delta))$ in the running time of $K$ calls is absorbed by $\wt O(\cdot)$. Since the output of a failed call is an arbitrary subset, every event on which sizes are bounded in expectation includes the success of the sampler calls. For an expectation conditioned on a success event ${\cal F}$ with $\Pr[{\cal F}]\geq 1/2$ that may depend on the outcome of a call, we use $\E[X\mid{\cal F}]\leq \E[X\,\mathbf{1}_{{\cal F}'}]/\Pr[{\cal F}]\leq 2\,\E[X\,\mathbf{1}_{{\cal F}'}]$, where ${\cal F}'\supseteq{\cal F}$ is an event determined before the call, so the expected running time of the call given its input applies. In the algorithm boxes, we write $\textsc{QSample}(q)$ when the failure parameter is fixed in the accompanying proof.

Throughout the paper, we will use the notation $\textsc{QLS}(A, s, \delta)$ to denote Algorithm~\ref{alg:qls}, the procedure of Theorem~\ref{thm:ls} that samples rows or columns of $A$ according to the leverage scores with size parameter $s$. Its guarantees and the cost of computing the scores are those of Theorem~\ref{thm:ls}, where $d$ is the number of columns (resp.\ rows) of $A$, and the running time is expected in the sense of Theorem~\ref{thm:general}. The sampling step takes $\wt O(\sqrt{ns})\cdot {\cal T}$ time, where ${\cal T}$ is the time to compute a single score. Similarly, we use $\textsc{QRLS}(A, k, \epsilon, \delta)$ to denote the procedure of Theorem~\ref{thm:ridge} that samples according to the rank-$k$ ridge leverage scores, and $\textsc{QRLS}_s(A, k, s, \delta)$ to denote its fixed-size variant in Corollary~\ref{thm:quantum_pcp}.

\section{A Quantum Recursive Sampling Framework for Coreset}
\label{sec:general}
Throughout this section, let us consider $A=\{a_1,\ldots,a_n\}\subset \R^d$ to be a set of $n$ points in $\R^d$, and $X$ to be a set. Let ${\rm cost}: \R^d\times X\rightarrow \R_{\geq 0}$ be a cost function, and for $x\in X$, let ${\rm cost}(A, x)=\sum_{i=1}^n {\rm cost}(a_i, x)$. The main objective of this section is to develop a framework for sampling a weighted subset of $A$ that approximates the cost of $A$. To do so, we prove that if the weights satisfy certain assumptions, then a generic recursive sampling framework could construct a coreset from these weights. The assumptions are listed in the following.

We use two pieces of notation throughout this section. First, for a finite set $A$, we write $A'\subset_{1/2} A$ for the random subset of $A$ that contains each point of $A$ independently with probability $1/2$. Applying this operation $T$ times gives a chain $A_0\subset_{1/2} A_1\subset_{1/2}\cdots\subset_{1/2} A_T=A$. Thus, a point of $A$ belongs to $A_t$ with probability $2^{-(T-t)}$, and $|A_t|$ is a sum of independent indicators with mean $n2^{-(T-t)}$. Equivalently, every $a_i\in A$ is assigned $T$ independent unbiased random bits, and $a_i\in A_t$ if and only if its first $T-t$ bits all equal one. Following~\cite{ag23}, we realize the chain in this way, and the random bits are simulated as described in the quantum computation model in the introduction. Second, given $A=\{a_1,\ldots,a_n\}$, a subset $A'\subseteq A$, and a weighted finite set $B$, we define the \emph{augmented weights}
\begin{align*}
    w'_i(A, A'; B) := & ~ \begin{cases}
        w_i(A, B), & \text{if $a_i\in A'$,} \\
        w_i(A, B\cup \{a_i\}), & \text{otherwise,}
    \end{cases}
\end{align*}
where $a_i$ joins $B$ with weight one, and we abbreviate $w'_i(A, A'):=w'_i(A, A'; A')$. In the algorithms below, $A'$ is the uniform half of $A$ and $B$ is a coreset of $A'$, and we read off the applicable case from the membership bit of $a_i$.

\begin{assumption}\label{assumption:recurse}
Given two finite subsets $A, B\subseteq \R^d$, let $w(A, B)\in \R^{|A|}$ be a nonnegative weight vector where $w_i(A, B)$ is the weight of $a_i$ with respect to $B$, and we assume that $w_i(A, B)$ depends on $A$ only through the point $a_i$, so that $w_i(A, B)=w_j(A', B)$ whenever $a_i$ is the $j$-th point of a finite set $A'\subseteq \R^d$, and in particular $w_i(A, A_S)=w_i(A_S, A_S)$ for $i\in S$. We assume $w$ satisfies the following conditions for every finite point set $A\subseteq \R^d$, with $n=|A|$ and $[n]$ indexing its points, where $B$, and $C$ in item 4, may carry weights:
\begin{itemize}
    \item Consistent total weights: for any subset $S\subseteq [n]$, $\sum_{i\in S} w_i(A_S, A_S)\leq {\rm sum}(w)$ where ${\rm sum}(w)$ is a finite upper bound on the total weights.
    \item Uniform sampling bound: for every subset $A'\subseteq A$, the augmented weights satisfy $w'_i(A, A')=w'_i(A, A'; A')\geq w_i(A, A)$ for all $i\in [n]$. 
    \item Importance sampling bound: for $\epsilon, \delta\in (0,1)$ there is a quantity $g(\epsilon, \delta, n, d)$ with the following property. Let $u\in \R^n$ satisfy $u_i\geq w_i(A, A)$ for all $i\in [n]$, and let $B$ be the weighted subset of $A$ obtained by including each $a_i$ independently with probability $q_i=\min\{1, g(\epsilon, \delta, n, d)\cdot u_i \}$ and reweighting each included point by $1/q_i$. Then $B$ has expected size at most $g(\epsilon, \delta, n, d)\cdot \|u\|_1$, and with probability at least $1-\delta$, $B$ is an $\epsilon$-coreset of $A$.
    \item Coreset preserves weights: let $B$ be an $\epsilon$-coreset of $A$, then $w_i(C, B)\in [(1+\epsilon)^{-1}, (1-\epsilon)^{-1}]\cdot w_i(C, A)$ for any fixed finite $C\subseteq \R^d$ and for every point $c_i$ of $C$.
\end{itemize}
\end{assumption}
We make two remarks on the importance sampling bound. First, for any $\alpha\geq g:=g(\epsilon, \delta, n, d)$, the vector $(\alpha/g)\cdot u$ is again an entrywise overestimate of $w(A, A)$, so the bound applies verbatim to the probabilities $q_i=\min\{1, \alpha\cdot u_i\}$, with expected size at most $\alpha\cdot \|u\|_1$. The algorithms below use $\alpha=c\cdot g$ for a constant $c$, and we may assume $g\cdot {\rm sum}(w)\geq 1$ by increasing $g$. Second, we take $g$ nondecreasing in $n$, hence $g(\epsilon, \delta, n, d)$ may be used for every subset of $A$ as well. In our instantiations, $g$ depends on $\delta$ only through $\log(1/\delta)$, as in Lemma~\ref{lem:importance_sample}.

\begin{algorithm}[!ht]
\caption{Quantum iterative sampling for coreset.}
\label{alg:quantum_iterate}
\begin{algorithmic}[1]
\Procedure{QIterateSample}{$A\in \R^{n\times d}, \epsilon, \delta$}
\State $c\gets 1000$
\State $\epsilon'\gets 0.01$, $\delta'\gets \delta/(2\lceil\log_2 n\rceil)$ \Comment{$\delta'\leq \delta/(2T)$ for the $T$ below}
\State $s \gets 4c\cdot g(\epsilon, \delta/2, n, d)\cdot {\rm sum}(w)$
\State $s'\gets 4c\cdot g(\epsilon', \delta', n, d)\cdot {\rm sum}(w)$
\State $T\gets \max\{1, \lceil\log_2(n/s')\rceil\}$
\State $A_0\subset_{1/2} A_1\subset_{1/2}\ldots\subset_{1/2} A_{T-1}\subset_{1/2}A_T=A$
\State $C_0\gets A_0$ \Comment{learned explicitly by $\textsc{QSample}(\mathbf{1}[a_i\in A_0], \delta'/(T+1))$}
\For{$t=1\to T-1$}
\State Implement a classical oracle for $w'_i(A_t, A_{t-1}; C_{t-1})$ for all $a_i\in A_t$ 
\State \Comment{$q_i=\min\{1,  c\cdot g(\epsilon', \delta', n, d)\cdot w'_i(A_t, A_{t-1}; C_{t-1})\}$}
\State $D_t\gets \textsc{QSample}(q, \delta'/(T+1))$
\State $C_t\gets D_t^\top A_t$  \Comment{$C_t\in \R^{\|D_t\|_0\times d}$}
\EndFor
\State Implement a classical oracle for $w_i'(A_T, A_{T-1}; C_{T-1})$ for all $a_i\in A_T$ 
\State \Comment{$q_i=\min\{1, c\cdot g(\epsilon, \delta/2, n, d)\cdot w_i'(A_T, A_{T-1}; C_{T-1}) \}$}
\State $D_T\gets \textsc{QSample}(q, \delta'/(T+1))$
\State $C_T\gets D_T^\top A_T$ \Comment{$C_T\in \R^{\|D_T\|_0\times d}$}
\State \Return $(C_T, D_T)$
\EndProcedure
\end{algorithmic}
\end{algorithm}

\begin{lemma}\label{lem:coreset_add}
Let $B$ be an $\epsilon$-coreset of $A$ and let $p\not \in A$, then $B\cup \{p\}$ is an $\epsilon$-coreset of $A\cup \{p\}$. 
\end{lemma}
\begin{proof}
Since $B$ is an $\epsilon$-coreset of $A$, we know that for any $x\in X$, ${\rm cost}(B, x)=(1\pm\epsilon)\cdot {\rm cost}(A, x)$. We note that
\begin{align*}
    {\rm cost}(B\cup \{p\}, x) = & ~ {\rm cost}(B, x) + {\rm cost}(\{p\}, x) \\
    \leq & ~ (1+\epsilon)\cdot {\rm cost}(A, x) + {\rm cost}(\{p\}, x) \\
    \leq & ~ (1+\epsilon)\cdot {\rm cost}(A\cup \{p\}, x),
\end{align*}
the lower bound can be established similarly.
\end{proof}

Theorem~\ref{thm:general} is stated for Algorithm~\ref{alg:quantum_iterate}, the iterative form of the recursive scheme of Algorithm~\ref{alg:quantum_recurse_main} in Section~\ref{sec:tech:grover}.

\begin{theorem}
\label{thm:general}
Let $A\in \R^{n\times d}$ with $n\geq 2$, let $\epsilon, \delta\in (0,1)$, and let $w$ satisfy Assumption~\ref{assumption:recurse}. Then, Algorithm~\ref{alg:quantum_iterate} is a quantum algorithm that, with probability at least $1-\delta$, constructs an $\epsilon$-coreset $C$ of $A$. More precisely, let ${\cal F}$ be the event that every intermediate sample $C_t$, $1\leq t\leq T-1$, of Algorithm~\ref{alg:quantum_iterate} is a $0.01$-coreset of $A_t$ and that each of its $T+1$ calls of $\textsc{QSample}$ returns an exact sample, with $T$ as in that algorithm. Then $\Pr[{\cal F}]\geq 1-\delta/2$, with probability at least $1-\delta$ the event ${\cal F}$ occurs and $C$ is an $\epsilon$-coreset of $A$, and conditioned on ${\cal F}$ the expected size of $C$ is $O(s)$, where $s:=4c\cdot g(\epsilon, \delta/2, n, d)\cdot {\rm sum}(w)$. Moreover, if a classical oracle for $w_i(Z, Y)$ can be implemented with
\begin{itemize}
    \item Preprocessing in time ${\cal T}_{\rm prep}(|Y|, d)$, linear in $|Y|$ for fixed $n$ and $d$,
    \item Query time ${\cal T}_{\rm query}(|Y|, d)$ for computing $w_i(Z, Y)$ and $w_i(Z, Y\cup\{z_i\})$ for any $i\in [|Z|]$, where $z_i$ denotes the $i$-th point of the finite set $Z\subseteq \R^d$, and we assume that this query time is bounded by ${\cal T}_{\rm query}(s', d)$ for every set $Y$ arising in the algorithm,
\end{itemize}
the algorithm runs, conditioned on ${\cal F}$, in expected time
\begin{align*}
    O(1+\log(n/s'))\cdot {\cal T}_{\rm prep}(s', d)+\wt O(\sqrt{n(s+s')}\cdot {\cal T}_{\rm query}(s', d)),
\end{align*}
where $s':=4c\cdot g(0.01, \delta', n, d)\cdot {\rm sum}(w)$ with $\delta'=\delta/(2\lceil\log_2 n\rceil)$, and the algorithm uses $\wt O(\sqrt{n(s+s')})$ queries to the rows or columns of $A$, likewise in expectation conditioned on ${\cal F}$. In all applications in this paper $s'=\wt O(s)$, because either $s=s'$ or $g$ scales as $\epsilon^{-2}$ up to logarithmic factors and $\epsilon<1$. The bounds then read $O(1+\log(n/s'))\cdot {\cal T}_{\rm prep}(s', d)+\wt O(\sqrt{ns}\cdot {\cal T}_{\rm query}(s', d))$ and $\wt O(\sqrt{ns})$ queries.
\end{theorem}

\begin{proof}
We divide the proof into three parts.

\paragraph{One level.} Let $A$ be a set of $n$ points, let $A'\subset_{1/2} A$, let $\epsilon_0\in [0, 1/3]$, and let $C'$ be an $\epsilon_0$-coreset of $A'$, where for $\epsilon_0=0$ we take $C'=A'$. We claim that for all $i$,
\begin{align}
    w'_i(A, A'; C') \in & ~ I_{\epsilon_0}\cdot w'_i(A, A'), \qquad I_{\epsilon_0}:=[(1+\epsilon_0)^{-1}, (1-\epsilon_0)^{-1}], \label{eq:one_level_approx}
\end{align}
and that
\begin{align}
    \E[\|w'(A, A')\|_1] \leq & ~ 2\, {\rm sum}(w). \label{eq:one_level_sum}
\end{align}
For~\eqref{eq:one_level_approx}, we consider two cases. If $a_i\in A'$, then $w'_i(A, A')=w_i(A, A')$, and by the coreset preserves weights condition of Assumption~\ref{assumption:recurse} with $B=C'$ and $C=A$, we have $w'_i(A, A'; C')=w_i(A, C')\in I_{\epsilon_0}\cdot w_i(A, A')=I_{\epsilon_0}\cdot w'_i(A, A')$. If $a_i\not\in A'$, then $C'\cup \{a_i\}$ is an $\epsilon_0$-coreset of $A'\cup\{a_i\}$ by Lemma~\ref{lem:coreset_add}, and thus the same condition gives $w'_i(A, A'; C')=w_i(A, C'\cup\{a_i\})\in I_{\epsilon_0}\cdot w_i(A, A'\cup \{a_i\})=I_{\epsilon_0}\cdot w'_i(A, A')$. 

For~\eqref{eq:one_level_sum}, we first condition on $|A'|=m$, and given this event, $A'$ is a uniformly random subset of $A$ of size $m$. Let $S$ be the diagonal indicator matrix of $A'$, so $SA=A'$. We show that
\begin{align*}
    \E[\|w'(A, SA)\|_1 \mid |S|=m] \leq & ~ \frac{n+1}{m+1}\cdot {\rm sum}(w).
\end{align*}
Let us denote $S^{(i)}$ as the diagonal indicator matrix for $S\cup \{i\}$. Then, note
\begin{align*}
    \sum_{i=1}^n w'_i(A, SA) = & ~ \sum_{i\in S} w_i(A, SA) + \sum_{i\not\in S} w_i(A, S^{(i)} A) \\
    = & ~  \|w(SA, SA)\|_1 + \sum_{i\not\in S} w_i(A, S^{(i)} A) \\
    \leq & ~ {\rm sum}(w) + \sum_{i\not\in S} w_i(A, S^{(i)} A),
\end{align*}
where the second step uses the convention of Assumption~\ref{assumption:recurse}, which we also use below, and the last step uses the consistent total weights assumption. To bound the second term, note that it is generated via the following random process: first selecting $S$, then selecting a random $i\not \in S$ and returning $w_i(A, S^{(i)}A)$. Since there are $n-m$ points not in $SA$, the expected value of this process is $\frac{1}{n-m} \E[\sum_{i\not\in S} w_i(A, S^{(i)}A)]$. The key observation is that this process is equivalent to another process: pick a random subset $S'\subset [n]$ of size $m+1$, then randomly pick a point $a_i\in S'A$ and return $w_i(A, S'A)$. In expectation, this is equal to the average weight over $S'A$. Since $S'A$ contains $m+1$ points and by the consistent total weights assumption, the average weight is at most $\frac{{\rm sum}(w)}{m+1}$. Therefore,
\begin{align*}
    \E[\sum_{i\not \in S} w_i(A, S^{(i)}A) \mid |S|=m] \leq & ~ (n-m)\cdot \frac{{\rm sum}(w)}{m+1},
\end{align*}
combining these results, we obtain the following expectation bound:
\begin{align*}
    \E[\sum_{i=1}^n w'_i(A, SA) \mid |S|=m] \leq & ~ {\rm sum}(w) + (n-m)\cdot \frac{{\rm sum}(w)}{m+1} 
    = ~ \frac{n+1}{m+1} \cdot {\rm sum}(w).
\end{align*}
Since $|A'|$ is a sum of $n$ independent unbiased bits, we have $\Pr[|A'|=m]=\binom{n}{m}2^{-n}$. Using $\frac{n+1}{m+1}\binom{n}{m}=\binom{n+1}{m+1}$, we obtain
\begin{align*}
    \E[\|w'(A, A')\|_1] \leq & ~ {\rm sum}(w)\cdot \E\Big[\frac{n+1}{|A'|+1}\Big] = {\rm sum}(w)\cdot 2^{-n}\sum_{m=0}^n \binom{n+1}{m+1} \\
    = & ~ (2-2^{-n})\cdot {\rm sum}(w)\leq 2\, {\rm sum}(w),
\end{align*}
which is~\eqref{eq:one_level_sum}.

Next, we derive the consequences of these two bounds. By the uniform sampling bound, we have $w'_i(A, A')\geq w_i(A, A)$. Hence,~\eqref{eq:one_level_approx} gives $w'_i(A, A'; C')\geq (1+\epsilon_0)^{-1}\, w_i(A, A)\geq (1-\epsilon_0)\, w_i(A, A)$ for all $i$, and on the event that $C'$ is an $\epsilon_0$-coreset of $A'$, it also gives
\begin{align*}
    \|w'(A, A'; C')\|_1 \leq & ~ \frac{1}{1-\epsilon_0}\cdot \|w'(A, A')\|_1 .
\end{align*}
Now let $\epsilon_1, \delta_1\in (0,1)$ and let $g_1\geq g(\epsilon_1, \delta_1, n, d)$. In the algorithm, $g$ is evaluated at the size of the input set, and this is admissible by the remarks following Assumption~\ref{assumption:recurse}. We sample $A$ with $q_i=\min\{1, c\cdot g_1\cdot w'_i(A, A'; C')\}$, where $c=1000$, and we write $q_i=\min\{1, g_1\cdot u_i\}$ with $u_i:=c\cdot w'_i(A, A'; C')\geq c(1-\epsilon_0)\, w_i(A, A)\geq w_i(A, A)$. By the importance sampling bound of Assumption~\ref{assumption:recurse}, the sample is an $\epsilon_1$-coreset of $A$ with probability at least $1-\delta_1$, and its expected size is at most $g_1\|u\|_1=c\, g_1 \|w'(A, A'; C')\|_1$. Taking the expectation over $A'$ and over the randomness that produced $C'$, on the event that $C'$ is an $\epsilon_0$-coreset of $A'$, the expected size is at most
\begin{align*}
    c\, g_1\cdot \frac{1}{1-\epsilon_0}\cdot 2\,{\rm sum}(w) \leq & ~ 4c\, g_1\cdot {\rm sum}(w)
\end{align*}
for $\epsilon_0\leq 1/3$. For $\epsilon_0=0.01$, the factor is $2/0.99<2.03$.

\paragraph{The chain of levels.} For $t\leq T-1$, let ${\cal E}_t$ be the event that $C_t$ is a $0.01$-coreset of $A_t$, and let ${\cal E}_T$ be the event that $C_T$ is an $\epsilon$-coreset of $A_T=A$. The set $C_0=A_0$ is an exact coreset of $A_0$, so ${\cal E}_0$ holds. By the choice of $T$, we have $\E|A_0|=n2^{-T}\leq s'$. Let ${\cal G}$ be the event that the $T+1$ calls of $\textsc{QSample}$ in the algorithm, for $A_0$ and for $D_1,\ldots,D_T$, all return exact samples. Each call is run with failure probability $\delta'/(T+1)$, which gives $\Pr[{\cal G}^c]\leq\delta'$. Moreover, on ${\cal G}$ every sampling step is a Bernoulli sample with the stated probabilities, so the one-level statement applies as written, whereas a failed call may return an arbitrary subset. For $1\leq t\leq T-1$, we apply the one-level statement to $A_{t-1}\subset_{1/2} A_t$ with $C'=C_{t-1}$, $\epsilon_0=0$ if $t=1$ and $\epsilon_0=0.01$ otherwise, and $(\epsilon_1, \delta_1)=(0.01, \delta')$. Conditioned on $A_{t-1}$, $C_{t-1}$, and ${\cal E}_{t-1}$, the level fails with probability at most $\delta'$, and $\E[|C_t|\,\mathbf{1}_{{\cal E}_{t-1}\cap{\cal G}}]\leq 4c\cdot g(0.01, \delta', n, d)\cdot {\rm sum}(w)=s'$. For $t=T$, we apply it with $(\epsilon_1, \delta_1)=(\epsilon, \delta/2)$, so, conditioned on ${\cal E}_{T-1}$, the final level fails with probability at most $\delta/2$, and $\E[|C_T|\,\mathbf{1}_{{\cal E}_{T-1}\cap{\cal G}}]\leq 4c\cdot g(\epsilon, \delta/2, n, d)\cdot {\rm sum}(w)=s$. Since $s'\geq 1$, we have $T\leq \lceil\log_2 n\rceil$ and hence $T\delta'\leq \delta/2$. Let ${\cal F}:={\cal G}\cap\bigcap_{t=1}^{T-1}{\cal E}_t$ be the event named in the statement. A union bound over the levels and the sampler calls gives $\Pr[{\cal F}\cap{\cal E}_T]\geq 1-(T-1)\delta'-\delta'-\delta/2\geq 1-\delta$ and $\Pr[{\cal F}]\geq 1-T\delta'\geq 1-\delta/2\geq 1/2$. Since ${\cal F}\subseteq {\cal E}_{t-1}\cap{\cal G}$ for every $t\in [T]$, the bounds above give $\E[|C_t|\mid {\cal F}]\leq \E[|C_t|\,\mathbf{1}_{{\cal E}_{t-1}\cap{\cal G}}]/\Pr[{\cal F}]\leq 2s'$ for $t\leq T-1$ and $\E[|C_T|\mid{\cal F}]\leq 2s$. All expectations in the running time analysis below are conditioned on ${\cal F}$. This proves the coreset guarantee and the size bound. Finally, the factor $O(\log((T+1)/\delta'))=O(\log(\log n/\delta))$ in the running time of the sampler calls is absorbed by $\wt O(\cdot)$.

\paragraph{Running time.}

\begin{itemize}
    \item Since we realize the chain by $T$ random bits per point as described at the beginning of this section, testing whether $a_i\in A_t$ costs $O(T)=O(\log n)$ per query, which the $\wt O$ absorbs. We learn the set $A_0$ explicitly by applying Lemma~\ref{lem:q_sampling_1_d} to $q_i=\mathbf{1}[a_i\in A_0]$, which takes $\wt O(\sqrt{n\cdot \E|A_0|})=\wt O(\sqrt{ns'})$ expected time and queries. Reading its points then takes $|A_0|$ row queries and $O(s'd)$ expected time, which we charge to the first preprocessing. Every preprocessing reads its input, so ${\cal T}_{\rm prep}(s', d)\geq s'd$.
    \item For each level $t\in [T]$, we first prepare a classical oracle for $w'_i(A_t, A_{t-1}; C_{t-1})$ in ${\cal T}_{\rm prep}(|C_{t-1}|, d)$ time. Moreover, $\E[|C_{t-1}|\mid{\cal F}]\leq 2s'$ for every $t\in [T]$, where for $t=1$ this holds because $\E|A_0|\leq s'$ and $\Pr[{\cal F}]\geq 1/2$. Since ${\cal T}_{\rm prep}$ is linear in its first argument by hypothesis, we have $\E[{\cal T}_{\rm prep}(|C_{t-1}|, d)\mid{\cal F}]\leq {\cal T}_{\rm prep}(2s', d)=O({\cal T}_{\rm prep}(s', d))$, and we write ${\cal T}_{\rm prep}(s', d)$ for this quantity. The oracles of Lemmas~\ref{lem:classical_ls} and~\ref{lem:classical_lewis} have linear preprocessing time. For the oracle of Lemma~\ref{lem:classical_ridge}, whose preprocessing time is superlinear, the same accounting applies since the proof of Theorem~\ref{thm:ridge} bounds every level size by $O(s')$ on the success event through Lemma~\ref{thm:ridge_via_unif}. A query for a point $a_i$ evaluates $w_i(A_t, C_{t-1})$ or $w_i(A_t, C_{t-1}\cup\{a_i\})$ according to the membership bit of $a_i$, which takes ${\cal T}_{\rm query}(s', d)$ time.
    \item Next, for $t\leq T-1$ we sample according to $q_i=\min\{1, c\cdot g(\epsilon', \delta', n, d)\cdot w'_i(A_t, A_{t-1}; C_{t-1})\}$ for $a_i\in A_t$ and $q_i=0$ for $a_i\notin A_t$, with 
    \begin{align*}
        \E[\sum_{i\in A_t} q_i\mid{\cal F}] \leq & ~ c\cdot g(\epsilon', \delta', n, d)\cdot \E[\sum_{i\in A_t} w'_i(A_t, A_{t-1}; C_{t-1})\,\mathbf{1}_{{\cal E}_{t-1}}]/\Pr[{\cal F}] \leq  ~ 1.02\,s'
    \end{align*}
    by the one-level statement, where we use the bound $2\,{\rm sum}(w)/0.99$ on the expectation there, the definition $s'=4c\,g(\epsilon', \delta', n, d)\,{\rm sum}(w)$, and $\Pr[{\cal F}]\geq 1/2$. Using Lemma~\ref{lem:q_sampling_1_d} and Jensen's inequality $\E\sqrt{n\sum_i q_i}\leq\sqrt{n\,\E\sum_i q_i}$, this step can be implemented in expected time
    \begin{align*}
        \wt O(\sqrt{n s'})\cdot {\cal T}_{\rm query}(s', d).
    \end{align*}
    \item Forming $C_t$ requires selecting and weighting $\|D_t\|_0$ points, which can be done in $O(s'd)$ expected time.
    \item Finally, we do a resampling with $\epsilon$ to form the final coreset. By the same computation as for the intermediate levels, we have $\E[\sum_{i} q_i\mid{\cal F}]\leq 1.02\,s$, and this step takes 
    \begin{align*}
        {\cal T}_{\rm prep}(s', d)+\wt O(\sqrt{ns}\cdot {\cal T}_{\rm query}(s', d))
    \end{align*}
    expected time. Summing the $T=O(1+\log(n/s'))$ oracle preparations gives the $O(1+\log(n/s'))\cdot {\cal T}_{\rm prep}(s', d)$ term. The $T-1$ intermediate sampling rounds together with the final one cost $\wt O((\sqrt{ns'}+\sqrt{ns})\cdot{\cal T}_{\rm query}(s',d))=\wt O(\sqrt{n(s+s')}\cdot{\cal T}_{\rm query}(s',d))$, where the $\wt O$ absorbs the factor $T=O(\log n)$. Each oracle query makes one row query, and reading $A_0$ and the sampled points costs $O(T\cdot\min\{s', n\}+\min\{s, n\})\leq \wt O(\sqrt{n(s+s')})$ further queries in expectation conditioned on ${\cal F}$, as $\min\{a, n\}\leq\sqrt{an}$. Therefore, the query count is $\wt O(\sqrt{n(s+s')})$. \qedhere
\end{itemize}
\end{proof}

Our main contribution is to prove that \emph{sensitivity sampling} satisfies Assumption~\ref{assumption:recurse}.

\begin{definition}\label{def:sensitivity}
Let $A=\{a_1,\ldots,a_n\}\subset \R^d$ and let ${\rm cost}: \R^d\times X\rightarrow \R_{\geq 0}$ be a cost function. We define the sensitivity of $a_i$ with respect to $B$, denoted by $s_i(A, B)$, as 
\begin{align*}
    s_i(A, B) = & ~ \max_{x\in X, {\rm cost}(B, x)\neq 0} \frac{{\rm cost}(a_i, x)}{{\rm cost}(B, x)}
\end{align*}
\end{definition}

We also need to define the dimension of a system $(A, X, {\rm cost})$:
\begin{definition}
\label{def:dim}
Given a point set $A=\{a_1,\ldots,a_n\}\subset \R^d$, a space $X$ and a cost function ${\rm cost}: \R^d\times X\rightarrow \R_{\geq 0}$, for a weight vector $v\in \R^n_{\geq 0}$, a subset $S\subseteq [n]$, $x\in X$ and $r\in [0,\infty)$ we define
\begin{align*}
    {\rm range}_{S, v}(x, r) = & ~ \{i\in S: v_i\cdot {\rm cost}(a_i, x)\leq r \}.
\end{align*}
The \emph{dimension} of $(A, X, {\rm cost})$ is the smallest integer ${\rm dim}$ such that for every subset $S\subseteq [n]$ with $|S|\geq 2$ and every weight vector $v\in \R^n_{\geq 0}$ we have
\begin{align*}
    |\{{\rm range}_{S, v}(x, r): x\in X, r\in [0,\infty) \} | \leq & ~ |S|^{\rm dim}.
\end{align*}
\end{definition}

\begin{lemma}[Sensitivity sampling with independent inclusions]
\label{lem:importance_sample}
Let $A=\{a_1,\ldots,a_n\}\subset \R^d$, let $w\in \R^n_{\geq 0}$, and write ${\rm cost}_w(A, x)=\sum_{i=1}^n w_i\cdot {\rm cost}(a_i, x)$. Let ${\rm dim}\geq 1$ be an integer that is at least the dimension of $(A, X, {\rm cost})$ of Definition~\ref{def:dim}. Let $u\in \R^n_{\geq 0}$ satisfy
\begin{align*}
    u_i \geq & ~ \sup_{x\in X:~{\rm cost}_w(A, x)\neq 0} \frac{w_i\cdot {\rm cost}(a_i, x)}{{\rm cost}_w(A, x)} \qquad \text{for all $i\in [n]$}
\end{align*}
(for unit weights, $u_i\geq s_i(A, A)$), let $t\geq \max\{\sum_{i=1}^n u_i, {\rm dim}, 2\}$, and let $\epsilon,\delta\in (0,1)$. Let $c_0$ be a sufficiently large absolute constant, let
\begin{align*}
    g := & ~ c_0\epsilon^{-2}\cdot ({\rm dim}\cdot \log(t/\epsilon)+\log(1/\delta)), \qquad q_i:=\min\{1, g\cdot u_i\},
\end{align*}
and let $S\subseteq [n]$ contain each $i$ independently with probability $q_i$, the point $a_i$ with $i\in S$ receiving the weight $w_i/q_i$. Then with probability at least $1-\delta$, for all $x\in X$,
\begin{align*}
    \Big| {\rm cost}_w(A, x) - \sum_{i\in S} \frac{w_i}{q_i}\cdot {\rm cost}(a_i, x) \Big| \leq & ~ \epsilon\cdot {\rm cost}_w(A, x).
\end{align*}
Moreover, $\E[|S|]=\sum_{i=1}^n q_i\leq g\sum_{i=1}^n u_i$, and $|S|\leq 2gt$ with probability at least $1-\delta$.
\end{lemma}

\begin{proof}
The counterpart of this statement for a prescribed number of independent draws from the distribution $u/\|u\|_1$ is Theorem 17 of~\cite{bflsz22} (Theorem 4.1 of~\cite{fl11}). The sampling primitive of this paper in Lemma~\ref{lem:q_sampling_1_d} includes each point independently with its own probability, so we prove the statement for that model. Throughout, $\log$ denotes the natural logarithm.

\emph{Size.} Since $\E|S|=\sum_i q_i\leq g\sum_i u_i\leq gt$, a Chernoff bound gives $\Pr[|S|\geq 2gt]\leq e^{-gt/3}\leq \delta$, where the last step uses $gt\geq g\geq 3\log(1/\delta)$.

\emph{Reduction to ranges.} If ${\rm cost}_w(A, x)=0$, both sides vanish. Otherwise, points with $q_i=1$ contribute equally to both sides, and points with $u_i=0$ contribute to neither. On the remaining set $L$, we define $h_x(i):=w_i\,{\rm cost}(a_i, x)/(q_i\,{\rm cost}_w(A, x))\leq 1/g$ and $\mu(R):=\sum_{i\in R}q_i$. Let $R_x(r):=\{i\in L: h_x(i)>r\}$, which is empty for $r\geq 1/g$, and let $f_x(i):=w_i\,{\rm cost}(a_i, x)/{\rm cost}_w(A, x)=q_ih_x(i)$, so $\sum_{i\in L}f_x(i)\leq 1$. By the layer-cake formula, we have $\sum_{i\in S\cap L}h_x(i)=\int_0^{1/g}|S\cap R_x(r)|\,\d r$ and $\sum_{i\in L}q_ih_x(i)=\int_0^{1/g}\mu(R_x(r))\,\d r$. Therefore, it suffices to show that $||S\cap R|-\mu(R)|\leq \frac{\epsilon}{2}\max\{\mu(R), g\}$ for every level set $R=R_x(r)$, since integrating this bound shows that the error is at most $\frac{\epsilon}{2}(\sum_{i\in L}f_x(i)+g\cdot\frac1g)\leq\epsilon$. By the hypothesis on ${\rm dim}$, these sets have at most $|M|^{\rm dim}$ traces on every $M\subseteq L$ with $|M|\geq 2$.

\emph{Symmetrization.} Let $S'$ be an independent copy of $S$ and let $m_R:=\max\{\mu(R), g\}$. Since ${\rm Var}|S'\cap R|\leq\mu(R)$, Chebyshev's inequality gives $\Pr[||S'\cap R|-\mu(R)|\leq\frac{\epsilon}{4}m_R]\geq 1-16/(\epsilon^2g)\geq 1/2$ once $\epsilon^2g\geq 32$. The first violating $R$ in a fixed enumeration is a function of $S$ alone, so it suffices to bound $|S\cap R|-|S'\cap R|$ on the event that $S'$ is well behaved, at the price of a factor $2$. Conditionally on $S\cup S'$ and $S\,\triangle\, S'$, the differences of the indicators on $S\,\triangle\, S'$ are independent uniform signs. Let $U:=S\cup S'$, $M:=S\,\triangle\,S'$, and $c_R:=|S\cap R|+|S'\cap R|\geq |M\cap R|$. Both $c_R$ and $M\cap R$ depend on $R$ only through the trace $R\cap U$. First, by Markov's inequality, $\Pr[|S'\cap R|\leq 2m_R]\geq 1/2$ for each $R$. Therefore, as above, if some level set has $|S\cap R|>4m_R$, then with probability at least $1/2$ some $R$ has $|S\cap R|>4m_R\geq 4g$ and $|S'\cap R|\leq 2m_R$. Such an $R$ has $c_R>4g$ and $\sum_{i\in M\cap R}\sigma_i=|S\cap R|-|S'\cap R|>|S\cap R|/2\geq c_R/3$, so Hoeffding's inequality gives conditional probability at most $e^{-c_R/18}\leq e^{-2g/9}\leq e^{-\epsilon^2 g/192}$ per trace. Hence, with $N\leq (4gt)^{\rm dim}$ traces and the condition displayed below, every level set has $|S\cap R|\leq 4m_R$ except with probability $2(\delta/16+\delta/32)=3\delta/16$, where the term $\delta/16$ accounts for $\Pr[|S|>2gt\text{ or }|S'|>2gt]\leq 2e^{-gt/3}$. Second, suppose this event holds. Then a level set $R$ with $||S\cap R|-\mu(R)|>\frac{\epsilon}{2}m_R$ and $||S'\cap R|-\mu(R)|\leq \frac{\epsilon}{4}m_R$ has $|S'\cap R|\leq 2m_R$, and hence $|M\cap R|\leq c_R\leq 6m_R$ and $|\sum_{i\in M\cap R}\sigma_i|>\frac{\epsilon}{4}m_R$. Since $\mu(R)$ is not determined by the trace $T=R\cap U$, we instead define $m^\star_T:=\max\{g, |M\cap T|/6\}$. Such an $R$ has $m_R\geq m^\star_T$, so $|\sum_{i\in M\cap T}\sigma_i|>\frac{\epsilon}{4}m^\star_T$ with $|M\cap T|\leq 6m^\star_T$, and Hoeffding's inequality gives conditional probability at most $2e^{-\epsilon^2 m^\star_T/192}\leq 2e^{-\epsilon^2 g/192}$ per trace. If $|U|\leq 1$, then the sign sums are at most $1$ in absolute value while both thresholds exceed $1$, so neither event occurs. There are at most $(4gt)^{\rm dim}$ traces when $|S|, |S'|\leq 2gt$, and a union bound over them gives $Ne^{-\epsilon^2 g/192}\leq \delta/32$. Hence, with the outer symmetrization factor $2$, the total failure probability is at most $2(3\delta/16+\delta/16+2\cdot\delta/32)=5\delta/8\leq\delta$ once
\begin{align*}
    \frac{\epsilon^2 g}{192} \geq & ~ {\rm dim}\cdot\log(4gt)+\log(32/\delta).
\end{align*}
Let $D:={\rm dim}\cdot\log(t/\epsilon)+\log(1/\delta)$, then $g=c_0\epsilon^{-2}D$. To check the condition, we expand $\log(4gt)$ and use $\log(a+b)\leq \log 2+\log^{+}a+\log^{+}b$ with $\log^{+}y:=\max\{0,\log y\}$, ${\rm dim}\leq t$, and $\ell\log^{+}b\leq \ell\log \ell+b$ for $\ell\geq 1$. A routine computation then shows that the right-hand side is at most $(\log(4c_0)/\log 2+12)\cdot D$, so the condition holds once $c_0\geq 192(\log(4c_0)/\log 2+12)$. Moreover, the same choice gives $\epsilon^2g\geq 32$ for the Chebyshev step, $g\geq 3\log(1/\delta)$ for the size bound, and $gt\geq 3\log(32/\delta)$ for the event $|S|, |S'|\leq 2gt$.
\end{proof}

\begin{theorem}
\label{thm:sensitivity}
Let $A=\{a_1,\ldots,a_n\}\subset \R^d$ and let ${\rm cost}: \R^d\times X\rightarrow \R_{\geq 0}$. Moreover, suppose the total sensitivity has a finite upper bound, i.e., there exists some ${\rm sum}(s)<\infty$ such that for any finite subset $C\subset \R^d$, $\sum_{i\in C} s_i(C, C)\leq {\rm sum}(s)$. Then, the sensitivity of $A$ with respect to $B$, $s(A, B)$ (Def.~\ref{def:sensitivity}) satisfies Assumption~\ref{assumption:recurse}.
\end{theorem}

\begin{proof}
We need to prove $s(A, B)$ satisfies the four items in Assumption~\ref{assumption:recurse}.
\begin{itemize}
    \item Consistent total weights: by assumption, we have that for any $S\subseteq [n]$, $\sum_{i\in S} s_i(A_S, A_S)\leq {\rm sum}(s)$ with ${\rm sum}(s)$ being finite.
    \item Uniform sampling bound: we analyze by cases. For the first case, where $a_i\in A'$, we have $w'_i(A, A')=s_i(A, A')$. Let $x_1, x_2$ be the two points that realize $s_i(A, A')$ and $s_i(A, A)$, respectively. Suppose ${\rm cost}(A', x_2)\neq 0$, then
    \begin{align*}
        \frac{{\rm cost}(a_i, x_1)}{{\rm cost}(A', x_1)} \geq & ~ \frac{{\rm cost}(a_i, x_2)}{{\rm cost}(A', x_2)} \\
        \geq & ~ \frac{{\rm cost}(a_i, x_2)}{{\rm cost}(A', x_2)+{\rm cost}(A\setminus A', x_2)} \\
        = & ~ \frac{{\rm cost}(a_i, x_2)}{{\rm cost}(A, x_2)}
    \end{align*}
    where we use the fact that ${\rm cost}$ is nonnegative, therefore increasing the denominator will only decrease the fraction. On the other hand, if ${\rm cost}(A', x_2)=0$, then it must be that ${\rm cost}(a_i, x_2)=0$ due to the nonnegativity of cost. Hence, $s_i(A, A)=0$, and consequently $s_i(A, A')=0$ as otherwise we could pick $x_1$ for $s_i(A, A)$.

    For the next case, where $a_i\not \in A'$, we have $w'_i(A, A')=s_i(A, A'\cup \{a_i\})$. Again, let $x_1, x_2$ be the two points that realize $s_i(A, A'\cup \{a_i\})$ and $s_i(A, A)$. The argument is similar: suppose ${\rm cost}(A', x_2)\neq 0$, then
    \begin{align*}
        \frac{{\rm cost}(a_i, x_1)}{{\rm cost}(A', x_1)+{\rm cost}(a_i, x_1)} \geq & ~ \frac{{\rm cost}(a_i, x_2)}{{\rm cost}(A', x_2)+{\rm cost}(a_i, x_2)} \\
        \geq & ~ \frac{{\rm cost}(a_i, x_2)}{{\rm cost}(A', x_2)+{\rm cost}(a_i, x_2)+{\rm cost}(A\setminus (A'\cup \{a_i\}), x_2)} \\
        = & ~ \frac{{\rm cost}(a_i, x_2)}{{\rm cost}(A, x_2)}.
\end{align*}
If ${\rm cost}(A', x_2)=0$ and ${\rm cost}(a_i, x_2)\neq 0$, then ${\rm cost}(A'\cup\{a_i\}, x_2)={\rm cost}(a_i, x_2)\neq 0$, so $x_2$ is admissible in the definition of $s_i(A, A'\cup\{a_i\})$ and attains the ratio ${\rm cost}(a_i, x_2)/({\rm cost}(A', x_2)+{\rm cost}(a_i, x_2))=1$. Since every sensitivity of a point of the reference set is at most $1$, this gives $s_i(A, A'\cup\{a_i\})=1\geq s_i(A, A)$. If instead ${\rm cost}(a_i, x_2)=0$, then $s_i(A, A)={\rm cost}(a_i, x_2)/{\rm cost}(A, x_2)=0\leq s_i(A, A'\cup\{a_i\})$. 
    \item Importance sampling bound: let ${\rm dim}$ be the dimension of $(A, X, {\rm cost})$ of Definition~\ref{def:dim}, which is the quantity required by Lemma~\ref{lem:importance_sample}. Note that ${\rm dim}\leq n$, since a set of $m\geq 2$ points has at most $2^m\leq m^n$ subsets and we assume $n\geq 2$. Let $u\in \R^n$ satisfy $u_i\geq s_i(A, A)$ for all $i$, and let $\ov u_i:=\min\{1, u_i\}$, which still satisfies $\ov u_i\geq s_i(A, A)$ since $s_i(A, A)\leq 1$ for every $i$. Moreover, $\sum_i \ov u_i\leq n$, and $\min\{1, g\cdot u_i\}=\min\{1, g\cdot \ov u_i\}$ for every $g\geq 1$. We apply Lemma~\ref{lem:importance_sample} to $\ov u$ with unit weights and $t:=n\geq \max\{\sum_i\ov u_i, {\rm dim}, 2\}$. Hence, with
\begin{align*}
    g(\epsilon, \delta, n, d) = & ~ c_0\epsilon^{-2}\cdot ({\rm dim}\cdot \log(n/\epsilon)+\log(1/\delta))\geq 1
\end{align*}
and $q_i=\min\{1, g(\epsilon, \delta, n, d)\cdot u_i\}=\min\{1, g(\epsilon, \delta, n, d)\cdot \ov u_i\}$, the weighted sample is an $\epsilon$-coreset of $A$ with probability at least $1-\delta$, and its expected size is at most $g(\epsilon, \delta, n, d)\cdot\|\ov u\|_1\leq g(\epsilon, \delta, n, d)\cdot\|u\|_1$. This verifies the importance sampling bound for every overestimate $u$, as required by Assumption~\ref{assumption:recurse}.
    \item Coreset preserves weights: let $B$ be an $\epsilon$-coreset of $A$. Then, we know that for any $x\in X$, ${\rm cost}(B, x)=(1\pm\epsilon)\cdot {\rm cost}(A, x)$. Now, let $C\subset \R^d$ be any fixed set of points, and let $x_1, x_2\in X$ be the points that achieve $s_i(C, A)$ and $s_i(C, B)$. We have:
    \begin{align*}
        w_i(C, B) = & ~ s_i(C, B) \\
        = & ~ \frac{{\rm cost}(c_i, x_2)}{{\rm cost}(B, x_2)} \\
        \leq & ~ (1-\epsilon)^{-1}\cdot \frac{{\rm cost}(c_i, x_2)}{{\rm cost}(A, x_2)} \\
        \leq & ~ (1-\epsilon)^{-1}\cdot \frac{\cost(c_i, x_1)}{{\rm cost}(A, x_1)} \\
        = & ~ (1-\epsilon)^{-1}\cdot s_i(C, A),
    \end{align*}
    where we use ${\rm cost}(B, x_2)\geq (1-\epsilon)\cdot {\rm cost}(A, x_2)$ in the third step. We could similarly establish that $s_i(C, B)\geq (1+\epsilon)^{-1}\cdot s_i(C, A)$. This proves the assertion. \qedhere
\end{itemize}
\end{proof}

In what follows, we demonstrate how to concretely implement sensitivity sampling for various cost functions, such as $\ell_p$ sensitivity and $k$-subspace sensitivity.

\subsection{\texorpdfstring{$\ell_2$}{l2} Sensitivity and Leverage Score}

Let $X=\R^d$ and ${\rm cost}(a_i, x)=(a_i^\top x)^2$. In this case, the $\ell_2$ sensitivity defined as
\begin{align*}
    s_i(A, B) = & ~ \max_{x\in \R^d, Bx\neq 0} \frac{(a_i^\top x)^2}{\|Bx\|_2^2}
\end{align*}
is, in fact, the leverage score $\tau_i(A)$. The leverage score has many favorable structures: for example, to obtain an $\epsilon$-coreset, it is sufficient to sample $O(\epsilon^{-2}d\log d)$ points, and one could sample with $w_i(A, A')$ instead of $w'_i(A, A')$.

\begin{algorithm}
\caption{Classical oracle for leverage score.}
\label{alg:classical_ls}
\begin{algorithmic}[1]
\State {\bf data structure} \textsc{LeverageScore}
\State {\bf members}
\State \hspace{4mm} $A\in \R^{n\times d}$
\State \hspace{4mm} $C\in \R^{s\times d}$
\State \hspace{4mm} $M\in \R^{O(\log n)\times d}$
\State \hspace{4mm} $N\in \R^{O(\log n)\times d}$
\State {\bf end members}
\State
\Procedure{Preprocess}{$A\in \R^{n\times d},  C\in \R^{s\times d}$}
\State $c\gets 1000$
\State Compute the compact SVD of $C$: $C=U\Sigma V^\top$ \Comment{$r={\rm rank}(C)\leq s$, $\Sigma\in \R^{r\times r}$ invertible, $V\in \R^{d\times r}$}
\State Let $G\in \R^{c\log n\times r}$ and $G^\prime\in \R^{c\log n\times d}$ be random Gaussian matrices
\State $M\gets G\Sigma^{-1}V^\top$ \Comment{$M\in \R^{c\log n\times d}$}
\State $N\gets G^\prime(I-VV^\top)$ \Comment{$N\in \R^{c\log n\times d}$, sketch of the orthogonal complement of ${\rm span}(C^\top C)$}
\EndProcedure
\State
\Procedure{Query}{$i\in [n]$}
\State {\bf if} $Na_i\neq 0$ {\bf then return} $\infty$ \Comment{$a_i\notin {\rm span}(C^\top C)$, so $q_i=1$}
\State \Return $\| Ma_i \|_2^2$
\EndProcedure
\State {\bf end data structure}
\end{algorithmic}
\end{algorithm}

\begin{lemma}
\label{lem:classical_ls}
Let $A=\{a_1,\ldots, a_n\}\subset \R^d$ and define ${\rm cost}: \R^d\times \R^d\rightarrow \R_{\geq 0}$ by ${\rm cost}(a_i, x)=(a_i^\top x)^2$, and let $w(A, B)$ be defined as
\begin{align*}
    w_i(A, B) = & ~ \begin{cases}
        a_i^\top (B^\top B)^\dagger a_i, & \text{if $a_i\in {\rm span}(B^\top B)$}, \\
        \infty, & \text{otherwise}.
    \end{cases}
\end{align*}
Then, the weights $w$ satisfy Assumption~\ref{assumption:recurse}. Moreover, there exists a randomized algorithm (Algorithm~\ref{alg:classical_ls}) that implements \textsc{Preprocess} and \textsc{Query} procedures, with
\begin{itemize}
    \item ${\cal T}_{\rm prep}(s, d) = \wt O(sd^{\omega-1})$.
    \item ${\cal T}_{\rm query}(s, d) = \wt O(d)$.
\end{itemize}
\end{lemma}

\begin{proof}
While leverage score is $\ell_2$ sensitivity and we could directly use Theorem~\ref{thm:sensitivity}, we include a proof that utilizes the structure of leverage score for completeness.
\begin{itemize}
    \item Consistent total weights: first note that 
    \begin{align*}
        \sum_{i=1}^n w_i(A, A) = & ~ \sum_{i=1}^n a_i^\top (A^\top A)^\dagger a_i \\
        = & ~ \tr[(A^\top A)^\dagger A^\top A] \\
        = & ~ {\rm rank}(A) \\
        \leq & ~ d
    \end{align*}
    hence we have ${\rm sum}(w)=d$. Let $S\subset [n]$ with $|S|\geq d$, then
    \begin{align*}
        \sum_{i\in S} w_i(A_S, A_S) = & ~ \sum_{i\in S} a_i^\top (A_S^\top A_S)^\dagger a_i \\
        = & ~ \tr[(A_S^\top A_S)^\dagger(A_S^\top A_S)] \\
        = & ~ {\rm rank}(A_S) \\
        \leq & ~ d.
    \end{align*}
    \item Uniform sampling bound: the proof closely follows that of Theorem 1 of~\cite{clm+15}, and we analyze by cases. Let $S$ be an indicator matrix with $A'=SA$ and let $S^{(i)}$ be the indicator matrix for $S\cup \{i\}$. We will show that $w'_i(A, A')=w_i(A, S^{(i)}A)$ via case analysis. If $a_i\in A'$, then $w'_i(A, A')=w_i(A, A')$ and $S=S^{(i)}$, consequently $w_i(A, SA)=w_i(A, S^{(i)}A)$. If $a_i\not\in A'$, then $w'_i(A, A')=w_i(A, A'\cup \{i\})=w_i(A, S^{(i)}A)$. This completes the proof. To show the overestimate, observe that $S^{(i)}$ is an indicator matrix for the sample and thus $S^{(i)}\preceq I_n$, we can then conclude
    \begin{align*}
        A^\top (S^{(i)} )^2 A \preceq & ~ A^\top A
    \end{align*}
    and
    \begin{align*}
        w'_i(A, A') = & ~ a_i^\top (A^\top (S^{(i)})^2 A)^\dagger a_i \\
        \geq & ~ a_i^\top (A^\top A)^\dagger a_i \\
        = & ~ w_i(A, A).
    \end{align*}
    \item Importance sampling bound: this is standard as $w_i(A, A)$ is the leverage score of matrix $A$. Let $g(\epsilon, \delta, n, d):=c\epsilon^{-2}\log(d/\delta)$ for a sufficiently large constant $c$. We include each row independently with probability $q_i=\min\{1, g\cdot u_i\}$ for overestimates $u_i\geq \tau_i(A)$ and reweight it by $1/q_i$. By the matrix Chernoff bound, with probability at least $1-\delta$ this gives $(1\pm\epsilon)A^\top A$ in the Loewner order, hence an $\epsilon$-coreset, and the expected size of the sample is $g\|u\|_1$. Lemma 4 of~\cite{clm+15} is the same statement with $\log d$ in place of $\log(d/\delta)$ and failure probability $d^{-\Omega(1)}$.
    \item Coreset preserves weights: because $B$ is an $\epsilon$-coreset of $A$, we know that for any $x\in \R^d$, $\|Bx\|_2^2 = (1\pm\epsilon) \cdot \|Ax\|_2^2$. This implies that $B^\top B$ is a spectral approximation to $A^\top A$ and ${\rm ker}(A)={\rm ker}(B)$. By the monotonicity of the pseudo-inverse on the common range, the same holds for $(B^\top B)^\dagger$ with respect to $(A^\top A)^\dagger$ with the factors $(1+\epsilon)^{-1}$ and $(1-\epsilon)^{-1}$. Let $C\subset \R^d$ be any fixed subset of $\R^d$. We conclude the proof by spectral approximation:
    \begin{align*}
       (1+\epsilon)^{-1}\cdot  c_i^\top (A^\top A)^\dagger c_i \leq c_i^\top (B^\top B)^\dagger c_i\leq (1-\epsilon)^{-1}\cdot  c_i^\top (A^\top A)^\dagger c_i.
    \end{align*}
\end{itemize}

Now, we turn to the runtime analysis of Algorithm~\ref{alg:classical_ls}. Let $C=U\Sigma V^\top$ be the compact SVD of $C$, with $r={\rm rank}(C)$, $\Sigma\in \R^{r\times r}$ invertible, and $V\in \R^{d\times r}$. Then we have $(C^\top C)^\dagger=(V\Sigma^2 V^\top)^\dagger = V \Sigma^{-2} V^\top$. By definition, for $a_i\in {\rm span}(V)$,
\begin{align*}
    w_i(A, C) = & ~ a_i^\top (C^\top C)^\dagger a_i \\
    = & ~ a_i^\top V \Sigma^{-2} V^\top a_i \\
    = & ~ \|\Sigma^{-1} V^\top a_i\|_2^2,
\end{align*}
and by a standard Johnson-Lindenstrauss trick~\citep{ss11}, it is sufficient to prepare the matrix $G \Sigma^{-1} V^\top$ for a JL matrix $G$. Then, with high probability, all $w_i(A, C)$ are approximated within a factor of $1\pm\epsilon$, and a simple scaling turns these estimates into overestimates for all $a_i\in {\rm span}(C^\top C)={\rm span}(V)$. By definition, the points outside this span have weight $\infty$, and thus $q_i=\min\{1, w_i\}=1$. This is the convention of~\cite{clm+15} under which Lemma~\ref{lem:clm_thm4} is stated, so its size bound accounts for these points. For such $a_i$ we have $(I-VV^\top)a_i\neq 0$, and for a Gaussian $G^\prime$ the vector $Na_i=G^\prime(I-VV^\top)a_i$ is nonzero with probability $1$, while every $a_i\in {\rm span}(V)$ satisfies $Na_i=0$ exactly. Therefore, the test in \textsc{Query} returns $\infty$ on the points outside the span and only on them. We assume that this exact zero test on $O(\log n)$ numbers can be performed, and with rational inputs we fingerprint $a_i$ against a rational basis of the orthogonal complement of the row space of $C$, computed by Gaussian elimination at the same $O(sd^{\omega-1})$ cost. It remains to analyze the runtime. 
\begin{itemize}
    \item Computing the compact SVD of $C$ takes $O(sd^{\omega-1})$ time.
    \item Computing $G\Sigma^{-1}$ takes $\wt O(s\log n)$ time and then we multiply $G\Sigma^{-1}$ with $V^\top$ which takes $\wt O(sd)$ time. Computing $N=G^\prime-(G^\prime V)V^\top$ takes $\wt O(rd)\leq \wt O(sd)$ time.
    \item For query, note that $M, N\in \R^{O(\log n)\times d}$, and thus testing $Na_i\neq 0$ and computing $\|Ma_i\|_2^2$ takes $\wt O(d)$ time.
\end{itemize}
This completes the proof of the assertion.
\end{proof}

\begin{remark}
If we faithfully execute the framework of Theorem~\ref{thm:sensitivity}, then we would need to compute the $w_i(A, C\cup \{a_i\})$ instead of $w_i(A, C)$. Instead, we only need to sample with $w_i(A, C)$, which is a key feature for leverage score and related notions. We take the same route for Lewis weights in Theorem~\ref{thm:lewis} and for ridge leverage scores in Theorem~\ref{thm:ridge}. We run the recursion of Algorithm~\ref{alg:quantum_iterate} with the one-sided values $w_i(A_t, C_{t-1})$ in place of $w'_i(A_t, A_{t-1}; C_{t-1})$. The coreset guarantee and the level sizes for this sampling rule are proved in Lemmas~\ref{lem:clm_thm4}, \ref{lem:lewis_level}, and~\ref{thm:ridge_via_unif}. We use Theorem~\ref{thm:general} only for its time and query accounting, and thus the oracles of Lemmas~\ref{lem:classical_ls}, \ref{lem:classical_lewis}, and~\ref{lem:classical_ridge} only need to evaluate $w_i(Z, Y)$. Nevertheless, for leverage scores the augmented value is available at no extra cost by the Sherman--Morrison formula of the next remark.
\end{remark}

\begin{lemma}[Theorem 4 of~\cite{clm+15}]
\label{lem:clm_thm4}
Let $A=\{a_1,\ldots,a_n\}\subset \R^d$, $m\in [n]$ and $\epsilon,\delta\in (0, 1)$. Form $SA$ by including each point of $A$ independently with probability $\frac{m}{n}$, and let $w$ be as in Lemma~\ref{lem:classical_ls}. Let $q_i=\min\{1, c\epsilon^{-2}\log(d/\delta)\cdot w_i(A, SA)\}$ for a sufficiently large constant $c$, and let $B$ contain each $a_i$ independently with probability $q_i$, weighted by $1/q_i$. Then $\E[\sum_{i=1}^n q_i]=O(\epsilon^{-2}\frac{nd}{m}\log(d/\delta))$, and with probability at least $1-\delta$, $B$ is an $\epsilon$-coreset of $A$, i.e., $(1-\epsilon)A^\top A\preceq B^\top B\preceq (1+\epsilon)A^\top A$.
\end{lemma}

\begin{remark}The statement combines two facts. First, by the monotonicity of the pseudo-inverse as in the proof of Lemma~\ref{lem:classical_ls}, we have $w_i(A, SA)\geq w_i(A, A)=\tau_i(A)$ for every $i$, and thus $q_i\geq \min\{1, c\epsilon^{-2}\log(d/\delta)\cdot\tau_i(A)\}$. Then, for every realization of $SA$, the spectral approximation follows from the matrix Chernoff bound for leverage score sampling with independent inclusions. This is the bound behind Lemma 4 of~\cite{clm+15}, and here its oversampling factor $c\epsilon^{-2}\log(d/\delta)$ replaces the $\log d$ and failure probability $d^{-\Omega(1)}$ printed there. Second, we have $\E[\sum_i\min\{1, w_i(A, SA)\}]=O(nd/m)$. Indeed, for $a_i\in {\rm span}(SA)$ the Sherman--Morrison formula gives $w_i(A, SA\cup\{a_i\})=w_i(A, SA)/(1+w_i(A, SA))$, and $w_i(A, SA\cup\{a_i\})=1$ otherwise, so in both cases $\min\{1, w_i(A, SA)\}\leq 2\,w'_i(A, SA)$. Moreover, $\E\|w'(A, SA)\|_1\leq 2\frac{n}{m}\cdot d$ by the exchange argument of the proof of Theorem~\ref{thm:general}, which applies to independent inclusions as well as to a uniformly random subset of exactly $m$ points.\end{remark}

Setting $m=n/2$, Lemma~\ref{lem:clm_thm4} itself is sufficient to prove Theorem~\ref{thm:general}, without resorting to $w_i(A, C\cup \{a_i\})$. The factor $2$ above is absorbed by the constant $c$. Our following theorem recovers the main result of~\cite{ag23}.

\begin{algorithm}
\caption{Quantum leverage score sampling with a size parameter.}
\label{alg:qls}
\begin{algorithmic}[1]
\Procedure{QLS}{$A\in \R^{n\times d}, s, \delta$}
\State $s'\gets c\cdot d\log(d\lceil\log_2 n\rceil/\delta)$ \Comment{$c$ a sufficiently large constant, and $s'=\wt O(d)$}
\State $(C_T, D_T)\gets \textsc{QIterateSample}(A, 0.01, \delta/4)$\label{ln:qls:rec} with the oracle of Algorithm~\ref{alg:classical_ls}, sampling with $w_i(A_t, C_{t-1})$ in place of $w'_i$ (Lemma~\ref{lem:clm_thm4}), and $O(s')$ points per level in expectation \Comment{Algorithm~\ref{alg:quantum_iterate}}
\State $\textsc{LeverageScore.Preprocess}(A, C_T)$ \Comment{Algorithm~\ref{alg:classical_ls} on $C_T$, ${\cal T}_{\rm prep}(s', d)=\wt O(d^\omega)$ time}
\State $\wt \tau_i\gets \textsc{LeverageScore.Query}(i)/((1-0.01)(1-\epsilon_{\rm JL}))$ \Comment{evaluated on demand, and $\tau_i(A)\leq \wt\tau_i\leq O(1)\cdot \tau_i(A)$}
\State $q_i\gets \min\{1, s\cdot \wt\tau_i/d\}$ \Comment{$\sum_{i=1}^n q_i\leq s\sum_i\wt\tau_i/d=O(s)$}
\State $D\gets \textsc{QSample}(q, \delta/(8(T+2)))$\label{ln:qls:D} \Comment{Lemma~\ref{lem:q_sampling_1_d}}
\State \Return $(D^\top A, D)$
\EndProcedure
\end{algorithmic}
\end{algorithm}
\noindent In Algorithm~\ref{alg:qls}, $T$ is the number of levels of the recursion in line~\ref{ln:qls:rec}.

\begin{theorem}
\label{thm:ls}
Let $A=\{a_1,\ldots,a_n\}\subset \R^d$ and $\epsilon, \delta\in (0, 1)$. Then, there exists a quantum algorithm that with probability $1-\delta$, constructs an $\epsilon$-coreset $B$ of $A$ of size $O(\epsilon^{-2}d\log(d/\delta))$, in expected time $\wt O(\epsilon^{-1} n^{0.5} d^{1.5} + d^\omega)$ and $\wt O(\epsilon^{-1}n^{0.5}d^{0.5})$ queries to points in $A$.

Furthermore, for a size parameter $s\geq d$, the procedure $\textsc{QLS}(A, s, \delta)$ of Algorithm~\ref{alg:qls} with probability at least $1-\delta$ computes overestimates $\wt\tau_i$ of the leverage scores with $\tau_i(A)\leq \wt\tau_i\leq O(1)\cdot\tau_i(A)$ for every $i\in [n]$, hence $\sum_{i=1}^n\wt\tau_i=O({\rm rank}(A))$, and returns a sample in which point $i$ is included independently with probability $q_i=\min\{1, s\cdot \wt\tau_i/d\}$ and carries the weight $1/q_i$. The sample has expected size $\sum_i q_i=O(s)$, and size $O(s+\log(1/\delta))$ with probability at least $1-\delta$. The procedure runs in expected time $\wt O(n^{0.5} s^{0.5} d + d^{\omega})$ and makes $\wt O(n^{0.5}s^{0.5})$ queries to points in $A$, the expectations being taken as in Theorem~\ref{thm:general}.
\end{theorem}

\begin{proof}

We first prove the statement about $\textsc{QLS}$ and then obtain the $\epsilon$-coreset by a choice of $s$. Let $T=O(\log n)$ be the number of levels of the constant-accuracy recursion and let $s'=O(d\log(dT/\delta))=\wt O(d)$ be its expected number of points per level.

\emph{Step 1: the constant-accuracy recursion.} We run $\textsc{QIterateSample}(A, 0.01, \delta/4)$ with the oracle of Algorithm~\ref{alg:classical_ls}. At level $t$, we sample with $q_i=\min\{1, c\,g\cdot w_i(A_t, C_{t-1})\}$ in place of the augmented weights, where $g=\Theta(\log(dT/\delta))$ is the oversampling factor of Lemma~\ref{lem:clm_thm4} at accuracy $0.01$ and failure probability $\delta/(8T)$. The per-level failure probability $\delta/(8\lceil\log_2 n\rceil)$ of the call is at most this. For the weights of Lemma~\ref{lem:classical_ls}, items 1 and 4 of Assumption~\ref{assumption:recurse} are proved there, and we replace items 2 and 3 by Lemma~\ref{lem:clm_thm4} with $m=n/2$, applied to the Bernoulli$(1/2)$ half $A_{t-1}$ of $A_t$. At level $t$, the oracle evaluates $w_i(A_t, C_{t-1})$. If $C_{t-1}$ is a $0.01$-coreset of $A_{t-1}$, then by item 4 this weight is within a constant factor of $w_i(A_t, A_{t-1})$, and the two are infinite for the same $i$ since ${\rm span}(C_{t-1})={\rm span}(A_{t-1})$. Hence, in this case, the sampling probabilities dominate those of Lemma~\ref{lem:clm_thm4} and the level produces a $0.01$-coreset $C_t$ of $A_t$ with probability at least $1-\delta/(8T)$. Moreover, $\sum_{i\in A_t}q_i$ then has expectation $O(s')$ over the uniform half $A_{t-1}$ by the second fact stated after Lemma~\ref{lem:clm_thm4}. Each oracle is prepared with a Gaussian sketch of $O(\log(n/\delta))$ rows, whose distortion bound holds for all rows of $A_t$ with probability at least $1-\delta/(8T)$. The $T+1$ calls of $\textsc{QSample}$ are run with failure probability $\delta/(8(T+2))$ each, in place of the parameter $\delta'/(T+1)$ printed in Algorithm~\ref{alg:quantum_iterate}, and either value closes the union bound below. By a union bound over the $T$ levels, the $T$ sketches, and the $T+1$ sampler calls, with probability at least $1-3\delta/8-\delta/8= 1-\delta/2$, every level succeeds and $C_T$ is a $0.01$-coreset of $A$. Conditioned on this event, every intermediate sample $C_t$, and hence every set on which an oracle is prepared, has expected size $O(s')$ as in Theorem~\ref{thm:general}. By the analysis of Theorem~\ref{thm:general} and Lemma~\ref{lem:classical_ls}, this step takes $O(\log n)\cdot{\cal T}_{\rm prep}(s', d)+\wt O(\sqrt{ns'}\cdot{\cal T}_{\rm query}(s', d))=\wt O(d^\omega+n^{0.5}d^{1.5})$ expected time and $\wt O(\sqrt{ns'})=\wt O(n^{0.5}d^{0.5})$ queries.

\emph{Step 2: the additional oracle.} Since $C_T$ is a $0.01$-coreset of $A$, we have $0.99\,A^\top A\preceq C_T^\top C_T\preceq 1.01\,A^\top A$, and thus ${\rm ker}(C_T)={\rm ker}(A)$. Hence, by the monotonicity of the pseudo-inverse on the common range, $\tau_i(A)/1.01\leq w_i(A, C_T)\leq \tau_i(A)/0.99$ for every $i$. This is item 4 of Assumption~\ref{assumption:recurse} for the weights of Lemma~\ref{lem:classical_ls}, with $B=C_T$ and $C=A$. When prepared on $C_T$, the procedure \textsc{Query} of Algorithm~\ref{alg:classical_ls} returns $(1\pm\epsilon_{\rm JL})\,w_i(A, C_T)$ for all $i$ simultaneously with probability $1-n^{-\Omega(1)}$. Here $\epsilon_{\rm JL}\leq 1/5$ is the distortion of its Gaussian sketch with $O(\log n)$ rows, and for $\delta<n^{-\Omega(1)}$ we take $O(\log(n/\delta))$ rows. Hence, $\wt\tau_i:=\textsc{Query}(i)/(0.99(1-\epsilon_{\rm JL}))$ satisfies
\begin{align*}
    \tau_i(A) \leq & ~ \wt\tau_i \leq \frac{1+\epsilon_{\rm JL}}{0.99^2(1-\epsilon_{\rm JL})}\cdot\tau_i(A) \leq 2\tau_i(A) \qquad \text{for all $i\in [n]$},
\end{align*}
and $\sum_i\wt\tau_i\leq 2\,{\rm rank}(A)\leq 2d$. Consequently, $\sum_i q_i\leq s\sum_i\wt\tau_i/d\leq 2s$, and by a Chernoff bound, $|D|\leq 2\max\{2s, 3\log(4/\delta)\}$ with probability $1-\delta/4$. Preparing this oracle takes ${\cal T}_{\rm prep}(s', d)=\wt O(s'd^{\omega-1})=\wt O(d^\omega)$ time. We then sample using Lemma~\ref{lem:q_sampling_1_d} with failure probability $\delta/(8(T+2))$, which takes $\wt O(\sqrt{ns})\cdot{\cal T}_{\rm query}(s', d)=\wt O(n^{0.5}s^{0.5}d)$ time and $\wt O(\sqrt{ns})$ queries. Altogether, $\textsc{QLS}$ succeeds with probability at least $1-\delta$, since it fails with probability $\delta/2$ in Step 1 and at most $\delta/4+\delta/8+\delta/8$ in the size bound, the sampler call, and the sketch of Step 2. It runs in expected time $\wt O(n^{0.5}d^{1.5}+d^\omega+n^{0.5}s^{0.5}d)=\wt O(n^{0.5}s^{0.5}d+d^\omega)$ and makes $\wt O(n^{0.5}(d^{0.5}+s^{0.5}))$ queries, which are the stated bounds for $s\geq d$.

\emph{Step 3: the $\epsilon$-coreset.} We run $\textsc{QLS}(A, s, \delta/2)$ with $s=C\epsilon^{-2}d\log(d/\delta)$ for a sufficiently large constant $C$, so $s\geq d$ and $q_i=\min\{1, s\wt\tau_i/d\}\geq\min\{1, C\epsilon^{-2}\log(d/\delta)\cdot\tau_i(A)\}$. We apply the matrix Chernoff bound for leverage score sampling with independent inclusions, Lemma 4 of~\cite{clm+15}, as in item 3 of the proof of Lemma~\ref{lem:classical_ls}. Therefore, $B:=D^\top A$ satisfies $(1-\epsilon)A^\top A\preceq B^\top B\preceq (1+\epsilon)A^\top A$ with probability at least $1-\delta/2$. By Step 2, $B$ has $O(s)=O(\epsilon^{-2}d\log(d/\delta))$ points, the expected time is $\wt O(n^{0.5}s^{0.5}d+d^\omega)=\wt O(\epsilon^{-1}n^{0.5}d^{1.5}+d^\omega)$, and the number of queries is $\wt O(n^{0.5}s^{0.5})=\wt O(\epsilon^{-1}n^{0.5}d^{0.5})$. \qedhere

\end{proof}

\subsection{\texorpdfstring{$\ell_p$}{lp} Sensitivity and Lewis Weights}

To preserve $\ell_p$ subspace, one could sample according to $\ell_p$ sensitivity: let us define ${\rm cost}(a_i, x)=|a_i^\top x|^p$ for $p\in (0, \infty)$, then the $\ell_p$ sensitivity is
\begin{align*}
    s_i(A, B) = & ~ \max_{x\in \R^d, Bx\neq 0} \frac{|a_i^\top x|^p}{\|Bx\|_p^p},
\end{align*}
and a computationally efficient proxy for $\ell_p$ sensitivity is $\ell_p$ Lewis weights, defined as the unique nonnegative weight vector $w_A\in \R^n$ with
\begin{align*}
    w_{A, i}^{2/p} = & ~ a_i^\top (A^\top W_A^{1-2/p}A)^{-1} a_i,
\end{align*}
where $W_A\in \R^{n\times n}$ is the diagonal matrix of $w_A$. Naturally, we define our weights as
\begin{align*}
    w_i(A, B) = & ~ (a_i^\top (B^\top W_B^{1-2/p}B)^{-1}a_i)^{p/2}.
\end{align*}
To implement recursive sampling according to Lewis weights, we need a stronger notion of approximation for $\epsilon$-coreset, as beyond sensitivity, the weights might not be preserved by an $\epsilon$-coreset. We explicitly define the notion of an \emph{$\epsilon$-approximator}, a weighted subset of points that preserves the Lewis quadratic form, and hence the weights. Note that an $\epsilon$-approximator is not necessarily an $\epsilon$-coreset.

\begin{definition}
\label{def:approximator}
Let $A=\{a_1,\ldots,a_n\}\subset \R^d$ and $\epsilon\in (0,1)$. We say a weighted subset $B$ of $A$ is an $\epsilon$-approximator of $A$ if
\begin{align*}
    (1-\epsilon)\cdot A^\top W_A^{1-2/p}A \preceq & ~ B^\top W_B^{1-2/p}B \preceq (1+\epsilon)\cdot A^\top W_A^{1-2/p}A,
\end{align*}
where $W_A$ and $W_B$ are the diagonal matrices of the $\ell_p$ Lewis weights of $A$ and of $B$.
\end{definition}

Note that the two-sided bound in Definition~\ref{def:approximator} forces ${\rm rowspan}(B)={\rm rowspan}(A)$. Moreover, by Lemma 2.6 of~\cite{wy23}, the pseudo-inverses satisfy $(1+\epsilon)^{-1}(A^\top W_A^{1-2/p}A)^\dagger\preceq (B^\top W_B^{1-2/p}B)^\dagger\preceq (1-\epsilon)^{-1}(A^\top W_A^{1-2/p}A)^\dagger$ on this row space. Hence, for any fixed $C$ and any $i$, we have $w_i(C, B)^{2/p}=(1\pm 2\epsilon)\cdot w_i(C, A)^{2/p}$ when $\epsilon\leq 1/2$. Raising both sides to the power $p/2$ gives
\begin{align*}
    w_i(C, B) = & ~ (1\pm 2p\epsilon)\cdot w_i(C, A) \qquad \text{whenever } \epsilon\leq \min\{1/4, 1/p\},
\end{align*}
where we use $(1+2\epsilon)^{p/2}\leq e^{p\epsilon}\leq 1+2p\epsilon$ for $p\epsilon\leq 1$ and $(1-2\epsilon)^{p/2}\geq 1-2p\epsilon$ for $\epsilon\leq 1/4$. Thus, an $\epsilon$-approximator preserves the weights up to a factor $1\pm O(p\epsilon)$, and an $O(1/p)$-approximator preserves them up to a constant factor. For Lewis weights, the analogue of Lemma~\ref{lem:clm_thm4} is the recursive algorithm of Lemma 6.2 of~\cite{cp15}. It computes an approximate inverse Lewis quadratic form of a uniform half $\wh A$ of $A$ recursively, and then samples the rows of $A$ with the relative weights given by this form.

For $p>2$, the relative weights $(a_i^\top (\wh A^\top W_{\wh A}^{1-2/p}\wh A)^{-1}a_i)^{p/2}$ computed from the half $\wh A$ are \emph{not} entrywise overestimates of the Lewis weights of $A$. In fact, they fall below them on small random instances. The following lemma isolates one level of this recursion in the form our algorithm uses, and its proof follows the proof of Lemma 6.2 of~\cite{cp15} step by step. In this lemma, the half is drawn by independent inclusions as in the chain of Algorithm~\ref{alg:quantum_iterate}, rather than as a subset of exactly $n/2$ rows. Moreover, the approximate inverse form is supplied by an approximator of the half, and the failure probability is made explicit. For $p<2$, we also use a modification that~\cite{cp15} note right after the proof of their Lemma 6.2. There, their Lemma 5.5 replaces their Lemma 5.6 for $p<2$, and their Lemma 5.3 replaces their Lemma 5.4 for $p<4$, so that for $p<2$ no polynomial factor in $d$ multiplies the weights. Our proof uses Lemma 5.5 in Step 1 and Lemma 5.3 in Steps 3 and 4 in these ranges. As a consequence, for $p<2$, the accuracy $\epsilon'$ of Step 2 needs no factor $1/\sqrt d$, and a level keeps $O_{p,\kappa}(\epsilon_1^{-2}d\log(d/\delta))$ rows in expectation. We write $M_A:=A^\top W_A^{1-2/p}A$ for the Lewis quadratic form of $A$. For a rank-deficient matrix, the Lewis weights are taken on its row space and $M_A^\dagger$ replaces $M_A^{-1}$.

\begin{lemma}[One level of the Lewis weight recursion]
\label{lem:lewis_level}
Let $p\in (0, \infty)$, let $A\in \R^{n\times d}$ have rows $a_1,\ldots,a_n$ (for a rank-deficient $A$, $d$ may be replaced by ${\rm rank}(A)$), and let $\epsilon_1, \delta\in (0, 1)$ and $\kappa\geq 1$. Let $\wh A\subset_{1/2}A$ consist of the rows of $A$ included independently with probability $1/2$, let $C$ be a $\frac14$-approximator of $\wh A$, and let $x_i:=(a_i^\top M_C^\dagger a_i)^{p/2}$ for $a_i\in {\rm rowspan}(C)$ and $x_i:=\infty$ otherwise. Let $\wh w_1,\ldots,\wh w_n$ satisfy $\kappa^{-1}x_i\leq \wh w_i\leq \kappa x_i$. Include each row $a_i$ independently with probability
\begin{align*}
    q_i = & ~ \min\{1, \beta\cdot \wh w_i\}, \qquad \beta:=c_{p,\kappa}\cdot \epsilon_1^{-2}\cdot d^{\theta_p}\cdot \log(d/\delta),
\end{align*}
scaled by $q_i^{-1/p}$, and let $A'$ be the resulting weighted subset. Here $\theta_p:=0$ for $p<2$ and $\theta_p:=p/2$ for $p\geq 2$. There is a constant $c_{p,\kappa}$ depending only on $p$ and $\kappa$ such that, for every outcome of $\wh A$ and $C$, with probability at least $1-\delta$ over the inclusions, $A'$ is an $\epsilon_1$-approximator of $A$, and moreover $\E[|A'|]=\E[\sum_{i=1}^n q_i]=O_{p,\kappa}(\epsilon_1^{-2}d^{\theta_p+1}\log(d/\delta))$, where the expectation is over $\wh A$ and the inclusions, for any $C$ and $\wh w$ satisfying the hypotheses (which may depend on $\wh A$).
\end{lemma}

\begin{proof}
If $A$ is rank-deficient, we replace $A$ by $AV$, with $V\in \R^{d\times r}$ as in the paragraph after Lemma~\ref{lem:flps_thm2}. Every form $M$ built from rows of $A$ satisfies $M=V(V^\top MV)V^\top$ and $M^\dagger=V(V^\top MV)^\dagger V^\top$. Hence the Lewis weights, the quantities $x_i$, $y_i$, $v_i$ below, and the approximator relation between $A'$ and $A$ are unchanged. Moreover, $d$ enters the argument only as an upper bound on the rank of matrices with rows in ${\rm rowspan}(A)$, and we therefore assume that $A$ has full column rank. Since $C$ is a $\frac14$-approximator of $\wh A$, we have ${\rm rowspan}(C)={\rm rowspan}(\wh A)$, and by Lemma 2.6 of~\cite{wy23}, $\frac45 M_{\wh A}^\dagger\preceq M_C^\dagger\preceq \frac43 M_{\wh A}^\dagger$ on this row space. Let $y_i:=(a_i^\top M_{\wh A}^\dagger a_i)^{p/2}$, and $y_i=\infty$ for $a_i\notin{\rm rowspan}(\wh A)$. Then $\kappa_1^{-1}y_i\leq \wh w_i\leq \kappa_1 y_i$ for $\kappa_1:=\kappa(4/3)^{p/2}$. We write $w_i$ for $w_{A,i}$ throughout, and all inverses of forms of $\wh A$ and of its extensions by one row are taken on ${\rm rowspan}(\wh A)$.

\emph{Step 1: the relative weights dominate the Lewis weights of $A$ up to a factor $d^{(p/2-1)\lor 0}$.} If $p\leq 2$, the monotonicity of Lewis weights in Lemma 5.5 of~\cite{cp15} gives $M_A\succeq M_{\wh A}$, hence $y_i\geq w_i$ for every $i$. If $p>2$ and $\wh A$ has full column rank, Lemma 5.6 of~\cite{cp15} gives $M_{\wh A}\preceq d^{1-2/p}M_A$. For $p>2$ in general, Lemmas 2.3 and 2.4 of~\cite{wy23} give the same bound by using $\|w_{\wh A}\|_1={\rm rank}(\wh A)\leq d$ and $\wh A\subseteq A$, hence $y_i\geq d^{-(p/2-1)}w_i$. In both cases, $y_i\geq d^{-((p/2-1)\lor 0)}w_i$. Since $d^{\theta_p}\cdot d^{-((p/2-1)\lor 0)}$ equals $1$ for $p<2$ and $d$ for $p\geq 2$, we obtain
\begin{align*}
    q_i \geq & ~ \min\{1, \gamma\cdot w_i\}, \qquad \gamma:=c_{p,\kappa}\kappa_1^{-1}\cdot \epsilon_1^{-2}\cdot d^{\mathbf{1}[p\geq 2]}\cdot \log(d/\delta).
\end{align*}
Rows outside ${\rm rowspan}(\wh A)$ have $q_i=1$.

\emph{Step 2: the carried weights are almost Lewis weights of $A'$.} As in~\cite{cp15}, we give the sampled copy $a'_j=q_i^{-1/p}a_i$ of row $i$ the weight $w'_j:=w_i/q_i$. Then $A'^\top W'^{1-2/p}A'$ is the Gram matrix of the rows of $W_A^{1/2-1/p}A$ included independently with probability $q_i$ and rescaled by $q_i^{-1/2}$, and the leverage scores of $W_A^{1/2-1/p}A$ are the Lewis weights $w_i$. Let $\epsilon':=\epsilon_1/K_p$ if $p<2$ and $\epsilon':=\epsilon_1/(K_p\sqrt d)$ if $p\geq 2$, for a constant $K_p$ depending only on $p$, which is chosen in Step 3 and accommodated by $c_{p,\kappa}$. In both cases, $\epsilon'^{-2}=K_p^2\epsilon_1^{-2}d^{\mathbf{1}[p\geq 2]}$, and hence $\gamma=\Omega(\epsilon'^{-2}\log(d/\delta))$. Using the matrix Chernoff bound for independent inclusion in Lemma 4 of~\cite{clm+15}, as in the proof of Lemma~\ref{lem:classical_ls}, we have $(1-\epsilon')\cdot M_A \preceq A'^\top W'^{1-2/p}A' \preceq (1+\epsilon')\cdot M_A$ with probability at least $1-\delta$. Hence $w'$ is a vector of $(1+2\epsilon')$-almost Lewis weights of $A'$ in the sense of Definition 5.1 of~\cite{cp15}.

\emph{Step 3: stability transfers the guarantee to the true Lewis weights of $A'$.} Let $\ov w'$ be the vector of Lewis weights of $A'$ and let $\eta:=\epsilon_1/(4(1\lor|1-2/p|))$. For $p<4$, Lemma 5.3 of~\cite{cp15} gives $\ov w'_j=(1+2\epsilon')^{\pm c_p}\, w'_j$ with $c_p=(p/2)/(1-|p/2-1|)$. For $p\geq 2$, Lemma 5.4 of~\cite{cp15} gives $\ov w'_j=(1+2\epsilon')^{\pm K_0 f_0(p)\sqrt d}\, w'_j$ for an absolute constant $K_0$ and a function $f_0$ of $p$ alone. By choosing $K_p$ large enough as a function of $p$, we have $\ov w'_j=(1\pm\eta)\, w'_j$ for all $j$, using Lemma 5.3 when $p<4$ and Lemma 5.4 when $p\geq 4$, and we also have $\epsilon'\leq \epsilon_1/4$. For $p<2$, the accuracy $\epsilon'=\epsilon_1/K_p$ does not depend on $d$, and this suffices because $c_p=1$. Indeed, $K_p\geq 8(1\lor|1-2/p|)$ gives $2\epsilon'\leq\eta$ and $\epsilon'\leq\epsilon_1/8$, so both values $(1+2\epsilon')^{\pm 1}$ lie in $[1-\eta, 1+\eta]$, that is, $\ov w'_j=(1\pm\eta)\, w'_j$. Hence $(\ov w'_j)^{1-2/p}=(1\pm\epsilon_1/2)\,(w'_j)^{1-2/p}$, and therefore
\begin{align*}
    M_{A'} = A'^\top \ov W'^{1-2/p}A' = & ~ (1\pm \epsilon_1/2)\cdot A'^\top W'^{1-2/p}A' = (1\pm\epsilon_1/2)(1\pm\epsilon_1/4)\cdot M_A,
\end{align*}
which gives $(1-\epsilon_1)M_A\preceq M_{A'}\preceq (1+\epsilon_1)M_A$. Thus, $A'$ is an $\epsilon_1$-approximator of $A$.

\emph{Step 4: expected size.} For $i\in [n]$, let $\wh A_i$ be $\wh A$ with the row $a_i$ appended if $a_i\notin \wh A$, and $\wh A_i=\wh A$ otherwise. Let $v_i:=w_{\wh A_i,i}$ be the Lewis weight of $a_i$ in $\wh A_i$. As in the proof of Theorem~\ref{thm:general}, we condition on $|\wh A|=m$, and given this event, $\wh A$ is a uniformly random subset of $m$ rows. As in Part 2 of the proof of Lemma 6.2 of~\cite{cp15}, consider the random process that draws such an $\wh A$ and then a uniformly random $i\notin \wh A$. It is the same as drawing a uniformly random subset of $m+1$ rows and then a uniformly random row in it. Since the Lewis weights of any matrix sum to its rank, which is at most $d$, this equivalence gives $\E[\sum_{i\notin\wh A}v_i\mid |\wh A|=m]\leq (n-m)\cdot \frac{d}{m+1}\leq \frac{n+1}{m+1}\cdot d$. We then average over $m$ using $\E[\frac{n+1}{|\wh A|+1}]=2-2^{-n}\leq 2$, which is computed in the proof of Theorem~\ref{thm:general}, and obtain $\E[\sum_{i\notin\wh A}v_i]\leq 2d$. Since $\sum_{i\in \wh A}v_i=\sum_{i\in\wh A}w_{\wh A,i}\leq d$ for every outcome, we have $\E[\sum_{i=1}^n v_i]\leq 3d$. It remains to show that $q_i\leq O_{p,\kappa}(\beta)\cdot v_i$ for every $i$, pointwise for every $C$ and $\wh w$ satisfying the hypotheses. If $a_i\in \wh A$, then $y_i=v_i$ and $q_i\leq \beta\kappa_1 v_i$. If $a_i\notin {\rm rowspan}(\wh A)$, then $v_i=1$ and $q_i=1=v_i$. The last row of $W^{1/2-1/p}\wh A_i$ lies outside the span of the other rows for every positive diagonal $W$, so its leverage score is $1$. Otherwise, let $\wt v$ be the Lewis weights of $\wh A_i$ and let $N:=\sum_{j\in\wh A}\wt v_j^{1-2/p}a_ja_j^\top$. Then $M_{\wh A_i}=N+v_i^{1-2/p}a_ia_i^\top$ and $v_i^{1-2/p}\cdot a_i^\top M_{\wh A_i}^{-1}a_i=v_i^{1-2/p}v_i^{2/p}=v_i$. By the Sherman--Morrison formula and the Cauchy--Schwarz inequality, we have for every $j\in\wh A$,
\begin{align*}
    a_j^\top M_{\wh A_i}^{-1}a_j \leq a_j^\top N^{-1}a_j = & ~ a_j^\top M_{\wh A_i}^{-1}a_j+\frac{v_i^{1-2/p}(a_j^\top M_{\wh A_i}^{-1}a_i)^2}{1-v_i}\leq \frac{a_j^\top M_{\wh A_i}^{-1}a_j}{1-v_i},
\end{align*}
which means that $\wt v$ restricted to $\wh A$ is a vector of $(1-v_i)^{-1}$-almost Lewis weights of $\wh A$. Likewise, $a_i^\top N^{-1}a_i=v_i^{2/p}/(1-v_i)$. Let $\eta_p:=1/2$ if $p<4$ and $\eta_p:=\min\{1/2, 1/(C'f_0(p)\sqrt d)\}$ if $p\geq 4$, for an absolute constant $C'$. Then $1-\eta_p\geq 1/2$ in both cases. If $v_i\leq \eta_p$, then the stability lemmas, Lemma 5.3 of~\cite{cp15} for $p<4$ and Lemma 5.4 for $p\geq 4$, give $\wt v_j=\Theta_p(1)\cdot w_{\wh A,j}$ for all $j\in\wh A$. In that case, $N=\Theta_p(1)\cdot M_{\wh A}$ in the Loewner order, so $a_i^\top M_{\wh A}^{-1}a_i\leq O_p(1)\cdot a_i^\top N^{-1}a_i\leq O_p(1)\cdot v_i^{2/p}$, and hence $y_i\leq O_p(1)\cdot v_i$ and $q_i\leq \beta\kappa_1\cdot O_p(1)\cdot v_i$. If $v_i>\eta_p$, then $q_i\leq 1<v_i/\eta_p$. For $p<4$, we have $v_i/\eta_p=2v_i\leq O_p(\beta)\cdot v_i$, and for $p\geq 4$, we have $v_i/\eta_p\leq O_p(\sqrt d)\cdot v_i\leq O_p(\beta)\cdot v_i$ since $\theta_p=p/2\geq 2$. Thus, in all cases $q_i\leq O_{p,\kappa}(\beta)\cdot v_i$, and by taking expectations, we have
\begin{align*}
    \E\Big[\sum_{i=1}^n q_i\Big] \leq & ~ O_{p,\kappa}(\beta)\cdot \E\Big[\sum_{i=1}^n v_i\Big] \leq O_{p,\kappa}(\beta d) = O_{p,\kappa}(\epsilon_1^{-2}d^{\theta_p+1}\log(d/\delta)). \qedhere
\end{align*}
\end{proof}

We also need the following result, due to~\cite{cp15} for $p\in (0,4)$ and to~\cite{flps22} for $p\geq 4$, that approximates $\ell_p$ Lewis weights in nearly exact leverage score time:

\begin{lemma}[Lemma 2.4 and Corollary 3.3 of~\cite{cp15} and Theorem 1 of~\cite{flps22}]
\label{lem:flps_thm2}
Let $A=\{a_1,\ldots,a_n\}\subset \R^d$ with $A\in \R^{n\times d}$ of full column rank, $p\in (0, \infty)$ and $\epsilon\in (0, 1)$. Then, there exists a deterministic algorithm that outputs a vector $\wt w_A\in \R^n$ such that for any $i\in [n]$, $\wt w_{A, i} =  (1\pm\epsilon)\cdot w_{A, i}$. Moreover, the runtime of this algorithm is
\begin{align*}
    O_p(nd^{\omega-1}\log(np/\epsilon)).
\end{align*}
\end{lemma}

The approximation notion in the lemma is entrywise closeness to the true Lewis weights, which is the notion (1.5) of~\cite{flps22}. For $p\in (0,4)$, the algorithm is the fixed-point iteration $w_i\gets (a_i^\top (A^\top W^{1-2/p}A)^{-1}a_i)^{p/2}$ of~\cite{cp15} with exact leverage scores, which needs $O_p(\log(\log n/\epsilon))$ steps by Lemma 3.5, Lemma 3.2, and Corollary 3.3 of that paper. For $p\geq 4$, and in fact for all $p>2$ by footnote 5 of~\cite{flps22}, Theorem 1 of~\cite{flps22} performs $O(p^3\log(np/\epsilon))$ leverage score computations of matrices $DA$ with $D$ diagonal. In both cases, each leverage score computation is for an $n\times d$ matrix and takes $O(nd^{\omega-1})$ time. Both sources assume full column rank, so for a rank-deficient $A$ we apply the lemma to $AV$, where $V\in \R^{d\times r}$ is an orthonormal basis of ${\rm rowspan}(A)$ and $r={\rm rank}(A)$. This matrix has the same Lewis weights as $A$, since the weights are defined through $a_i^\top (A^\top W^{1-2/p}A)^\dagger a_i$ and $(A^\top W^{1-2/p}A)^\dagger=V(V^\top A^\top W^{1-2/p}AV)^{-1}V^\top$ by Lemma 2.5 of~\cite{wy23}.

\begin{algorithm}
\caption{Classical oracle for Lewis weights.}
\label{alg:classical_lewis}
\begin{algorithmic}[1]
\State {\bf data structure} \textsc{LewisWeights}
\State {\bf members}
\State \hspace{4mm} $A\in \R^{n\times d}$
\State \hspace{4mm} $C\in \R^{s\times d}$
\State \hspace{4mm} $M\in \R^{O_p(\log(n/\delta))\times d}$
\State \hspace{4mm} $N\in \R^{O(\log n)\times d}$
\State {\bf end members}
\State
\Procedure{Preprocess}{$A\in \R^{n\times d},  C\in \R^{s\times d}, \delta\in (0,1)$}
\State $\gamma\gets \min\{1/8, 1/(8p)\}$, $m\gets c\gamma^{-2}\log(n/\delta)$ for an absolute constant $c$
\State Compute an orthonormal basis $V_C\in \R^{d\times r}$ of ${\rm rowspan}(C)$, $r={\rm rank}(C)$
\State Generate $\wt W_C$ via Lemma~\ref{lem:flps_thm2} on $CV_C$ with accuracy $\gamma$ \Comment{$\wt W_C\in \R^{s\times s}$}
\State Compute the compact SVD of $\wt W_C^{1/2-1/p}C$: $\wt W_C^{1/2-1/p}C=U\Sigma V^\top$ \Comment{$\Sigma\in \R^{r\times r}$ invertible, $V\in \R^{d\times r}$, ${\rm span}(V)={\rm rowspan}(C)$}
\State Let $G\in \R^{m\times r}$ and $G^\prime\in \R^{c\log n\times d}$ be random Gaussian matrices
\State $M\gets G\Sigma^{-1}V^\top$ \Comment{$M\in \R^{m\times d}$}
\State $N\gets G^\prime(I-VV^\top)$ \Comment{sketch of the orthogonal complement of ${\rm rowspan}(C)$}
\EndProcedure
\State
\Procedure{Query}{$i\in [n]$}
\State {\bf if} $Na_i\neq 0$ {\bf then return} $\infty$ \Comment{$a_i\notin {\rm rowspan}(C)$, so $q_i=1$}
\State \Return $\| Ma_i \|_2^p$
\EndProcedure
\State {\bf end data structure}
\end{algorithmic}
\end{algorithm}

\begin{lemma}
\label{lem:classical_lewis}
Let $A=\{ a_1,\ldots, a_n \}\subset \R^d$, $p\in (0, \infty)$, $\delta\in (0, 1)$, $C\in \R^{s\times d}$, and define $w(A, C)$ as
\begin{align*}
    w_i(A, C)^{2/p} = & ~ \begin{cases}
        a_i^\top (C^\top W_C^{1-2/p} C)^{\dagger} a_i, & \text{if $a_i\in {\rm rowspan}(C)$}, \\
        \infty, & \text{otherwise},
    \end{cases}
\end{align*}
where $W_C$ is the diagonal matrix of the $\ell_p$ Lewis weights of $C$, taken on its row space. Then $\sum_{i=1}^n w_i(A, A)={\rm rank}(A)\leq d$, so ${\rm sum}(w)=d$, and there exists a randomized algorithm (Algorithm~\ref{alg:classical_lewis}) that implements \textsc{Preprocess} and \textsc{Query} procedures such that, with probability at least $1-\delta$ over the randomness of \textsc{Preprocess}, every query $i\in [n]$ returns a value $\wh w_i$ with $\frac12\cdot w_i(A, C)\leq \wh w_i\leq 2\cdot w_i(A, C)$ (and $\wh w_i=\infty$ exactly when $w_i(A,C)=\infty$), with
\begin{itemize}
    \item ${\cal T}_{\rm prep}(s, d) = \wt O_p(sd^{\omega-1})$.
    \item ${\cal T}_{\rm query}(s, d) = \wt O_p(d)$.
\end{itemize}
\end{lemma}

\begin{proof}

\begin{itemize}
    \item Consistent total weights: observe that we can alternatively define $w_A$ as $w_{A, i}=\tau_i(W_A^{1/2-1/p}A)$, i.e., it is the leverage score of $W_A^{1/2-1/p}A$. Since the sum of the leverage scores is the rank, which is at most $d$, we have ${\rm sum}(w)=d$.
    \item Approximator preserves weights: this follows from Definition~\ref{def:approximator}, as shown in the paragraph after it. Therefore, to preserve the weights up to a constant factor, we work with $O(1/p)$-approximators, and in the end we do a final resampling using the approximated Lewis weights.
\end{itemize}

To analyze Algorithm~\ref{alg:classical_lewis}, we first consider a variant where the Johnson--Lindenstrauss transformation is not applied. Let $V_C$ be an orthonormal basis of ${\rm rowspan}(C)$. As shown in the paragraph after Lemma~\ref{lem:flps_thm2}, the matrix $CV_C$ has full column rank and the same Lewis weights as $C$, so Lemma~\ref{lem:flps_thm2} applies to it. We compute $\wt W_C$ using Lemma~\ref{lem:flps_thm2} with accuracy $\gamma=\min\{1/8, 1/(8p)\}$, which is a $1\pm\gamma$ entrywise approximation to $W_C$. Then we know that $\wt W_C^{1-2/p}$ is a $(1\pm 2\gamma)^{|1-2/p|}$ entrywise approximation to $W_C^{1-2/p}$, and this approximation guarantee propagates to $C^\top \wt W_C^{1-2/p} C$ and, by Lemma 2.6 of~\cite{wy23}, to its pseudo-inverse on ${\rm rowspan}(C)$. So far, we have established that for any $a_i\in {\rm rowspan}(C)$, $a_i^\top (C^\top \wt W_C^{1-2/p} C)^{\dagger} a_i = (1\pm 2\gamma)^{|1-2/p|}\cdot a_i^\top (C^\top W_C^{1-2/p} C)^{\dagger} a_i$. Our final output is the $(p/2)$-th power of the quantity, hence the approximate weight is a $(1\pm 2\gamma)^{|p/2-1|}$ approximation to the true quantity $w_i(A, C)$. Rows $a_i\notin {\rm rowspan}(C)$ have $w_i(A, C)=\infty$, and the test $Na_i\neq 0$ in \textsc{Query} detects exactly these rows with probability $1$. This is the same exact zero test as in the proof of Lemma~\ref{lem:classical_ls}, and $V$ can be taken equal to $V_C$.

When applying the Johnson--Lindenstrauss transformation, we are effectively approximating $a_i^\top (C^\top \wt W_C^{1-2/p}C)^{\dagger}a_i=\|\Sigma^{-1}V^\top a_i\|_2^2$ within a factor $1\pm\gamma$, simultaneously for all $n$ rows and with probability at least $1-\delta$, by a Gaussian $G$ with $m=O(\gamma^{-2}\log(n/\delta))=O_p(\log(n/\delta))$ rows~\citep{ss11}. Combining the two approximations gives $\wh w_i=\|Ma_i\|_2^p=\big((1\pm 2\gamma)^{|1-2/p|}(1\pm\gamma)\big)^{p/2}\cdot w_i(A, C)$, and for $\gamma=\min\{1/8, 1/(8p)\}$, the multiplicative factor lies in $[1/2, 2]$ for every $p>0$. For $p<2$ the factor lies in $[1-2\gamma, 1+2\gamma]$, and for $p\geq 2$ it lies in $[e^{-3\gamma p}, e^{3\gamma p/2}]\subseteq [e^{-3/8}, e^{3/16}]$. Let us analyze the runtime.
\begin{itemize}
    \item \textsc{Preprocess}: computing $V_C$ takes $O(sd^{\omega-1})$ time. To compute $\wt W_C$, we set the accuracy in Lemma~\ref{lem:flps_thm2} to $\gamma=\Theta(1/(1\lor p))$, and it runs in time $\wt O_p(sd^{\omega-1})$. Computing the compact SVD takes $O(sd^{\omega-1})$ time, and applying the random Gaussian matrices $G$ and $G^\prime$ takes $\wt O_p(sd)$ time.
    \item \textsc{Query}: note that $M\in \R^{\wt O_p(1)\times d}$ and $N\in \R^{O(\log n)\times d}$, hence answering one query takes time $\wt O_p(d)$.
\end{itemize}
This completes the proof.
\end{proof}

Lemmas~\ref{lem:lewis_level} and~\ref{lem:classical_lewis} give an approach to compute an $O(1/p)$-approximator for $A$, but our ultimate goal is to compute an $\epsilon$-coreset for $A$, which has a different objective. The following result states that sampling according to the appropriate scaling of overestimates of Lewis weights indeed yields an $\epsilon$-coreset:

\begin{lemma}[Theorems 1.3 and A.2 of~\cite{wy23}]
\label{lem:wy23_thm13}
Let $A=\{a_1,\ldots,a_n\}\subset \R^d$, $\epsilon,\delta\in (0, 1)$ and $p\in (0, \infty)$ and let $u\in \R^n$ be an overestimate of $w_A$ with $\|u\|_1\leq O(d)$. Consider the sampling scheme where each point is sampled with probability $q_i=\min\{1, \alpha\cdot u_i\}$ where
\begin{itemize}
    \item $\alpha=\epsilon^{-2}(\log^3 d+\log(1/\delta))$ for $p\in (0, 1)$.
    \item $\alpha=\epsilon^{-2}\log(n/\delta)$ for $p=1$.
    \item $\alpha=\epsilon^{-2}(\log^2 d\log n+\log(1/\delta))$ for $p\in (1, 2)$.
    \item $\alpha=\epsilon^{-2}d^{p/2-1}(\log^2 d\log n+\log(1/\delta))$ for $p\geq 2$.
\end{itemize}
Let $S\in\R^{n\times n}$ be the diagonal matrix with $S_{ii}=q_i^{-1/p}$ for the sampled points and $S_{ii}=0$ otherwise. Then, with probability at least $1-\delta$, $SA$ is an $\epsilon$-coreset of $A$, and the number of samples has expectation $\sum_i q_i\leq \alpha\cdot \|u\|_1$ and is $O(\alpha\cdot \|u\|_1)$ with probability at least $1-\delta$.
\end{lemma}

Theorem 1.3 of~\cite{wy23} is stated for $p>2$ and for \emph{one-sided} Lewis weights, which Definition 2.2 there defines as vectors $u$ with $u_i\geq (a_i^\top (A^\top U^{1-2/p}A)^{-1}a_i)^{p/2}$ for all $i$, where $U={\rm diag}(u)$. When $p>2$, every entrywise overestimate $u$ of $w_A$ is one-sided, since $1-2/p>0$ gives $U^{1-2/p}\succeq W_A^{1-2/p}$, and hence $a_i^\top (A^\top U^{1-2/p}A)^{-1}a_i\leq a_i^\top (A^\top W_A^{1-2/p}A)^{-1}a_i=w_{A,i}^{2/p}\leq u_i^{2/p}$. For $p<2$, Theorem A.2 of~\cite{wy23} requires only that each point be sampled with probability at least $\min\{1, \alpha\cdot w_{A,i}\}$, and $q_i=\min\{1,\alpha\cdot u_i\}$ satisfies this for any overestimate $u$. For $p=2$, the statement is standard leverage score sampling. The number of samples is a sum of independent Bernoulli variables with mean $\sum_i q_i\leq \alpha\|u\|_1$. Since $\alpha\geq \log(1/\delta)$ in every case, by a Chernoff bound, the number of samples is at most $2\alpha\|u\|_1+O(\log(1/\delta))=O(\alpha\|u\|_1)$ with probability at least $1-\delta$.

\begin{theorem}
\label{thm:lewis}
Let $A=\{a_1,\ldots,a_n\}\subset \R^d$, $\epsilon,\delta\in (0, 1)$ and $p\in (0, \infty)$. There exists a quantum algorithm that with probability at least $1-\delta$, constructs an $\epsilon$-coreset of $A$ with size $O_p(\alpha\cdot d)$, for $\alpha$ given in Lemma~\ref{lem:wy23_thm13}. The runtimes for generating the coreset, in expectation conditioned on the success event of probability at least $1-\delta$ (as in Theorem~\ref{thm:general}), are
\begin{itemize}
    \item $\wt O_p(\epsilon^{-1}n^{0.5}d^{1.5}+d^{\omega})$ for $p\in (0, 2)$.
    \item $\wt O_p(d^{p/2+\omega})+\wt O_p(n^{0.5}d^{p/4+1}(\epsilon^{-1}+d^{0.5}))$ for $p\geq 2$.
\end{itemize}
The numbers of queries to the points in $A$ for generating the coreset, in the same expectation, are
\begin{itemize}
    \item $\wt O_p(\epsilon^{-1}n^{0.5}d^{0.5})$ for $p\in (0,2)$.
    \item $\wt O_p(n^{0.5}d^{p/4}(\epsilon^{-1}+d^{0.5}))$ for $p\geq 2$.
\end{itemize}

\end{theorem}

\begin{proof}
Our strategy will be to first construct an $O(1/p)$-approximator of $A$, which in turn gives an $O(1)$-approximation to $w_A$, then we will sample according to these approximations, in conjunction with Lemma~\ref{lem:q_sampling_1_d}. We assume that $A$ has full column rank. Otherwise, every statement below holds on ${\rm rowspan}(A)$ with $d$ replaced by ${\rm rank}(A)$. Note that the intermediate sets $A_t$ may be rank-deficient even when $A$ is not, but Lemma~\ref{lem:lewis_level} does not require full column rank and is applied to them with $d\geq {\rm rank}(A_t)$. Set $\epsilon_1:=\min\{1/4, 1/(4p)\}$, let $s'$ be the quantity chosen in Stage 1 below, and let $T:=\lceil\log_2(n/s')\rceil=O(\log n)$.
\begin{itemize}
    \item Stage 1: constructing an $\epsilon_1$-approximator, hence an $O(1/p)$-approximator, of $A$. We run the recursion of Algorithm~\ref{alg:quantum_iterate} on the chain $A_0\subset_{1/2}A_1\subset_{1/2}\cdots\subset_{1/2}A_T=A$ with $C_0=A_0$, using the oracle of Algorithm~\ref{alg:classical_lewis} with the same accuracy $\epsilon_1$ at every level, so $s=s'$. We replace the sampling step of every level by the rule of Lemma~\ref{lem:lewis_level} and use the relative weights without augmentation, since that lemma handles the rows outside $A_{t-1}$ directly and takes the place of the conditions of Assumption~\ref{assumption:recurse}. We use Theorem~\ref{thm:general} only to account for the time and the queries. At level $t$, we sample $A_t$ with
    \begin{align*}
        q_i = & ~ \min\{1, \beta\cdot \wh w_i\},\qquad \beta:=c_{p,2}\cdot \epsilon_1^{-2}\cdot d^{\theta_p}\cdot \log(8Td/\delta),
    \end{align*}
    where $\wh w_i$ is the output of \textsc{Query}$(i)$ after \textsc{Preprocess}$(A_t, C_{t-1}, \delta/(8T))$, $c_{p,2}$ is the constant of Lemma~\ref{lem:lewis_level} for $\kappa=2$, and $\theta_p$ is as in that lemma, that is, $\theta_p=0$ for $p<2$ and $\theta_p=p/2$ for $p\geq 2$. We show by induction on $t$ that with probability at least $1-\delta/4$, $C_t$ is an $\epsilon_1$-approximator of $A_t$ for every $t\leq T$. The base case is trivial since $C_0=A_0$. Suppose $C_{t-1}$ is an $\epsilon_1$-approximator of $A_{t-1}$, the Bernoulli$(1/2)$ half of $A_t$. Since $\epsilon_1\leq 1/4$, Lemma~\ref{lem:classical_lewis} gives $\wh w_i=2^{\pm1}\cdot w_i(A_t, C_{t-1})$ for all $a_i\in A_t$ with probability at least $1-\delta/(8T)$, and Lemma~\ref{lem:lewis_level} with $\kappa=2$ and failure probability $\delta/(8T)$ shows that $C_t$ is an $\epsilon_1$-approximator of $A_t$. Thus, the $2T$ events have total failure probability $\delta/4$. For $t\in [T]$, let ${\cal H}_t$ be the event that $C_{t-1}$ is an $\epsilon_1$-approximator of $A_{t-1}$ and the oracle guarantee of level $t$ holds. On ${\cal H}_t$, the hypotheses of Lemma~\ref{lem:lewis_level} are met with $\kappa=2$ and $\wh A=A_{t-1}\subset_{1/2}A_t$. In this case, Step 4 of the proof of that lemma bounds $\sum_{i\in A_t}q_i$ pointwise by $O_{p,2}(\beta)\sum_{i\in A_t}v_i$, where $v_i$ depends only on $A_t$ and $A_{t-1}$ and $\E[\sum_{i\in A_t}v_i\mid A_t]\leq 3d$. Therefore, when the sampler call of level $t$ returns an exact sample, an event included in ${\cal S}$ below, we have $\E[|C_t|\,\mathbf{1}_{{\cal H}_t}]=\E[\sum_{i\in A_t}q_i\,\mathbf{1}_{{\cal H}_t}]\leq s'$ for
    \begin{align*}
        s' := & ~ O_p(\epsilon_1^{-2}d^{\theta_p+1}\log(Td/\delta)) = \wt O_p(d^{\theta_p+1}),
    \end{align*}
    which is $\wt O_p(d)$ for $p<2$, and $\E|A_0|=n2^{-T}\leq s'$ by the choice of $T$. Let ${\cal S}$ be the intersection of the $2T$ events of this stage, the success of the $T+2$ calls of the sampler of Lemma~\ref{lem:q_sampling_1_d} in the two stages, and the three events of Stage 2. The sampler is called for $A_0$, for the $T$ levels, and for the final sample, each time with failure probability $\delta/(8(T+2))$, so these calls fail with probability at most $\delta/8$ in total. By the last item below, the event ${\cal S}$ has probability at least $1-6\delta/8\geq 1/4$, and it is contained in every ${\cal H}_t$. Consequently, we have $\E[|C_t|\mid {\cal S}]\leq s'/\Pr[{\cal S}]\leq 4s'$ and $\E[\sum_{i\in A_t}q_i\mid {\cal S}]\leq 4s'$ for every $t\leq T$. As in Theorem~\ref{thm:general}, we condition all expectations in the running time analysis on ${\cal S}$. Each level prepares one oracle in ${\cal T}_{\rm prep}(|C_{t-1}|, d)=\wt O_p(|C_{t-1}|d^{\omega-1})$ time, which is linear in $|C_{t-1}|$ and hence has conditional expectation $\wt O_p(s'd^{\omega-1})$. It then samples using Lemma~\ref{lem:q_sampling_1_d} in $\wt O(\sqrt{n\sum_{i\in A_t}q_i})\cdot {\cal T}_{\rm query}$ time. Therefore, by Jensen's inequality, the expected time of Stage 1 conditioned on ${\cal S}$ is
    \begin{align*}
        \wt O_p(d^{\theta_p+\omega}) + \wt O_p(n^{0.5} d^{\theta_p/2+1.5}).
    \end{align*}
    We let $\wt B:=C_T$ denote the resulting approximator. Note that the expected size of $\wt B$ conditioned on ${\cal S}$ is $\wt O_p(d^{\theta_p+1})$. In particular, for $p<2$, Stage 1 takes $\wt O_p(d^{\omega}+n^{0.5}d^{1.5})$ expected time, and $\wt B$ has $\wt O_p(d)$ rows in expectation.
    \item Stage 2: constructing an $\epsilon$-coreset of $A$. Observe that $\wt B$ gives an $O(1)$-approximation to $w_A$. Since $\wt B$ is an $\epsilon_1$-approximator of $A$ with $\epsilon_1\leq \min\{1/4, 1/p\}$, by the consequence of Definition~\ref{def:approximator}, for any $a_i$ we have
    \begin{align*}
        (a_i^\top (\wt B^\top W_{\wt B}^{1-2/p} \wt B)^{-1} a_i)^{p/2} = & ~ (1\pm 2p\epsilon_1)\cdot (a_i^\top (A^\top W_{A}^{1-2/p} A)^{-1} a_i)^{p/2} \\
        = & ~ (1\pm 1/2)\cdot w_{A, i},
    \end{align*}
    and after appropriately rescaling this yields the desired oversampling vector $u$. Concretely, we run \textsc{Preprocess}$(A, \wt B, \delta/8)$ of Algorithm~\ref{alg:classical_lewis} and set $u_i:=4\cdot \wh w_i$, where $\wh w_i=2^{\pm 1}\cdot w_i(A, \wt B)$ is the output of \textsc{Query}$(i)$. Then we have $u_i\geq 4\cdot \frac12\cdot \frac12\cdot w_{A,i}=w_{A,i}$ for all $i$, and
    \begin{align*}
        \|u\|_1 = & ~ 4\sum_{i=1}^n \wh w_i \leq 8\sum_{i=1}^n (a_i^\top (\wt B^\top W_{\wt B}^{1-2/p} \wt B)^{-1} a_i)^{p/2} \\
        \leq & ~ 12\cdot \sum_{i=1}^n (a_i^\top (A^\top W_{A}^{1-2/p} A)^{-1} a_i)^{p/2} \\
        = & ~ 12d,
    \end{align*}
    and we could invoke Lemma~\ref{lem:wy23_thm13} with $\delta/8$ in place of $\delta$ to generate the desired $\epsilon$-coreset. Since $\wt B$ has expected size $\wt O_p(d^{\theta_p+1})$ conditioned on ${\cal S}$ and the cost of \textsc{Preprocess} is linear in $|\wt B|$, generating $\wt W_{\wt B}$ takes $\wt O_p(d^{\theta_p+\omega})$ expected time. Computing the SVD and applying the Gaussian matrices takes $\wt O_p(d^{\theta_p+\omega})$ expected time as well. Each query takes $\wt O_p(d)$ time. Moreover, $\alpha\|u\|_1=\wt O_p(\epsilon^{-2}d^{p/2\lor 1})$ on ${\cal S}$, so sampling according to $q_i=\min\{1, \alpha\cdot u_i\}$ with Lemma~\ref{lem:q_sampling_1_d} takes $\wt O_p(\sqrt{n\alpha\|u\|_1})\cdot \wt O_p(d)=\wt O_p(\epsilon^{-1}n^{0.5}d^{(p/2\lor 1)/2+1})$ time. All in all, we obtain the following runtime for generating the $\epsilon$-coreset, where in each line the first term is the preprocessing cost of the two stages, the second term is the sampling cost of Stage 2, and the third term is the sampling cost of Stage 1:
    \begin{itemize}
        \item For $p\in (0, 2)$, it takes time $\wt O_p(d^{\omega}+\epsilon^{-1}n^{0.5}d^{1.5}+n^{0.5}d^{1.5})$, which is $\wt O_p(\epsilon^{-1}n^{0.5}d^{1.5}+d^{\omega})$ since $\epsilon<1$.
        \item For $p\geq 2$, it takes time $\wt O_p(d^{p/2+\omega}+\epsilon^{-1}n^{0.5}d^{p/4+1}+n^{0.5}d^{p/4+1.5})$.
    \end{itemize}
    The query complexity follows from Theorem~\ref{thm:general} and Lemma~\ref{lem:q_sampling_1_d}. Stage 1 makes $\wt O_p(\sqrt{ns'})=\wt O_p(n^{0.5}d^{\theta_p/2+0.5})$ queries to the points of $A$, and Stage 2 makes $\wt O_p(\sqrt{n\alpha\|u\|_1})=\wt O_p(\epsilon^{-1}n^{0.5}d^{(p/2\lor 1)/2})$ queries. For $p\in (0,2)$, both counts are at most $\wt O_p(\epsilon^{-1}n^{0.5}d^{0.5})$, since $\theta_p=0$ and $\epsilon<1$.
    \item Failure probability and size. The three events of Stage 2 in ${\cal S}$ are the oracle guarantee, the coreset guarantee of Lemma~\ref{lem:wy23_thm13}, and its size guarantee, each with failure probability $\delta/8$. Together with the failure probabilities of the level events and the sampler calls computed in Stage 1, a union bound shows that ${\cal S}$ has probability at least $1-\delta/4-\delta/8-3\delta/8\geq 1-\delta$. On ${\cal S}$, the output is an $\epsilon$-coreset of $A$ of size $O(\alpha\|u\|_1)=O_p(\alpha d)$. This is because $\|u\|_1\leq 12d$ holds once $\wt B$ is an approximator and the Stage 2 oracle guarantee holds, and both are part of ${\cal S}$.
\end{itemize}
This concludes the proof.
\end{proof}

\begin{remark}[Row sparsity]\label{rem:row_sparsity}
Suppose that every row of $A$ has at most $r\leq d$ nonzero entries (this $r$ is not the rank of $C$ in the oracles of this section) and that a row query returns these entries with their positions. Then the procedure \textsc{Query} of Algorithms~\ref{alg:classical_ls}, \ref{alg:classical_lewis}, and~\ref{alg:classical_ridge_ls} takes $\wt O_p(r)$ time instead of $\wt O_p(d)$, since it uses $a_i$ only through its inner products with the $O_p(\log(n/\delta))$ rows of $M$ and $N$, which are held in QRAM (for rational inputs, the fingerprint of the zero test in the proof of Lemma~\ref{lem:classical_ls} can be taken as one more inner product, with a random combination of the basis of the orthogonal complement of the row space of $C$). The other work per row, namely reading the points of $A_0$, of the samples $C_t$, and of the output, and computing each $q_i$ and weight from the value returned by \textsc{Query}, takes $O(r)$ time, and the membership tests of the chain read no row. Therefore, the query counts of Theorems~\ref{thm:ls}, \ref{thm:lewis}, and~\ref{thm:ridge} do not change, and since each of their running times is $\wt O_p(Qd)$ for the query count $Q$ plus additive preprocessing terms, such as $d^\omega$, $d^{p/2+\omega}$, and $dk^{\omega-1}$, the running times become $\wt O_p(Qr)$ plus the same additive terms. For example, Theorem~\ref{thm:lewis} for $p\in (0,2)$ and Theorem~\ref{thm:ls} for $p=2$ give time $\wt O_p(\epsilon^{-1}n^{0.5}d^{0.5}r+d^\omega)$, which matches the leading term $\wt O(r\sqrt{nd}/\epsilon)$ of Corollary 11 of~\cite{lj25} (arXiv v1). This term comes from computing their sparsifier, which takes $\wt O(r\sqrt{nd}/\epsilon+d^\omega+dr^2)$ time by their Theorem 10 applied to $\ell_p$ losses (their Section 4.2), and their Corollary 11 adds $\poly(d,1/\epsilon)$ time to solve the regression on the sparsifier. Table~\ref{tab:coreset} compares our time with this sparsifier time. For simplicity, we state all our bounds for dense points, that is, $r=d$.
\end{remark}

\subsection{\texorpdfstring{$k$}{k}-Subspace Sensitivity and Ridge Leverage Score}
\label{sec:ridge}

Let ${\cal F}_k$ be the set of all $k$-dimensional subspaces in $\R^d$. We can define the cost with respect to a subspace by identifying $X={\cal F}_k$ and ${\rm cost}: \R^d\times {\cal F}_k\rightarrow \R_{\geq 0}$ as
\begin{align*}
    {\rm cost}(a_i, x) = & ~ \|a_i^\top (I-P_x)\|_2^2,
\end{align*}
where $P_x$ is the orthogonal projection onto $x$. Then, the $k$-subspace sensitivity is
\begin{align*}
    s_i(A, B) = & ~ \max_{x\in X} \frac{\|a_i^\top (I-P_x)\|_2^2}{\|B(I-P_x) \|_F^2}.
\end{align*}
Similar to $\ell_p$ sensitivity, $k$-subspace sensitivity is controlled by the \emph{ridge leverage score}, defined as
\begin{align*}
    \ov \tau_i(A) = & ~ a_i^\top (A^\top A+\lambda_{A_k} I)^{\dagger}a_i
\end{align*}
where $\lambda_{A_k}=\|A-A_k\|_F^2/k$. The matrix in this definition is invertible unless ${\rm rank}(A)\leq k$. For every $i\in [n]$, we have
\begin{align*}
    s_i(A, A) \leq & ~ (1+1/k)\cdot \ov\tau_i(A).
\end{align*}
To see this, let $Q=I-P_x$ and $M=A^\top A+\lambda_{A_k}I$, and suppose $Qa_i\neq 0$. Applying the Cauchy--Schwarz inequality in the semi-inner product induced by $M$ with the unit vector $y=Qa_i/\|Qa_i\|_2$, we have $\|a_i^\top Q\|_2^2\leq \ov\tau_i(A)\cdot(\|Ay\|_2^2+\lambda_{A_k})\leq (1+1/k)\,\ov\tau_i(A)\,\|AQ\|_F^2$. Here we use $\|Ay\|_2\leq \|AQ\|_F$ and $\lambda_{A_k}=\|A-A_k\|_F^2/k\leq \|AQ\|_F^2/k$, where the latter holds because $AP_x$ has rank at most $k$. The ridge leverage score alone is not an overestimate of the sensitivity. For example, for $A=I_d$ and $k=1$, we have $s_i(A, A)=1/(d-1)>1/d=\ov\tau_i(A)$. However, the value $2\ov\tau_i(A)$ is an overestimate, and the factor $2$ is absorbed by the constant $c$ in the sampling probabilities of our framework. It is worth noting that the results of~\cite{cmm17} used below are sampling guarantees for ridge leverage scores, and they do not pass through this bound. We then define the weights similarly to leverage scores:
\begin{align*}
    w_i(A, B) = & ~ \begin{cases}
        a_i^\top (B^\top B+\lambda_{B_k}I)^\dagger a_i, & \text{if $a_i\in {\rm span}(B^\top B+\lambda_{B_k}I)$}, \\
        \infty, & \text{otherwise}.
    \end{cases}
\end{align*}
We will explicitly define the notion of $\epsilon$-approximator:
\begin{definition}
Let $A=\{a_1,\ldots,a_n\}\subset \R^d$, $\epsilon\in (0, 1)$ and $1\leq k\leq d$. We say $B$ is an $\epsilon$-approximator of $A$ if
\begin{itemize}
    \item $B$ is an $\epsilon$-coreset of $A$.
    \item The following additive-multiplicative spectral approximation guarantee holds:
    \begin{align*}
        (1-\epsilon) B^\top B - \epsilon \lambda_{A_k} I \preceq A^\top A \preceq (1+\epsilon) B^\top B + \epsilon \lambda_{A_k} I.
    \end{align*}
\end{itemize}
\end{definition}

The following two results due to~\cite{cmm17} illustrate that an $\epsilon$-approximator indeed preserves all weights, and uniform sampling gives sufficiently good approximation.

\begin{lemma}[Lemma 12 of~\cite{cmm17}\footnote{Note that while the original Lemma in~\cite{cmm17} states the result in terms of ridge leverage score, their proof essentially shows that $B^\top B+\lambda_{B_k}I$ is a $1\pm 4\epsilon$ spectral approximation of $A^\top A+\lambda_{A_k}I$, which gives the desired approximator guarantee.}]
\label{lem:cmm_lem12}
Let $A=\{a_1,\ldots,a_n\}\subset \R^d$ and $\epsilon \in (0, 1/2]$. If $B$ is an $\epsilon$-approximator of $A$, then for any fixed $C$ and for all $i\in [n]$, $w_i(C, B)= (1\pm 4\epsilon)\cdot w_i(C, A)$.
\end{lemma}

\begin{lemma}[Theorem 14 of~\cite{cmm17}]
\label{thm:ridge_via_unif}
Let $A=\{a_1,\ldots,a_n\}\subset \R^d$, $k\in [d]$ and $\epsilon,\delta\in (0, 1)$. Form $SA$ by including each point of $A$ independently with probability $\frac{1}{2}$. Let $q_i=\min\{1, c\epsilon^{-2}\log(k/\delta)\cdot w_i(A, SA)\}$ for a sufficiently large constant $c$, and let $B$ contain each $a_i$ independently with probability $q_i$, weighted by $1/q_i$. Then, with probability at least $1-\delta$, $B$ is an $\epsilon$-approximator of $A$ of size $O(\epsilon^{-2}k\log(k/\delta))$.
\end{lemma}
\begin{remark}\cite{cmm17} state Theorem 14 for a constant accuracy, with the sampling probabilities $\min\{1, c\log(k/\delta)\cdot\min\{1, w_i(A, SA)\}\}$, and conclude that $B$ satisfies the hypotheses of their Lemma 12 for some constant $\epsilon$. However, their proof gives the $\epsilon$-dependence stated here. By their Theorem 11, which is monotonicity, we have $\min\{1, w_i(A, SA)\}\geq \ov\tau_i(A)$. Therefore, these overestimates are fed to their Lemmas 21 and 22, which are the independent-sampling forms of their Theorem 5 and Lemma 20 (the ingredient of their Theorem 6) stated for general $\epsilon$ with the probabilities $\min\{1, c\epsilon^{-2}\log(k/\delta)\wt\tau_i\}$. These probabilities coincide with our $q_i$ since $c\epsilon^{-2}\log(k/\delta)\geq 1$, so these lemmas yield the additive-multiplicative spectral approximation and the projection-cost preservation at accuracy $\epsilon$. We bound the number of sampled points through their Lemma 13 as in their proof. The reweighted columns contribute at most $3k/u=O(k\log(k/\delta))$ and the remaining ones at most $c\epsilon^{-2}\log(k/\delta)\cdot 12k$. This holds in expectation conditioned on the event that the reweighted half satisfies the hypotheses of their Lemma 12, an event of probability $1-\delta/3$ in their proof, and by a Chernoff bound it also holds with probability $1-\delta$.\end{remark}

\begin{algorithm}
\caption{Classical oracle for ridge leverage score.}
\label{alg:classical_ridge_ls}
\begin{algorithmic}[1]
\State {\bf data structure} \textsc{RidgeLeverageScore}
\State {\bf members}
\State \hspace{4mm} $A\in \R^{n\times d}$
\State \hspace{4mm} $C\in \R^{s\times d}$
\State \hspace{4mm} $M\in \R^{O(\log n)\times d}$
\State \hspace{4mm} $N\in \R^{O(\log n)\times d}$
\State {\bf end members}
\State
\Procedure{Preprocess}{$A\in \R^{n\times d},  C\in \R^{s\times d}$}
\State $c\gets 1000$
\State Compute the compact SVD of $C$: $C=U\Sigma V^\top$ \Comment{$r={\rm rank}(C)\leq s$, $\Sigma\in \R^{r\times r}$ invertible, $V\in \R^{d\times r}$}
\State $\lambda\gets \frac{1}{k}\sum_{i=k+1}^{r} \sigma_i^2$ \Comment{$\lambda=\lambda_{C_k}=\|C-C_k\|_F^2/k$, and $\lambda=0$ iff $r\leq k$}
\State Let $G\in \R^{c\log n\times r}$ and $G^\prime\in \R^{c\log n\times d}$ be random Gaussian matrices
\State $M\gets G(\Sigma^2+\lambda I)^{-1/2}V^\top$ \Comment{$M\in \R^{c\log n\times d}$}
\State $N\gets G^\prime(I-VV^\top)$ \Comment{$N\in \R^{c\log n\times d}$, sketch of the orthogonal complement of ${\rm span}(C^\top C)$}
\EndProcedure
\State
\Procedure{Query}{$i\in [n]$}
\State {\bf if} $\lambda=0$ {\bf and} $Na_i\neq 0$ {\bf then return} $\infty$ \Comment{${\rm rank}(C)\leq k$ and $a_i\notin {\rm span}(C^\top C)$, so $q_i=1$}
\State {\bf if} $\lambda>0$ {\bf then return} $\| Ma_i \|_2^2+\lambda^{-1}\|Na_i\|_2^2$ \Comment{sketch of $a_i^\top V(\Sigma^2+\lambda I)^{-1}V^\top a_i+\lambda^{-1}\|(I-VV^\top)a_i\|_2^2$}
\State \Return $\| Ma_i \|_2^2$ \Comment{$\lambda=0$ and $a_i\in {\rm span}(C^\top C)$}
\EndProcedure
\State {\bf end data structure}
\end{algorithmic}
\end{algorithm}

\begin{lemma}
\label{lem:classical_ridge}
Let $A=\{ a_1,\ldots, a_n \}\subset \R^d$, $k\leq d$, $\epsilon, \delta\in (0, 1)$, and define $w(A, B)$ as
\begin{align*}
    w_i(A, B) = & ~ \begin{cases}
        a_i^\top (B^\top B+\lambda_{B_k}I)^\dagger a_i, & \text{if $a_i\in {\rm span}(B^\top B+\lambda_{B_k}I)$}, \\
        \infty, & \text{otherwise}.
    \end{cases}
\end{align*}
Then $\sum_{i=1}^n w_i(A, A)\leq 2k$, and there exists a randomized algorithm (Algorithm~\ref{alg:classical_ridge_ls}) that implements \textsc{Preprocess} and \textsc{Query} procedures with
\begin{itemize}
    \item ${\cal T}_{\rm prep}(s, d) = \wt O(ds^{\omega-1})$.
    \item ${\cal T}_{\rm query}(s, d) = \wt O(d)$.
\end{itemize}
\end{lemma}

\begin{proof}
Let $A=U\Sigma V^\top$ be the SVD of $A$. Then,
\begin{align*}
    \sum_{i=1}^n w_i(A, A) = & ~ \sum_{i=1}^n a_i^\top (A^\top A+\lambda_{A_k}I)^\dagger a_i \\
    = & ~ \tr[A^\top A (A^\top A+\lambda_{A_k}I)^\dagger] \\
    = & ~ \tr[V \Sigma^2 V^\top (V\Sigma^2 V^\top+\lambda_{A_k} V V^\top)^\dagger] \\
    = & ~ \tr[\Sigma^2 (\Sigma^2+\lambda_{A_k}I)^\dagger]\\
    = & ~ \sum_{i=1}^n \frac{\sigma_i^2}{\sigma_i^2+\frac{\|A-A_k\|_F^2}{k}} \\
    \leq & ~ k + \sum_{i=k+1}^n \frac{\sigma_i^2}{\sigma_i^2+\frac{\|A-A_k\|_F^2}{k}} \\
    \leq & ~ k + \sum_{i=k+1}^n \frac{\sigma_i^2}{\frac{\|A-A_k\|_F^2}{k}} \\
    = & ~ k + k\cdot \frac{\|A-A_k\|_F^2}{\|A-A_k\|_F^2} \\
    \leq & ~ 2k
\end{align*}
where for the fifth step, we upper bound $\frac{\sigma_i^2}{\sigma_i^2+\frac{\|A-A_k\|_F^2}{k}}$ by 1 for $i\leq k$. The runtime analysis is identical to that of Lemma~\ref{lem:classical_ls}, with the following additions for the correctness of Algorithm~\ref{alg:classical_ridge_ls}. Let $C=U\Sigma V^\top$ be the compact SVD of the current approximator $C$, where $r={\rm rank}(C)$ and $V\in \R^{d\times r}$. Let $\lambda=\lambda_{C_k}=\frac{1}{k}\sum_{i>k}\sigma_i^2=\|C-C_k\|_F^2/k$ be the value computed by \textsc{Preprocess}. If $\lambda>0$, then $C^\top C+\lambda I$ is invertible and
\begin{align*}
    (C^\top C+\lambda I)^{-1} = & ~ V(\Sigma^2+\lambda I)^{-1}V^\top+\lambda^{-1}(I-VV^\top),
\end{align*}
and hence $w_i(A, C)=\|(\Sigma^2+\lambda I)^{-1/2}V^\top a_i\|_2^2+\lambda^{-1}\|(I-VV^\top)a_i\|_2^2$. Both terms are sketched by Gaussian matrices with $O(\log n)$ rows. Thus, \textsc{Query} returns a $1\pm\epsilon$ approximation of $w_i(A, C)$, which we turn into an overestimate by scaling. The second term vanishes when $r=d$, but it cannot be dropped whenever $r<d$. In our applications, the approximator has $s=\wt O(k)$ rows, typically far fewer than $d$. If $\lambda=0$, i.e. $r\leq k$, then the weight is $a_i^\top (C^\top C)^\dagger a_i$ for $a_i\in {\rm span}(C^\top C)$ and $\infty$ otherwise. In this case, \textsc{Query} handles the two subcases in the same way as Algorithm~\ref{alg:classical_ls}, using the same exact zero test on $Na_i$. Finally, computing $N=G^\prime-(G^\prime V)V^\top$ takes $\wt O(rd)\leq \wt O(sd)$ time and each query takes $\wt O(d)$ time. Therefore, ${\cal T}_{\rm prep}$ and ${\cal T}_{\rm query}$ are as stated.
\end{proof}

One of the key features of the $\epsilon$-approximator for $k$-subspace approximation is that it is also an $\epsilon$-coreset. 

\begin{theorem}
\label{thm:ridge}
Let $A=\{a_1,\ldots,a_n\}\subset \R^d$, $\epsilon,\delta\in (0, 1)$ and $k\in [d]$. There exists a quantum algorithm $\textsc{QRLS}(A, k, \epsilon, \delta)$ that with probability at least $1-\delta$, constructs an $\epsilon$-approximator of $A$ (in particular an $\epsilon$-coreset of $A$ for the $k$-subspace cost) with size $ O(\epsilon^{-2}k\log(k/\delta))$, in expected time $\wt O(\epsilon^{-1}n^{0.5}dk^{0.5}+dk^{\omega-1})$ and, in expectation, $\wt O(\epsilon^{-1}n^{0.5}k^{0.5})$ queries to the points in $A$ (the expectations as in Theorem~\ref{thm:general}).
\end{theorem}

\begin{proof}
The proof is almost identical to the proof of Theorem~\ref{thm:ls}, except that the sizes are $s=\wt O(\epsilon^{-2}k)$ and $s'=\wt O(k)$. Let $T=O(\log n)$ be the number of levels. We run Algorithm~\ref{alg:quantum_iterate} as $\textsc{QIterateSample}(A, \epsilon, \delta/2)$ with the oracle of Lemma~\ref{lem:classical_ridge}. Its $T+1$ calls of $\textsc{QSample}$ are run with failure probability $\delta/(4T(T+1))$ each, which is at most $\delta/(4T)$ in total. This replaces the parameter $\delta'/(T+1)$ printed in Algorithm~\ref{alg:quantum_iterate}, although that parameter would serve as well. We replace items 2 and 3 of Assumption~\ref{assumption:recurse} by Lemma~\ref{thm:ridge_via_unif}, which we apply at accuracy $0.01$ and failure probability $\delta/(4T)$ at each of the $T-1$ intermediate levels, and at accuracy $\epsilon$ and failure probability $\delta/4$ at the final level. By Lemma~\ref{lem:cmm_lem12}, the oracle at level $t$ evaluates $w_i(A_t, C_{t-1})=\Theta(1)\cdot w_i(A_t, A_{t-1})$. The constant factors of Lemmas~\ref{lem:cmm_lem12} and~\ref{lem:classical_ridge} are absorbed by the constant $c$ of Lemma~\ref{thm:ridge_via_unif}, and therefore its size bound applies at every level. As in Step 2 of the proof of Theorem~\ref{thm:ls}, the remaining $\delta/2$ covers the Gaussian sketches of the $T$ oracle preparations of Algorithm~\ref{alg:classical_ridge_ls}, each with $O(\log(n/\delta))$ rows and failure probability at most $\delta/(4T)$. It also covers the event $|A_0|\leq 2\max\{s', 3\log(4/\delta)\}$, which has probability at least $1-\delta/4$ by a Chernoff bound since $\E|A_0|\leq s'$. The intersection of these events has probability at least $1-\delta$. On this intersection, the final level outputs an $\epsilon$-approximator of $A$ of size $O(\epsilon^{-2}k\log(k/\delta))$, and every intermediate sample, hence every set on which an oracle is prepared, has $O(s')$ points. Therefore, on this event each preparation costs ${\cal T}_{\rm prep}(s', d)=\wt O(dk^{\omega-1})$, since $s'=\wt O(k)$ and $k\leq d$.
Hence, when we plug these choices into Lemma~\ref{lem:classical_ridge} and Theorem~\ref{thm:general}, the hypothesis of the latter that ${\cal T}_{\rm prep}$ be linear is not needed, and we obtain a runtime of
\begin{align*}
    & ~ \wt O(\epsilon^{-1}n^{0.5}dk^{0.5}+dk^{\omega-1}). \qedhere
\end{align*}
\end{proof}

\section{Quantum Column Subset Selection and Low-Rank Approximation}
\label{sec:cmm}

In this section, we present the first application of the generic sampling framework developed in Section~\ref{sec:general}. In particular, when the cost is the $k$-subspace cost defined as ${\rm cost}(A, x)=\|(I-P_x)A\|_F^2$ where $x\in {\cal F}_k$, then an $\epsilon$-coreset of $A$ can be used to compute a Frobenius norm low-rank approximation. Let $A\in \R^{d\times n}$, we let the set of points be $\{a_1,\ldots,a_n\}\subset \R^d$, and the goal is to compute a weighted subset of columns of $A$.
\begin{lemma}[Lemma 3 of~\cite{cem+15}]
Let $A=\{a_1,\ldots,a_n\}\subset \R^d$, $\epsilon\in (0, 1)$, $k\in [\min\{n, d\}]$ and let $B\subset A$ be an $(\epsilon/3)$-coreset of $A$ with respect to the $k$-subspace cost. Then, the projection onto the top-$k$ left singular vectors of $B$, denoted by $P_{B_k}$, satisfies
\begin{align*}
    \|A-P_{B_k}P_{B_k}^\top A\|_F^2 \leq & ~ (1+\epsilon) \|A-A_k\|_F^2.
\end{align*}
\end{lemma}
\begin{remark}For a $\gamma$-coreset $B$, the two-sided guarantee gives $(1-\gamma)\|A-P_{B_k}P_{B_k}^\top A\|_F^2\leq \|B-P_{B_k}P_{B_k}^\top B\|_F^2\leq \|B-P_{A_k}P_{A_k}^\top B\|_F^2\leq (1+\gamma)\|A-A_k\|_F^2$, which yields the factor $(1+\gamma)/(1-\gamma)$. This factor is at most $1+3\gamma$ for $\gamma\leq 1/3$, and the statement above is this bound with $\gamma=\epsilon/3$. Moreover, the factor $(1+\gamma)/(1-\gamma)$ is attained up to any constant by two orthogonal columns of squared norms $a>b$ with coreset weights $1-\gamma$ and $1+\gamma$, once $(1+\gamma)b>(1-\gamma)a$.\end{remark}

\cite{cmm17} showed that sampling according to ridge leverage scores gives a projection-cost preserving sample, and in fact a stronger $\epsilon$-approximator, which is an $\epsilon$-coreset. We discuss this in Section~\ref{sec:ridge}, where we also show that the $k$-subspace sensitivity is at most $(1+1/k)$ times the ridge leverage score. We hence summarize the result below.

\begin{corollary}
\label{thm:quantum_pcp}
Let $A\in \R^{d\times n}$, $\epsilon,\delta\in (0, 1)$, $k\leq \min\{n, d\}$ be a positive integer. There exists a quantum algorithm $\textsc{QLowRankCMM}(A, k, \epsilon, \delta)$ (the procedure $\textsc{QRLS}(A, k, \epsilon, \delta)$ of Theorem~\ref{thm:ridge} run on the columns of $A$) that constructs an $\epsilon$-coreset $C$ of $A$ for the $k$-subspace cost with probability at least $1-\delta$. The size of the coreset is at most $O(k\log (k/\delta)/\epsilon^{2})$, the runtime is $\wt O(n^{0.5}dk^{0.5}\epsilon^{-1}+dk^{\omega-1})$,  and the number of queries to the columns of $A$ is $\wt O(\epsilon^{-1}n^{0.5}k^{0.5})$.

Moreover, for any $s\geq k$ there is a fixed-size variant $\textsc{QRLS}_s(A, k, s, \delta)$ that, with probability at least $1-\delta$, returns constant-factor overestimates $\wt \tau_i$ of the rank-$k$ ridge leverage scores $\ov \tau_i(A)$ of the columns of $A$ with $\sum_i \wt \tau_i=O(k)$, together with a sample of the columns in which column $i$ is included independently with probability $q_i=\min\{1, s\cdot \wt \tau_i/k\}$ (expected size $O(s)$), in $\wt O(n^{0.5}d(k^{0.5}+s^{0.5})+dk^{\omega-1})$ time and $\wt O(n^{0.5}(k^{0.5}+s^{0.5}))$ queries to the columns of $A$.
\end{corollary}

\begin{proof}
The coreset statement follows from applying Theorem~\ref{thm:ridge} to the columns of $A$. The fixed-size variant is obtained as for $\textsc{QLS}$ in Theorem~\ref{thm:ls}, with the oracle of Lemma~\ref{lem:classical_ridge}. The recursion of Theorem~\ref{thm:general} at constant accuracy and failure probability $\delta/2$ yields an $O(1)$-approximator $C_T$ of size $s'=\wt O(k)$ in $\wt O(n^{0.5}dk^{0.5}+dk^{\omega-1})$ time and $\wt O(n^{0.5}k^{0.5})$ column queries. By Lemma~\ref{lem:cmm_lem12}, we have $w_i(A, C_T)=\Theta(1)\cdot \ov\tau_i(A)$ for every $i$, and Algorithm~\ref{alg:classical_ridge_ls} evaluates this weight up to a constant factor in $\wt O(d)$ time per query. After scaling by a constant, we obtain overestimates $\wt \tau_i$ with $\sum_i \wt \tau_i=O(\sum_i \ov \tau_i(A))=O(k)$ by Lemma~\ref{lem:classical_ridge}. Finally, Lemma~\ref{lem:q_sampling_1_d} with failure probability $\delta/2$ draws the sample, with $\sum_i q_i=O(s)$, in $\wt O(\sqrt{ns}\cdot d)$ time and $\wt O(\sqrt{ns})$ column queries.
\end{proof}

We note that in addition, $C$ is a column subset selection of $A$:
\begin{definition}[Rank-$k$ Column Subset Selection]\label{def:rank_k_subset_selection}
For $n' \leq n$, a subset of $A$'s columns $C \in \R^{d \times n'}$ (possibly with repetitions) is a $(1 + \epsilon)$ factor column subset selection if there  exists a rank-$k$ matrix $X \in \R^{n' \times n}$ with 
\begin{align*}
    \| A - C X \|_F^2 \leq (1 + \epsilon) \| A - A_k \|^2_F.
\end{align*}
\end{definition}
Note that for column subset selection alone, sampling by ridge leverage scores only needs an oversampling factor of $1/\epsilon$, rather than $1/\epsilon^2$:
\begin{lemma}[Theorem 7 of~\cite{cmm17}]
\label{lem:cmm_thm7}
Let $A\in \R^{d\times n}$, $k\leq \min\{n,d\}$, $\epsilon,\delta\in (0,1)$, and let $\wt \tau_i\geq \ov \tau_i(A)$ be overestimates of the rank-$k$ ridge leverage scores of the columns of $A$. Let $C$ consist of $t=O(\epsilon^{-1}(\log k+\log(1/\delta))\sum_i \wt \tau_i)$ columns of $A$, drawn independently with replacement, each with probability proportional to $\wt \tau_i$. Then, with probability at least $1-\delta$, $\|A-(CC^\dagger A)_k\|_F^2\leq (1+\epsilon)\|A-A_k\|_F^2$, and since $(CC^\dagger A)_k$ has rank $k$ and lies in the column span of $C$, $C$ is a $(1+\epsilon)$ factor rank-$k$ column subset selection of $A$ (Definition~\ref{def:rank_k_subset_selection}).
\end{lemma}
When the $\wt\tau_i$ are constant-factor approximations of $\ov\tau_i(A)$, by Lemma~\ref{lem:classical_ridge} we have $\sum_i\wt\tau_i=O(k)$, and hence $t=O(\epsilon^{-1}k\log(k/\delta))$. \cite{cmm17} state Theorem 7 for $t$ draws with replacement, and their Appendix B writes out independent sampling only for the $\epsilon^{-2}$ forms of their Theorem 5 and Lemma 20. The following lemma transfers Theorem 7 to the independent sampling of $\textsc{QRLS}_s$.

\begin{lemma}[Bernoulli form of Theorem 7 of~\cite{cmm17}]
\label{lem:cmm_thm7_bernoulli}
Under the hypotheses of Lemma~\ref{lem:cmm_thm7}, let $t$ be as there, with the constant in $t$ large enough that $t\geq 4\log(1/\delta)$, and let $p_i=\wt\tau_i/\sum_j\wt\tau_j$. Let $B$ be a random subset of the columns of $A$ in which column $i$ is included independently with probability $q_i\geq \min\{1, 2tp_i\}$. Then $\E|B|=\sum_i q_i$, and with probability at least $1-2\delta$, $B$ is a $(1+\epsilon)$ factor rank-$k$ column subset selection of $A$ (Definition~\ref{def:rank_k_subset_selection}). In particular, the inclusion probabilities $\min\{1, s\,\wt\tau_i/k\}$ of $\textsc{QRLS}_s(A, k, s, \delta')$ satisfy the hypothesis whenever $s\geq 2tk/\sum_j\wt\tau_j=O(\epsilon^{-1}k(\log k+\log(1/\delta)))$, a threshold that does not depend on $\sum_j\wt\tau_j$.
\end{lemma}
\begin{proof}
We first note that adding columns preserves column subset selection. Indeed, if every column of $C$ occurs in $C'$ and ${\rm rank}(X)\leq k$, then summing the rows of $X$ belonging to the draws of each column of $C'$ gives $X'$ with $C'X'=CX$ and ${\rm rank}(X')\leq k$. Let $X_1, X_2,\ldots$ be independent draws from the distribution $(p_i)_{i\in [n]}$, let $N\sim {\rm Poisson}(2t)$ be independent of them, and let $S_N=\{X_1,\ldots,X_N\}\subseteq [n]$. By Poisson splitting, the events $\{i\in S_N\}$ are independent with $\Pr[i\in S_N]=1-e^{-2tp_i}\leq \min\{1, 2tp_i\}\leq q_i$. Therefore, adding each $i\notin S_N$ independently with probability $r_i:=\big(q_i-(1-e^{-2tp_i})\big)/e^{-2tp_i}\in [0,1]$ gives a set $B\supseteq S_N$ with the law of the statement, and $\E|B|=\sum_i q_i$. By Lemma~\ref{lem:cmm_thm7}, with probability at least $1-\delta$, the columns $a_{X_1},\ldots,a_{X_t}$ are a $(1+\epsilon)$ factor rank-$k$ column subset selection of $A$. If this event and the event $\{N\geq t\}$ both hold, then $S_N\supseteq\{X_1,\ldots,X_t\}$, and hence, by monotonicity, $B\supseteq S_N$ is also such a selection. Applying Markov's inequality to $e^{-\theta N}$ with $\theta=\log 2$, we have $\Pr[N<t]\leq \Pr[N\leq t]\leq e^{-t}2^{t}=(2/e)^t\leq \delta$, since $t\geq 4\log(1/\delta)\geq \log(1/\delta)/\log(e/2)$. A union bound gives the probability $1-2\delta$. For $A\neq 0$, let $\sigma_1\geq \cdots\geq \sigma_r>0$ be the nonzero singular values of $A$ and let $\lambda=\|A-A_k\|_F^2/k$. Then, as in the proof of Lemma~\ref{lem:classical_ridge}, we have $\sum_i\wt\tau_i\geq\sum_i\ov\tau_i(A)=\sum_{j=1}^r\sigma_j^2/(\sigma_j^2+\lambda)$, and this sum is at least $1/2$. Indeed, if $r\leq k$, then $\lambda=0$ and every term equals $1$. If $r>k$ and $\sigma_k^2\geq \lambda$, then the $k$-th term is at least $1/2$. If $r>k$ and $\sigma_k^2<\lambda$, then $\sigma_j^2+\lambda<2\lambda$ for $j>k$, and thus these terms sum to at least $\sum_{j>k}\sigma_j^2/(2\lambda)=k/2$. Consequently, $t\geq c\,\epsilon^{-1}\log(1/\delta)\sum_i\wt\tau_i\geq 4\log(1/\delta)$ once the constant $c$ in $t$ is at least $8$. For $A=0$, every column subset is a $(1+\epsilon)$ factor selection. For the last claim, $\min\{1, s\,\wt\tau_i/k\}\geq \min\{1, 2tp_i\}$ for all $i$ as soon as $s/k\geq 2t/\sum_j\wt\tau_j$, and by the definition of $t$, we have $2tk/\sum_j\wt\tau_j=O(\epsilon^{-1}(\log k+\log(1/\delta)))\cdot k$.

\end{proof}
The weights $1/q_i$ that $\textsc{QRLS}_s$ attaches to the sampled columns play no role in the column subset selection property, since the matrix $X$ in Definition~\ref{def:rank_k_subset_selection} is unconstrained apart from its rank.
 We utilize this fact to further derive an algorithm for outputting a low-rank approximation of $A$, which could subsequently be generalized to tensors. We state a tool for solving a bilinear multiple response regression.

\begin{lemma}[Generalized Low-Rank Approximation~\citep{ft07}]
Let $A\in \R^{d\times n}$, $B\in \R^{d\times p}$ and $C\in \R^{q\times n}$, let $k\leq \min\{n,d\}$ be a positive integer. The following bilinear regression problem
\begin{align*}
    \min_{X\in \R^{p\times q}:{\rm rank}(X)\leq k} \|A-BXC\|_F^2
\end{align*}
is minimized by $X_*=B^\dagger [P_B AP_C]_k C^\dagger$ where $P_B, P_C$ are the projection matrices onto $B, C$ respectively.
\end{lemma}

\begin{algorithm}[!ht]
\caption{Quantum low-rank approximation.}
\label{alg:quantum_lra}
\begin{algorithmic}[1]
\Procedure{QLowRank}{$A\in \R^{d\times n}, k,\epsilon$}
\State $k_1\gets O(\epsilon^{-1}k\log k)$ \Comment{Lemma~\ref{lem:cmm_thm7_bernoulli} at $\delta=0.0005$}
\State $k_2\gets O(k_1\log k_1+\epsilon^{-1}k_1)$
\State $k_2^\prime\gets O(\epsilon^{-2}k_1\log k_1)$
\State $k_3\gets O(\epsilon^{-2}k_2\log k_2)$
\State $C\gets \textsc{QRLS}_s(A, k, k_1, 0.001)$\label{ln:lra:C} \Comment{$C\in \R^{d\times k_1}$, a $(1+\epsilon)$ rank-$k$ column subset selection (Lemma~\ref{lem:cmm_thm7_bernoulli})}
\State $S\gets \textsc{QLS}(C, k_2, 0.001)$ \Comment{$S\in \R^{k_2\times d}$, Theorem~\ref{thm:ls}.}
\State $T_1\gets \textsc{QLS}(C, k_2^\prime, 0.001)$ \Comment{$T_1\in \R^{k_2^\prime\times d}$, Theorem~\ref{thm:ls}.}
\State $T_2\gets \textsc{QLS}(SA, k_3, 0.001)$ \Comment{$T_2\in \R^{n\times k_3}$, Theorem~\ref{thm:ls}.}
\State Solve $\min_{Y\in \R^{k_1\times k_2}:{\rm rank}(Y)\leq k} \|T_1CYSAT_2-T_1AT_2 \|_F^2$ by the lemma of~\cite{ft07}:
\State $\wh M\gets (T_1C)^\dagger [P_{T_1C} T_1 AT_2P_{SAT_2}]_k (SAT_2)^\dagger$ \Comment{$\wh M\in \R^{k_1\times k_2}$ and ${\rm rank}(\wh M)\leq k$.}
\State Write $\wh M$ into factored form $\wh M=\wh X\wh Y$ \Comment{$\wh X\in \R^{k_1\times k}, \wh Y\in \R^{k\times k_2}$.}
\State \Return $C\wh X$, $\wh YSA$ in factored form
\EndProcedure
\end{algorithmic}
\end{algorithm}

\begin{theorem}
\label{thm:quantum_lra}
Let $A\in \R^{d\times n}$ and $\epsilon\in (0,0.1)$, and let $k\leq \min\{d,n\}$ be a positive integer. Then, there exists a quantum algorithm (Algorithm~\ref{alg:quantum_lra}) that outputs a pair of matrices $M\in \R^{d\times k}, N\in \R^{n\times k}$ of rank at most $k$, represented in factored form (Remark~\ref{rem:factored_output}), such that
\begin{align*}
    \|A-MN^\top\|_F^2 \leq & ~ (1+\epsilon)\cdot \|A-A_k\|_F^2
\end{align*}
holds with probability at least $0.99$.
Moreover, Algorithm~\ref{alg:quantum_lra} runs in time
\begin{align*}
    & ~ \wt O(\epsilon^{-0.5}n^{0.5}dk^{0.5}+dk^{\omega-1}+\epsilon^{-2.5}d^{0.5}k^{1.5}+\epsilon^{-4}n^{0.5}k^{1.5}+\poly(k/\epsilon)),
\end{align*}
 and uses $\wt O(\epsilon^{-2}n^{0.5}k^{0.5})$ queries to the columns and $\wt O(\epsilon^{-1.5}d^{0.5}k^{0.5})$ queries to the rows of $A$.
\end{theorem}

\begin{remark}\label{rem:factored_output}
The factors are returned implicitly. The factor $M=C\wh X$ is specified by the $k_1$ sampled columns of $A$ together with $\wh X$, and $N^\top=\wh YSA$ is specified by the row indices and weights of $S$ together with $\wh Y$. Any entry of $M$ or $N$ can be computed on demand in $\wt O(\poly(k/\epsilon))$ time, and this is the same output convention as our kernel low-rank approximation result. Materializing the support $SA$ of the second factor costs an additional $\wt O(nk_2)=\wt O(\epsilon^{-2}nk)$ time, and forming $M$ and $N$ as explicit dense matrices by the naive products $C\wh X$ and $(\wh YSA)^\top$ costs $\wt O(\epsilon^{-1}dk^2+\epsilon^{-2}nk^2)$. The cost of the first product would dominate the leading term unless $n\gtrsim \epsilon^{-1}k^3$, which our assumption on the dimensions gives, and that of the second product unless $d\gtrsim \epsilon^{-1.5}n^{0.5}k^{1.5}$. In contrast, no cost linear in $n$ is incurred in factored form.
\end{remark}

\begin{proof}

Write $\OPT:=\|A-A_k\|_F^2$. We first prove correctness, in three steps.

\emph{Step 1: a good column subset.} Suppose that $\textsc{QRLS}_s(A, k, k_1, 0.001)$ succeeds. By Corollary~\ref{thm:quantum_pcp}, the scores $\wt\tau_i$ it computes are constant-factor overestimates of $\ov\tau_i(A)$ with $\sum_i\wt\tau_i=O(k)$, and $C$ includes column $i$ independently with probability $\min\{1, k_1\wt\tau_i/k\}$. We apply Lemma~\ref{lem:cmm_thm7_bernoulli} with $\delta=0.0005$, and by its last claim, the hypothesis $k_1\geq 2tk/\sum_j\wt\tau_j$ holds when the constant in $k_1=O(\epsilon^{-1}k\log k)$ is large enough. Consequently, with probability at least $1-2\delta=0.999$, $C$ is a $(1+\epsilon)$ factor rank-$k$ column subset selection in the sense of Definition~\ref{def:rank_k_subset_selection}, meaning that there is a rank-$k$ matrix $X^\star\in \R^{k_1\times n}$ with $\|A-CX^\star\|_F^2\leq (1+\epsilon)\OPT$.

\emph{Step 2: a good rank-$k$ solution in the row span of $SA$ exists.} Let $Z:=CX^\star\in \R^{d\times n}$, which has rank at most $k$ and whose column span is contained in that of $C$. Therefore, for every row index $i\in [d]$, the leverage score of the $i$-th row of $Z$ is at most that of the $i$-th row of $C$, since both are the $i$-th diagonal entries of the orthogonal projectors onto the respective column spans and the projector for $Z$ is dominated by the one for $C$ in the positive semidefinite order. The matrix $S$ samples rows independently with probabilities $\min\{1, c(\log k_1+\epsilon^{-1})\cdot \wt\tau_i\}$, where $\wt\tau_i=\Theta(\tau_i(C))$ are computed by $\textsc{QLS}$ in Theorem~\ref{thm:ls}. For $c$ large enough, these probabilities dominate $\min\{1, c^\prime(\log k+\epsilon^{-1})\tau_i(Z)\}$, and since $\sum_i \wt\tau_i=O(k_1)$, the expected number of sampled rows is $k_2=O(k_1\log k_1+\epsilon^{-1}k_1)$. The proof of Lemma~\ref{lem:leverage_score_optimal_solution} only uses that the sampling probabilities dominate the leverage scores of the design, with the sample size governed by their sum. Thus, we can apply this lemma to the design $Z$ with these probabilities in place of the leverage scores of $Z$, and for $Y_S:=(SZ)^\dagger SA$ this gives
\begin{align*}
    \|ZY_S-A\|_F^2\leq (1+\epsilon)\cdot \min_{Y}\|ZY-A\|_F^2\leq (1+\epsilon)\cdot \|Z-A\|_F^2\leq (1+\epsilon)^2\OPT,
\end{align*}
where the second inequality takes $Y=I$. Since $ZY_S=C\cdot X^\star (SCX^\star)^\dagger\cdot SA=C\wh Y_0 SA$ with $\wh Y_0:=X^\star (SCX^\star)^\dagger\in \R^{k_1\times k_2}$ of rank at most $k$, we have
\begin{align*}
    \min_{Y\in \R^{k_1\times k_2}:{\rm rank}(Y)\leq k}\|A-CYSA\|_F^2\leq (1+\epsilon)^2\OPT.
\end{align*}

\emph{Step 3: finding such a solution.} Write $f(Y):=\|CYSA-A\|_F^2$, $g(Y):=\|T_1(CYSA-A)\|_F^2$, and $h(Y):=\|T_1(CYSA-A)T_2\|_F^2$. The matrix $T_1$ samples $k_2^\prime=O(\epsilon^{-2}k_1\log k_1)$ rows of $C$ according to leverage scores, $C$ has rank at most $k_1$, and $CYSA$ is always of the form $CX$. Therefore, by Lemma~\ref{lem:leverage_score_weak_affine} with design $C$ and response $A$, there is a constant $c_1$ such that $g(Y)=(1\pm\epsilon)f(Y)+c_1$ for all $Y$. Similarly, $T_2$ samples $k_3=O(\epsilon^{-2}k_2\log k_2)$ columns of $SA$ according to the leverage scores of the rows of $(SA)^\top$, and $SA$ has rank at most $k_2$. Applying Lemma~\ref{lem:leverage_score_weak_affine} with design $(SA)^\top$ and response $(T_1A)^\top$, we obtain a constant $c_2$ such that $\|(WSA-T_1A)T_2\|_F^2=(1\pm\epsilon)\|WSA-T_1A\|_F^2+c_2$ for all $W$, and taking $W=T_1CY$ gives $h(Y)=(1\pm\epsilon)g(Y)+c_2$ for all $Y$. Let $\wh M$ minimize $h$ over $\{Y:{\rm rank}(Y)\leq k\}$. By the lemma of~\cite{ft07}, we have $\wh M=(T_1C)^\dagger [P_{T_1C}T_1AT_2P_{SAT_2}]_k (SAT_2)^\dagger$, which the algorithm computes, and ${\rm rank}(\wh M)\leq k$. Comparing with $\wh Y_0$ of Step 2, we have $h(\wh M)\leq h(\wh Y_0)$, and since the constant $c_2$ cancels, we obtain $(1-\epsilon)g(\wh M)\leq (1+\epsilon)g(\wh Y_0)$. Let $\wh B=CC^\dagger A-A$ be the residual of the unconstrained problem $\min_X\|CX-A\|_F^2$. The constant $c_1=\|T_1\wh B\|_F^2-\|\wh B\|_F^2$ does not cancel, but it is small. Since $T_1$ samples with rescaling, $\|T_1\wh B\|_F^2$ is an unbiased estimator of $\|\wh B\|_F^2$, and hence, by Markov's inequality, $\|T_1\wh B\|_F^2\leq 1000\|\wh B\|_F^2$ with probability at least $0.999$. Consequently, $|c_1|\leq 999\|\wh B\|_F^2\leq 999 f(\wh Y_0)$, where the last inequality holds because the unconstrained minimum is at most $f(\wh Y_0)$. Substituting $g=(1\pm\epsilon)f+c_1$ on both sides, we have
\begin{align*}
    (1-\epsilon)^2 f(\wh M)\leq (1+\epsilon)^2 f(\wh Y_0)+2\epsilon\cdot |c_1|\leq \big((1+\epsilon)^2+1998\epsilon\big)\cdot f(\wh Y_0),
\end{align*}
which implies $f(\wh M)\leq (1+O(\epsilon))\cdot (1+\epsilon)^2\OPT$, and rescaling $\epsilon$ by a constant gives $\|A-C\wh M SA\|_F^2\leq (1+\epsilon)\OPT$. Moreover, the output $M=C\wh X$, $N^\top=\wh YSA$ with $\wh X\wh Y=\wh M$ has rank at most $k$. The call to $\textsc{QRLS}_s$, the three $\textsc{QLS}$ calls, Lemma~\ref{lem:cmm_thm7_bernoulli} applied with $\delta=0.0005$, Lemma~\ref{lem:leverage_score_optimal_solution}, the two applications of Lemma~\ref{lem:leverage_score_weak_affine}, and the Markov step each fail with probability at most $0.001$, and hence, by a union bound, the success probability is at least $0.99$.

For the running time, by the fixed-size variant of Corollary~\ref{thm:quantum_pcp} with $s=k_1$, generating $C$ takes $\wt O(\epsilon^{-0.5}n^{0.5}dk^{0.5}+dk^{\omega-1})$ time. Generating $S$ takes $\wt O(\sqrt{dk_2}k_1+k_1^\omega)=\wt O(\epsilon^{-2}d^{0.5}k^{1.5}+\epsilon^{-\omega}k^\omega)$ time, and generating $T_1$ takes $\wt O(\sqrt{dk_2^\prime}k_1+k_1^\omega)=\wt O(\epsilon^{-2.5}d^{0.5}k^{1.5}+\epsilon^{-\omega}k^\omega)$ time, using $k_1=\wt O(\epsilon^{-1}k)$, $k_2=\wt O(\epsilon^{-2}k)$, and $k_2^\prime=\wt O(\epsilon^{-3}k)$. We never materialize $SA$, since the algorithm only accesses $SAT_2$, a $\poly(k/\epsilon)$-size submatrix of $A$, and the factor $\wh YSA$ is returned implicitly as in Remark~\ref{rem:factored_output}. Generating $T_2$ takes $\wt O(\sqrt{nk_3} k_2+k_2^\omega)=\wt O(\epsilon^{-4}n^{0.5}k^{1.5}+\epsilon^{-2\omega}k^{\omega})$ time, using $k_3=\wt O(\epsilon^{-4}k)$. Finally, computing $T_1 C$, $SAT_2$, their pseudoinverses, and their projections takes $\poly(k/\epsilon)$ time, since forming these matrices is simply selecting and rescaling entries, and the resulting matrices are of size $\poly(k/\epsilon)$. The matrix $T_1AT_2$ is obtained in the same way, hence $\wh M$ can be computed in $\poly(k/\epsilon)$ time. Summing up these bounds gives the running time stated in the theorem. For the query counts: generating $C$ uses $\wt O(n^{0.5}(k^{0.5}+k_1^{0.5}))=\wt O(\epsilon^{-0.5}n^{0.5}k^{0.5})$ queries to columns of $A$, the Grover searches for $S$ and $T_1$ use $\wt O(\sqrt{dk_2}+\sqrt{dk_2^\prime})=\wt O(\epsilon^{-1.5}d^{0.5}k^{0.5})$ queries to rows, and the search for $T_2$ uses $\wt O(\sqrt{nk_3})=\wt O(\epsilon^{-2}n^{0.5}k^{0.5})$ queries to columns.
\end{proof}
\section{Quantum Kernel Low-Rank Approximation}
\label{sec:kernel}
Given a set of points $\{ x_1,\ldots,x_n \}\subset \R^d$ and a positive definite kernel function ${\sf K}: \R^d\times \R^d\rightarrow \R$, the kernel low-rank approximation problem asks to find a pair $M, N\in \R^{n\times k}$ such that $\|K-MN^\top\|_F^2 \leq (1+\epsilon)\cdot \|K-K_k\|_F^2$, where $K\in \R^{n\times n}$ is the kernel matrix induced by ${\sf K}$, with $K_{i, j} = {\sf K}(x_i, x_j)$. Note that explicitly forming the matrix $K$ takes $\Omega(n^2)$ evaluations of ${\sf K}(\cdot, \cdot)$, which is usually too expensive to be afforded. Since ${\sf K}$ is positive definite, there exists a feature mapping $\phi: \R^d\rightarrow \R^{d'}$ such that $K=\Phi \Phi^\top$ where $\Phi\in \R^{n\times d'}$ with the $i$-th row being $\phi(x_i)$. \cite{mm17} gives a low-rank approximation for $\Phi$ using $\wt O(\epsilon^{-1}nk)$ evaluations of ${\sf K}(\cdot, \cdot)$ and an additional $\wt O(\epsilon^{-2}nk^2)$ time (their Table 1 and Theorem 17). \cite{mw17} show that the above guarantee for $K$ can be achieved with $\wt O(\epsilon^{-2.5}nk)$ kernel evaluations and an additional $\wt O(\epsilon^{-2(\omega-1)}nk^{\omega-1})$ time, and~\cite{bcw20} with $\wt O(\epsilon^{-1}nk)$ kernel evaluations and an additional $\wt O(\epsilon^{-(\omega-1)} nk^{\omega-1})$ time\footnote{Note that~\cite{mw17,bcw20} phrase their algorithm as a low-rank approximation for PSD matrix $A$, and their runtime is stated in terms of reads to $A$. Observe that a read to an entry of $A$ could be translated to one kernel evaluation.}. In this section, we will present a quantum algorithm based on the techniques developed in Sections~\ref{sec:general} and~\ref{sec:cmm}, that computes a low-rank approximation for the kernel matrix using \emph{a sublinear number of kernel evaluations and additional operations.} 

Before diving into our main result, we introduce some notations. We will extensively use $KD_1$ or $D_2^\top K D_1$ to denote a weighted sampling of $K$, in particular,
\begin{itemize}
    \item If $D\in \R^{n\times t}$, we use $D^\top K_i\in \R^{t}$ to denote the vector $v$ with $v_j:=D(j)\cdot {\sf K}(x_i, x_j)$, where $j\in D$ is the $j$-th sample of $D$, and $D(j)$ is the corresponding weight.
    \item If $D\in \R^{n\times t}$, we use ``$K D$ in factored form'' to denote a data structure that when the $i$-th row is queried, computes $v\in \R^t$ where $v_j:=D(j)\cdot {\sf K}(x_i, x_j)$ for $j\in D$.
    \item If $D_1\in \R^{n\times t_1}$ and $D_2\in \R^{n\times t_2}$, we use ``$D_2^\top KD_1$ in factored form'' to denote a data structure that supports queries to either row or column, where for $i$-th row, it computes a vector $v^{\rm row}\in \R^{t_1}$ where $v^{\rm row}_j:=D_1(j)D_2(i)\cdot {\sf K}(x_i, x_j)$ for $j\in D_1$ and $i\in D_2$. Similarly the operation applies to the column.
    \item Sometimes given $KD\in \R^{n\times t_1}$ in factored form, we will compose it with another matrix $M\in \R^{t_1\times t_2}$. We then use ``$KDM$ in factored form'' to denote a data structure that supports row queries, such that when the $i$-th row is queried, it returns $M^\top v$ where $v_j:=D(j)\cdot {\sf K}(x_i, x_j)$ for $j\in D$.
\end{itemize}

\begin{algorithm}[!ht]
\caption{Quantum kernel low-rank approximation.}
\label{alg:quantum_bcw}
\begin{algorithmic}[1]
\Procedure{QLowRankKernel}{$\{x_1,\ldots,x_n\}\in (\R^d)^n,{\sf K}:\R^d\times \R^d\rightarrow \R, k, \epsilon, \delta$}
\State $c\gets 1000$
\State $t\gets c\sqrt{\frac{nk}{\epsilon}}\log(n/\delta)$ 
\State $k'\gets 100(k/\epsilon)\log(12k/(\epsilon\delta))$\label{ln:bcw:kp} \Comment{$k'=\wt O(k/\epsilon)$}
\State $D_1\gets \textsc{QNystr{\"o}mKernel}(\{x_1,\ldots,x_n\}, {\sf K}, k', \delta/6)$ with the probabilities of its last sampling step replaced by $p_i=\min\{1, \sqrt{\epsilon n/k}\cdot 16 q_i\log(12n/\delta)\}$\label{ln:bcw:D1} \Comment{$D_1\in \R^{n\times t}$}
\State $D_2\gets \textsc{QNystr{\"o}mKernel}(\{x_1,\ldots,x_n\}, {\sf K}, k', \delta/6)$ with the probabilities of its last sampling step replaced by $p_i=\min\{1, \sqrt{\epsilon n/k}\cdot 16 q_i\log(12n/\delta)\}$ \label{ln:bcw:D2} \Comment{$D_2\in \R^{n\times t}$}
\State $C\gets KD_1$ in factored form \Comment{$C\in \R^{n\times t}$.}
\State $R\gets D_2^\top K D_1$ in factored form \Comment{$R\in \R^{t\times t}$.}
\State $\epsilon_0\gets 0.01$
\State $\wt R\gets \textsc{QLowRankCMM}(R^\top, k/\epsilon, \epsilon_0, \delta/6)$\label{ln:bcw:Rt} \Comment{Corollary~\ref{thm:quantum_pcp} on the columns of $R^\top$}
\State $Z\gets \text{top-$k/\epsilon$ left singular vectors of $\wt R$}$\label{ln:bcw:Z} \Comment{$Z\in \R^{t\times k/\epsilon}$, orthonormal columns}
\State \Comment{Solve the regression $\min_{W\in \R^{n\times k/\epsilon}} \|C-WZ^\top\|$.}
\State Implement oracle for $p_i=\min\{1, \frac{\epsilon k'}{k}\cdot \|z_i\|_2^2\}$, $i\in [t]$\label{ln:bcw:pZ} \Comment{$\sum_{i=1}^t p_i\leq k'$}
\State $D_3\gets \textsc{QSample}(p, 0.001)$\label{ln:bcw:D3} \Comment{$D_3\in \R^{t\times k'}$ samples rows of $Z$}
\State \Comment{Solve the surrogate regression $\min_W \|CD_3-WZ^\top D_3\|$.}
\State $W\gets CD_3(Z^\top D_3)^\dagger$ in factored form\label{ln:bcw:W} \Comment{$W=K\,(D_1D_3(Z^\top D_3)^\dagger)\in \R^{n\times k/\epsilon}$}
\State \Comment{Solve the regression $\min_{Y: {\rm rank}(Y)\leq k} \|K-W Y W^\top \|_F^2$.}
\State $D_4\gets \textsc{QLS}(W, k'/\epsilon^2, \delta/6)$ \Comment{$D_4\in \R^{n\times k'/\epsilon^2}$, sample rows.}
\State $D_5\gets \textsc{QLS}(W, k'/\epsilon^2, \delta/6)$ \Comment{$D_5\in \R^{n\times k'/\epsilon^2}$, sample columns.}
\State Compute $D_4^\top W$ and $W^\top D_5$ \Comment{$D_4^\top W\in \R^{k'/\epsilon^2\times k/\epsilon}$, $W^\top D_5\in \R^{k/\epsilon\times k'/\epsilon^2}$.}
\State $P_{D_4^\top W}\gets D_4^\top W (W^\top D_4 D_4^\top W)^\dagger W^\top D_4, P_{W^\top D_5}\gets D_5^\top W (W^\top D_5 D_5^\top W)^\dagger W^\top D_5$\label{ln:bcw:P} \Comment{projectors onto ${\rm col}(D_4^\top W)$ and ${\rm row}(W^\top D_5)$}
\State Compute $D_4^\top K D_5$ \Comment{$D_4^\top K D_5\in \R^{k'/\epsilon^2\times k'/\epsilon^2}$.}
\State Compute $[P_{D_4^\top W} (D_4^\top K D_5) P_{W^\top D_5}]_k$ \Comment{Of rank $k$.}
\State $Y_*\gets (D_4^\top W)^\dagger [P_{D_4^\top W} (D_4^\top K D_5) P_{W^\top D_5}]_k (W^\top D_5)^\dagger$ \Comment{$Y_*\in \R^{k/\epsilon\times k/\epsilon}$ of rank $k$.}
\State $U_*\gets \text{top-$k$ left singular vectors of $Y_*$}$ \Comment{$U_*\in \R^{k/\epsilon\times k}$.}
\State $D_6\gets \textsc{QLS}(WU_*, c(k\log k+k/\epsilon), \delta/6)$ \Comment{$D_6\in \R^{n\times O(k\log k+k/\epsilon)}$.}
\State \Comment{Solve the regression $\min_{N\in \R^{n\times k}} \|D_6^\top K-D_6^\top WU_* N^\top \|_F^2$.}
\State $N^\top\gets (D_6^\top W U_*)^\dagger (D_6^\top K)$\label{ln:bcw:N} \Comment{$N\in \R^{n\times k}$ in factored form}
\State \Return $M:=WU_*$ and $N$ in factored form
\EndProcedure
\end{algorithmic}
\end{algorithm}

\begin{algorithm}[!ht]
\caption{Quantum generalized ridge leverage score sampling via recursive Nystr{\"o}m method.}
\label{alg:quantum_nystrom}
\begin{algorithmic}[1]
\Procedure{QNystr{\"o}mKernel}{$\{x_1,\ldots,x_n\}\in (\R^d)^n, {\sf K}: \R^d\times \R^d\rightarrow \R, s, \delta$}
\State $c\gets 100$
\State $T\gets \max\{1, \lceil\log_2(n/s)\rceil\}$, $\delta'\gets \delta/(2T)$ \Comment{$\delta'$ is the failure parameter of each level of the recursion.}
\State Let $S_0\subset_{1/2} S_1\subset_{1/2} \ldots \subset_{1/2} S_T=[n]$ \Comment{Starting from $[n]$, uniformly sampling half of the indices, and $\E|S_0|\leq s$.}
\State Set $k$ to be the largest integer with $ck\log(2k/\delta)\leq s$
\State Let $D_0\in \R^{n\times |S_0|}$ be the sampling matrix of $S_0$ (weights $1$)
\For{$t=1\to T+1$}
\State $M_{t-1}\gets \{D_{t-1}(i) D_{t-1}(j)\cdot{\sf K}(x_i, x_j) \}_{(i,j)\in D_{t-1}\times D_{t-1}}$ \Comment{$M_{t-1}=C_{t-1}C_{t-1}^\top$ for $C_{t-1}=D_{t-1}^\top K^{1/2}$, of size $\wt O(s)\times \wt O(s)$.}
\State $\lambda\gets \frac{1}{k}\sum_{i>k} \sigma_i(M_{t-1})$ \Comment{$\lambda=\|C_{t-1}-(C_{t-1})_k\|_F^2/k$.}
\State $\wh M\gets (M_{t-1}+\lambda I)^{-1}$
\State $S\gets S_{\min\{t, T\}}$ \Comment{$S=[n]$ at the additional level $t=T+1$.}
\State \Comment{$D_{t-1}^\top K_i:=\{D_{t-1}(j)\,{\sf K}(x_i, x_j)\}_{j\in D_{t-1}}$ for $i\in S$, where $D_{t-1}(j)$ = weight of $x_j$ in $D_{t-1}$}
\State Implement oracle for $q_i\gets \frac{5   }{\lambda}\cdot ({\sf K}(x_i, x_i)- (D_{t-1}^\top K_i)^\top  \wh M D_{t-1}^\top K_i)$ for $i\in S$ \Comment{$q_i=5\,w_i(\Phi_S, C_{t-1})$ for $\Phi=K^{1/2}$, by the Woodbury identity (Lemma~\ref{lem:qnystrom}).}
\State \Comment{$p_i=\min\{1, 16q_i\log(2k/\delta')\}$ for $t\leq T$, and $p_i=\min\{1, 16q_i\log(2k/\delta)\}$ at $t=T+1$ (as in Algorithm~\ref{alg:qls})}
\State $\wt D_t\gets \textsc{QSample}(p, \delta/(2(T+2)))$ \Comment{$\wt D_t\in \R^{|S|\times \wt O(s)}$.}
\State $D_t\gets D_{S}\cdot \wt D_t$ \Comment{$D_t\in \R^{n\times \wt O(s)}$, weights $p_i^{-1/2}$.}
\EndFor
\State \Return $D_{T+1}$ and the oracle $q$ of the last level
\EndProcedure
\end{algorithmic}
\end{algorithm}

\begin{lemma}[Guarantee of Algorithm~\ref{alg:quantum_nystrom}]
\label{lem:qnystrom}
Let $K$ be the kernel matrix of $X=\{x_1,\ldots,x_n\}$ and let $\Phi:=K^{1/2}$, a symmetric matrix with $\Phi^\top\Phi=K$ whose rows $\phi_i$ satisfy $\phi_i^\top\phi_j={\sf K}(x_i,x_j)$. Let $s\geq 1$, $\delta\in (0,1)$, and let $k$ be the integer set by $\textsc{QNystr{\"o}mKernel}(X, {\sf K}, s, \delta)$. With probability at least $1-\delta$, the procedure returns a sample $D$ of $[n]$ in which $i$ is included independently with probability $p_i=\min\{1, 16\log(2k/\delta)\,q_i\}$ and weight $p_i^{-1/2}$, where the values $q_i$ of its last oracle satisfy
\begin{align*}
\ov\tau_i(\Phi)\leq q_i\leq 6\,\ov\tau_i(\Phi)\qquad \text{for every $i\in [n]$},
\end{align*}
$\ov\tau_i(\Phi)=\phi_i^\top (K+\lambda_{\Phi_k}I)^{-1}\phi_i$ being the rank-$k$ ridge leverage scores of the rows of $\Phi$ (Definition~\ref{def:ridge_lg}), and in particular $\sum_i q_i\leq 12k$, and the sample has expected size $O(k\log(k/\delta))=O(s)$. If the probabilities of the last sampling step are replaced by $\min\{1, \beta\cdot 16\log(2n/\delta)\,q_i\}$ for some $\beta\geq 1$, the same holds with expected size $O(\beta k\log(n/\delta))$. The procedure runs in expected time $\wt O(\sqrt{ns}\,(s{\cal T}_{\sf K}+s^2)+s^2{\cal T}_{\sf K}+s^\omega)$ and makes $\wt O(\sqrt{ns}\cdot s)$ kernel evaluations, and the rescaled last step costs $\wt O(\sqrt{n\beta s}\,(s{\cal T}_{\sf K}+s^2))$ additional time and $\wt O(\sqrt{n\beta s}\cdot s)$ additional kernel evaluations.
\end{lemma}
\begin{proof}
As in the proof of Theorem~\ref{thm:ridge} and in $\textsc{QLS}$ of Algorithm~\ref{alg:qls}, Algorithm~\ref{alg:quantum_nystrom} runs Algorithm~\ref{alg:quantum_iterate} on the rows of $\Phi$ with the weights of Lemma~\ref{lem:classical_ridge} at rank $k$ and constant accuracy at every level, followed by one additional oracle level. The chain $S_0\subset_{1/2}\cdots\subset_{1/2}S_T=[n]$ is the chain of Algorithm~\ref{alg:quantum_iterate} with $\E|S_0|=n2^{-T}\leq s$, and $C_{t-1}:=D_{t-1}^\top\Phi$ is the weighted sample of level $t-1$, with $C_0=\Phi_{S_0}$. We never form the rows of $\Phi$, since every quantity the algorithm uses is an inner product $\phi_i^\top\phi_j={\sf K}(x_i,x_j)$.

\emph{The oracle.} Note that $M_{t-1}=D_{t-1}^\top KD_{t-1}=C_{t-1}C_{t-1}^\top$, so the value $\lambda$ computed at level $t$ is $\frac1k\sum_{i>k}\sigma_i(C_{t-1})^2=\lambda_{(C_{t-1})_k}$. For $\lambda>0$, the Woodbury identity gives
\begin{align*}
(C_{t-1}^\top C_{t-1}+\lambda I)^{-1} = & ~ \frac{1}{\lambda}\Big(I-\Phi^\top D_{t-1}(M_{t-1}+\lambda I)^{-1}D_{t-1}^\top\Phi\Big).
\end{align*}
Since $\Phi\phi_i=K_i$ is the $i$-th column of $K$, the weight in Lemma~\ref{lem:classical_ridge} of a row $\phi_i$ with $i\in S$ is
\begin{align*}
w_i(\Phi_S, C_{t-1}) = & ~ \phi_i^\top (C_{t-1}^\top C_{t-1}+\lambda I)^{-1}\phi_i \\
= & ~ \frac{1}{\lambda}\Big({\sf K}(x_i,x_i)-(D_{t-1}^\top K_i)^\top (M_{t-1}+\lambda I)^{-1}D_{t-1}^\top K_i\Big),
\end{align*}
so the oracle of Algorithm~\ref{alg:quantum_nystrom} returns $q_i=5\,w_i(\Phi_S, C_{t-1})$ exactly. Once $\wh M$ is formed, each query takes $\wt O(s)$ kernel evaluations and $\wt O(s^2)$ time. If $\lambda=0$, i.e., ${\rm rank}(M_{t-1})\leq k$, the oracle follows the two cases of Algorithm~\ref{alg:classical_ridge_ls}, which we omit from the display. It returns $\infty$ if ${\sf K}(x_i,x_i)>(D_{t-1}^\top K_i)^\top M_{t-1}^\dagger D_{t-1}^\top K_i$, i.e., $\phi_i\notin{\rm span}(C_{t-1}^\top C_{t-1})$, and $5\,(D_{t-1}^\top K_i)^\top (M_{t-1}^\dagger)^2 D_{t-1}^\top K_i=5\,\phi_i^\top(C_{t-1}^\top C_{t-1})^\dagger\phi_i$ otherwise.

\emph{The recursion.} As in the proof of Theorem~\ref{thm:ridge}, we use Lemma~\ref{thm:ridge_via_unif} at accuracy $0.01$ and failure probability $\delta'=\delta/(2T)$ at each of the $T$ levels in place of items 2 and 3 of Assumption~\ref{assumption:recurse}. Since the oracle is exact, the factor $1\pm 0.04$ of Lemma~\ref{lem:cmm_lem12} is absorbed by the literal constant $5$. Moreover, the oversampling of Lemma~\ref{thm:ridge_via_unif} is absorbed by the constant $16$, which, like $c$, stands for a sufficiently large constant. Hence, with probability at least $1-T\delta'=1-\delta/2$, every $C_t$ is a $0.01$-approximator of $\Phi_{S_t}$ of expected size $O(k\log(k/\delta'))=O(s+k\log\log n)=\wt O(s)$, and $C_T$ is a $0.01$-approximator of $\Phi$.

\emph{The additional level.} By Lemma~\ref{lem:cmm_lem12} with $\epsilon=0.01$ and $C=\Phi$, we have $w_i(\Phi, C_T)=(1\pm 0.04)\,w_i(\Phi,\Phi)=(1\pm 0.04)\,\ov\tau_i(\Phi)$ for every $i$, and hence the last oracle returns $q_i=5\,w_i(\Phi, C_T)\in [4.8\,\ov\tau_i(\Phi), 5.2\,\ov\tau_i(\Phi)]$. Thus, $\ov\tau_i(\Phi)\leq q_i\leq 6\,\ov\tau_i(\Phi)$ for every $i$, and $\sum_iq_i\leq 5.2\sum_i\ov\tau_i(\Phi)\leq 10.4\,k\leq 12k$ by Lemma~\ref{lem:classical_ridge}. The final sample is drawn by Lemma~\ref{lem:q_sampling_1_d} with $\sum_ip_i\leq 16\log(2k/\delta)\sum_iq_i=O(k\log(k/\delta))$, and with the rescaling this bound becomes $O(\beta k\log(n/\delta))$. The sampler is called at most $T+2$ times, for $S_0$, the $T$ levels, and the additional level, each time with failure probability $\delta/(2(T+2))$, so the failure probability is at most $\delta/2+\delta/2=\delta$.

\emph{Running time.} The bounds follow from Theorem~\ref{thm:general} as in the proof of Theorem~\ref{thm:ridge}, using the oracle costs computed above. Each of the $T+1=O(\log n)$ oracle preparations forms $M_{t-1}$ from $\wt O(s^2)$ kernel evaluations, and computes its eigendecomposition and $\wh M$ in $\wt O(s^\omega)$ time. Each of the $T+2$ sampler calls makes $\wt O(\sqrt{ns})$ queries in expectation, except that the rescaled last call makes $\wt O(\sqrt{n\beta s})$ queries in expectation. Each query costs $\wt O(s)$ kernel evaluations and $\wt O(s^2)$ time. Here the expectations are understood as in Theorem~\ref{thm:general}.
\end{proof}

\begin{lemma}[Ridge leverage scores of the rows of $KD$]
\label{lem:row_rls}
Let $K\in \R^{n\times n}$ be positive semidefinite with ${\rm rank}(K)>m$, let $\rho_i:=\ov\tau_i(K^{1/2})$ be the rank-$m$ ridge leverage scores of the rows of $K^{1/2}$, $\lambda:=\frac1m\sum_{j>m}\sigma_j(K)=\lambda_{(K^{1/2})_m}$ and $\lambda':=\|K-K_m\|_F^2/m$. Let $D\in \R^{n\times t}$ select column $i$ of $K$ independently with probability $p_i\geq \min\{1, \beta\sqrt{n/m}\,\rho_i\}$ and weight $p_i^{-1/2}$, where $\beta\geq c\log(n/\delta)$ for a sufficiently large constant $c$, let $C:=KD$, and let $c_i^\top$ be the $i$-th row of $C$. Then with probability at least $1-\delta$, for every $i\in [n]$ and every $\mu\geq 0.9\lambda'$,
\begin{align*}
c_i^\top (C^\top C+\mu I)^{-1}c_i\leq & ~ 3\sqrt{n/m}\cdot \rho_i .
\end{align*}
\end{lemma}
\begin{proof}
Let $\Psi:=K^{1/2}D$. Then $C=K^{1/2}\Psi$, $C^\top C=\Psi^\top K\Psi$, $c_i=\Psi^\top\psi_i$, and $\rho_i=\psi_i^\top (K+\lambda I)^{-1}\psi_i$ with $\psi_i:=K^{1/2}e_i$. Let $\lambda_s:=\lambda\sqrt{m/n}$ and $\epsilon_1:=1/10$. We will use two facts about the spectrum. First, the Cauchy--Schwarz inequality $(\sum_{j>m}\sigma_j(K))^2\leq (n-m)\sum_{j>m}\sigma_j(K)^2$ gives $\lambda^2\leq \frac{n-m}{m}\lambda'\leq \frac{n}{m}\lambda'$, and hence $\lambda_s^2\leq \lambda'$. Second, for every $x\geq 0$, we have
\begin{align}\label{eq:row_rls_scalar}
\frac{x}{x^2+\lambda'}\leq & ~ \frac{2\sqrt{n/m}}{x+\lambda},
\end{align}
because $x(x+\lambda)\leq 2x^2$ if $x\geq \lambda$, and if $x<\lambda$, then $x(x+\lambda)\leq 2\lambda x\leq (\lambda/\sqrt{\lambda'})(x^2+\lambda')\leq \sqrt{n/m}\,(x^2+\lambda')$ by $x^2+\lambda'\geq 2x\sqrt{\lambda'}$.

\emph{The sample $\Psi$ approximates $K$ at the ridge parameter $\lambda_s$.} Note that $(K+\lambda_sI)^{-1}\preceq (\lambda/\lambda_s)(K+\lambda I)^{-1}$, so the $\lambda_s$-ridge leverage scores $\tau_i:=\psi_i^\top (K+\lambda_sI)^{-1}\psi_i$ of the columns of $K^{1/2}$ satisfy $\tau_i\leq (\lambda/\lambda_s)\rho_i=\sqrt{n/m}\,\rho_i$, and hence $\sum_i\tau_i\leq \sqrt{n/m}\sum_i\rho_i\leq 2\sqrt{nm}\leq 2n$ by Lemma~\ref{lem:classical_ridge}. The proof of Lemma 21 of~\cite{cmm17} is a matrix Bernstein bound in which the rank enters only through the trace of the variance proxy, which is the sum of the ridge leverage scores, and through the logarithm in the oversampling factor. We apply this proof to the columns of $K^{1/2}$ with the ridge parameter $\lambda_s$ in place of $\|A-A_k\|_F^2/k$, the overestimates $\sqrt{n/m}\,\rho_i\geq \tau_i$, the trace bound $2n$ in place of $2k$, and the oversampling factor $\beta\geq c\log(n/\delta)$, and obtain that with probability at least $1-\delta$,
\begin{align}\label{eq:psi_spec}
(1-\epsilon_1)\Psi\Psi^\top-\epsilon_1\lambda_sI\preceq & ~ K\preceq (1+\epsilon_1)\Psi\Psi^\top+\epsilon_1\lambda_sI .
\end{align}
Assuming this event occurs, we show that the claim holds for $\mu=\lambda'$, and for $\mu\geq 0.9\lambda'$ the claim then follows from $(C^\top C+\mu I)^{-1}\preceq \frac{1}{0.9}(C^\top C+\lambda' I)^{-1}$.

\emph{From $\Psi^\top K\Psi$ to $\Psi^\top\Psi$.} Let $T:=\Psi^\top\Psi\in \R^{t\times t}$ and $G:=\Psi\Psi^\top\in \R^{n\times n}$. Conjugating the left inequality of~\eqref{eq:psi_spec} by $\Psi$ gives $\Psi^\top K\Psi\succeq (1-\epsilon_1)T^2-\epsilon_1\lambda_sT$, and since $2\lambda_s\nu\leq \nu^2+\lambda_s^2\leq \nu^2+\lambda'$ for every eigenvalue $\nu$ of $T$, we have
\begin{align*}
C^\top C+\lambda' I=\Psi^\top K\Psi+\lambda' I\succeq & ~ (1-\epsilon_1)T^2-\tfrac{\epsilon_1}{2}(T^2+\lambda' I)+\lambda' I\succeq (1-\tfrac{3\epsilon_1}{2})(T^2+\lambda' I).
\end{align*}
Hence, $c_i^\top (C^\top C+\lambda' I)^{-1}c_i\leq (1-\frac{3\epsilon_1}{2})^{-1}\,\psi_i^\top\Psi(T^2+\lambda' I)^{-1}\Psi^\top\psi_i=(1-\frac{3\epsilon_1}{2})^{-1}\,\psi_i^\top G(G^2+\lambda' I)^{-1}\psi_i$. The last step uses the identity $\Psi(T^2+\lambda' I)^{-1}\Psi^\top=G(G^2+\lambda' I)^{-1}$, which holds since both sides equal $U\Sigma^2(\Sigma^4+\lambda' I)^{-1}U^\top$ for a compact singular value decomposition $\Psi=U\Sigma V^\top$.

\emph{From $G$ to $K$.} Applying~\eqref{eq:row_rls_scalar} to the eigenvalues of $G$ gives $G(G^2+\lambda' I)^{-1}\preceq 2\sqrt{n/m}\,(G+\lambda I)^{-1}$, and since $\lambda_s\leq \lambda$, the right inequality of~\eqref{eq:psi_spec} gives $K+\lambda I\preceq (1+\epsilon_1)G+(\lambda+\epsilon_1\lambda_s)I\preceq (1+\epsilon_1)(G+\lambda I)$, i.e., $(G+\lambda I)^{-1}\preceq (1+\epsilon_1)(K+\lambda I)^{-1}$. Putting everything together, we have
\begin{align*}
c_i^\top (C^\top C+\lambda' I)^{-1}c_i\leq & ~ \frac{2(1+\epsilon_1)}{1-3\epsilon_1/2}\sqrt{n/m}\cdot \psi_i^\top (K+\lambda I)^{-1}\psi_i\leq 2.6\sqrt{n/m}\cdot \rho_i ,
\end{align*}
and $2.6/0.9\leq 3$ gives the claim for $\mu\geq 0.9\lambda'$.
\end{proof}

\begin{theorem}
\label{thm:quantum_psd}
There exists a quantum algorithm (Algorithm~\ref{alg:quantum_bcw}, run with $\delta=0.001$) that given any set of points $X=\{x_1,\ldots, x_n\}\subset \R^d$ and a positive definite kernel function ${\sf K}: \R^d\times \R^d\rightarrow \R$ and any positive integer $k\leq n$ and $\epsilon\in (0,1)$ with $k/\epsilon\leq n$ (we write $m:=k/\epsilon$ and treat it as an integer, and otherwise replace it by $\lceil k/\epsilon\rceil$ throughout), runs in expected
\begin{align*} 
& ~ \wt O\big(n^{0.75}k^{1.25}\epsilon^{-1.25}({\cal T}_{\sf K}+k\epsilon^{-1})+n^{0.5}k^{1.5}\epsilon^{-2.5}({\cal T}_{\sf K}+1) \\
& ~ \qquad +n^{0.5}k^{\omega-0.5}\epsilon^{0.5-\omega}+\poly(k/\epsilon)\cdot({\cal T}_{\sf K}+1)\big)
\end{align*}
time, where ${\cal T}_{\sf K}$ is the time to evaluate ${\sf K}$ on any pair of points $x_i, x_j$, and makes, in expectation, $\wt O(\epsilon^{-2.5}n^{0.5}k^{1.5}+\epsilon^{-1.25}n^{0.75}k^{1.25}+\poly(k/\epsilon))$ queries to the points in $X$, and returns a pair of rank-$k$ matrices $M, N\in \R^{n\times k}$ (given implicitly in factor form) such that 
\begin{align*}
    \|K-MN^\top\|_F^2 \leq & ~ (1+\epsilon) \|K-K_k\|_F^2
\end{align*}
holds with probability at least $0.99$.
\end{theorem}

\begin{remark}[PSD low-rank approximation]\label{rem:psd_case}
A positive semidefinite matrix $K\in \R^{n\times n}$ given by entry access is the kernel matrix of the linear kernel on the rows of any factor $\Phi$ with $K=\Phi\Phi^\top$, and one entry read is one kernel evaluation. Since the algorithm accesses the points only through kernel evaluations, $\Phi$ is never formed, and hence, for $k$ and $\epsilon$ as in Theorem~\ref{thm:quantum_psd}, the theorem applies with ${\cal T}_{\sf K}=O(1)$. The expected time is then $\wt O(n^{0.75}k^{2.25}\epsilon^{-2.25}+n^{0.5}k^{1.5}\epsilon^{-2.5}+n^{0.5}k^{\omega-0.5}\epsilon^{0.5-\omega}+\poly(k/\epsilon))$. Moreover, the proof of Theorem~\ref{thm:quantum_psd} counts the kernel evaluations, which here are entry reads, and gives $\wt O(\epsilon^{-1.25}n^{0.75}k^{1.25}+\epsilon^{-2.5}n^{0.5}k^{1.5}+\poly(k/\epsilon))$ of them in expectation. The leading terms are the PSD rows of Tables~\ref{tab:optimization} and~\ref{tab:query_complexity}.
\end{remark}

\begin{proof}
Our algorithm could be interpreted as a quantum implementation of a generalization of~\cite{bcw20}, where they only tackle the case ${\sf K}(x_i, x_j)=x_i^\top x_j$, in which the kernel matrix $K$ is given directly.
Throughout the proof, we write $\OPT_k:=\|K-K_k\|_F^2$, and we assume without loss of generality that $\epsilon\leq 1/2$, since $\epsilon$ is rescaled by a constant at the end of the proof.

\paragraph{What is imported from~\cite{bcw20}.} Algorithm~\ref{alg:quantum_bcw} follows Algorithm 3 of~\cite{bcw20}, which samples a column sketch $C$ and a row sketch $R$ of $K$, computes an orthonormal $Z$ that spans a rank-$m$ spectral low-rank approximation of $R$, solves the spectral regression $\min_W\|C-WZ^\top\|_2$ by their Algorithm 2, and applies their Algorithm 1 to the column span of $W$. The proof of their Theorem 4.16 for Algorithm 1 uses only the following two properties of the orthonormal basis $Q$ it receives: (Q1) $QQ^\top$ is an $(\epsilon,k)$-spectral-Frobenius projection for $K$, i.e., $\|K-QQ^\top K\|_2^2\leq (\epsilon/k)\cdot\OPT_k$, which enters only once, through Lemma 4.14 of~\cite{bcw20} (due to~\cite{cw17}), to give
\begin{align}\label{eq:Q_lra}
\min_{X:{\rm rank}(X)\leq k}\|K-QXQ^\top\|_F^2\leq \|K-QQ^\top K_kQQ^\top\|_F^2\leq (1+\epsilon)\cdot\OPT_k ,
\end{align}
and (Q2) the leverage scores of $Q$ are its squared row norms, which are used to sample rows and columns of $K$ for the rank-constrained regression and for the final regression. In our algorithm, the column span of $W$ plays the role of $Q$, the leverage scores of $W$ are the squared row norms of $Q$, and we sample by them with $\textsc{QLS}$ instead of reading them off $Q$. Therefore, the proof has two parts, where Part 1 establishes (Q1) for an orthonormal basis $Q$ of the column span of $W$, and Part 2 solves the regressions in the basis $W$ from sampled rows and columns.

\paragraph{Part 1: the column span of $W$.} We use the following two properties of $D_1$ and $D_2$ at rank $m$ and accuracy $0.1$. They are the projection-cost preservation statements behind steps 2 and 3 of Algorithm 3 of~\cite{bcw20}, which are their Lemmas 4.11 and 4.12, due to~\cite{mw17}. \cite{bcw20} draw $t$ columns and $t$ rows with replacement by constant-factor approximations of the rank-$m$ ridge leverage scores of $K^{1/2}$, computed by the Nystr{\"o}m method of~\cite{mm17} in linear time. Here $D_1$ and $D_2$ include every index independently, with the probabilities of Lemma~\ref{lem:qnystrom} rescaled by $\sqrt{\epsilon n/k}=\sqrt{n/m}$. Throughout, the numerical constants $c$, $16$, and $100$ of Algorithms~\ref{alg:quantum_bcw} and~\ref{alg:quantum_nystrom} stand for sufficiently large absolute constants, chosen such that each sampling lemma we invoke applies at the stated accuracy with failure probability at most $0.001$. They enter the bounds only through the constants of the $O(\cdot)$ notation, whereas the constant $5$ is literal, since the bounds $q_i\leq 6\rho_i$ and $\sum_iq_i\leq 12m$ below use it.
\begin{itemize}
\item[(P1)] For every orthogonal projection $X\in \R^{n\times n}$ of rank at most $m$: $\|C-XC\|_F^2=(1\pm 0.1)\|K-XK\|_F^2$ and
\begin{align*}
\tfrac{9}{10}\|K-XK\|_2^2-\tfrac{1}{10m}\|K-K_m\|_F^2\leq & ~ \|C-XC\|_2^2 \\
\leq & ~ \tfrac{11}{10}\|K-XK\|_2^2+\tfrac{1}{10m}\|K-K_m\|_F^2 .
\end{align*}
\item[(P2)] For every orthogonal projection $X\in \R^{t\times t}$ of rank at most $m$: $\|R-RX\|_F^2=(1\pm 0.1)\|C-CX\|_F^2$ and
\begin{align*}
\tfrac{9}{10}\|C-CX\|_2^2-\tfrac{1}{10m}\|C-C_m\|_F^2\leq & ~ \|R-RX\|_2^2 \\
\leq & ~ \tfrac{11}{10}\|C-CX\|_2^2+\tfrac{1}{10m}\|C-C_m\|_F^2 .
\end{align*}
\end{itemize}
\emph{Proof of (P1) and (P2).} Let $\rho_i:=\ov\tau_i(K^{1/2})$, $\lambda:=\frac1m\sum_{j>m}\sigma_j(K)$, and $\lambda':=\|K-K_m\|_F^2/m$ be as in Lemma~\ref{lem:row_rls}, and assume ${\rm rank}(K)>m$. Otherwise $K=K_m$ and $\lambda=\lambda'=0$, so $\ov\tau_i(K)=(P_K)_{ii}=\rho_i$ for the orthogonal projector $P_K$ onto ${\rm range}(K)$, and, for the column sample $C=KD_1$, the score $\ov\tau_i(C^\top)$ is the $i$-th diagonal entry of the projector onto ${\rm range}(C)\subseteq{\rm range}(K)$, hence it is at most $\rho_i$. Thus, the overestimates $\wt\tau_i$ and $\hat\tau_i$ used below remain valid also when ${\rm rank}(K)\leq m$, with the pseudo-inverse convention of Section~\ref{sec:ridge}. We apply Lemma~\ref{lem:qnystrom} with $s=k'$, which makes its rank parameter $m$, and with $\beta=\sqrt{\epsilon n/k}=\sqrt{n/m}$. If both calls succeed, then $D_1$ and $D_2$ include index $i$ independently with probability $p_i=\min\{1, \sqrt{n/m}\cdot 16\log(12n/\delta)\,q_i\}$ and weight $p_i^{-1/2}$, where $\rho_i\leq q_i\leq 6\rho_i$, and hence $\sum_iq_i\leq 12m$ and $\sum_ip_i=O(\sqrt{nm}\log(n/\delta))=O(t)$. We use two facts from~\cite{bcw20} and~\cite{cmm17}. First, the rank-$m$ ridge leverage scores of the columns of $K$ satisfy $\ov\tau_i(K)\leq 2\sqrt{n/m}\,\rho_i$, which is Lemma 4.10 of~\cite{bcw20}, due to~\cite{mw17}. In the eigenbasis $K=\sum_j\sigma_jv_jv_j^\top$, these two scores are $\sum_jv_{ji}^2\sigma_j^2/(\sigma_j^2+\lambda')$ and $\sum_jv_{ji}^2\sigma_j/(\sigma_j+\lambda)$, and the ratio of the $j$-th terms is at most $2$ if $\sigma_j\geq \lambda$. In the other case, the ratio is at most $\lambda/\sqrt{\lambda'}$ by $\sigma_j^2+\lambda'\geq 2\sigma_j\sqrt{\lambda'}$, and $\lambda^2\leq \frac nm\lambda'$ by the Cauchy--Schwarz inequality. Second, let each column $a_i$ of a matrix $A$ be included independently with probability $p_i:=\min\{1, c'\epsilon_1^{-2}\log(k/\delta)\,\wt\tau_i\}$ and weight $p_i^{-1/2}$, where $\wt\tau_i\geq \ov\tau_i(A)$ are overestimates of the rank-$k$ ridge leverage scores of the columns and $c'$ is a sufficiently large constant. Then by Lemma 21 of~\cite{cmm17}, with probability at least $1-\delta$, the sample $B$ has $O(\epsilon_1^{-2}\log(k/\delta)\sum_i\wt\tau_i)$ columns and satisfies $(1-\epsilon_1)BB^\top-\epsilon_1\lambda_{A_k}I\preceq AA^\top\preceq (1+\epsilon_1)BB^\top+\epsilon_1\lambda_{A_k}I$. On the same event, it also satisfies $\|B-XB\|_F^2=(1\pm\epsilon_1)\|A-XA\|_F^2$ for every orthogonal projection $X$ of rank at most $k$ by their Theorem 6, whose proof uses $X$ only through ${\rm rank}(X)\leq k$ and, as they remark after it, applies to independent sampling with their Lemmas 21 and 22 in place of Theorem 5 and Lemma 20. For $\epsilon_1=1/11$, conjugating the spectral inequality by $I-X$ and taking spectral norms gives
\begin{align*}
\tfrac{9}{10}\|A-XA\|_2^2-\tfrac{1}{10}\lambda_{A_k}\leq \|B-XB\|_2^2\leq \tfrac{11}{10}\|A-XA\|_2^2+\tfrac{1}{10}\lambda_{A_k},
\end{align*}
since $\frac{1}{1-\epsilon_1}=\frac{11}{10}$, $\frac{\epsilon_1}{1-\epsilon_1}=\frac{1}{10}$, $\frac{1}{1+\epsilon_1}\geq \frac{9}{10}$, and $\frac{\epsilon_1}{1+\epsilon_1}\leq \frac{1}{10}$. (P1) is the case $A=K$, $B=C=KD_1$, $k=m$, and $\lambda_{A_m}=\frac1m\|K-K_m\|_F^2$, with the overestimates $\wt\tau_i:=2\sqrt{n/m}\,q_i\geq \ov\tau_i(K)$, for which $p_i=\min\{1, 8\log(12n/\delta)\,\wt\tau_i\}$ and $\sum_i\wt\tau_i\leq 24\sqrt{nm}$, so the sample has $O(t)$ columns. (P2) is the case $A=C^\top$, whose columns are the rows $c_i$ of $C$, $B=R^\top=(D_2^\top C)^\top$, $k=m$, and $\lambda_{A_m}=\frac1m\|C-C_m\|_F^2=:\lambda_C$, applied conditionally on $D_1$. Note that we may condition on $D_1$ since $D_2$ is independent of $D_1$. The overestimates are $\hat\tau_i:=3\sqrt{n/m}\,q_i$, where $q_i$ now denotes the last-oracle values of the second call, which also lie in $[\rho_i, 6\rho_i]$. The Frobenius part of (P1), applied to the projection $X$ onto the top-$m$ left singular vectors of $C$, gives $\|C-C_m\|_F^2\geq \frac{9}{10}\|K-XK\|_F^2\geq \frac{9}{10}\|K-K_m\|_F^2$, i.e., $\lambda_C\geq 0.9\lambda'$. Hence, by Lemma~\ref{lem:row_rls} applied to $D_1$ with $\beta=16\log(12n/\delta)$, with probability at least $0.999$ over $D_1$ we have $\ov\tau_i(C^\top)=c_i^\top (C^\top C+\lambda_CI)^{-1}c_i\leq 3\sqrt{n/m}\,\rho_i\leq \hat\tau_i$ for every $i$. Then $p_i=\min\{1, \frac{16}{3}\log(12n/\delta)\,\hat\tau_i\}$ and $\sum_i\hat\tau_i\leq 36\sqrt{nm}$, so the sample has $O(t)$ rows.

Minimizing the Frobenius parts over $X$ gives $\|C-C_m\|_F^2\leq 1.1\|K-K_m\|_F^2\leq 1.1\,\OPT_k$ and $\|R-R_m\|_F^2\leq 1.1\|C-C_m\|_F^2\leq 1.21\,\OPT_k$.

\emph{The matrix $Z$.} Corollary~\ref{thm:quantum_pcp} is Theorem~\ref{thm:ridge} applied to the columns of $R^\top=D_1^\top KD_2\in \R^{t\times t}$, and hence the output of the corollary is an $\epsilon_0$-approximator. That is, $\wt R\in \R^{t\times \wt O(k/\epsilon)}$ is a weighted subset of the columns of $R^\top$ which is an $\epsilon_0$-coreset of them for the $m$-subspace cost and satisfies
\begin{align*}
(1-\epsilon_0)\wt R\wt R^\top-\epsilon_0\lambda I\preceq R^\top R\preceq (1+\epsilon_0)\wt R\wt R^\top+\epsilon_0\lambda I,\qquad \lambda:=\|R-R_m\|_F^2/m .
\end{align*}
The matrix $Z\in \R^{t\times m}$ consists of the top-$m$ left singular vectors of $\wt R$. Its rows are indexed by the samples of $D_1$, as are the columns of $C$, so $C-WZ^\top$ is well defined. We derive two consequences. First, by Lemma 3 of~\cite{cem+15} as stated in Section~\ref{sec:cmm}, with $\gamma=\epsilon_0$ in the remark following it, we have
\begin{align}\label{eq:Z_frob}
\|R-RZZ^\top\|_F^2=\|(I-ZZ^\top)R^\top\|_F^2\leq (1+3\epsilon_0)\|R-R_m\|_F^2 .
\end{align}
Second, a unit vector $v$ orthogonal to the columns of $Z$ satisfies $v^\top\wt R\wt R^\top v\leq \sigma_{m+1}(\wt R)^2$, and the upper spectral inequality gives $v^\top R^\top Rv\leq (1+\epsilon_0)\sigma_{m+1}(\wt R)^2+\epsilon_0\lambda$. Moreover, the lower spectral inequality gives $(1-\epsilon_0)\sigma_{m+1}(\wt R)^2\leq \sigma_{m+1}(R)^2+\epsilon_0\lambda$ by the monotonicity of eigenvalues under the positive semidefinite order. Since $\|R-RZZ^\top\|_2^2=\max\{v^\top R^\top Rv: \|v\|_2=1, Z^\top v=0\}$, we have
\begin{align}\label{eq:Z_spec}
\|R-RZZ^\top\|_2^2\leq \frac{1+\epsilon_0}{1-\epsilon_0}\big(\sigma_{m+1}(R)^2+\epsilon_0\lambda\big)+\epsilon_0\lambda\leq 1.03\,\sigma_{m+1}(R)^2+0.03\,\lambda .
\end{align}
To bound $\sigma_{m+1}(R)^2$, we use the upper inequalities of (P2) and (P1). Let $X_C$ be the projection onto the top-$m$ right singular vectors of $C$ and $X_K$ the projection onto the top-$m$ left singular vectors of $K$. Then we have
\begin{align*}
& ~ \sigma_{m+1}(R)^2\leq \|R-RX_C\|_2^2\leq \tfrac{11}{10}\sigma_{m+1}(C)^2+\tfrac{1}{10m}\|C-C_m\|_F^2, \\
& ~ \sigma_{m+1}(C)^2\leq \|C-X_KC\|_2^2\leq \tfrac{11}{10}\sigma_{m+1}(K)^2+\tfrac{1}{10m}\|K-K_m\|_F^2,
\end{align*}
and $\sigma_{m+1}(K)^2\leq \frac{1}{m-k}\sum_{i=k+1}^{m+1}\sigma_i(K)^2\leq \frac{2\epsilon}{k}\OPT_k$ because $m\geq 2k$. Combining these with the Frobenius tail bounds above gives $\sigma_{m+1}(C)^2\leq 2.3\frac{\epsilon}{k}\OPT_k$, $\sigma_{m+1}(R)^2\leq 2.7\frac{\epsilon}{k}\OPT_k$, and $\lambda\leq 1.21\frac{\epsilon}{k}\OPT_k$. Therefore, by~\eqref{eq:Z_spec} and~\eqref{eq:Z_frob},
\begin{align}\label{eq:Z_bounds}
\|R-RZZ^\top\|_2^2\leq 2.9\,\frac{\epsilon}{k}\OPT_k,\qquad \|R-RZZ^\top\|_F^2\leq 1.3\,\OPT_k ,
\end{align}
and by the lower inequalities of (P2) with $X=ZZ^\top$,
\begin{equation}\label{eq:C_Z}
\begin{aligned}
\|C-CZZ^\top\|_2^2\leq & ~ \tfrac{10}{9}\Big(\|R-RZZ^\top\|_2^2+\tfrac{1}{10m}\|C-C_m\|_F^2\Big)\leq 3.4\,\frac{\epsilon}{k}\OPT_k, \\
\|C-CZZ^\top\|_F^2\leq & ~ \tfrac{10}{9}\|R-RZZ^\top\|_F^2\leq 1.5\,\OPT_k .
\end{aligned}
\end{equation}

\emph{The matrix $W$.} The matrix $D_3$ includes the $i$-th row of $Z$ independently with probability $p_i=\min\{1,\alpha\|z_i\|_2^2\}$, where $\alpha:=\epsilon k'/k=100\log(12k/(\epsilon\delta))$, and rescales it by $p_i^{-1/2}$. This is the sampling matrix $S$ of Algorithm 2 of~\cite{bcw20}, which samples by the leverage scores of $Z^\top$ oversampled by a logarithmic factor, and these leverage scores are the squared row norms of $Z$. Moreover, $W=CD_3(Z^\top D_3)^\dagger$ is its output $\wh W=CS(Z^\top S)^\dagger$. The proof of Theorem 4.18 of~\cite{bcw20}, via their Lemma 4.19, equations (4.18)--(4.21), and Lemma 4.25, uses only that $Z$ has orthonormal columns and that $S$ is drawn in this way. It shows that, with probability at least $0.999$,
\begin{align}\label{eq:W_spec}
\|C-WZ^\top\|_2^2\leq \wt O(1)\cdot\Big(\|C-CZZ^\top\|_2^2+\frac{1}{m}\|C-CZZ^\top\|_F^2\Big)\leq \wt O(1)\cdot \frac{\epsilon}{k}\OPT_k ,
\end{align}
where the second step follows from~\eqref{eq:C_Z}. Their statement assumes in addition that $ZZ^\top$ is an $(O(1),m)$-spectral-Frobenius projection for $C$, but this assumption serves only to rewrite the first bound in terms of $\|C-C_m\|_F^2$, and we use the first bound directly. The event of~\eqref{eq:W_spec} is the matrix Bernstein bound of their Lemma 4.22, which is Lemma 21 of~\cite{cmm17}, together with Markov bounds on expectations. The failure probability of the matrix Bernstein bound enters through the oversampling factor $\alpha\geq 100\log(1000m)$, so a failure probability of $0.001$ costs a constant factor in the $\wt O(1)$.

Let $Q$ be an orthonormal basis of the column span of $W$, a span that contains $WZ^\top y$ for every $y$, so $\|(I-QQ^\top)Cy\|_2\leq \|Cy-WZ^\top y\|_2$ and hence $\|C-QQ^\top C\|_2^2\leq \|C-WZ^\top\|_2^2$. Applying the lower inequality of (P1) to $X=QQ^\top$, whose rank is at most $m$, we have
\begin{align}\label{eq:Q_sf}
\|K-QQ^\top K\|_2^2\leq \tfrac{10}{9}\Big(\|C-QQ^\top C\|_2^2+\tfrac{1}{10m}\|K-K_m\|_F^2\Big)\leq \wt O(1)\cdot\frac{\epsilon}{k}\OPT_k .
\end{align}
This is (Q1) with $\epsilon$ replaced by $\epsilon\cdot{\rm polylog}(n,k/\epsilon)$, so we run the algorithm with $\epsilon/{\rm polylog}(n,k/\epsilon)$ in place of $\epsilon$, and the $\wt O(\cdot)$ in the running time absorbs this change. Thus, we obtain $\|K-QQ^\top K\|_2^2\leq (\epsilon/k)\cdot\OPT_k$ and hence~\eqref{eq:Q_lra}.

\paragraph{Part 2: the rank-constrained regression in the basis $W$.} We will focus on solving the rank-constrained regression $\min_{Y: {\rm rank}(Y)\leq k} \|K-WYW^\top\|_F^2$, which is the major divergence between our approach and that of~\cite{bcw20}. In~\cite{bcw20}, since they could afford linear in $n$ time, they compute an orthonormal basis for $W$ denoted by $Q$, and instead solve the regression $\min_{X: {\rm rank}(X)\leq k} \|K-QXQ^\top\|_F^2$. Let $Q$ be the orthonormal basis of the column span of $W$ from Part 1, and let $T\in \R^{{\rm rank}(W)\times k/\epsilon}$ be the change-of-basis matrix such that $QT=W$. The matrix $T$ has full row rank, so $WT^\dagger=QTT^\dagger=Q$, and we observe that $X$ could be recovered via the following procedures:
\begin{itemize}
    \item Solve 
    \begin{align}\label{eq:exact_rank_regression}
        \min_{Y: {\rm rank}(Y)\leq k} \|K-WYW^\top\|_F^2
    \end{align}
    and let $Y'$ denote the optimal solution of the above regression.
    \item Set $X_*:=T Y' T^\top$.
\end{itemize}
To see that $X_*$ is optimal for the rank-constrained regression against $Q$, note
\begin{align*}
    QX_* Q^\top = & ~ Q T Y' T^\top Q^\top \\
    = & ~ WY' W^\top,
\end{align*}
and if there exists a solution $X'$ with lower cost, then 
\begin{align*}
    \|K-W T^{\dagger} X' T^{\dagger\top} W^\top\|_F^2 = & ~ \|K - Q X' Q^\top\|_F^2 \\
    < & ~ \|K - Q X_* Q^\top \|_F^2 \\
    = & ~ \|K - W Y' W^\top \|_F^2,
\end{align*}
contradicting the definition of $Y'$. Both~\cite{bcw20} and Algorithm~\ref{alg:quantum_bcw} construct leverage score sampling matrices according to the leverage scores of $W$ (in the context of~\cite{bcw20}, they sample according to the row norms of $Q$, which are the leverage scores of $W$), then we solve the surrogate regression
\begin{align}\label{eq:approx_rank_regression}
    \min_{Y: {\rm rank}(Y)\leq k} \|D_4^\top K D_5 - D_4^\top W Y W^\top D_5\|_F^2, 
\end{align}
so it suffices to show that the optimal solution of Eq.~\eqref{eq:approx_rank_regression} is a good approximation to the optimal solution of Eq.~\eqref{eq:exact_rank_regression}. Let $f(Y):=\|K-WYW^\top\|_F^2$, $g(Y):=\|D_4^\top(K-WYW^\top)\|_F^2$, and $h(Y):=\|D_4^\top(K-WYW^\top)D_5\|_F^2$, and let $\OPT:=\min_{Y:{\rm rank}(Y)\leq k}f(Y)$. By the change of basis above, $\OPT$ equals the left-hand side of~\eqref{eq:Q_lra}, hence $\OPT\leq (1+\epsilon)\cdot\OPT_k$. The matrix $D_4$ samples $k'/\epsilon^2=\Omega(\epsilon^{-2}m\log(m/\delta))$ rows by constant-factor approximations of the leverage scores of $W$, whose rank is at most $m$. The set $\{WYW^\top\}$ is not an affine space, but it is contained in $\{WH: H\in \R^{m\times n}\}$. Therefore, by Lemma~\ref{lem:leverage_score_weak_affine} with design $W$, response $K$, and failure probability $0.001$, together with Remark~\ref{rem:delta_dependence}, there is a constant $c_1$ with $g(Y)=(1\pm\epsilon)f(Y)+c_1$ for all $Y$, where $c_1=\|D_4^\top K_*\|_F^2-\|K_*\|_F^2$ with $K_*:=K-WH_*$ and $H_*:=\arg\min_{H}\|K-WH\|_F^2$. Similarly, $D_5$ samples $k'/\epsilon^2$ columns by the leverage scores of the rows of $W$. By Lemma~\ref{lem:leverage_score_weak_affine} with design $W$, response $(D_4^\top K)^\top$, and failure probability $0.001$, there is a constant $c_2$ with $\|(VW^\top-D_4^\top K)D_5\|_F^2=(1\pm\epsilon)\|VW^\top-D_4^\top K\|_F^2+c_2$ for all $V\in \R^{k'/\epsilon^2\times m}$, and taking $V=D_4^\top WY$ gives $h(Y)=(1\pm\epsilon)g(Y)+c_2$ for all $Y$. Let $Y_*$ minimize $h$ over $\{Y:{\rm rank}(Y)\leq k\}$, i.e., $Y_*$ solves Eq.~\eqref{eq:approx_rank_regression}, and recall that $Y'$ minimizes $f$ over the same set, so $f(Y')=\OPT$. Then $h(Y_*)\leq h(Y')$, and $c_2$ cancels, so we have $(1-\epsilon)g(Y_*)\leq (1+\epsilon)g(Y')$. Substituting $g=(1\pm\epsilon)f+c_1$ on both sides, we obtain
\begin{align}\label{eq:f_Y_*}
(1-\epsilon)^2f(Y_*)\leq (1+\epsilon)^2\,\OPT+2\epsilon\,|c_1| .
\end{align}
The constant $c_1$ does not cancel, but it is small. To see this, we first observe that $\|K_*\|_F^2=\min_H\|K-WH\|_F^2\leq \OPT$, because the unconstrained problem has the larger solution space. Moreover, $\|D_4^\top K_*\|_F^2$ is an unbiased estimator of $\|K_*\|_F^2$,
\begin{align*}
    \E[\|D_4^\top K_*\|_F^2] = & ~ \E[\tr[K_*^\top D_4 D_4^\top K_*]] = \tr[K_*^\top \E[D_4 D_4^\top] K_*] = \tr[K_*^\top I_n K_*] = \|K_*\|_F^2,
\end{align*}
since $D_4$ includes the $i$-th row with probability $q_i$ and rescales it by $q_i^{-1/2}$. By Markov's inequality, with probability at least $0.999$, we have $\|D_4^\top K_*\|_F^2\leq 1000\|K_*\|_F^2$, hence $|c_1|\leq 999\|K_*\|_F^2\leq 999\,\OPT$ and, by~\eqref{eq:f_Y_*},
\begin{align*}
    f(Y_*) \leq & ~ \frac{(1+\epsilon)^2+1998\epsilon}{(1-\epsilon)^2} \cdot\OPT = (1+O(\epsilon))\cdot \OPT\leq (1+O(\epsilon))\cdot\OPT_k .
\end{align*}
 We apply the lemma of~\cite{ft07} stated in Section~\ref{sec:cmm}, which is Theorem 4.15 of~\cite{bcw20}, with $B=D_4^\top W$, $C=W^\top D_5$, and $A=D_4^\top KD_5$ in its notation. It shows that the minimizer $Y_*$ of Eq.~\eqref{eq:approx_rank_regression} admits the closed form computed in Algorithm~\ref{alg:quantum_bcw}.

We now have a rank-$k$ solution in factored form, which is $WY_*W^\top$ with $f(Y_*)\leq (1+O(\epsilon))\cdot\OPT_k$. Compute the top-$k$ left singular vectors of $Y_*$, denoted as $U_*\in \R^{m\times k}$, and write $Y_*=U_* V_*$ with $V_*:=U_*^\top Y_*$, which is possible because ${\rm rank}(Y_*)\leq k$. We set $M:=WU_*\in \R^{n\times k}$ and plug the decomposition into the regression, so the right low-rank factor could be found by solving
\begin{align*}
    \min_{N\in \R^{n\times k}} \|K-MN^\top\|_F^2 \leq & ~ \|K-WU_*V_*W^\top\|_F^2 \\
    \leq & ~ (1+O(\epsilon)) \|K-K_k\|_F^2.
\end{align*}
To solve the regression, we employ leverage score sampling on the rows of $M$, whose rank is at most $k$. By Lemma~\ref{lem:leverage_score_optimal_solution} with failure probability $0.001$, together with Remark~\ref{rem:delta_dependence}, it suffices to sample $O(k\log k+k/\epsilon)$ rows, and the solution $N_*^\top=(D_6^\top M)^\dagger(D_6^\top K)$ of the sketched regression satisfies
\begin{align*}
    \|K-MN_*^\top\|_F^2 \leq & ~ (1+\epsilon) \min_{N\in \R^{n\times k}} \|K-MN^\top\|_F^2 \\
    \leq & ~ (1+O(\epsilon)) \|K-K_k\|_F^2.
\end{align*}
Finally, by properly scaling $\epsilon$, we conclude the proof of correctness.

\paragraph{Failure probability.} The six calls are $\textsc{QNystr{\"o}mKernel}$ for $D_1$ and $D_2$, $\textsc{QLowRankCMM}$ for $\wt R$, and $\textsc{QLS}$ for $D_4$, $D_5$, and $D_6$, and each call is made with failure parameter $\delta/6$, for a total of $\delta=0.001$. Note that the property (P1), the event of Lemma~\ref{lem:row_rls} for $D_1$, and the property (P2) each hold with probability at least $0.999$ for sufficiently large constants. The event~\eqref{eq:W_spec} for $D_3$ has probability at least $0.999$. Moreover, the two applications of Lemma~\ref{lem:leverage_score_weak_affine}, the application of Lemma~\ref{lem:leverage_score_optimal_solution}, the Markov step, and the sampler call for $D_3$ each fail with probability at most $0.001$. Therefore, by a union bound, the success probability is at least $1-0.001-0.003-0.001-0.005=0.99$.

Next, we analyze the runtime of Algorithm~\ref{alg:quantum_bcw}, item by item as follows:
\begin{itemize}
    \item Forming the generalized ridge leverage score sampling matrices $D_1$ and $D_2$ (Algorithm~\ref{alg:quantum_nystrom}) involves selecting $O(k'^2)$ entries from $K$, which could be implemented by $k'^2$ evaluations of the kernel function. In the loop, we compute the SVD of a $k'\times k'$ matrix, which takes $O(k'^\omega)$ time, and forming $\wh M$ also takes $O(k'^\omega)$ time. Next, we need to analyze the complexity of implementing the sampling oracle, for any fixed $i$, we form $D_{t-1}^\top K_i$ by forming a vector of length $k'$ through $k'$ kernel evaluations and an extra $k'^2$ time for the quadratic form. The last sampling step uses the probabilities $p_i=\min\{1,\sqrt{\epsilon n/k}\cdot 16 q_i\log(12n/\delta)\}$, whose sum is
    \begin{align*}
        \sum_{i=1}^n p_i = & ~ O( \sqrt{nk/\epsilon}\log(n/\delta))=O(t)
    \end{align*}
    by Lemma~\ref{lem:qnystrom}. Therefore, by Lemma~\ref{lem:q_sampling_1_d}, the last sampling step costs $\wt O(\sqrt{nt})\cdot (k'{\cal T}_{\sf K}+k'^2)$ time. Each of the $O(\log n)$ intermediate levels samples with $\sum_ip_i=\wt O(k')$ and costs $\wt O(\sqrt{nk'})\cdot (k'{\cal T}_{\sf K}+k'^2)=\wt O(n^{0.5}k^{1.5}\epsilon^{-1.5}{\cal T}_{\sf K}+n^{0.5}k^{2.5}\epsilon^{-2.5})$. The first term of this cost is within the second term of the theorem, and the second term of this cost is at most $n^{0.75}k^{2.25}\epsilon^{-2.25}$ because $k/\epsilon\leq n$, which is needed for $t\leq n$. Thus, the overall runtime of this part is
    \begin{align*}
        & ~ \wt O(\sqrt{n\sum_{i}p_i})\cdot (k' {\cal T}_{\sf K}+k'^2)+k'^2 {\cal T}_{\sf K}+k'^\omega \\
        = & ~ \wt O(n^{0.75}k^{1.25}\epsilon^{-1.25}({\cal T}_{\sf K}+k\epsilon^{-1})) + k^2\epsilon^{-2} ({\cal T}_{\sf K}+k^{\omega-2}\epsilon^{2-\omega}).
    \end{align*}
    \item For matrices $C$ and $R$, we do not explicitly compute the data structure for them.
    \item To form the coreset $\wt R$ of the columns of $R^\top=D_1^\top KD_2$ using Corollary~\ref{thm:quantum_pcp}, we need to show that the generic quantum sampling algorithm could be implemented even though the input is given in factor form. Observe that the algorithm requires uniformly sampling columns of the input matrix, which is oblivious to the input. To form the initial coreset $C_0$, we need to query a total of $\wt O(\sqrt{nk/\epsilon})\times \wt O(k/\epsilon)$ entries of $K$, which can be done in $\wt O(n^{0.5}k^{1.5}\epsilon^{-1.5})$ kernel evaluations. Then we compute the SVD of this matrix, in time $\wt O(n^{0.5}k^{\omega-0.5}\epsilon^{0.5-\omega})$. Subsequently we need to implement the classical ridge leverage score data structure (Algorithm~\ref{alg:classical_ridge_ls}), which can be done in time $\wt O(n^{0.5}k^{\omega-0.5}\epsilon^{0.5-\omega})$ and then apply the random Gaussian matrix takes $\wt O(n^{0.5}k^{1.5}\epsilon^{-1.5})$ time. To implement each query, we can form the query vector by $\wt O(n^{0.5}k^{0.5}\epsilon^{-0.5})$ kernel evaluations and an additional $\wt O(n^{0.5}k^{0.5}\epsilon^{-0.5})$ time. The total runtime is
    \begin{align*}
        \wt O(n^{0.75}k^{1.25}\epsilon^{-1.25}+n^{0.5}k^{1.5}\epsilon^{-1.5})\cdot ({\cal T}_{\sf K}+1)+\wt O(n^{0.5}k^{\omega-0.5}\epsilon^{0.5-\omega}).
    \end{align*}
    \item Form matrix $Z$ by computing SVD of $\wt R$, since $\wt R\in \R^{\sqrt{nk/\epsilon}\times k/\epsilon}$, this step could be done in time $O(\epsilon^{0.5-\omega}n^{0.5}k^{\omega-0.5})$.
    \item Form the sampling matrix $D_3$ involves sampling the $t$ rows of $Z$ according to $p_i=\min\{1,(\epsilon k'/k)\|z_i\|_2^2\}$, where each oracle call returns the squared norm of a stored row of $Z$ in $O(k/\epsilon)$ time, and the sum of $p_i$'s is
    \begin{align*}
        \sum_{i=1}^t p_i \leq & ~ \frac{\epsilon k'}{k} \cdot \sum_{i=1}^t \|z_i\|_2^2 = \frac{\epsilon k'}{k}\cdot \|Z\|_F^2 = k'
    \end{align*}
    because $Z$ has $k/\epsilon$ orthonormal columns. Thus, by Lemma~\ref{lem:q_sampling_1_d} applied to the $t$ row indices of $Z$, the overall runtime of this step is
    \begin{align*}
        \wt O(\sqrt{tk'}\cdot k/\epsilon) = & ~ \wt O((nk/\epsilon)^{1/4}(k/\epsilon)^{1/2}\cdot k/\epsilon) = \wt O(n^{0.25}k^{1.75}\epsilon^{-1.75}),
    \end{align*}
    which is dominated by $n^{0.5}k^{1.5}\epsilon^{-2.5}$ since $k\leq n$.
    \item Form matrix $W$, we only need to explicitly compute $(Z^\top D_3)^\dagger\in \R^{k'\times k/\epsilon}$, which is a small matrix and could be computed in time $\wt O(k^\omega/\epsilon^\omega)$. Note that again, we won't explicitly compute the data structure for $W$, and a row of $W$ is instead formed on demand from $k'$ kernel evaluations.
    \item Form the leverage score sampling matrix $D_4$ and $D_5$ with respect to $W\in \R^{n\times k/\epsilon}$ and sample $k'/\epsilon^2$ rows/columns. The argument is similar to forming that of $\wt R$, except we use Algorithm~\ref{alg:classical_ls}, whose approximators are $\wt O(k/\epsilon)\times k/\epsilon$ matrices formed from rows of $W$. Since we need to oversample $k'/\epsilon^2=\wt O(k/\epsilon^3)$ rows and columns, we could scale the scores accordingly and make the sum of probabilities be at most $\wt O(k/\epsilon^3)$. To implement the oracle, we form the $\wt O(k/\epsilon)$ rows of each approximator with $\wt O(k^2\epsilon^{-2})$ kernel evaluations, $k'$ for each row. The subsequent SVDs and applications of the JL matrices take $\wt O(k^{\omega}\epsilon^{-\omega})$ time, which is the additive term $d^\omega$ of Theorem~\ref{thm:ls} with $d=k/\epsilon$. We multiply the $O(\log n)\times k/\epsilon$ sketching matrices $M$ and $N$ of Algorithm~\ref{alg:classical_ls} once by $((Z^\top D_3)^\dagger)^\top$, in $\wt O(k'k/\epsilon)$ time. Since the $i$-th row of $W$ is $w_i=((Z^\top D_3)^\dagger)^\top v_i$ with $v_i:=D_3^\top D_1^\top K_i\in \R^{k'}$, a query on this row is answered by forming $v_i$ with $k'$ kernel evaluations and applying the two precomposed $O(\log n)\times k'$ matrices to it in $\wt O(k')$ time. Without this precomposition, a query would cost $\wt O(k'k/\epsilon)$ time. The final Grover search makes $\wt O(\sqrt{nk'/\epsilon^2})$ queries, and the $O(\log n)$ intermediate levels make fewer. Therefore, the total runtime is
    \begin{align*}
        \wt O(\sqrt{nk'/\epsilon^2})\cdot k'({\cal T}_{\sf K}+1)+\poly(k/\epsilon) = & ~ \wt O(n^{0.5}k^{1.5}\epsilon^{-2.5})\cdot ({\cal T}_{\sf K}+1)+\poly(k/\epsilon).
    \end{align*}
    \item Form matrix $D_4^\top W$ and $W^\top D_5$ could be done via selecting entries, in time $\wt O(k^2\epsilon^{-4})\cdot {\cal T}_{\sf K}$, and forming the projection matrices $P_{D_4^\top W}$ and $P_{W^\top D_5}$ takes time $\poly(k/\epsilon)$.
    \item Form the matrix $D_4^\top K D_5$ is selecting its $(k'/\epsilon^2)^2=\wt O(k^2\epsilon^{-6})$ entries from $K$, in time $\wt O(k^2\epsilon^{-6})\cdot {\cal T}_{\sf K}$, and this is the largest number of kernel evaluations among the $\poly(k/\epsilon)$ terms.
    \item Compute $[P_{D_4^\top W}(D_4^\top K D_5) P_{W^\top D_5}]_k$ involves multiplying $\poly(k/\epsilon)$ size matrices and computing an SVD of a $k'/\epsilon^2\times k'/\epsilon^2$ matrix, which takes $\wt O((k'/\epsilon^2)^\omega)=\wt O(k^{\omega}\epsilon^{-3\omega})$ time, the largest of the $\poly(k/\epsilon)$ arithmetic terms. Forming $Y_*$ and its top-$k$ singular vectors takes $\poly(k/\epsilon)$ time.
    \item Form the sampling matrix $D_6$ involves performing leverage score sampling according to matrix $WU_*\in \R^{n\times k}$ with the smaller target row count $O(k\log k+k/\epsilon)\leq k'/\epsilon^2$, so the runtime is subsumed by the time to form $D_4$ and $D_5$. Finally, forming the matrix $N$ only requires computing $(D_6^\top WU_*)^\dagger$, which takes $\poly(k/\epsilon)\cdot {\cal T}_{\sf K}$ time.
\end{itemize}
Hence, the overall running time of Algorithm~\ref{alg:quantum_bcw} is the one stated in the theorem. By the same accounting, the number of kernel evaluations, and hence the number of queries to the points of $X$, is $\wt O(n^{0.75}k^{1.25}\epsilon^{-1.25}+n^{0.5}k^{1.5}\epsilon^{-2.5}+\poly(k/\epsilon))$. The first of the three terms comes from $D_1$, $D_2$, and $\wt R$, the second from $D_4$, $D_5$, and $D_6$, and the third from the explicitly formed $\poly(k/\epsilon)$-size submatrices. The $\poly(k/\epsilon)$ term is at least the $\wt O(k^2\epsilon^{-6})$ entries of $D_4^\top KD_5$.
\end{proof}

\section{Quantum \texorpdfstring{$(k, p)$}{(k, p)}-Subspace Approximation}
\label{sec:subspace_approx}
In this section, we consider a generalized version of the $k$-subspace cost studied in Section~\ref{sec:ridge}, which we call the $(k, p)$-subspace cost~\citep{wy24}: let ${\cal F}_k$ be the set of all $k$-dimensional subspaces, then
\begin{align*}
    {\rm cost}(A, x) = & ~ \left(\sum_{i=1}^n\|a_i^\top (I - P_x)\|_2^p\right)^{1/p}.
\end{align*}
By defining the matrix $(p, 2)$-norm as
\begin{align*}
    \|Y\|_{p,2} = & ~ \left(\sum_{i=1}^n \|e_i^\top Y\|_2^p\right)^{1/p},
\end{align*}
we have
\begin{align*}
    \mathrm{cost}(A, x) = & ~ \| A (I-P_x)\|_{p,2}.
\end{align*}
The $k$-subspace cost function we studied in Section~\ref{sec:ridge} is just the $(k, 2)$-subspace cost, and~\cite{wy24} have shown that, similar to the $k$-subspace cost, one could sample according to the powers of the ridge leverage score. We recall their main result in the following.

\begin{lemma}[Theorems 3.3 and 3.6 of~\cite{wy24}, one-round form]
\label{lem:wy24}
Let $A\in \R^{n\times d}$, $\epsilon\in (0,1)$, $\delta\in (0,1)$ and $\alpha=\Theta(\epsilon^2)/(\log^3 n+\log(1/\delta))$. Let $S$ be the $\ell_p$ sampling matrix of~\cite{wy24} with probabilities $\{q_i\}_{i=1}^n$ (row $i$ is kept independently with probability $q_i$ and rescaled by $q_i^{-1/p}$), where
\begin{align*}
    q_i \geq & ~ \begin{cases}
        \min\{1, n^{p/2-1}\ov\tau_i(A, \lambda_{A_k})^{p/2}/\alpha\}, & \text{if $p> 2$,} \\
        \min\{1, \ov\tau_i(A, \lambda_{A_{k'}})^{p/2}/\alpha \}, & \text{if $1\leq p< 2$},
    \end{cases}
\end{align*}
with $\lambda_{A_k}=\|A-A_k\|_F^2/k$ and, for $p<2$, $\lambda_{A_{k'}}=\|A-A_{k'}\|_F^2/k'$ for a rank parameter $k'\geq k$. Write $L:=O(\log(n/(\epsilon\delta)))$ and $\OPT:=\min_{x\in {\cal F}_k}\|A(I-P_x)\|_{p,2}^p$. Then, with probability at least $1-\delta$, for every $x\in {\cal F}_k$,
\begin{align*}
    \|SA(I-P_x)\|_{p,2}^p = & ~ (1\pm\epsilon) \|A(I-P_x)\|_{p,2}^p \pm O(\epsilon)\,L^{p/2}\cdot \OPT \quad (p>2), \\
    \text{resp.}\quad \|SA(I-P_x)\|_{p,2}^p = & ~ (1\pm\epsilon) \|A(I-P_x)\|_{p,2}^p \pm O(\epsilon)\,L^{p}\cdot \OPT \quad (1\leq p<2),
\end{align*}
(the statement of Theorem 3.6 in arXiv v3 shows the exponent $p/2$ on its second norm where its proof display and Definition 1.1 have $p$, and we follow the proof) under the following hypotheses:
\begin{itemize}
    \item For $p>2$ (Theorem 3.3 of~\cite{wy24}): $n\geq n'$ for some $n'=O(k^{p/2}/\epsilon)$. If $q_i=\min\{1, n^{p/2-1}\wh\tau_i^{p/2}/\alpha\}$ for values $\wh\tau_i\geq \ov\tau_i(A,\lambda_{A_k})$, then the expected number of sampled rows is $\sum_i q_i\leq n^{1-2/p}\alpha^{-2/p}\sum_i\wh\tau_i$. For $\wh\tau_i=\ov\tau_i(A,\lambda_{A_k})$, this is at most $2kn^{1-2/p}\alpha^{-2/p}$, since these scores sum to at most $2k$ by Lemma~\ref{lem:classical_ridge}. This sharpens the bound $O(kn^{1-2/p}/\alpha)$ in the proof of Theorem 3.5 of~\cite{wy24}, which takes $1/\alpha$ outside the minimum, and we prove it after the lemma.
    \item For $1\leq p<2$ (Theorem 3.6 of~\cite{wy24}): there is a rank-$k'$ subspace $\wt F$ with $\|a_i^\top (I-P_{\wt F})\|_2^p\leq O(1/n)\|A(I-P_{\wt F})\|_{p,2}^p$ for all $i$ and $\|A(I-P_{\wt F})\|_{p,2}^p\leq O(\OPT)$ (a ``flat'' constant-factor solution). If $q_i=\min\{1,\wh\tau_i^{p/2}/\alpha\}$ for values $\wh\tau_i\geq \ov\tau_i(A,\lambda_{A_{k'}})$, then the expected number of sampled rows is $\sum_i q_i\leq n^{1-p/2}(\sum_i\wh\tau_i)^{p/2}/\alpha$ by H\"older's inequality. For $\wh\tau_i=\ov\tau_i(A,\lambda_{A_{k'}})$, this is $O(k'^{p/2}n^{1-p/2}/\alpha)$ (proof of Theorem 3.7 of~\cite{wy24}, from $\sum_i\ov\tau_i(A,\lambda_{A_{k'}})\leq 2k'$). \cite{wy24} establish the precondition with $k'=\wt O(k)$ by computing a constant-factor solution and splitting each heavy row into reweighted copies (their Lemmas 2.3 and 2.4).
\end{itemize}
Since $\|A(I-P_x)\|_{p,2}^p\geq \OPT$ for every $x$, replacing $\epsilon$ by $\epsilon/(CL^{p/2})$ (resp.\ $\epsilon/(CL^{p})$) for a suitable constant $C$ turns the guarantee into $\|SA(I-P_x)\|_{p,2}^p=(1\pm\epsilon)\|A(I-P_x)\|_{p,2}^p$ for all $x\in {\cal F}_k$, hence $\|SA(I-P_x)\|_{p,2}=(1\pm\epsilon)\|A(I-P_x)\|_{p,2}$ since $p\geq 1$. This rescaling costs a factor $O(L^{p})$ (resp.\ $O(L^{2p})$) in $1/\alpha$. For $p>2$, the bound on the number of sampled rows in the first item depends on $\alpha$ only through $\alpha^{-2/p}$, so it grows only by a factor $O(L^2)$. The rescaling also requires, for $p>2$, the hypothesis $n\geq O(k^{p/2}L^{p/2}/\epsilon)$ in place of $n\geq O(k^{p/2}/\epsilon)$, and the $\wt O(\cdot)$ below hides these factors. The full procedures of~\cite{wy24} (Theorems 3.5 and 3.7, $O(\log\log n)$ rounds of this sampling) run in $\wt O(\nnz(A)+d^\omega)$ time and output coresets of size $k^{p/2}\epsilon^{-p}(\log(n/\delta))^{O(p^2)}$ for $p>2$ and $k\epsilon^{-4/p}(\log(n/\delta))^{O(1)}$ for $p<2$.
\end{lemma}
To see the bound on $\sum_i q_i$ for $p>2$ in Lemma~\ref{lem:wy24}, note that $\min\{1,y^{p/2}\}\leq y$ for every $y\geq 0$. Indeed, $y^{p/2}\leq y$ for $y\leq 1$ as $p/2\geq 1$, and $\min\{1,y^{p/2}\}\leq 1<y$ for $y>1$. We apply this with $y=n^{1-2/p}\alpha^{-2/p}\wh\tau_i$, for which $y^{p/2}=n^{p/2-1}\wh\tau_i^{p/2}/\alpha$, and sum over $i$.

Note that the one-round bounds on $\sum_i q_i$ cannot be replaced by the $n$-independent output sizes. To see this, consider $p=4$ and $k\leq \sqrt n$, and let $A\in \R^{n\times d}$ have $N=k\sqrt n$ rows forming an equal-norm tight frame of rank $k$ and $n-N$ zero rows. Then we have $\lambda_{A_k}=0$, and the zero rows have $q_i=0$. Each frame row has $\ov\tau_i=k/N=1/\sqrt n$ and $n^{p/2-1}\ov\tau_i^{p/2}=1$, so $q_i=1$ for all $N$ rows and $\sum_i q_i=k\sqrt n=kn^{1-2/p}$.

To speed up their algorithm, we note that the dominating runtime part is to sample from the rescaled leverage score distribution, and we could use Theorem~\ref{thm:ridge} with an inflated sample size.

\begin{theorem}
\label{thm:quantum_lp}
Under the hypotheses of Lemma~\ref{lem:wy24} ($n\geq \wt O(k^{p/2}/\epsilon)$ for $p>2$, none beyond those of Theorem~\ref{thm:ridge} for $p=2$, and the flatness precondition at a rank $k'=\wt O(k)$ for $p\in [1,2)$) and with $\delta$ a constant, there exists a quantum algorithm that outputs, with probability at least $1-O(\delta)$, a weighted row sample $SA$ with $\|SA(I-P_x)\|_{p,2}^p=(1\pm\epsilon)\|A(I-P_x)\|_{p,2}^p$ for all $x\in {\cal F}_k$ (hence also $\|SA(I-P_x)\|_{p,2}=(1\pm\epsilon)\|A(I-P_x)\|_{p,2}$), of expected size $\wt O(\epsilon^{-4/p}kn^{1-2/p})$ for $p\geq 2$ and $\wt O(\epsilon^{-2}k^{p/2}n^{1-p/2})$ for $p\in [1,2)$, and such that

\begin{itemize}
    \item For $p\geq 2$, it runs in expected time $\wt O(n^{1-1/p}dk^{0.5}\epsilon^{-2/p}+dk^{\omega-1})$ and uses, in expectation, $\wt O(n^{1-1/p}k^{0.5}\epsilon^{-2/p})$ queries to the points in $A$.
    \item For $p\in [1,2)$, it runs in expected time $\wt O(n^{1-p/4}dk^{p/4}\epsilon^{-1}+dk^{\omega-1})$ and uses, in expectation, $\wt O(n^{1-p/4}k^{p/4}\epsilon^{-1})$ queries to the points in $A$.
\end{itemize}
Here $\wt O(\cdot)$ hides factors $(\log(n/\epsilon))^{O(p)}$ in the hypothesis $n\geq \wt O(k^{p/2}/\epsilon)$, and factors $(\log(n/\epsilon))^{O(1)}$ elsewhere.

\end{theorem}

\begin{proof}
We apply Lemma~\ref{lem:wy24} with $\epsilon$ rescaled as stated there, which makes its guarantee multiplicative, and with $k'=\wt O(k)$ for $p<2$, as in the hypothesis. Then, with the probabilities $q_i$ chosen below, the sum of the sampling probabilities is $O(kn^{1-2/p}\alpha^{-2/p})$ for $p>2$ and $O(k'^{p/2}n^{1-p/2}/\alpha)$ for $p\in [1, 2)$. For constant $\delta$, the rescaled $\alpha$ satisfies $1/\alpha=O(\epsilon^{-2}L^{p}\log^3 n)$ for $p>2$, resp.\ $1/\alpha=O(\epsilon^{-2}L^{2p}\log^3 n)=\wt O(\epsilon^{-2})$ for $p<2$. Hence, for $p>2$, we have $\alpha^{-2/p}=O(\epsilon^{-4/p}L^2(\log n)^{6/p})=\wt O(\epsilon^{-4/p})$. In both cases, the hidden factor is $(\log(n/\epsilon))^{O(1)}$. For $p=2$, the sum is $O(k/\alpha)$ by the definition of $q_i$, and the guarantee is that of Section~\ref{sec:ridge}. Sampling with Lemma~\ref{lem:q_sampling_1_d} costs $\wt O(\sqrt{n\sum_i q_i})$ queries with per-query cost $d$, so for $p>2$ the number of queries is $\wt O(\sqrt{n\cdot kn^{1-2/p}\epsilon^{-4/p}})=\wt O(n^{1-1/p}k^{0.5}\epsilon^{-2/p})$, and for $p\in [1,2)$ it is $\wt O(\sqrt{n\cdot k'^{p/2}n^{1-p/2}\epsilon^{-2}})=\wt O(n^{1-p/4}k'^{p/4}\epsilon^{-1})=\wt O(n^{1-p/4}k^{p/4}\epsilon^{-1})$. The ridge leverage scores at rank $k$, resp.\ $k'$, are supplied by the oracle of Lemma~\ref{lem:classical_ridge} (Algorithm~\ref{alg:classical_ridge_ls}) built on the constant-accuracy output of Theorem~\ref{thm:ridge} with rank parameter $\min\{k,d\}$, resp.\ $\min\{k',d\}$. By Theorem~\ref{thm:ridge}, this output is an approximator, not only a coreset, and for $k'\geq d$ the ridge parameter $\lambda_{A_{k'}}$ vanishes and the scores coincide with those at rank $d$. By Lemma~\ref{lem:cmm_lem12}, the \textsc{Query} of the oracle returns a constant-factor approximation of $\ov\tau_i(A,\lambda_{A_k})$, resp.\ $\ov\tau_i(A,\lambda_{A_{k'}})$, in $\wt O(d)$ time, and scaling it by a constant gives a value $\wh\tau_i$ between the score and a constant multiple of it. We set $q_i=\min\{1,n^{p/2-1}\wh\tau_i^{p/2}/\alpha\}$, resp.\ $q_i=\min\{1,\wh\tau_i^{p/2}/\alpha\}$, which is admissible since Lemma~\ref{lem:wy24} only requires $q_i\geq \min\{1,\cdot\}$. Since $\sum_i\wh\tau_i=O(k)$, resp.\ $O(k')$, by Lemma~\ref{lem:classical_ridge}, the bounds of Lemma~\ref{lem:wy24} give the sums of the $q_i$ stated above. Here $k\leq d$, as ${\cal F}_k$ consists of $k$-dimensional subspaces of $\R^d$, and we write $k'=k$ for $p\geq 2$. The preprocessing takes $\wt O(d\min\{k',d\}^{\omega-1})$ time, and so does the additive term of Theorem~\ref{thm:ridge}. Since $k'=\wt O(k)$, both are $\wt O(dk^{\omega-1})$. The $\wt O(n^{0.5}dk'^{0.5})$ cost of Theorem~\ref{thm:ridge} is dominated by the sampling term, since $1-1/p\geq 1/2$, resp.\ since $n^{0.5}k'^{0.5}\leq n^{1-p/4}k'^{p/4}$ for $k'\leq n$. Up to logarithmic factors, the additive term $dk^{\omega-1}$ is also at most the sampling term whenever $n\geq k^2$. Indeed, the sampling term is at least $n^{0.5}dk^{0.5}$, where for $p<2$ we use $n^{1/2-p/4}\geq k^{1/2-p/4}$, and for $n\geq k^2$ we have $k^{\omega-1}\leq k^{1.5}\leq n^{0.5}k^{0.5}$ since $\omega<2.5$. Hence, for $n\geq k^2$, the first term of each running time is its leading term, with no assumption relating $n$ and $d$. Since the algorithm performs one round of sampling, its output has the stated expected size $m$ rather than the $n$-independent size of~\cite{wy24}. To reach the latter, one can run the algorithm of their Theorems 3.5 and 3.7 on the $m$ weighted sampled rows, where for the $(p,2)$-norm a weight $w$ rescales a row by $w^{1/p}$. As stated there, this takes $\wt O(md+d^\omega)$ time. Their proofs do not itemize the running time, and their Section 1.4.2 notes that ridge leverage scores can be approximated up to constant factors in $\wt O(\nnz(A)+d^\omega)$ time. We implement every step of their algorithm as follows and obtain $\wt O(md+dk^{\omega-1})$ time. Their algorithm runs $r=O(\log\log m)$ rounds. For $p<2$, a round first computes a constant-factor solution $\wt F$ of the current matrix by their Lemma 2.3 and flattens the matrix with respect to $\wt F$ by their Lemma 2.4, which multiplies the number of rows by at most $3/2$. The subspace $\wt F$ is spanned by $\wt O(k)$ rows, and we take the rank parameter of their Theorem 3.6 to be an upper bound $k''=\wt O(k)$ on its dimension, as that theorem requires. We set $k''=k$ for $p\geq 2$. Then the round applies their Theorem 3.3, resp.\ 3.6, with probabilities computed from constant-factor approximations of the ridge leverage scores of the current matrix, and keeps a subset of its rows. Hence every matrix that appears has at most $(3/2)^r m=\wt O(m)$ rows. As above, we approximate the ridge leverage scores of all rows of such a matrix $B$ by writing $B$ into the QRAM in $O(|B|d)$ time, running Theorem~\ref{thm:ridge} on $B$ at constant accuracy with rank parameter $\min\{k'',d\}$, and querying the oracle of Lemma~\ref{lem:classical_ridge} built on its output for every row. This takes $\wt O(|B|^{0.5}dk''^{0.5}+dk^{\omega-1}+|B|d)=\wt O(|B|d+dk^{\omega-1})$ time. For $p<2$, their Lemma 2.3 computes $\wt F$ in $\wt O(|B|d+k^\omega)$ time, and $k^\omega\leq dk^{\omega-1}$. Their Lemma 2.4 uses the residual norms $\rho_i=\|b_i^\top(I-P_{\wt F})\|_2$ of the rows $b_i$ of $B$, whose direct computation takes $\wt O(|B|dk)$ time. Instead, we compute an orthonormal basis $V$ of $\wt F$ and a sketch $G'(I-VV^\top)$ as in Algorithm~\ref{alg:classical_ridge_ls}, with a Gaussian matrix $G'$ of $O(\log(m/\delta))$ rows, in $\wt O(dk^{\omega-1})$ time. This gives estimates $\wh\rho_i\in[\rho_i/2,2\rho_i]$ for all rows in $\wt O(|B|d)$ time, and we run the splitting rule of their Lemma 2.4 with $\wh\rho_i$ in place of $\rho_i$. That is, let $\wh R=\sum_i\wh\rho_i^p$. If $\wh R=0$, then all $\rho_i$ vanish and $B$ is already flat. Otherwise, each row with $\wh\rho_i^p\geq 2\wh R/|B|$ is replaced by $l_i=\lceil \wh\rho_i^p|B|/(2\wh R)\rceil$ copies of $l_i^{-1/p}b_i$. This preserves $\|B(I-P_F)\|_{p,2}$ for every subspace $F$ and adds at most $\sum_i\wh\rho_i^p|B|/(2\wh R)=|B|/2$ rows. Since $\rho_i^p\leq 2^p\wh\rho_i^p$, each copy satisfies $\|l_i^{-1/p}b_i^\top(I-P_{\wt F})\|_2^p=\rho_i^p/l_i\leq 2^{p+1}\wh R/|B|$, and each row that is not split satisfies $\rho_i^p<2^{p+1}\wh R/|B|$. Let $B'$ be the flattened matrix and $R=\sum_i\rho_i^p=\|B'(I-P_{\wt F})\|_{p,2}^p$. Since $\wh R\leq 2^pR$ and $|B'|\leq 3|B|/2$, every row $b'$ of $B'$ satisfies $\|b'^\top(I-P_{\wt F})\|_2^p\leq 2^{2p+1}R/|B|\leq 3\cdot 4^p\|B'(I-P_{\wt F})\|_{p,2}^p/|B'|$, where $3\cdot 4^p<48$. This is the first half of the flatness precondition of their Theorem 3.6. The second half, $\|B'(I-P_{\wt F})\|_{p,2}^p\leq O(\min_{x\in{\cal F}_k}\|B'(I-P_x)\|_{p,2}^p)$, follows from their Lemma 2.3, since flattening preserves all these costs. If each of these subroutines is run with failure probability $\delta/r$ in each round, the reduction fails with probability $O(\delta)$. Summing over the $r$ rounds, it takes $\wt O(md+dk^{\omega-1})$ time, which is at most the sampling time plus $\wt O(dk^{\omega-1})$ whenever $n\geq k^{p/2}\epsilon^{-2}$ for $p\geq 2$, resp.\ $n\geq k\epsilon^{-4/p}$ for $p\in [1,2)$, since then we have $md\leq n^{1-1/p}dk^{0.5}\epsilon^{-2/p}$, resp.\ $md\leq n^{1-p/4}dk^{p/4}\epsilon^{-1}$, up to the hidden logarithmic factors. These conditions hold under our assumption that $n\gg\poly(k,1/\epsilon)$, and composing the two $(1\pm\epsilon)$ guarantees costs a constant factor in $\epsilon$. Finally, the constant-accuracy approximator of Theorem~\ref{thm:ridge}, the Gaussian sketches of the oracle of Lemma~\ref{lem:classical_ridge} built on it, the guarantee of Lemma~\ref{lem:wy24}, and the sampler of Lemma~\ref{lem:q_sampling_1_d} each fail with probability at most $\delta$. For the sketches, the distortion bound of the oracle holds for all $n$ rows since, as in the proof of Theorem~\ref{thm:ridge}, they are taken with $O(\log(n/\delta))$ rows. By a union bound, the success probability is at least $1-4\delta=1-O(\delta)$, and the running time is an expected time in the sense of the convention following Lemma~\ref{lem:q_sampling_1_d}. For $p\in [1,2)$, we assume the flatness precondition of Lemma~\ref{lem:wy24}. The step of~\cite{wy24} that establishes it changes the sampling distribution, and we leave a sublinear-time implementation of it open.
\end{proof}
\section{Quantum Tensor Low-Rank Approximation}
\label{sec:tensor}
In this section, we provide a quantum algorithm for computing the Frobenius norm low-rank approximation of a 3rd order tensor $A\in \R^{n\times n \times n}$. The goal is to find a rank-$k$ tensor $B:=\sum_{i=1}^k u_i\otimes v_i\otimes w_i$ for $u_i,v_i,w_i\in \R^n$, such that $\|A-B\|_F^2 \leq (1+\epsilon)\cdot {\rm OPT}$ where ${\rm OPT}:=\inf_{B:{\rm rank}(B)=k} \|A-B\|_F^2$. The first caveat is that such an optimal rank-$k$ solution might not even exist. We provide algorithms with constant relative error $(2+\epsilon)$, specifically a bicriteria algorithm with output rank $k^2$, and a fixed-parameter algorithm with exact output rank $k$ when an optimal rank-$k$ solution exists.

\subsection{Preliminary}

 Given a 3rd order tensor $A\in \R^{n\times n\times n}$, we define the rank of $A$ as the smallest integer $k$ such that $A=\sum_{i=1}^k u_i\otimes v_i\otimes w_i$ where $u_i,v_i, w_i\in \R^n$. We use $\otimes$ to denote the Kronecker product of two matrices, i.e., for $A\in \R^{a\times b}, B\in \R^{c\times d}$, $A\otimes B\in \R^{ac\times bd}$ and $A\otimes B=\begin{bmatrix}
A_{1,1} B & A_{1,2} B& \ldots & A_{1, b} B \\
\vdots & \vdots & \ldots & \vdots \\
A_{a,1} B & A_{a,2}  B & \ldots & A_{a, b} B
\end{bmatrix}$. We use $\odot$ to denote a product of two matrices defined as for $A\in \R^{a\times b}, B\in \R^{a\times d}$, $A\odot B\in\R^{a\times bd}$ where $A\odot B=\begin{bmatrix}
    A_{1,*}\otimes B_{1,*} \\
    A_{2,*}\otimes B_{2,*} \\
    \vdots \\
    A_{a,*}\otimes B_{a,*}
\end{bmatrix}$, i.e., the matrix formed by computing tensor product between corresponding rows of $A$ and $B$. Given a tensor $A\in \R^{n_1\times n_2\times n_3}$ and three matrices $B_1\in \R^{n_1\times d_1}, B_2\in \R^{n_2\times d_2}$ and $B_3\in \R^{n_3\times d_3}$, we define the $(\cdot,\cdot,\cdot)$ operator as
\begin{align*}
    A(B_1, B_2, B_3)_{i,j,l} = & ~ \sum_{i'=1}^{n_1}\sum_{j'=1}^{n_2}\sum_{l'=1}^{n_3} A_{i',j',l'} (B_1)_{i',i} (B_2)_{j',j} (B_3)_{l',l}, \forall (i,j,l)\in [d_1]\times [d_2]\times [d_3],
\end{align*}
consequently, $A(B_1,B_2,B_3)\in \R^{d_1\times d_2\times d_3}$. One could also set any of the $B_i$'s as $I_{n_i}$ and for example, $A(B_1,I_{n_2}, I_{n_3})\in \R^{d_1\times n_2\times n_3}$. When the dimension of the identity matrix is clear from context, we abbreviate it as $I$ for notational simplicity. For $A\in \R^{n_1\times n_2\times n_3}$, we use $A_1\in \R^{n_1\times n_2n_3},A_2\in \R^{n_2\times n_1n_3}$ and $A_3\in \R^{n_3\times n_1n_2}$ to denote the three matrices such that the $[3]\setminus \{i\}$ dimensions are flattened. We use the convention $(A_1)_{i,(j-1)n_3+l}=(A_2)_{j,(i-1)n_3+l}=(A_3)_{l,(i-1)n_2+j}=A_{i,j,l}$. Under this convention, for the design matrices $B_m$ in Algorithm~\ref{alg:tensor_from_low_rank_to_CURT} and $Z_j$ in the proof of Theorem~\ref{thm:quantum_fpt_two} below, the $i$-th row is the Kronecker product of the $i$-th columns of the two remaining factors in increasing mode order. For factor matrices $U, V, W\in \R^{n\times k}$ with columns $U_i, V_i, W_i$, we write $U\otimes V\otimes W:=\sum_{i=1}^k U_i\otimes V_i\otimes W_i\in \R^{n\times n\times n}$ for the tensor with these factors, as in~\cite{swz19}. Note that this is not the Kronecker product of the three matrices. 

We also state an algorithm due to~\cite{swz19} for sampling according to leverage scores of $U\odot V$:

\begin{lemma}
\label{lem:tensor_leverage}
Given two matrices $U\in \R^{k\times n_1}$ and $V\in \R^{k\times n_2}$, there exists an algorithm 
\begin{align*}
    \textsc{TensorLeverageScore}(U, V, n_1, n_2, k, \epsilon, R_{\rm sample})
\end{align*}
that takes 
\begin{align*}
    O((n_1+n_2)\cdot \poly(\log(n_1n_2), k, \epsilon^{-1})\cdot R_{\rm sample})
\end{align*}
time to generate a weighted sampling matrix $D\in \R^{n_1n_2\times R_{\rm sample}}$ according to a constant-factor approximation of the leverage score distribution of the columns of $U\odot V$, with probability at least $1-1/\poly(n_1n_2)$ minus the failure probability of its internal sketch, which can be made any constant at a constant factor in the $\poly$ term. (This is Lemma C.32 and Algorithm 10 of~\cite{swz19}. When $U\odot V$ is rank-deficient, the matrix $R^{-1}$ used there is read as the pseudo-inverse of a rank-revealing factorization, which samples by the leverage scores of the row space of $U\odot V$.)
\end{lemma}

To obtain our fixed-parameter tractable algorithm for rank-$k$ tensor low-rank approximation, we require the following result from~\cite{swz19}:
\begin{lemma}
\label{lem:poly_solver}
Let $\max\{t_i, d_i\}\leq n$, given a $t_1\times t_2\times t_3$ tensor $A$ and three matrices: $T_1\in \R^{t_1\times d_1}, T_2\in \R^{t_2\times d_2}$ and $T_3\in \R^{t_3\times d_3}$, if for any $\delta>0$ there exists a solution to
\begin{align*}
    \min_{X_1, X_2, X_3} \|\sum_{i=1}^k (T_1X_1)_i\otimes (T_2X_2)_i\otimes (T_3X_3)_i-A \|_F^2 := & ~ \OPT,
\end{align*}
and each entry of $X_i$ could be expressed with $O(n^\delta)$ bits (and each entry of $A$ and of $T_1, T_2, T_3$ is a rational number of $O(n^\delta)$ bits), then there exists an algorithm that takes $n^{O(\delta)}\cdot 2^{O(d_1k+d_2k+d_3k)}$ time and outputs three matrices $\wh X_1, \wh X_2$ and $\wh X_3$ such that $\|\sum_{i=1}^k (T_1\wh X_1)_i\otimes (T_2\wh X_2)_i\otimes (T_3\wh X_3)_i-A \|_F^2 =  \OPT$.
\end{lemma}

\begin{remark}\label{rem:poly_solver_reading}
We read the proof of Theorem C.45 of~\cite{swz19} as follows. It binary-searches the cost, with range and precision fixed by the bit-size lower bound of their Theorem B.11, and at each step it decides whether the polynomial system ``cost $\leq$ guess'' has a solution in the ball $\|x\|_2^2\leq 2^{O(n^\delta)}$. Therefore, its output has cost at most that of \emph{any} point of that ball with entries of $O(n^\delta)$ bits, up to an additive $2^{-\Omega(n^\delta)}$. The lemma as printed states the case in which such a point is a minimizer, and we use it in the general form, as Lemma~\ref{lem:swz19_fpt} does.
\end{remark}

\subsection{A \texorpdfstring{$(2+\epsilon)$}{(2+eps)} Bicriteria Algorithm with Output Rank \texorpdfstring{$k^2$}{k\^2}}
\label{sec:tensor:two}

In the first stage, we fix a $k$-dimensional span $Q$ for the first factors. We then treat the second and third modes \emph{jointly} as one matrix low-rank approximation problem on the $n\times kn$ matrix obtained by contracting $A$ with $Q$, and we solve it using the column coreset of Corollary~\ref{thm:quantum_pcp}. A rank-$k$ factorization of this matrix unfolds to a tensor of CP rank $k^2$, with only $k$ distinct first factors.

\begin{algorithm}[!ht]
\caption{Quantum $(2+\epsilon)$ bicriteria tensor low-rank approximation with output rank $k^2$.}
\label{alg:quantum_bicriteria_two}
\begin{algorithmic}[1]
\Procedure{QBicriteriaTensorLowRankTwo}{$A\in \R^{n\times n\times n}, k, \epsilon$}
\State $C_1\gets \textsc{QLowRankCMM}(A_1, k, \epsilon/3, 0.001)$ \Comment{$C_1\in \R^{n\times s_1}$, $s_1=\wt O(k/\epsilon^2)$, by Corollary~\ref{thm:quantum_pcp}.}
\State $Q\gets$ top-$k$ left singular vectors of $C_1$ \Comment{$Q\in \R^{n\times k}$, $Q^\top Q=I_k$.}
\State $S_0\gets$ leverage score sampling matrix of the rows of $Q$ with $m=O(k\log k+k/\epsilon^2)$ rows \Comment{Lemma~\ref{lem:leverage_score_optimal_solution}.}
\State $\wt Q^\top\gets (S_0Q)^\dagger S_0$ \Comment{$\wt Q^\top\in \R^{k\times n}$.}
\State Let $\wt M\in \R^{n\times kn}$ be the mode-$2$ unfolding of $A(\wt Q, I, I)$, i.e., $\wt M_{j,(t,l)}=\sum_{i\in {\rm supp}(S_0)}\wt Q_{i,t}A_{i,j,l}$\label{ln:bic2:M} \Comment{accessed implicitly}
\State $C_M\gets \textsc{QLowRankCMM}(\wt M, k, \epsilon/3, 0.001)$ \Comment{$C_M\in \R^{n\times s}$, $s=\wt O(k/\epsilon^2)$ columns of $\wt M$.}
\State $V\gets$ top-$k$ left singular vectors of $C_M$ \Comment{$V\in \R^{n\times k}$.}
\State $S_1\gets$ leverage score sampling matrix of the rows of $V$ with $m_1=O(k\log k+k/\epsilon)$ rows
\State $Y\gets (S_1V)^\dagger S_1\wt M$ \Comment{$Y\in \R^{k\times kn}$.}
\State For $i, t\in [k]$ let $y_{i,t}\in \R^n$ be the vector $(Y_{i,(t,l)})_{l\in [n]}$
\State \Return $Q$, $V$, $\{y_{i,t}\}_{i,t\in [k]}$, representing $B=\sum_{i,t\in [k]} Q_t\otimes V_i\otimes y_{i,t}$
\EndProcedure
\end{algorithmic}
\end{algorithm}

\begin{theorem}
\label{thm:quantum_bicriteria_two}
Given a 3rd order tensor $A\in \R^{n\times n\times n}$, a positive integer $k\leq n$ and $\epsilon\in (0, 0.1)$, let $\OPT:=\inf_{{\rm rank}\text{-}k~A_k}\|A-A_k\|_F^2$. Algorithm~\ref{alg:quantum_bicriteria_two} outputs $Q, V\in \R^{n\times k}$ and vectors $y_{i,t}\in \R^n$, $i,t\in [k]$, such that the tensor $B=\sum_{i,t\in [k]}Q_t\otimes V_i\otimes y_{i,t}$, of CP rank at most $k^2$ with only $k$ distinct first factors, satisfies
\begin{align*}
    \|A-B\|_F^2\leq & ~ (2+\epsilon)\cdot \OPT
\end{align*}
with probability at least $0.9$. It runs in time
\begin{align*}
    \wt O(\epsilon^{-1}n^2k^{0.5}+\epsilon^{-3}n^{1.5}k^2+\epsilon^{-3}nk^3+\epsilon^{2-2\omega}nk^{\omega-1}) = \wt O(\epsilon^{-1}n^2k^{0.5}+n^{1.5}\poly(k/\epsilon)),
\end{align*}
using $\wt O(\epsilon^{-1}nk^{0.5})$ queries to mode-$1$ fibers of $A$ (the columns of $A_1$) and $\wt O(\epsilon^{-3}n^{0.5}k^2)$ queries to mode-$2$ and mode-$3$ fibers.
\end{theorem}

\begin{proof}
We run Algorithm~\ref{alg:quantum_bicriteria_two} with accuracy parameter $\epsilon_0:=\epsilon/13$ and prove $\|A-B\|_F^2\leq (2+13\epsilon_0)\cdot \OPT$. This rescaling and the accuracy $\epsilon/3$ of the two coreset calls change the running time and query bounds only by constant factors, which are absorbed by the $\wt O(\cdot)$. We write $\epsilon$ for $\epsilon_0$ throughout the proof, and fix $\eta>0$ and a rank-$k$ tensor $B^\star=\sum_{i=1}^k u_i\otimes v_i\otimes w_i$ with $\|A-B^\star\|_F^2\leq (1+\epsilon)\OPT'$, where $\OPT':=\OPT+\eta$. Such a $B^\star$ exists by the definition of $\OPT$ as an infimum, even when $\OPT=0$ is not attained, and if the infimum is attained, one may take $\eta=0$. Since the output $B$ of the algorithm does not depend on $B^\star$, the bound $\|A-B\|_F^2\leq (2+13\epsilon)(\OPT+\eta)$ for every $\eta>0$ gives the claim, and hence below we write $\OPT$ for $\OPT'$. The unfolding $B^\star_1$ of $B^\star$ has matrix rank at most $k$, so we have $\|A_1-[A_1]_k\|_F^2\leq \|A_1-B^\star_1\|_F^2=\|A-B^\star\|_F^2\leq (1+\epsilon)\OPT$.

\emph{Step 1: the first-factor span.} By Corollary~\ref{thm:quantum_pcp}, $C_1\subseteq A_1$ is an $(\epsilon/3)$-coreset of the columns of $A_1$ for the $k$-subspace cost. Therefore, by Lemma 3 of~\cite{cem+15}, as stated in Section~\ref{sec:cmm}, the projection $QQ^\top$ onto the top-$k$ left singular vectors of $C_1$ satisfies
\begin{align*}
    e_1:=\|A_1-QQ^\top A_1\|_F^2\leq (1+\epsilon)\|A_1-[A_1]_k\|_F^2\leq (1+\epsilon)^2\OPT.
\end{align*}

\emph{Step 2: Pythagoras.} Since $Q$ has orthonormal columns, $(I-QQ^\top)A_1$ is orthogonal to $QY$ for every $Y\in \R^{k\times n^2}$, and hence we have
\begin{align}\label{eq:tensor_pythagoras}
    \|A_1-QY\|_F^2 = & ~ e_1+\|Q^\top A_1-Y\|_F^2 \qquad \text{for all } Y\in \R^{k\times n^2}.
\end{align}

\emph{Step 3: reshaping.} The matrix $Q^\top A_1\in \R^{k\times n^2}$ is the mode-$1$ unfolding of the $k\times n\times n$ tensor $A(Q, I, I)$, and we denote the mode-$2$ unfolding of this tensor by $M\in \R^{n\times kn}$, i.e., $M_{j,(t,l)}=\sum_{i}Q_{i,t}A_{i,j,l}$. Unfoldings preserve Frobenius norms. The tensor $B^\star(Q, I, I)=\sum_{i=1}^k (Q^\top u_i)\otimes v_i\otimes w_i$ has mode-$2$ unfolding $M^\star=\sum_{i=1}^k v_i\big((Q^\top u_i)\otimes w_i\big)^\top$, whose rank is at most $k$, and we have
\begin{align*}
    \|M-M^\star\|_F^2=\|Q^\top (A_1-B^\star_1)\|_F^2\leq \|A_1-B^\star_1\|_F^2\leq (1+\epsilon)\OPT,
\end{align*}
since $Q^\top$ is a contraction. Conversely, let $V\in \R^{n\times k}$ and $Y\in \R^{k\times kn}$, and for $i\in [k]$, let $Y^{(i)}\in \R^{k\times n}$ be the $i$-th row of $Y$ reshaped with $Y^{(i)}_{t,l}=Y_{i,(t,l)}$. Then $VY$ is the mode-$2$ unfolding of the $k\times n\times n$ tensor $\sum_{i,t\in [k]}e_t\otimes V_i\otimes (Y^{(i)}_{t,*})^\top$, and applying $Q$ to the first mode gives $B=\sum_{i,t\in [k]}Q_t\otimes V_i\otimes y_{i,t}$ with $y_{i,t}=(Y^{(i)}_{t,*})^\top$. The mode-$1$ unfolding of $B$ is $Q\cdot Y'$, where $Y'$ is the mode-$1$ unfolding of the $k\times n\times n$ tensor above. Therefore, by Eq.~\eqref{eq:tensor_pythagoras}, we have
\begin{align}\label{eq:tensor_total}
    \|A-B\|_F^2 = & ~ e_1+\|M-VY\|_F^2.
\end{align}
Moreover, $B$ has CP rank at most $k^2$ and only $k$ distinct first factors.

\emph{Step 4: implicit access.} A column $(t,l)$ of $M$ depends on the whole frontal slice $A_{*,*,l}$, which is too expensive to read. Instead, $S_0$ samples $m=O(k\log k+k/\epsilon^2)$ rows of $Q$ according to the leverage scores of $Q$, and we apply Lemma~\ref{lem:leverage_score_optimal_solution} with accuracy parameter $\epsilon^2$, design $Q$ of rank $k$, and response $A_1$. Thus, $Y_0:=(S_0Q)^\dagger S_0A_1=\wt Q^\top A_1$ satisfies $\|QY_0-A_1\|_F^2\leq (1+\epsilon^2)\min_Y\|QY-A_1\|_F^2=(1+\epsilon^2)e_1$, and by Eq.~\eqref{eq:tensor_pythagoras} with $Y=Y_0$, we have $\|Q^\top A_1-\wt Q^\top A_1\|_F^2\leq \epsilon^2 e_1$. Note that the matrix $\wt M$ of the algorithm is the mode-$2$ unfolding of $A(\wt Q, I, I)$, hence
\begin{align*}
    \|\wt M-M\|_F^2=\|(\wt Q^\top-Q^\top)A_1\|_F^2\leq \epsilon^2 e_1\leq \epsilon^2(1+\epsilon)^2\OPT.
\end{align*}
Since $\wt Q^\top$ is supported on the $m$ rows in ${\rm supp}(S_0)$, an entry $\wt M_{j,(t,l)}$ is computed from the $m$ entries $A_{i,j,l}$, $i\in {\rm supp}(S_0)$. Hence, a column $(t,l)$ of $\wt M$ is computed from the $m$ mode-$2$ fibers $A_{i,*,l}$, a row $j$ from the $m$ mode-$3$ fibers $A_{i,j,*}$, and all three fiber types are admissible queries, see Section~\ref{sec:query}. We write the $mk$ entries of $\wt Q^\top$ to QRAM once, and then a column of $\wt M$ can be formed inside the sampler of Lemma~\ref{lem:q_sampling_1_d}. Every bound below is routed through $M$ and never through $\wt Q^\top(A_1-B^\star_1)$, because $\wt Q^\top$ is not a contraction.

\emph{Step 5: the matrix problem.} Corollary~\ref{thm:quantum_pcp} applied to $\wt M$, with $d=n$ rows and $kn$ columns, returns an $(\epsilon/3)$-coreset $C_M$ of the columns of $\wt M$ for the $k$-subspace cost. Let $V$ be the top-$k$ left singular vectors of $C_M$. By Lemma 3 of~\cite{cem+15}, we have
\begin{align*}
    \|\wt M-VV^\top \wt M\|_F^2\leq (1+\epsilon)\|\wt M-\wt M_k\|_F^2\leq (1+\epsilon)\|\wt M-M^\star\|_F^2,
\end{align*}
using ${\rm rank}(M^\star)\leq k$. Next, $S_1$ samples $m_1=O(k\log k+k/\epsilon)$ rows of $V$ according to its leverage scores, and by Lemma~\ref{lem:leverage_score_optimal_solution} with design $V$ and response $\wt M$, the matrix $Y=(S_1V)^\dagger S_1\wt M$ satisfies
\begin{align*}
    \|VY-\wt M\|_F^2\leq (1+\epsilon)\min_{Y'}\|VY'-\wt M\|_F^2=(1+\epsilon)\|\wt M-VV^\top\wt M\|_F^2\leq (1+\epsilon)^2\|\wt M-M^\star\|_F^2.
\end{align*}
By the triangle inequality and the bounds of Steps 3 and 4, for $\epsilon\leq 0.1$ we have
\begin{align*}
    \|\wt M-M^\star\|_F\leq & ~ \|\wt M-M\|_F+\|M-M^\star\|_F \\
    \leq & ~ \big(\epsilon(1+\epsilon)+\sqrt{1+\epsilon}\big)\sqrt{\OPT}\leq (1+1.6\epsilon)\sqrt{\OPT},
\end{align*}
so $\|\wt M-M^\star\|_F^2\leq (1+3.5\epsilon)\OPT$, and consequently $\|VY-\wt M\|_F^2\leq (1+\epsilon)^2(1+3.5\epsilon)\OPT\leq (1+6.4\epsilon)\OPT$. Then
\begin{align*}
    \|M-VY\|_F\leq & ~ \|M-\wt M\|_F+\|\wt M-VY\|_F \\
    \leq & ~ \big(\epsilon(1+\epsilon)+\sqrt{1+6.4\epsilon}\big)\sqrt{\OPT}\leq (1+4.3\epsilon)\sqrt{\OPT},
\end{align*}
hence $\|M-VY\|_F^2\leq (1+10.5\epsilon)\OPT$ for $\epsilon\leq 0.1$. By Eq.~\eqref{eq:tensor_total} and Step 1, we have
\begin{align*}
    \|A-B\|_F^2\leq (1+\epsilon)^2\OPT+(1+10.5\epsilon)\OPT\leq (2+13\epsilon)\OPT.
\end{align*}
The failure events are the two calls of Corollary~\ref{thm:quantum_pcp}, each with probability $0.001$, and the two applications of Lemma~\ref{lem:leverage_score_optimal_solution}, each run with failure probability at most $0.04$ by Remark~\ref{rem:delta_dependence}, so a union bound gives success probability at least $0.9$.

\emph{Running time and queries.} By Corollary~\ref{thm:quantum_pcp} with $d=n$ and $n^2$ columns, computing $C_1$ takes $\wt O(\epsilon^{-1}n^2k^{0.5}+nk^{\omega-1})$ time and $\wt O(\epsilon^{-1}nk^{0.5})$ queries to the columns of $A_1$, i.e., mode-$1$ fibers of $A$. The SVD of $C_1\in \R^{n\times s_1}$ takes $\wt O(ns_1^{\omega-1})=\wt O(\epsilon^{2-2\omega}nk^{\omega-1})$ time. The leverage scores of $Q$ are its squared row norms, which are computed in $O(nk)$ time, and $(S_0Q)^\dagger$ costs $\poly(k/\epsilon)$. By Theorem~\ref{thm:general} with the oracle of Lemma~\ref{lem:classical_ridge}, the call $\textsc{QLowRankCMM}(\wt M, k, \epsilon/3, 0.001)$ accesses $\wt O(\sqrt{kn\cdot s})=\wt O(\epsilon^{-1}n^{0.5}k)$ columns of $\wt M$, where $s=\wt O(k/\epsilon^2)$. Each column is formed from $m$ mode-$2$ fibers in $O(nm)=\wt O(\epsilon^{-2}nk)$ time, and this gives a total of $\wt O(\epsilon^{-3}n^{1.5}k^2)$ time and $\wt O(\epsilon^{-3}n^{0.5}k^2)$ mode-$2$ fiber queries. The oracle preprocessing of this call takes $\wt O(nk^{\omega-1})$ time per level for the SVDs of the $n\times \wt O(k)$ approximators, each oracle query on an already formed column takes $\wt O(n)$ time, which is dominated by forming the column, and the SVD of $C_M\in \R^{n\times s}$ takes $\wt O(\epsilon^{2-2\omega}nk^{\omega-1})$ time. For these SVDs, we materialize the $\wt O(k)$ approximator columns per level and the $s$ coreset columns of $\wt M$, which costs $\wt O(s\cdot nm)=\wt O(\epsilon^{-4}nk^2)$ time and $\wt O(sm)$ mode-$2$ fiber queries. This is at most the sampling cost above when $n\geq \epsilon^{-2}$. When $n<\epsilon^{-2}$, at most $\min\{s,kn\}$ columns are materialized, which takes $\wt O(\epsilon^{-2}n^2k^2)\leq \wt O(\epsilon^{-3}n^{1.5}k^2)$ time and $\wt O(\epsilon^{-2}nk^2)$ queries. Computing $Y$ requires the $m_1$ sampled rows of $\wt M$, and each of them is formed from $m$ mode-$3$ fibers in $O(nkm)$ time. This takes $\wt O(m_1\cdot nkm)=\wt O(\epsilon^{-3}nk^3)$ time and $\wt O(\epsilon^{-3}k^2)$ mode-$3$ fiber queries, plus $\wt O(m_1k\cdot kn)$ time for the product $(S_1V)^\dagger S_1\wt M$. Summing up gives the stated time, and the queries are $\wt O(\epsilon^{-1}nk^{0.5})$ mode-$1$ fibers and $\wt O(\epsilon^{-3}n^{0.5}k^2)$ mode-$2$ and mode-$3$ fibers.
\end{proof}

\begin{remark}[A limitation of this route]\label{rem:towards_one_eps}

The route of Theorem~\ref{thm:quantum_bicriteria_two} cannot reach $(1+\epsilon)$, since its first stage fixes the span of the first factors to a $k$-dimensional subspace $\mathrm{span}(Q)$. Consider $k=1$, $n=m+1$, $A_{1,1,1}=1$, and $A_{2,j,j}=c/\sqrt m$ for $j=2,\ldots,n$, where $1<c\leq \sqrt m$ and all other entries are zero. Then $\OPT=c^2$, attained by $e_1\otimes e_1\otimes e_1$, and approximating the diagonal slice instead costs at least $1+c^2(1-1/m)\geq c^2$. Once $c^2(1-\epsilon)>1+\epsilon$, the first stage returns $Q=e_2$ with residual $\|A_1-QQ^\top A_1\|_F^2=1$, and $Q^\top A_1$ reshapes to $(c/\sqrt m)\cdot I_m$ padded with a zero row and column, whose best rank-$r$ approximation has error $c^2(1-r/m)$. Therefore, every output tensor of CP rank $r$ with first factors in $\mathrm{span}(Q)$ costs at least $1+c^2(1-r/m)$, which is a ratio of at least $1/c^2+1-r/m$ to $\OPT$. This ratio tends to $1+1/c^2$ as $m\to\infty$, and if in addition $c^2\downarrow (1+\epsilon)/(1-\epsilon)$, it tends to $2-2\epsilon/(1+\epsilon)$, which exceeds $1+\epsilon$ for every $\epsilon\in (0,0.1]$. This is a limitation of the route and not of the problem, and whether $(1+\epsilon)$ is reachable in sublinear time remains open, see Section~\ref{sec:conclusion}.

\end{remark}

\subsection{Quantum Tensor Low-Rank Approximation: Fixed-Parameter Tractable Algorithm}
\label{sec:tensor:fpt_two}

Reusing the first stage of Section~\ref{sec:tensor:two}, we fix a $k$-dimensional span $Q$ for the first factors and form the $k\times n\times n$ tensor $A(\wt Q, I, I)$ explicitly. We then solve the remaining rank-$k$ problem to $(1+\epsilon)$ using the classical fixed-parameter algorithm of~\cite{swz19} cited in Table~\ref{tab:optimization}. The output has exact rank $k$ and ratio $(2+\epsilon)$. We import the classical algorithm in the following form.

\begin{lemma}[Competitor-relative form of the proof method of Theorem C.2 (I) in~\cite{swz19}]
\label{lem:swz19_fpt}
Let $A'\in \R^{n\times n\times n}$ have entries of $O(n^\delta)$ bits (fixed-point, i.e., integers over a common power-of-two denominator, which is how we read the rounding of~\cite{swz19} to ``an integer expressed using $O(n^{\delta'_2})$ bits''), let $k\geq 1$, $\epsilon\in (0,1)$ and $n^{\delta'_2}=C_0\,n^\delta 2^{O(k^2/\epsilon)}$ for a sufficiently large constant $C_0$ (the parameter of~\cite{swz19}), and let $B'=\sum_{i=1}^k u'_i\otimes v'_i\otimes w'_i$ be any fixed rank-$k$ tensor whose factor matrices $U', V', W'\in \R^{n\times k}$ satisfy $\max\{\|U'\|_F, \|V'\|_F, \|W'\|_F\}\leq 2^{O(n^{\delta'_2})}$ ($B'$ is not given to the algorithm). There is a classical randomized algorithm that runs in $({\rm nnz}(A')+n\poly(k/\epsilon)+2^{O(k^2/\epsilon)})\cdot n^{O(\delta)}$ time in the unit cost RAM model with $O(\log n)$-bit words and outputs $U, V, W\in \R^{n\times k}$ with entries of $n^{\delta'_2}$ bits such that $\|U\otimes V\otimes W-A'\|_F^2\leq (1+\epsilon)\|B'-A'\|_F^2+2^{-\Omega(n^{\delta'_2})}\max\{1, \|A'\|_F^2\}$ with probability at least $9/10$.
\end{lemma}
Theorem C.2 of~\cite{swz19} is stated relative to $\OPT(A')=\inf_{{\rm rank}\text{-}k}\|A_k-A'\|_F^2$ under attainment hypotheses, and Theorem C.1 there is stated relative to the minimum over rank-$k$ tensors. We import the proof method of Part (I) of Theorem C.2, in which the factors of a fixed rank-$k$ tensor are rounded to $O(n^{\delta'_2})$-bit integers. The rounding gives the reduced problem a near-optimal solution of bounded bit size, which is the hypothesis of the polynomial system verifier of Lemma~\ref{lem:poly_solver} (Theorem C.45 of~\cite{swz19}), read as in Remark~\ref{rem:poly_solver_reading}. The sketched regressions of Theorem C.1 are then run with those fixed factors, and they yield a rank-$k$ output within $(1+\epsilon)$ of the cost of the rounded tensor. In~\cite{swz19}, the fixed tensor is an optimal $A_k$, and the rounding loss is absorbed through a lower bound on $\OPT(A')$. We instead fix an arbitrary bounded-norm competitor and keep the rounding loss as the additive term, as~\cite{swz19} themselves do in their Theorem 1.2 and in Part (IV) of Theorem C.2, where, as they explain, an additive $2^{-\poly(n)}$ is needed to write down the solution in the unit cost RAM model, since in general its entries may be irrational. We use the lemma with the competitor exhibited in the proof below and absorb the additive term there. Note that the rank-$k$ infimum of the tensor we hand to the algorithm need not be attained.

\begin{algorithm}[!ht]\caption{Quantum $(2+\epsilon)$ FPT rank-$k$ low-rank approximation.}\label{alg:quantum_fpt_two}
\begin{algorithmic}[1]
\Procedure{\textsc{QFPTLowRankTwo}}{$A,k,\epsilon$} \Comment{Theorem~\ref{thm:quantum_fpt_two}}
\State $C_1\gets \textsc{QLowRankCMM}(A_1, k, \epsilon/3, 0.001)$ and $Q\gets$ top-$k$ left singular vectors of $C_1$ \Comment{As in Algorithm~\ref{alg:quantum_bicriteria_two}.}
\State $Q'\gets$ $Q$ with entries rounded to $cn^{\delta'_2}$ bits after the binary point, $S_0\gets$ leverage score sampling matrix of the rows of $Q'$ with $m=O(\epsilon^{-2}k\log k)$ expected rows and sampling probabilities floored at $n^{-2}$, and $\wt Q^\top\gets (S_0Q')^\dagger S_0$, entries rounded to $c\,n^\delta+\lceil \log_2(100/\epsilon)\rceil$ bits after the binary point ($c\geq 2$)
\State Form $\wt A:=A(\wt Q, I, I)\in \R^{k\times n\times n}$ explicitly: $\wt A_{t,j,l}=\sum_{i\in {\rm supp}(S_0)}\wt Q_{i,t}A_{i,j,l}$ \Comment{Reads the $m$ slabs $A_{i,*,*}$, $i\in {\rm supp}(S_0)$.}
\For{$j=1,2,3$}
\State $(X_j, V_j, W_j)\gets$ the classical algorithm of Lemma~\ref{lem:swz19_fpt} on $\wt A$ (whose entries are fixed-point numbers of $O(n^\delta)$ bits, formed from the rounded $\wt Q^\top$, and which is padded by zeros to $n\times n\times n$), with $X_j\in \R^{k\times k}$ the first $k$ rows of the returned $n\times k$ first factor\label{ln:fpt2:X} \Comment{$V_j, W_j\in \R^{n\times k}$}
\State $\wh c_j\gets \|G(\wt A_{(2)}-(T_j)_{(2)})\|_F^2$ for a Rademacher matrix $G\in \R^{s\times n}$ with i.i.d.\ $\pm 1$ entries scaled by $s^{-1/2}$, $s\geq 48000\,\epsilon^{-2}$\label{ln:fpt2:c} \Comment{$T_j:=\sum_{i=1}^k (X_j)_i\otimes (V_j)_i\otimes (W_j)_i$}
\EndFor
\State $j^\star\gets \arg\min_j \wh c_j$
\State \Return $U:=Q'X_{j^\star}\in \R^{n\times k}$, $V_{j^\star}$, $W_{j^\star}$
\EndProcedure
\end{algorithmic}
\end{algorithm}
\noindent\emph{Note on Algorithm~\ref{alg:quantum_fpt_two}.} In Line~\ref{ln:fpt2:c}, $(\cdot)_{(2)}$ denotes the mode-$2$ unfolding, which is an $n\times kn$ matrix.

\begin{theorem}\label{thm:quantum_fpt_two}
Let $k\geq 1$ be an integer, let $A\in \R^{n\times n\times n}$ have entries of $O(n^\delta)$ bits for some $\delta>0$, where we assume $n^\delta\geq \log n$ as in~\cite{swz19}, let $\OPT:=\inf_{{\rm rank}\text{-}k~A_k}\|A-A_k\|_F^2>0$ be attained by $A_k=U^\star\otimes V^\star\otimes W^\star$ with $\max\{\|U^\star\|_F, \|V^\star\|_F, \|W^\star\|_F\}\leq 2^{O(n^{\delta'_2})}$, where $n^{\delta'_2}=C_0\,n^\delta 2^{O(k^2/\epsilon)}$ for a sufficiently large constant $C_0$ is the parameter of Lemma~\ref{lem:swz19_fpt}, and assume $\OPT\geq 2^{-n^\delta}\|A\|_F^2$. For $\epsilon\in (0, 0.1)$ with $\epsilon\geq 2^{-n^\delta}$, Algorithm~\ref{alg:quantum_fpt_two} outputs $n\times k$ matrices $U, V, W$ with
\begin{align*}
    \|U\otimes V\otimes W-A\|_F^2\leq & ~ (2+\epsilon)\cdot \OPT
\end{align*}
with probability at least $0.99$, in $(n^2k^2/\epsilon^2+n\poly(k/\epsilon)+2^{O(k^2/\epsilon)})\cdot n^{O(\delta)}$ time in the unit-cost RAM model with $O(\log n)$-bit words of Lemma~\ref{lem:swz19_fpt} (the word-level reading of the QRAM model of the introduction, and bit-level counts differ by an $O(\log n)$ factor absorbed in $n^{O(\delta)}$), using $\wt O(\epsilon^{-1}nk^{0.5})$ queries to mode-$1$ fibers and $\wt O(\epsilon^{-2}nk)$ queries to mode-$3$ fibers of $A$. Without the hypothesis $\OPT\geq 2^{-n^\delta}\|A\|_F^2$, carrying the rounding perturbation $\eta=2^{-\Omega(n^\delta)}\|A\|_F$ and the additive term of Lemma~\ref{lem:swz19_fpt} through the two triangle inequalities of the proof and using $2ab\leq \epsilon a^2+b^2/\epsilon$ gives, after a further constant rescaling of $\epsilon$, $\|U\otimes V\otimes W-A\|_F^2\leq (2+\epsilon)\OPT+\epsilon^{-1}2^{-\Omega(n^\delta)}\|A\|_F^2+2^{-\Omega(n^{\delta'_2})}\max\{1, \|A\|_F^2\}$ (the $\epsilon^{-1}$ on the last term and the polynomial factor relating $\|\wt A'\|_F$ to $\|A\|_F$ are absorbed using $\epsilon\geq 2^{-n^\delta}$). The cases in which no optimal rank-$k$ solution exists or $\OPT=0$ are not treated.
\end{theorem}

\begin{proof}
We run the algorithm with accuracy parameter $\epsilon_0:=\epsilon/14$ and prove the bound $(2+14\epsilon_0)\OPT$. The rescaling changes the running time only by constants, and the hypotheses $\epsilon\geq 2^{-n^\delta}$ and $n^{\delta'_2}=C_0\,n^\delta 2^{O(k^2/\epsilon)}$ transfer to $\epsilon_0$ up to the constant $14$, which is absorbed by the $O(\cdot)$'s. Throughout the proof, we write $\epsilon$ for $\epsilon_0$ and let $B^\star:=A_k$.

\emph{Steps 1--4 of the proof of Theorem~\ref{thm:quantum_bicriteria_two}} apply verbatim. By Step 1, we have $e_1:=\|A_1-QQ^\top A_1\|_F^2\leq (1+\epsilon)^2\OPT$. Let $\wt A=A(\wt Q, I, I)$ and $A(Q, I, I)$ be the tensors whose mode-$1$ unfoldings are $\wt Q^\top A_1$ and $Q^\top A_1$, and Step 4 gives $\|\wt A-A(Q, I, I)\|_F^2\leq \epsilon^2e_1$. The rank-$k$ tensor $B^\star(Q, I, I)=\sum_{i=1}^k (Q^\top u_i)\otimes v_i\otimes w_i$ satisfies $\|A(Q, I, I)-B^\star(Q, I, I)\|_F^2\leq \OPT$, since $Q^\top$ is a contraction. \emph{Magnitude and rounding of $\wt Q^\top$.} The matrix $S_0$ samples row $i$ of $Q$ with probability $p_i=\max\{\min\{1, \alpha\tau_i(Q)\}, n^{-2}\}$, where $\alpha=O(\epsilon^{-2}\log k)$, and rescales it by $p_i^{-1/2}$. The floor only increases the probabilities, and both Lemma~\ref{lem:leverage_score_optimal_solution} and Lemma~\ref{lem:leverage_score_weak_affine} tolerate this. By Lemma~\ref{lem:leverage_score_weak_affine} with $m=O(\epsilon^{-2}k\log k)$ expected rows, we have $\|(S_0Q)^\dagger\|_{\rm op}\leq 2$ with probability at least $0.998$. Since every entry of $S_0$ is at most $p_i^{-1/2}\leq n$, on this event every entry of $\wt Q^\top=(S_0Q)^\dagger S_0$ has magnitude at most $2n$. We round the entries of $\wt Q^\top$ to $c\,n^\delta+\lceil\log_2(100/\epsilon)\rceil$ bits after the binary point, for a constant $c\geq 2$, before forming $\wt A$. By the hypotheses $\epsilon\geq 2^{-n^\delta}$ and $n^\delta\geq \log n$, the rounded entries and the entries of $\wt A'$ have $O(n^\delta)$ bits, and $\|\wt A'\|_F\leq 2^{O(n^\delta)}$. The rounding replaces $\wt A$ by $\wt A'$ with $\|\wt A'-\wt A\|_F\leq \sqrt{km}\cdot 2^{-c\,n^\delta}(\epsilon/100)\|A\|_F\leq \epsilon\sqrt{\OPT}/100$, where we use the hypothesis $\OPT\geq 2^{-n^\delta}\|A\|_F^2$ and $km\leq n^2\leq 2^{2n^\delta}$. The exact basis $Q$ is likewise irrational in general, so the algorithm rounds its entries to $O(n^{\delta'_2})$ bits after the binary point, forms $S_0$ and $\wt Q^\top$ from the rounded $Q'$, and returns $U:=Q'X_{j^\star}$. We write $\|X_{j^\star}\|_F\|V_{j^\star}\|_F\|W_{j^\star}\|_F\leq 2^{Cn^{\delta'_2}}$ for the outputs of the verifier, whose entries have $n^{\delta'_2}$ bits. Let $C'$ be the constant in the pseudo-inverse perturbation bound $\|(S_0Q')^\dagger S_0A_1-(S_0Q)^\dagger S_0A_1\|_F\leq 2^{C'n^\delta}\|Q'-Q\|_F$, which holds on the subspace-embedding event $\sigma_{\min}(S_0Q)\geq 1/2$ of Lemma~\ref{lem:leverage_score_weak_affine}. We choose the constant $c$ larger than $C+1$ and $C'+1$, which is possible since $C$ and $C'$ do not depend on $c$. Then, $\|Q'-Q\|_F\leq \sqrt{nk}\,2^{-cn^{\delta'_2}}$ perturbs the returned tensor and $\wt Q^\top A_1$ by at most $2^{-n^{\delta'_2}/2}\leq 0.01\epsilon\sqrt{\OPT}$, where we use $\log(nk)\leq n^\delta\leq n^{\delta'_2}$, $\epsilon\geq 2^{-n^\delta}$, and $\OPT\geq 2^{-n^\delta}\|A\|_F^2$. We charge both perturbations to the allowance $0.01\epsilon\sqrt{\OPT}$ of the rounding of $\wt Q^\top$, and we carry this allowance through the two triangle inequalities below. Therefore, using $e_1\leq (1+\epsilon)^2\OPT$, we have
\begin{align*}
    \|\wt A'-B^\star(Q, I, I)\|_F\leq \epsilon(1+\epsilon)\sqrt{\OPT}+\sqrt{\OPT}+0.01\epsilon\sqrt{\OPT}\leq (1+1.62\epsilon)\sqrt{\OPT},
\end{align*}
so the optimal rank-$k$ cost of $\wt A'$ is at most $(1+3.6\epsilon)\OPT$. Moreover, $B^\star(Q, I, I)$ is a rank-$k$ witness of this cost whose factors $Q^\top u_i, v_i, w_i$ have norms at most those of $U^\star, V^\star, W^\star$.

\emph{The classical step.} We apply Lemma~\ref{lem:swz19_fpt} to $A':=\wt A'$ with the competitor $B':=B^\star(Q, I, I)$. Here $A'$ is a $k\times n\times n$ tensor with $O(n^\delta)$-bit entries, padded by zeros to $n\times n\times n$, so ${\rm nnz}(A')\leq kn^2$. The factor matrices of $B'$ are the padded $Q^\top U^\star$ and $V^\star, W^\star$, and their Frobenius norms are at most those of $U^\star, V^\star, W^\star$, and hence at most $2^{O(n^{\delta'_2})}$ by hypothesis. Each of the three runs returns $U'_j, V_j, W_j\in \R^{n\times k}$, and we let $X_j\in \R^{k\times k}$ be the first $k$ rows of $U'_j$ and let $T_j:=\sum_{i=1}^k (X_j)_i\otimes (V_j)_i\otimes (W_j)_i$. Dropping the other rows of $U'_j$ does not increase the cost, since the corresponding slabs of the padded tensor are zero. For each run, the guarantee of Lemma~\ref{lem:swz19_fpt} holds with probability at least $9/10$, and in that case $\|\wt A'-T_j\|_F^2\leq (1+\epsilon)\|\wt A'-B^\star(Q, I, I)\|_F^2+2^{-\Omega(n^{\delta'_2})}\max\{1, \|\wt A'\|_F^2\}\leq (1+\epsilon)(1+3.6\epsilon)\OPT+\epsilon\OPT\leq (1+6\epsilon)\OPT$. We absorb the additive term as follows. Since the tensor $A\neq 0$ has $O(n^\delta)$-bit entries, we have $\|A\|_F^2\geq 2^{-O(n^\delta)}$ and $\OPT\geq 2^{-n^\delta}\|A\|_F^2\geq 2^{-O(n^\delta)}$. Moreover, $\|\wt A'\|_F^2\leq 2^{O(n^\delta)}$ and $\epsilon\geq 2^{-n^\delta}$. Therefore, $2^{-\Omega(n^{\delta'_2})}\max\{1, \|\wt A'\|_F^2\}\leq 2^{-\Omega(n^{\delta'_2})+O(n^\delta)}\leq \epsilon\OPT$, once the constant $C_0$ in $n^{\delta'_2}=C_0\,n^\delta 2^{O(k^2/\epsilon)}$ is taken large enough relative to the constants in the $O(n^\delta)$ bounds above. We do not use any lower bound on the competitor's own cost $\|\wt A'-B^\star(Q, I, I)\|_F^2$, and no such bound is available. This cost can vanish, e.g., when $Q$ spans exactly the column span of $U^\star$ and the residual is orthogonal to it.

\emph{Selection.} With probability at least $1-10^{-3}$, at least one of the three independent runs succeeds. Let $c_j:=\|\wt A'-T_j\|_F^2$ and $M_j:=(\wt A'-T_j)_{(2)}\in \R^{n\times kn}$. Then, the estimate $\wh c_j=\|GM_j\|_F^2$ has expectation $\|M_j\|_F^2=c_j$ and variance at most $2c_j^2/s$, since each row $g$ of $\sqrt s\,G$ gives $\|g^\top M_j\|_2^2$ with mean $c_j$ and variance at most $2c_j^2$, and the $s$ rows are independent. Thus, by Chebyshev's inequality, $|\wh c_j-c_j|\leq (\epsilon/4)c_j$ fails for a given $j$ with probability at most $32/(s\epsilon^2)$, and hence it holds for all three $j$ with probability at least $1-2\cdot 10^{-3}$ once $s\geq 48000\,\epsilon^{-2}$. On these events, the selected candidate $T:=T_{j^\star}$ satisfies $c_{j^\star}\leq \frac{1+\epsilon/4}{1-\epsilon/4}\min_j c_j\leq (1+\epsilon)(1+6\epsilon)\OPT\leq (1+7.7\epsilon)\OPT$ for $\epsilon\leq 0.1$. Here $X:=X_{j^\star}$, $V:=V_{j^\star}$, $W:=W_{j^\star}$ are the matrices returned by the algorithm. By the triangle inequality once more, we have
\begin{align*}
    \|A(Q, I, I)-T\|_F\leq & ~ \|A(Q, I, I)-\wt A\|_F+\|\wt A-\wt A'\|_F+\|\wt A'-T\|_F \\
    \leq & ~ \big(\epsilon(1+\epsilon)+0.01\epsilon+\sqrt{1+7.7\epsilon}\big)\sqrt{\OPT},
\end{align*}
hence $\|A(Q, I, I)-T\|_F^2\leq (1+11\epsilon)\OPT$ for $\epsilon\leq 0.1$.

\emph{Lifting.} The output $U\otimes V\otimes W$ with $U=Q'X$ is the tensor $\sum_{i=1}^k (Q'X_i)\otimes V_i\otimes W_i$ of rank at most $k$, and its mode-$1$ unfolding is $Q'\cdot T_{(1)}$, where $T_{(1)}$ is the mode-$1$ unfolding of $T$. By Step 2 with $Y=T_{(1)}$, we have
\begin{align*}
    \|QT_{(1)}-A_1\|_F^2=e_1+\|Q^\top A_1-T_{(1)}\|_F^2= & ~ e_1+\|A(Q, I, I)-T\|_F^2 \\
    \leq & ~ (1+\epsilon)^2\OPT+(1+11\epsilon)\OPT\leq (2+13.1\epsilon)\OPT
\end{align*}
for $\epsilon\leq 0.1$. By the rounding of $Q$ above, $\|(Q'-Q)T_{(1)}\|_F\leq 0.01\epsilon\sqrt{\OPT}$, and hence, by the triangle inequality, we have $\|U\otimes V\otimes W-A\|_F^2\leq (2+13.1\epsilon)\OPT+0.04\epsilon\OPT\leq (2+14\epsilon)\OPT$.
The failure events are the coreset call at $0.001$, the applications of Lemma~\ref{lem:leverage_score_optimal_solution} and Lemma~\ref{lem:leverage_score_weak_affine} to $S_0$ at $0.004$ and $0.002$, respectively, by Remark~\ref{rem:delta_dependence}, the event of probability $10^{-3}$ that all three classical runs fail, and the three cost estimates at $2\cdot 10^{-3}$. By a union bound, the success probability is at least $0.99$.

\emph{Running time and queries.} Computing $C_1$ takes $\wt O(\epsilon^{-1}n^2k^{0.5}+nk^{\omega-1})$ time and $\wt O(\epsilon^{-1}nk^{0.5})$ mode-$1$ fiber queries, and as in Theorem~\ref{thm:quantum_bicriteria_two}, computing $Q$, $S_0$, $\wt Q^\top$ costs $\wt O(\epsilon^{2-2\omega}nk^{\omega-1})+\poly(k/\epsilon)$. Forming $\wt A$ reads the $m$ slabs $A_{i,*,*}$, $i\in {\rm supp}(S_0)$, i.e., $mn=\wt O(\epsilon^{-2}nk)$ mode-$3$ fibers, and computes $kn^2$ entries at $O(m)$ arithmetic operations each, which takes $O(n^2km)=\wt O(\epsilon^{-2}n^2k^2)$ time. Each of the three classical runs takes $({\rm nnz}(\wt A)+n\poly(k/\epsilon)+2^{O(k^2/\epsilon)})n^{O(\delta)}=(kn^2+n\poly(k/\epsilon)+2^{O(k^2/\epsilon)})n^{O(\delta)}$ time. Each cost estimate $\wh c_j$ is computed exactly as $\|G\wt A'_{(2)}-(GV_j)Z_j\|_F^2$, where $Z_j\in \R^{k\times kn}$ is the matrix whose $i$-th row is ${\rm vec}((X_j)_i(W_j)_i^\top)$. Here $\wt A'$ is the tensor written as $\wt A$ in Algorithm~\ref{alg:quantum_fpt_two}, and we factor the scalar $s^{-1/2}$ out of $G$, so all intermediate quantities are fixed-point numbers. The product $G\wt A'_{(2)}$ costs $O(skn^2)=O(\epsilon^{-2}kn^2)$ operations on $O(n^\delta)$-bit numbers. Computing $GV_j$, $(GV_j)Z_j$, and the final norm costs $O(\epsilon^{-2}nk^2)$ operations on $n^{\delta'_2}$-bit numbers, each $\poly(n^{\delta'_2})=n^{O(\delta)}2^{O(k^2/\epsilon)}$ word operations. The second of these takes $ab\cdot n^{O(\delta)}$ time with $a=\epsilon^{-1}nk$ and $b=\epsilon^{-1}k\,2^{O(k^2/\epsilon)}$, which is at most $(a^2+b^2)\cdot n^{O(\delta)}=(\epsilon^{-2}n^2k^2+2^{O(k^2/\epsilon)})\cdot n^{O(\delta)}$, since $b^2=2^{O(k^2/\epsilon)}$ due to $\log(k/\epsilon)=O(k^2/\epsilon)$. Finally, computing $U=Q'X_{j^\star}$ takes $O(nk^2)$ time, and summing up these costs gives the stated bounds.
\end{proof}

\subsection{Tensor CURT Decomposition: Fixed-Parameter Tractable and Bicriteria}

\begin{algorithm}[!ht]\caption{Converting a tensor low-rank approximation to a CURT decomposition.}\label{alg:tensor_from_low_rank_to_CURT}
\begin{algorithmic}[1]
\Procedure{\textsc{FromLowRankToCURT}}{$A,U_B,V_B,W_B,n,k,\epsilon$} \Comment{Lemma~\ref{lem:tensor_from_low_rank_to_CURT}}
\State $d_1\leftarrow d_2 \leftarrow d_3 \leftarrow O(k\log k + k/\epsilon)$.
\State $\epsilon_0\leftarrow 0.01$.
\State Let $B_1 := V_B^\top \odot W_B^\top \in \mathbb{R}^{k\times n^2}$ (implicit, and only $B_1D_1$ is formed, in $O(kd_1)$ time)
\State $D_1\leftarrow$\textsc{TensorLeverageScore}($V_B^\top, W_B^\top, n,n,k,\epsilon_0,d_1$)
\State Form $\wh{U} = A_1 D_1 (B_1 D_1)^\dagger \in \mathbb{R}^{n\times k}$.
\State Let $B_2 := \wh{U}^\top \odot W_B^\top \in \mathbb{R}^{k\times n^2}$ (implicit, and only $B_2D_2$ is formed, in $O(kd_2)$ time)
\State $D_2\leftarrow$\textsc{TensorLeverageScore}($\wh{U}^\top, W_B^\top,n,n,k,\epsilon_0,d_2$).
\State Form $\wh{V} = A_2 D_2 (B_2 D_2)^\dagger \in \mathbb{R}^{n\times k}$
\State Let $B_3 := \wh{U}^\top \odot \wh{V}^\top \in \mathbb{R}^{k\times n^2}$ (implicit, and only $B_3D_3$ is formed, in $O(kd_3)$ time)
\State $D_3\leftarrow$\textsc{TensorLeverageScore}($\wh{U}^\top, \wh{V}^\top,n,n,k,\epsilon_0,d_3$)
\State $C\leftarrow A_1 D_1$, $R\leftarrow A_2 D_2$, $T\leftarrow A_3 D_3$\label{ln:curt:CRT} \Comment{$D_i=\wt D_i\Sigma_i$: selection $\wt D_i$, weights $\Sigma_i$}
\State $U\leftarrow \sum_{i=1}^k ( (B_1 D_1)^\dagger )_i \otimes ( (B_2 D_2)^\dagger )_i \otimes ( (B_3 D_3)^\dagger )_i$
\State \Return $C$, $R$, $T$ and $U$
\EndProcedure
\end{algorithmic}
\end{algorithm}
\noindent\emph{Note on Algorithm~\ref{alg:tensor_from_low_rank_to_CURT}.} Line~\ref{ln:curt:CRT}: the returned factors are the unweighted selections $C=A_1\wt D_1$, $R=A_2\wt D_2$, $T=A_3\wt D_3$, and the weights $(\Sigma_1)_{ii}(\Sigma_2)_{jj}(\Sigma_3)_{ll}$ are folded into $U_{ijl}$.

\begin{lemma}[A modification of Theorem C.40 in~\cite{swz19}]\label{lem:tensor_from_low_rank_to_CURT}
Given a $3$rd order tensor $A\in \mathbb{R}^{n\times n \times n}$, let $k\geq 1$, and let $U_B,V_B,W_B\in \mathbb{R}^{n\times k}$ be any three matrices (a competitor). Then there is a classical algorithm (Algorithm~\ref{alg:tensor_from_low_rank_to_CURT}) which takes $\wt O(n \poly( k/\epsilon) )$
time and outputs three matrices $C\in \mathbb{R}^{n\times c}$ with columns from $A$, $R\in \mathbb{R}^{n\times r}$ with rows from $A$, $T\in \mathbb{R}^{n\times t}$ with tubes from $A$, and a tensor $U\in \mathbb{R}^{c\times r\times t}$ of CP rank at most $k$ such that $c=r=t=O(k\log k +k/\epsilon)$, and
\begin{align*}
\left\| \sum_{i=1}^c \sum_{j=1}^r \sum_{l=1}^t U_{i,j,l} \cdot C_i \otimes R_j \otimes T_l - A \right\|_F^2 \leq(1+\epsilon) \left\|\sum_{i=1}^k (U_B)_i\otimes (V_B)_i\otimes (W_B)_i-A\right\|_F^2
\end{align*}
holds with probability at least $0.99$ (three applications of Lemma~\ref{lem:leverage_score_optimal_solution}, each run with failure probability $0.002$ at a constant factor in $c, r, t$ by Remark~\ref{rem:delta_dependence}, and three \textsc{TensorLeverageScore} calls with internal sketch failure $0.001$ each).
\end{lemma}
The proof of Theorem C.40 of~\cite{swz19} shows this competitor-relative form, which is the form we need below, and we follow it with only three changes. We form $A_iD_i$ by reading the $d_i$ sampled fibers in $O(nd_i)$ time, in place of the ${\rm nnz}(A)$ pass. The sampler of Lemma~\ref{lem:tensor_leverage} contributes the polylogarithmic factor. The three regressions are unconstrained $k$-row regressions, and we solve them by argmin on leverage-score column samples of their own designs, so Lemma~\ref{lem:leverage_score_optimal_solution} applies within its hypotheses.

\begin{theorem}[Bicriteria Tensor CURT Decomposition]
\label{thm:quantum_bicriteria_tensor_CURT}
Given a 3rd order tensor $A\in \R^{n\times n\times n}$ and a positive integer $k\leq n$, $\epsilon\in (0,0.1)$, there exists an algorithm which takes $\wt{O}(\epsilon^{-1}n^2k^{0.5}+n^{1.5}\poly(k/\epsilon))$ time, $\wt O(\epsilon^{-1}nk^{0.5}+\epsilon^{-3}n^{0.5}k^2+\poly(k/\epsilon))$ queries to the rows, columns and tubes of $A$ (the additive term reads the sampled fibers of the conversion) and outputs three matrices $C, R, T\in \R^{n\times m}$ with $m=O(r\log r+r/\epsilon)=\wt O(k^2/\epsilon)$, where $r=k^2$ bounds the CP rank of the bicriteria solution of Theorem~\ref{thm:quantum_bicriteria_two}, and $U \in \R^{m \times m \times m}$ of CP rank at most $r$ such that
\begin{align*}
    \left\| \sum_{i=1}^m \sum_{j=1}^m \sum_{l=1}^m U_{i,j,l} \cdot C_i \otimes R_j \otimes T_l - A \right\|_F^2  \leq & ~ (2+\epsilon)\cdot \inf_{{\rm rank}\text{-}k~A_k} \|A-A_k\|_F^2
\end{align*}
with probability at least $0.89$.
\end{theorem}
\begin{proof}
It directly follows from combining Theorem~\ref{thm:quantum_bicriteria_two} and Lemma~\ref{lem:tensor_from_low_rank_to_CURT}. We invoke Lemma~\ref{lem:tensor_from_low_rank_to_CURT} with rank parameter $r=k^2$ and the bicriteria solution $B=\sum_{i,t\in [k]}Q_t\otimes V_i\otimes y_{i,t}$ as the competitor, whose factor matrices are the $n\times k^2$ matrices with columns $Q_t$, $V_i$, and $y_{i,t}$ indexed by the pairs $(i,t)$. The three fiber counts of the lemma are then all $O(r\log r+r/\epsilon)=m$. Moreover, the lemma's $\wt O(n\poly(k/\epsilon))$ time and $\poly(k/\epsilon)$ fiber reads are absorbed, and the two failure probabilities add to at most $0.11$. Finally, the product $(1+\epsilon)(2+\epsilon)$ of the two guarantees is $(2+\epsilon)$ after rescaling $\epsilon$ by a constant.
\end{proof}

\begin{theorem}[Fixed-Parameter Tractable Tensor CURT Decomposition]
\label{thm:quantum_fpt_curt}
Let $\epsilon\in (0,0.1)$, and let $A\in \R^{n\times n\times n}$, $k$, $\delta$ and $\OPT$ satisfy the hypotheses of Theorem~\ref{thm:quantum_fpt_two} with $\epsilon/4$ in place of $\epsilon$, in particular $\epsilon\geq 2^{2-n^\delta}$. There is a quantum algorithm that outputs $C\in \R^{n\times c}$, $R\in \R^{n\times r}$ and $T\in \R^{n\times t}$ with $c=r=t=O(k\log k+k/\epsilon)$, consisting of columns, rows and tubes of $A$, and $U\in \R^{c\times r\times t}$ of CP rank at most $k$, such that
\begin{align*}
    \Big\|\sum_{i=1}^c\sum_{j=1}^r\sum_{l=1}^t U_{i,j,l}\, C_i\otimes R_j\otimes T_l-A\Big\|_F^2\leq & ~ (2+\epsilon)\cdot \OPT
\end{align*}
with probability at least $0.98$, where $\OPT$ is the infimum of Theorem~\ref{thm:quantum_fpt_two}. The algorithm runs in $(n^2k^2/\epsilon^2+n\poly(k/\epsilon)+2^{O(k^2/\epsilon)})\cdot n^{O(\delta)}$ time in the unit-cost RAM model with $O(\log n)$-bit words of Theorem~\ref{thm:quantum_fpt_two}, and uses $\wt O(\epsilon^{-1}nk^{0.5})$ queries to mode-$1$ fibers, $\wt O(\epsilon^{-2}nk)$ queries to mode-$3$ fibers and $\poly(k/\epsilon)$ further queries to the rows, columns and tubes of $A$.
\end{theorem}
\begin{proof}
We run Algorithm~\ref{alg:quantum_fpt_two} with accuracy $\epsilon/4$, and then feed its output $U, V, W\in \R^{n\times k}$ to Algorithm~\ref{alg:tensor_from_low_rank_to_CURT}, also run with accuracy $\epsilon/4$, as the competitor of Lemma~\ref{lem:tensor_from_low_rank_to_CURT}. Note that the lemma holds for every fixed competitor, and the randomness of the conversion is independent of $U, V, W$. Therefore, the output has cost at most $(1+\epsilon/4)(2+\epsilon/4)\cdot\OPT=(2+\tfrac34\epsilon+\epsilon^2/16)\cdot\OPT\leq (2+\epsilon)\cdot\OPT$, since $\epsilon^2/16\leq \epsilon/4$ for $\epsilon\leq 4$. The fiber counts are $O(k\log k+4k/\epsilon)=O(k\log k+k/\epsilon)$, and the two failure probabilities add to at most $0.02$. Theorem~\ref{thm:quantum_fpt_two} at accuracy $\epsilon/4$ contributes its stated time and query bounds with $\epsilon/4$ in place of $\epsilon$, and this substitution leaves each term in its stated $O(\cdot)$, $\poly(\cdot)$, or $\wt O(\cdot)$ class. The conversion performs $\wt O(n\poly(k/\epsilon))$ arithmetic operations on numbers of $\poly(k)\,n^{\delta'_2}=\poly(n^{\delta'_2})$ bits, since the inputs have $O(n^{\delta'_2})$ and $O(n^\delta)$ bits and the pseudo-inverses grow the bit size by $\poly(k)$ factors, and each operation costs $\poly(n^{\delta'_2})$ word operations. By the AM--GM inequality, we have $n\poly(k/\epsilon)\poly(n^{\delta'_2})\leq n^2+\poly(k/\epsilon)^2\poly(n^{\delta'_2})^2$, and using $n^{\delta'_2}=O(n^\delta 2^{O(k^2/\epsilon)})$ and $\log(k/\epsilon)=O(k^2/\epsilon)$, the right-hand side is $(n^2+2^{O(k^2/\epsilon)})\cdot n^{O(\delta)}$, which, with the polylogarithmic factor absorbed in $n^{O(\delta)}$ since $n^\delta\geq \log n$, is within the stated time. Finally, the conversion reads $d_1+d_2+d_3=O(k\log k+k/\epsilon)$ sampled fibers of $A$, which accounts for the additive query term.
\end{proof}

\section{Improved Quantum Coreset Algorithm for \texorpdfstring{$(k,p)$}{(k,p)}-Clustering and Application}
\label{sec:clustering}

In this section, we give an improved quantum coreset construction for $(k, p)$-clustering. We observe that the size of the coreset obtained in the main result of~\cite{xclj23} (their Theorem 3.1) scales linearly with $d$, which causes an additional $d^{0.5}$ factor in the final runtime.

We begin by recalling the $(k, p)$-clustering problem in $\R^d$: let $A=\{a_1,\ldots,a_n\}\subset \R^d$, $X=(\R^d)^k$ and ${\rm cost}(a_i, x)=\min_{j\in [k]} \|a_i-x_j\|_2^p$, where $p\geq 1$ is the power of the distance, and $x_j$ is one of the centers in $x$. When $p=1$, this is the well-studied $k$-median problem, and when $p=2$, this captures the $k$-means problem. To construct a coreset, we implement the sensitivity sampling framework in quantum sublinear time.

We need the following quantum algorithm, due to~\cite{xclj23}, that computes an $(\alpha,\beta)$-bicriteria approximation.

\begin{definition}[Bicriteria Approximation]
Let $A\subset \R^d$, assume $\OPT$ is the optimal cost of the $(k,p)$-clustering problem for $A$, we say a set $x\subset \R^d$ is an $(\alpha,\beta)$-bicriteria approximation if $|x|\leq \alpha k$ and $\cost(A, x)\leq \beta \OPT$.
\end{definition}

\begin{lemma}[Lemma 3.7 of~\cite{xclj23}, with a size cap]
\label{lem:bicriteria}
Let $A\subset \R^d$ with $n\geq 2$ points, there exists a quantum algorithm that outputs a set $x$ with at most $c_{\rm bi}k\lceil\log_2 n\rceil^2$ points on every outcome, where $c_{\rm bi}:=354$, such that $x$ is an $(O(\log^2 n), 2^{O(p)})$-bicriteria approximation to the $(k,p)$-clustering problem for $A$ with probability at least $5/6$. The algorithm uses $\wt O(\sqrt{nk})$ queries to $A$, takes $\wt O(\sqrt{nk}d)$ time and $\poly(k\log n)$ additional preprocessing time.
\end{lemma}

Note that $\cost(A, x)$ does not increase when centers are added to $x$.

The size cap is our modification of the algorithm of~\cite{xclj23}, and Lemma~\ref{lem:discrete_constant} needs it. Their logarithms are to base $2$, as their Appendix A uses that $\lceil\log n\rceil$ halvings reduce $n$ points to at most one, and we write them as $\log_2$. Their Algorithm 1 takes the union of three repetitions of the following procedure. A loop adds a sample of $13k\lceil\log_2 n\rceil$ points to the output in each iteration and shrinks a set $D_t\subseteq A$ of remaining points. The loop stops once an estimate $\wt r_t$ of $|D_t|$ with $\big|\wt r_t-|D_t|\big|\leq |D_t|/2$ (their Line 8) satisfies $\wt r_t\leq 39k\lceil\log_2 n\rceil$, or after $3\lceil\log_2 n\rceil$ iterations, and then their Line 11 adds all points of $D_t$ to the output. The proof in their Appendix A shows that, with probability at least $5/6$, one of the three repetitions has cost at most $2^{O(p)}\OPT$, and that this repetition stops its loop by the first condition with all subroutines successful, so that $|D_t|\leq 2\wt r_t\leq 78k\lceil\log_2 n\rceil$ for it. In the other repetitions, the loop may stop by the second condition, which we call a late stop, or with a wrong estimate, and then $D_t$ can contain almost all of $A$. Hence, their output has no worst-case size bound. We make two changes. First, we keep at most $13k\lceil\log_2 n\rceil$ points of each sample, which changes nothing when their sampler succeeds. Second, we implement their Line 11 by the repeated Grover search described after Lemma~\ref{lem:q_sampling_1_d}, with $q_i=1$ on $D_t$ and $q_i=0$ elsewhere, computed by their membership oracle of $D_t$, and with the failure probability $1/\poly(nk)$ of their subroutines. This search finds the points of $D_t$ one at a time, and we stop it as soon as it has found $78k\lceil\log_2 n\rceil+1$ points. Since $\lceil\log_2 n\rceil\geq 1$ for $n\geq 2$, each repetition then outputs at most $3\lceil\log_2 n\rceil\cdot 13k\lceil\log_2 n\rceil+78k\lceil\log_2 n\rceil+1\leq 118k\lceil\log_2 n\rceil^2$ points, and each run outputs at most $3\cdot 118k\lceil\log_2 n\rceil^2=c_{\rm bi}k\lceil\log_2 n\rceil^2$ points. The stop does not change the output of the repetition used by their proof, since $|D_t|\leq 78k\lceil\log_2 n\rceil$ there, and by the monotonicity above, the points of the other repetitions do not increase the cost of the union. Therefore, the success probability is still at least $5/6$. The stop also bounds the expected cost of their Line 11 by $\wt O(\sqrt{nk})$ membership queries on every outcome of the loop, while without it this bound holds only when $|D_t|=\wt O(k)$. Indeed, let $J\leq 78k\lceil\log_2 n\rceil+1$ be the number of searches that find a point. For $j\leq J$, the $j$-th search starts with $|D_t|-j+1\geq J-j+1$ points of $D_t$ not yet found and takes $O(\sqrt{n/(|D_t|-j+1)})$ queries in expectation. Hence, these $J$ searches take $O(\sum_{i=1}^{J}\sqrt{n/i})=O(\sqrt{nk\log n})$ queries in expectation, and at most one final search that finds no point takes $\wt O(\sqrt n)$ queries.

We also need a quantum approximate nearest neighbor oracle, which would be crucial to approximately find the center a point belongs to.

\begin{lemma}[Lemma 3.4 of~\cite{xclj23}]
\label{lem:q_ann}
Let $A\subset \R^d$ and $x\subset \R^d$ with $|x|=m$, given two parameters $\delta>0$, $c_\tau\in [2.5,3)$, there exists a quantum oracle that given $a_i\in A$, returns $\tau(a_i)\in x$, using $\poly(m\log(n/\delta))$ preprocessing time. With probability at least $1-\delta$, $\tau:A\rightarrow x$ is a mapping such that for all $a_i\in A$,
\begin{align*}
    \|a_i-\tau(a_i)\|_2\leq & ~ c_\tau\cdot \dist(a_i, x),\quad\text{hence}\quad \|a_i-\tau(a_i)\|_2^p\leq c_\tau^p\cdot \cost(a_i, x)<3^p\cost(a_i, x).
\end{align*}
Each query to $\tau$ takes $O(d\poly\log(mn/\delta))$ time.
\end{lemma}
Note that $\tau$ is also a \emph{partition oracle}, as we could assign $a_i$ to $\tau(a_i)$, which is one of the $m$ clusters.

We need two other ingredients: one being estimating $\cost(A, x)=\sum_{i=1}^n \cost(a_i, x)$ and the other being estimating the number of points falling in each cluster. 

\begin{lemma}[Lemma 6 of~\cite{lcw19}, with quantum maximum finding~\citep{dh96}]
\label{lem:q_sum}
Let $C=\{c_1,\ldots,c_n\}$ be a collection of nonnegative numbers given by an oracle, let $c=\sum_{i=1}^n c_i$, and let $\epsilon,\delta\in (0,1)$. Suppose a number $M>0$ is known with $c_i\in [0, M]$ for all $i\in [n]$, and each $c_i/M$ is specified with $O(\log(n/\epsilon))$ bits. There exists a quantum algorithm that outputs an approximation $\wt c$ where $\wt c=(1\pm\epsilon)\cdot c$ with probability at least $1-\delta$, using $\wt O(\sqrt n\log(1/\delta)/\epsilon)$ queries to $C$. The same bounds hold for arbitrary nonnegative reals $c_i$ if $M:=\max_{i\in [n]} c_i$ is supplied, or is computed by Lemma~\ref{lem:dh_min} below on the values $-c_i$ with $\wt O(\sqrt n\log(1/\delta))$ additional queries and each of the two steps run with failure probability $\delta/2$.
\end{lemma}
Lemma 6 of~\cite{lcw19} estimates the mean of values $F(i)\in [0,1]$ given with $l$ bits of precision within a factor $1\pm\epsilon$ with probability $2/3$ using $O(\sqrt n/\epsilon)$ queries, and its extra factor $l\log l$ is absorbed in $\wt O$. We apply it to the values $c_i/M$ and take the median of $O(\log(1/\delta))$ repetitions. When $M=\max_i c_i$, we have $M\leq c$, so truncating each $c_i/M$ to $l=\lceil\log_2(2n/\epsilon)\rceil$ bits changes $c$ by at most $n2^{-l}M\leq\epsilon c/2$, which is absorbed by rescaling $\epsilon$. Moreover, $M=0$ forces $c=0$.

\begin{lemma}[Theorem 4.4 of~\cite{xclj23}]
\label{lem:multidim_count}
Let $A\in (\R^d)^n, x\in (\R^d)^m$ and $\tau:A\rightarrow x$, let $C_j=\{a\in A: \tau(a)=x_j \}$, let $\epsilon\in (0,1/3),\delta>0$, then there exists a quantum algorithm that with probability at least $1-\delta$, outputs a list of  estimates $\wt n_j$ for all $j\in [m]$ where $\wt n_j=(1\pm\epsilon)\cdot |C_j|$, using $\wt O(\sqrt{nm/\epsilon}\log(1/\delta))$ queries to $\tau$ and an additional $\wt O((\sqrt{nm/\epsilon}+m/\epsilon)\log(n/\delta))$ time.
\end{lemma}

\begin{lemma}[Theorem 4.2 of~\cite{xclj23}, multidimensional approximate summation]
\label{lem:multidim_sum}
Let $1\leq m\leq n$ be integers, $\tau:A\rightarrow [m]$, and let $f:A\rightarrow \{0,1,\ldots,M\}$ be integer-valued, both computable by one oracle call per point. Let $\epsilon\in (0,1/3)$ and $\delta>0$. There exists a quantum algorithm that, with probability at least $1-\delta$, outputs estimates $\wt s_j=(1\pm\epsilon)\cdot \sum_{a\in A:\tau(a)=j}f(a)$ for all $j\in [m]$ simultaneously, using $\wt O(\sqrt{nm/\epsilon}\log(1/\delta)\log M)$ queries and $\wt O((\sqrt{nm/\epsilon}+m/\epsilon)\log(n/\delta)\log M)$ additional time. (Their proof applies Lemma~\ref{lem:multidim_count} to each of the $O(\log M)$ binary digits of $f$, which gives relative accuracy on every cell, however light.)
\end{lemma}

Our algorithm, given in Section~\ref{sec:one_shot_optimal}, is a quantum implementation of the one-shot coreset construction of~\cite{hlw24}. It has two main advantages: (1) Seeded with a discrete constant-factor solution, it computes a coreset with points only from $A$. (2) Its sampling probabilities are relatively easy to compute.

\begin{remark}[Black-box use of a one-shot sampler]\label{rem:one_shot_blackbox}
Our Grover-based sampling step uses a one-shot sampler only through its sampling probabilities. Given \emph{any} one-shot clustering sensitivity sampler that, after computing approximate centers, produces a coreset of size $s$ and evaluates one sampling probability in time ${\cal T}$, the procedure of Lemma~\ref{lem:q_sampling_1_d} performs the sampling step in $\wt O(\sqrt{ns})\cdot {\cal T}$ time. Consequently, any improvement to classical one-shot clustering coresets, in the size $s$ or the evaluation time ${\cal T}$, immediately improves this step of our quantum algorithm.
\end{remark}

\subsection{One-Shot Sensitivity Sampling in Sublinear Time}
\label{sec:one_shot_optimal}

By Remark~\ref{rem:one_shot_blackbox}, the sampling step of a one-shot sampler costs $\wt O(\sqrt{ns})$ evaluations of its sampling probabilities. In this section, we show that the cell statistics that a one-shot sampler needs can be estimated quantumly, and we prove that the coreset construction of~\cite{hlw24} can be implemented in this way, with the same coreset size up to polylogarithmic factors, yielding Theorem~\ref{thm:one_shot_main}.

\begin{corollary}[Cell statistics estimation]
\label{cor:cell_mass}
Let $c: A\rightarrow [M]$ be a partition oracle computable in time ${\cal T}_{\rm cell}$ per point, and let $v:A\rightarrow \R_{\geq 0}$ be a value function computable together with $c(a)$ in the same time, for instance $v(a)=\dist^p(a, x')$ for a center set $x'$. For each cell $c\in [M]$, let $n_c:=|\{a\in A: c(a)=c\}|$ and $\sigma_c:=\sum_{a\in A: c(a)=c} v(a)$. Suppose that the values satisfy $v(a)\in \{0\}\cup [F, RF]$ for all $a\in A$, for known $F>0$ and $R\geq 1$. Then for any $\delta>0$, with probability at least $1-\delta$, all $n_c$ and $\sigma_c$ can be estimated within a factor $1\pm 1/4$ using $\wt O(\sqrt{nM}\log(2R)\log(1/\delta))$ queries and $\wt O(\sqrt{nM}\cdot {\cal T}_{\rm cell}\log(2R)\log(1/\delta))$ time.
\end{corollary}
\begin{proof}
The masses are estimated by Lemma~\ref{lem:multidim_count} with the partition oracle $c$ in place of $\tau$ and accuracy $1/4$, using $\wt O(\sqrt{nM})$ queries. For the cost sums, let $u:=F/16$ and $f(a):=\lfloor v(a)/u\rceil\in \{0,1,\ldots,\lceil 16R\rceil\}$, where we round to the nearest integer. If $v(a)=0$, then $f(a)=0$, and if $v(a)\geq F$, the rounding changes $v(a)$ by at most $u/2\leq v(a)/32$. Therefore, $u\sum_{a:c(a)=c}f(a)\in [\frac{15}{16}\sigma_c, \frac{17}{16}\sigma_c]$ for every cell. By Lemma~\ref{lem:multidim_sum} with accuracy $1/8$ and $\log\lceil 16R\rceil=O(\log(2R))$, we can estimate all the integer sums within $1\pm 1/8$ in $\wt O(\sqrt{nM}\log(2R)\log(1/\delta))$ queries. Note that $(1+1/8)(17/16)<5/4$ and $(1-1/8)(15/16)>3/4$.
\end{proof}

\begin{remark}[Instantiation with the construction of~\citet{hlw24}]
\label{rem:hlw_instantiation}
We verified the construction of~\cite{hlw24} against their full version. Their Algorithm 1 performs a \emph{single} round of importance sampling. Given an $O(1)$-approximate solution $A^\star$ with $k$ centers, each cluster is partitioned into rings by the ratio of $\dist^p(a, A^\star)$ to the cluster's average cost $\Delta_i$. The rings are merged by cost into $O_p(\log^2(k/\epsilon))$ groups, as in their Lemma A.5, and each group $G$ contributes $\Gamma_G$ samples drawn with probability proportional to $\dist^p(a, A^\star)$ within $G$. Their algorithm runs in $O(nk)$ classical time. Two adjustments are needed. First, their coreset is \emph{not} points-only as written, since Lines 2--3 of their Algorithm 1 add the centers $A^\star$ with weights $|P_i\setminus \mathcal{G}|$, which represent the discarded inner and light rings. We repair this by taking $A^\star$ to be a \emph{discrete} constant-factor approximation, i.e., one whose centers are input points, and their algorithm permits this at the cost of a $2^{O(p)}$ factor in the approximation constant. Second, their sampler draws $\Gamma_G$ points \emph{with replacement} within each group, while our primitive performs independent per-point sampling, and converting to independent sampling with $q_a=\min\{1, \Gamma_G\cdot \dist^p(a, A^\star)/{\rm cost}(G, A^\star)\}$ is a standard Poissonization step. Lemma~\ref{lem:one_shot_robust} below shows that their analysis tolerates both adjustments, together with the constant-factor statistics estimates and the approximate center weights provided by our quantum access.
\end{remark}

\begin{lemma}[Robustness of the one-shot sampler of~\citet{hlw24}]\label{lem:one_shot_robust}
Let $p\geq 1$ be a constant, $A\subset \R^d$, and let $A^\star$ be a $2^{O(p)}$-approximate solution with induced clusters $P_1,\ldots,P_k$ and average costs $\Delta_i$. Suppose that the ring and group decomposition of~\citet{hlw24} is constructed from estimates $\wh \Delta_i\in [\Delta_i/2, 2\Delta_i]$ and ring-cost estimates correct up to a factor of $2$, that each group $G$ is sampled by independent Bernoulli variables with probabilities $q_a=\min\{1, \Gamma_G\cdot \dist^p(a, A^\star)/\wh c_G\}$ and weights $1/q_a$, where $\wh c_G\in [{\rm cost}(G, A^\star)/2,\; 2\,{\rm cost}(G, A^\star)]$, and that each center $a_i^\star$ receives an estimated weight $\wh w(a_i^\star)=|P_i\setminus \mathcal{G}|\pm \epsilon|P_i|$. If the sample budgets $\Gamma_G$ are as in Theorem A.11 of~\citet{hlw24}, inflated by a factor of $2^{O(p)}$, then with probability at least $0.99$ the output is an $O(\epsilon)$-coreset of $A$ of size $k+2^{O(p)}\sum_{G}\Gamma_G$. Moreover, for every dimension $d$ and every integer $n\geq \max\{2,|A|,k\}$, and without any preprocessing of the input, the budgets may be taken as $\Gamma_G=O_p(k^{\frac{2p+2}{p+2}}\epsilon^{-2}\log n\cdot\log^4(2dk/\epsilon))$ for the main groups and $\Gamma_G=O_p(k\epsilon^{-2}\log n\cdot\log^3(2dk/\epsilon))$ for the outer groups, so the coreset has size $\wt O_p(k^{\frac{2p+2}{p+2}}\epsilon^{-2})$. Finally, for an \emph{arbitrary} center set $A^\star\in (\R^d)^k$ with induced clusters $P_1,\ldots,P_k$ and the same estimates and budgets, the output $(S, w)$ satisfies, with probability at least $0.99$, $|\sum_{s\in S}w_s\dist^p(s, C)-{\rm cost}(A, C)|\leq O(\epsilon)\cdot({\rm cost}(A, C)+{\rm cost}(A, A^\star))$ for all $C\in (\R^d)^k$. On the same event, the output has size $k+2^{O(p)}\sum_{G}\Gamma_G$.
\end{lemma}

\begin{proof}
We reduce to the analysis of Theorem A.11 of~\citet{hlw24}, given in their Appendices A.5--A.7, and we track every place where that analysis interacts with the three modifications.

\emph{Step 1: the perturbed decomposition.} The constructed rings $\wh R_{ij}$ partition each $P_i$, and membership gives the two-sided bound $2^j\wh \Delta_i\leq \dist^p(a, A^\star)<2^{j+1}\wh \Delta_i$. Every object of their analysis is defined relative to the decomposition the algorithm actually built. These objects include the huge and tiny subsets $H(G,C)$ and $T(G,C)$, the far subsets $F(G,C)$, the cost vectors of their Definition A.14, and the coverings, so we run their argument with respect to the hatted decomposition throughout. First, their Eq.~(7) and Observation A.7 convert ring membership into distance bounds, and they hold verbatim with $\wh \Delta_i$ in place of $\Delta_i$. Second, rings placed in a common group have true costs within a factor of $8$ of one another instead of $2$, so the constants of their Observation A.4 inflate from $(2k, 4k)$ to $(8k, 32k)$. Third, the covering-number bounds are unaffected, because their Lemma B.7 is applied with radii $r_i=2^{j+1}\wh \Delta_i$ and their Definition B.8 partitions by the ratio $\dist^p(a_i^\star, C)/\wh \Delta_i$ with the same number of levels. Finally, the identity $|P_i|\cdot \Delta_i={\rm cost}(P_i, A^\star)$ is used only to bound the total cost of inner rings and of discarded light groups, and there $\wh \Delta_i\leq 2\Delta_i$ loses a factor of $2$. Each effect inflates constants by at most $2^{O(p)}$, which is absorbed by rescaling $\epsilon$.

\emph{Step 2: Bernoulli sampling.} We write the estimator of a cost vector $y$ as $\sum_{a\in G}\xi_a y_a/q_a$ with $\xi_a\sim {\rm Ber}(q_a)$ independent. This estimator is exactly unbiased regardless of the accuracy of $\wh c_G$, because each weight is the reciprocal of the probability actually used. The symmetrization step of their Appendix A.6, which reduces the problem to a Gaussian process, requires only independence and unbiasedness. Moreover, the chaining argument that follows in their Lemmas A.17, A.21, and A.22 is carried out pointwise in the realization $S_G$. Therefore, the randomness of the sampler enters only at two further points. First, their event $\xi_G$ in their Eq.~(10) asserts that the per-cluster weight sums concentrate. Under Bernoulli sampling, the summands are independent, and by the hatted Observation A.4 they are bounded by $\max_a 1/q_a\leq 2^{O(p)}k|P_i\cap G|/\Gamma_G$, so Bernstein's inequality yields the same failure probability $k\cdot \exp(-\Omega(\epsilon^2\Gamma_G/k))$ as their Lemma A.13. Second, their Eq.~(14) and the end of their Lemma A.22 use the deterministic sample-size bound $|P_i\cap S_G|\leq \Gamma_G$, and we replace it by the event $|S_G|\leq 4\Gamma_G$. Since $\mathbb{E}[|S_G|]=\sum_{a\in G} q_a\leq 2\Gamma_G$, by the Chernoff bound this event fails with probability at most $\exp(-\Omega(\Gamma_G))$. On this event, the same chains hold with constants doubled, and the covering numbers are evaluated at $4\Gamma_G$ instead of $\Gamma_G$. This costs only a constant factor. Their Definitions A.8, A.10 and A.18 impose their conditions point by point, through sets and errors that depend only on $G$, $C$ and the point, so coverings of the parts of a partition of a set $S\subseteq G$ combine into a covering of $S$ at the same $\alpha$. Their covering number $\mathcal N_G(\Gamma,\alpha)$ is a maximum over the sets $S\subseteq G$ with $|S|\leq\Gamma$. Hence, splitting a set of size at most $m\Gamma$ into $m$ parts of size at most $\Gamma$ gives $\mathcal N_G(m\Gamma,\alpha)\leq \mathcal N_G(\Gamma,\alpha)^m$ for every integer $m\geq 1$, and likewise for the outer and layered covering numbers. With $m=4$, each entropy integral at most doubles, while the factor $1/\sqrt{\Gamma_G}$ in their chaining bound comes from the budget and is unchanged, so the error bound grows by a constant factor. The same inequality shows that their Eqs.~(8) and~(9) still hold after $\Gamma_G$ is multiplied by an integer. For the budgets of Step~4 nothing is lost, since the bounds of their Lemma B.2 do not depend on $\Gamma_G$. The complement is handled in the same way as they handle $\bar \xi_G$ in their Eqs.~(14)--(15). The crude deterministic bound grows by the factor $|S_G|/\Gamma_G$, and the expectation of this factor restricted to the complement is $\exp(-\Omega(\Gamma_G))\leq \epsilon/(4k)$ because $\Gamma_G=\Omega(k\epsilon^{-2}\log(k\epsilon^{-1}))$.

\emph{Step 3: center weights.} Item 3 of their Lemma A.12 is linear in the center weights, so replacing $|P_i\setminus \mathcal{G}|$ by $\wh w(a_i^\star)$ changes its left-hand side by at most $\sum_{i\in [k]} \epsilon|P_i|\cdot \dist^p(a_i^\star, C)$. By averaging the relaxed triangle inequality over $P_i$, we have $|P_i|\cdot \dist^p(a_i^\star, C)\leq 2^p\cdot ({\rm cost}(P_i, C)+{\rm cost}(P_i, A^\star))$. Therefore, the total change is at most $2^p\epsilon\cdot {\rm cost}(A, C+A^\star)$, and their Item 3 holds with $\epsilon$ replaced by $(1+2^p)\epsilon$.

We sum the three items over the $O_p(\log^2(k/\epsilon))$ groups and apply Markov's inequality at threshold $500\epsilon$ to the expectation bound $4\epsilon$ of their proof of Theorem A.11, which gives failure probability $0.008$. Note that they use the threshold $40\epsilon$ and probability $0.9$ instead. The remaining $0.002$ of the allowed failure probability $0.01$ is charged to the events $|S_G|\leq 4\Gamma_G$ of Step~2, which the size bound uses. Over the $O_p(\log^2(k/\epsilon))$ groups, the failure probabilities of these events sum to at most $\sum_G\exp(-\Omega(\Gamma_G))\leq 0.002$, once the constant factor in $\Gamma_G=\Omega(k\epsilon^{-2}\log(k\epsilon^{-1}))$ is large enough. After rescaling $\epsilon$ by $2^{O(p)}$, we obtain the first claim. Their proof bounds the expected error by $4\epsilon\cdot {\rm cost}(P, C+A^\star)=4\epsilon\,({\rm cost}(P, C)+{\rm cost}(P, A^\star))$, and hence the error is at most $40\epsilon\cdot{\rm cost}(P, C+A^\star)$ with probability $0.9$, and at most $500\epsilon\cdot{\rm cost}(P, C+A^\star)$ except with probability $0.008$, for all $C$ simultaneously. This is because their Lemma A.12 normalizes every group by ${\rm cost}(G, C+A^\star)$ or ${\rm cost}(P^G, C+A^\star)$ and its third item by ${\rm cost}(P, C+A^\star)$. Moreover, their proof invokes the approximation guarantee of $A^\star$ only in its last line, to write ${\rm cost}(P, C+A^\star)\leq O(1)\,{\rm cost}(P, C)$. The ring, group, and budget definitions and the statements of the three items use $A^\star$ only through the induced clusters, the averages $\Delta_i$, the costs ${\rm cost}(\cdot, A^\star)$, and the covering numbers. The bounds on these covering numbers in their Lemma B.2, which Step 4 uses, are uniform in $A^\star$. Items 2 and 3 are stated there in this normalization, with proofs cited from Lemma 13 of~\cite{clss22} and Lemma 4 of~\cite{css21}, which are their references [14] and [18]. This gives the relative form of the last claim for an arbitrary $A^\star$. The size bound of the last claim also holds on the same event, since the events $|S_G|\leq 4\Gamma_G$ of Step~2 do not use the approximation guarantee of $A^\star$.

\emph{Step 4: sample budgets without their Assumption 1.} The budgets of their Theorem B.1 are proved under their Assumption 1, which asks for at most $\poly(k/\epsilon)$ distinct points, dimension $O(\epsilon^{-2}\log\|P\|_0)$, and unweighted input. The third condition holds for our input. The first is used only to turn $\log\|P\|_0$ into $O(\log(k/\epsilon))$ (it is also cited in the proof of their Eq.~(35), a step of their proof of Lemma B.2, but only for this conversion). The second is used only where the $d$-branch of their Lemma B.2 bounds the small-$\alpha$ part of an entropy integral, in their Eqs.~(26)--(27) for the main-group budget of their Eq.~(8) and likewise for the outer-group budget of their Eq.~(9). Their Lemma B.2 states both branches with no condition on $d$, and we apply it with $\log\|P\|_0\leq \log n$. This also covers the sets $A^\star\cup S$ that their proof embeds, which have at most $k+n\leq 2n$ points, since $\log(2n)\leq 2\log n$ for $n\geq 2$. The main-group $d$-branch of their Lemma B.2 comes from their Lemma B.7, which assumes $d\geq \log k$. We may assume this by appending zero coordinates up to dimension $d':=\max\{d,\lceil\log k\rceil\}$. This changes no distance, and a covering for all center sets in $(\R^{d'})^k$ is also one for all center sets in $(\R^d)^k$, so we apply their Lemma B.2 in $\R^{d'}$. The padding is used only in the analysis, and the algorithm is unchanged. Let $L:=\log(2d'k/\epsilon)\leq 2\log(2dk/\epsilon)$ and $\alpha_0:=\epsilon/\sqrt{d'}$. We split each integral at $\alpha_0$ instead of $\epsilon$, using the $d$-branch only below $\alpha_0$. For main groups, the part below $\alpha_0$ is $O_p(\sqrt{kd'}\cdot \alpha_0\sqrt L)=O_p(\epsilon\sqrt{kL})$. Above $\alpha_0$, their Eq.~(28), started at $\alpha_0$ and with $\log n$ in place of $\log(k\epsilon^{-1})$, together with $\log\epsilon^{-1}\leq L$, $\log(\epsilon^{-1}\alpha^{-1})\leq 2L$ and $\int_{\alpha_0}^{(8p)^p\epsilon^{-p+1}}\alpha^{-1}d\alpha=O_p(L)$, gives $O_p(\sqrt{k^{\frac{2p+2}{p+2}}L^4\log n})$. Their Lemma B.2 is stated only for $0<\alpha\leq 1$, while their Eq.~(28) uses it up to $(8p)^p\epsilon^{-p+1}$. For $p\leq 2$ we need it only on $[\alpha_0,1]$. Indeed, for $a\in P_i$ let $c=\dist(a,C)$ and $r=\dist(a,A^\star)=\dist(a,a_i^\star)$. Since $|\dist(a,C)-\dist(a_i^\star,C)|\leq r$, the mean value theorem gives $|\dist^p(a,C)-\dist^p(a_i^\star,C)|\leq pr(c+r)^{p-1}\leq p2^{p-1}r\max\{c,r\}^{p-1}\leq p2^{p-1}(\sqrt{c^pr^p}+r^p)$, where the last step uses $(c/r)^{p/2-1}\leq 1$ for $c\geq r$ and $p\leq 2$. Their relative covering error in Definition A.8 is $\sqrt{c^pr^p}+r^p$ times $\sqrt{({\rm cost}(G,C)+{\rm cost}(G,A^\star))/{\rm cost}(G,A^\star)}\geq 1$, so the zero vector is an $\alpha$-covering for every $\alpha\geq p2^{p-1}$. Covering numbers are nonincreasing in $\alpha$, so on $[1,p2^{p-1}]$ the integrand is at most its value at $\alpha=1$, and this piece contributes $O_p(\sqrt{k^{\frac{2p+2}{p+2}}L^2\log n})$. For $p>2$ we follow their Eq.~(28), which applies the bound of their Lemma B.2 on $(1,(8p)^p\epsilon^{-p+1}]$, beyond its stated range. For $p>2$ our main-group budget therefore rests on this step of their proof of Theorem B.1. For outer groups, the covering number is $1$ for $\alpha\geq 1$. Indeed, their Definition A.10 asks for $|\dist^p(a,C)-v_a|\leq\alpha(\dist^p(a,C)+\dist^p(a,A^\star))$ point by point, and the zero vector satisfies this for every $\alpha\geq 1$. Below $\alpha_0$, the $d$-branch of their outer bound gives $\int_0^{\alpha_0}O_p(\sqrt{kd'\log\alpha^{-1}})\,d\alpha=O_p(\sqrt{kd'}\cdot\alpha_0\sqrt L)=O_p(\epsilon\sqrt{kL})$. On $[\alpha_0,1]$, its other branch gives $\int_{\alpha_0}^1O_p(\alpha^{-1}\sqrt{k\log n\log\alpha^{-1}})\,d\alpha=O_p(\sqrt{kL\log n}\cdot\log\alpha_0^{-1})=O_p(\sqrt{kL^3\log n})$, since $\log\alpha_0^{-1}\leq L$. Their outer bound cites Lemma 21 of~\cite{clss22}. In the full version (arXiv:2202.12793), the $\alpha^{-2}$-branch is Lemma 21 and the $d$-branch is Lemma 22. Both lemmas are stated only for integer $p$ and $\alpha\leq 1/2$, with $\log(1/(\alpha\epsilon))$ and $\log(4p/(\alpha\epsilon))$ respectively in place of $\log(1/\alpha)$. This changes the two parts only by factors $O_p(1)$, since $\log(1/(\alpha\epsilon))\leq 2L$ for $\alpha\geq\alpha_0$ and $\log(4p/(\alpha\epsilon))\leq\log(4p)+2\log\alpha^{-1}$ for $\alpha\leq\alpha_0\leq\epsilon$. The proof of Lemma 22 places a net around every input point and chooses $k$ net points, so its bound carries an additive $k\log\|P\|_0\leq k\log n$ that the statement omits. Below $\alpha_0$ this adds $O(\alpha_0\sqrt{k\log n})\leq O(\epsilon\sqrt{k\log n})$. The covering number is nonincreasing in $\alpha$, so the range $(1/2,1]$ adds at most $O_p(\sqrt{kL\log n})$. The outer budget is unchanged. For non-integer $p$ we use the outer bound of their Lemma B.2 as stated, as their proof of Theorem B.1 does. These bounds do not depend on $\Gamma_G$, so their self-referential Eqs.~(8) and~(9) hold for $\Gamma_G=O_p(k^{\frac{2p+2}{p+2}}\epsilon^{-2}L^4\log n)$ and $\Gamma_G=O_p(k\epsilon^{-2}L^3\log n)$, respectively, since both dominate their term $k\epsilon^{-2}\log(k/\epsilon)$. Compared with their Theorem B.1, the main-group budget replaces $\log(k/\epsilon)$ by $\log n$ and $\log^4\epsilon^{-1}$ by $L^4$. Summing over the $O_p(\log^2(k/\epsilon))$ groups gives the stated coreset size.
\end{proof}

We obtain a discrete constant-factor solution without constructing a coreset, using the bicriteria centers of Lemma~\ref{lem:bicriteria}, the cell counts of Lemma~\ref{lem:multidim_count}, and the following two facts.

\begin{lemma}[Theorem 1 of~\cite{dh96}]
\label{lem:dh_min}
Given query access to $N$ values, there is a quantum algorithm that outputs an index of a minimum value with probability at least $1/2$ using $O(\sqrt N)$ queries, and repeating it $O(\log(1/\delta))$ times and returning the smallest value found succeeds with probability at least $1-\delta$. In particular, for $c\in \R^d$ a nearest point of $A$ to $c$ can be found with probability at least $1-\delta$ using $\wt O(\sqrt n\log(1/\delta))$ queries to the points of $A$ and $\wt O(\sqrt n d\log(1/\delta))$ time.
\end{lemma}

\begin{lemma}[Cell counts as a proxy for the clustering cost]
\label{lem:cell_proxy}
Let $x'=(x'_1,\ldots,x'_m)\in (\R^d)^m$, let $\gamma\geq 1$, let $\tau:A\rightarrow x'$ satisfy $\|a-\tau(a)\|_2^p\leq \gamma\cdot \cost(a, x')$ for all $a\in A$, let $C_j=\{a\in A:\tau(a)=x'_j\}$ and $n_j=|C_j|$, and define, for any finite $C\subset \R^d$, the weighted cost $W(C)=\sum_{j=1}^m n_j\cdot \dist^p(x'_j, C)$. Then for every finite $C\subset \R^d$,
\begin{align*}
    W(C)\leq & ~ 2^{p-1}\big(\gamma\cdot \cost(A, x')+\cost(A, C)\big) \qquad \text{and} \\
    \cost(A, C)\leq & ~ 2^{p-1}\big(\gamma\cdot \cost(A, x')+W(C)\big).
\end{align*}
\end{lemma}
\begin{proof}
Fix $a\in C_j$. By the triangle inequality and the relaxed triangle inequality $(u+v)^p\leq 2^{p-1}(u^p+v^p)$ for $u,v\geq 0$ and $p\geq 1$, we have
\begin{align*}
    \dist^p(x'_j, C)\leq & ~ (\|x'_j-a\|_2+\dist(a, C))^p\leq 2^{p-1}(\|a-\tau(a)\|_2^p+\dist^p(a, C)) \\
    \leq & ~ 2^{p-1}(\gamma\cdot \cost(a, x')+\dist^p(a, C)).
\end{align*}
Since the cells partition $A$, summing over $a\in C_j$ and then over $j\in [m]$ gives $W(C)=\sum_{j}\sum_{a\in C_j}\dist^p(x'_j, C)\leq 2^{p-1}(\gamma\cdot \cost(A, x')+\cost(A, C))$. Symmetrically, for $a\in C_j$ we have $\dist^p(a, C)\leq (\|a-x'_j\|_2+\dist(x'_j, C))^p\leq 2^{p-1}(\gamma\cdot \cost(a, x')+\dist^p(x'_j, C))$, and summing gives the second inequality.
\end{proof}

\begin{lemma}[Discrete constant-factor solution in $\wt O(n^{0.5}kd+d\cdot\poly(k))$ time]
\label{lem:discrete_constant}
Let $A=\{a_1,\ldots,a_n\}\subset \R^d$, $k\geq 1$, $p\geq 1$ a constant and $\delta\in (0,1/2)$. There is a quantum algorithm that, with probability at least $1-\delta$, outputs a set $A^\star\subseteq A$ of at most $k$ input points with $\cost(A, A^\star)\leq 2^{O(p)}\cdot \min_{x\in (\R^d)^k}\cost(A, x)$, using $\wt O(n^{0.5}k\log(1/\delta))$ queries to the points of $A$, $\wt O(n^{0.5}kd\log(1/\delta))+\wt O(d\cdot \poly(k))$ time, and $\poly(k\log(n/\delta))$ additional preprocessing time.
\end{lemma}
\begin{proof}
Let $\OPT=\min_{x\in (\R^d)^k}\cost(A, x)$. We run each step below with failure probability at most $\delta/4$.
\begin{enumerate}
    \item Generate $x'\in (\R^d)^m$, $m=O(k\log^2 n)$, with $\cost(A, x')\leq 2^{O(p)}\cdot \OPT$ by Lemma~\ref{lem:bicriteria}. We amplify the success probability from $5/6$ to $1-\delta/4$ by $O(\log(1/\delta))$ independent runs, and keep the candidate of least estimated cost, using Lemma~\ref{lem:q_sum} at constant accuracy with a constant loss absorbed into $2^{O(p)}$. By the size cap of Lemma~\ref{lem:bicriteria}, every candidate has at most $c_{\rm bi}k\lceil\log_2 n\rceil^2$ centers on every outcome, so the selected one does too, even if it comes from a failed run. Each $\cost(a_i, x^\prime)$ is a minimum over $m=\wt O(k)$ centers, so each candidate takes $\wt O(\sqrt n kd)$ time, and hence the step takes $\wt O(\sqrt{n}kd\log(1/\delta))$ time and $\wt O(\sqrt{n}k\log(1/\delta))$ queries in total.
    \item Generate the oracle $\tau$ of Lemma~\ref{lem:q_ann} on $A, x'$ with $c_\tau\in [2.5,3)$, and also generate estimates $\wt n_j\in [\frac34 n_j, \frac54 n_j]$ of the cell sizes $n_j=|\{a\in A:\tau(a)=x'_j\}|$ by Lemma~\ref{lem:multidim_count} with accuracy $1/4$. This step takes $\wt O(\sqrt{nm}d)=\wt O(\sqrt{nk}d)$ time and $\wt O(\sqrt{nk})$ queries.
    \item Consider the weighted instance $\wt W(C)=\sum_{j=1}^m \wt n_j\cdot \dist^p(x'_j, C)$ on the $m$ weighted points $(x'_j, \wt n_j)$. Compute $\wt C\in (\R^d)^k$ with $\wt W(\wt C)\leq 2^{O(p)}\cdot \min_{C\in (\R^d)^k}\wt W(C)$ by applying Lemma~\ref{lem:approx_centers} to the weighted set, in $d\cdot \poly(k)$ time and no queries to $A$. Note that Lemma~\ref{lem:approx_centers} is stated for unweighted inputs, but the weights $\wt n_j$ here are polynomially bounded, and any polynomial-time constant-factor approximation for weighted $(k,p)$-clustering suffices.
    \item For each $c\in \wt C$, find a nearest point $a(c)\in A$ by Lemma~\ref{lem:dh_min}, in $\wt O(\sqrt nd)$ time per center, and output $A^\star=\{a(c): c\in \wt C\}$.
\end{enumerate}
Correctness. Let $W$ be as in Lemma~\ref{lem:cell_proxy} for this $x'$ and $\tau$, with $\gamma:=c_\tau^p<3^p$ by Lemma~\ref{lem:q_ann}. Then $\wt W(C)\in [\frac34 W(C), \frac54 W(C)]$ for every $C$. Let $C^\star$ attain $\OPT$. By Step 3 and Lemma~\ref{lem:cell_proxy}, we have
\begin{align*}
    \wt W(\wt C)\leq 2^{O(p)}\wt W(C^\star)\leq 2^{O(p)}W(C^\star)\leq 2^{O(p)}\cdot 2^{p-1}(\gamma\cdot \cost(A, x')+\OPT)\leq 2^{O(p)}\cdot \OPT,
\end{align*}
where we use $\cost(A, x')\leq 2^{O(p)}\OPT$ and $\gamma<3^p$. The second inequality of Lemma~\ref{lem:cell_proxy} then gives $\cost(A, \wt C)\leq 2^{p-1}(\gamma\cdot \cost(A, x')+\frac43 \wt W(\wt C))\leq 2^{O(p)}\cdot \OPT$. Finally, for every $a\in A$ and $c\in \wt C$, we have $\|a-a(c)\|_2\leq \|a-c\|_2+\|c-a(c)\|_2\leq 2\|a-c\|_2$, since $a(c)$ is a nearest input point to $c$. Hence, $\dist(a, A^\star)\leq 2\dist(a, \wt C)$ and $\cost(A, A^\star)\leq 2^p\cdot \cost(A, \wt C)\leq 2^{O(p)}\cdot \OPT$. A union bound over the four steps gives the failure probability, and summing the costs gives the time and query bounds, where the preprocessing of Lemma~\ref{lem:q_ann} and of Lemma~\ref{lem:bicriteria} is $\poly(k\log(n/\delta))$.
\end{proof}

\begin{lemma}[Center replacement for near-center points]
\label{lem:center_replacement}
Let $p\geq 1$, $\epsilon\in (0,1]$, $a, a^\star\in \R^d$ and let $C\subset \R^d$ be finite. Then
\begin{align*}
    |\dist^p(a, C)-\dist^p(a^\star, C)|\leq \epsilon\cdot \dist^p(a, C)+(3p/\epsilon)^p\cdot \|a-a^\star\|_2^p.
\end{align*}
\end{lemma}
\begin{proof}
Let $x:=\dist(a, C)$, $y:=\dist(a^\star, C)$, and $\eta:=\|a-a^\star\|_2$. By the triangle inequality, we have $|x-y|\leq \eta$, so the mean value theorem gives $|x^p-y^p|\leq p\,|x-y|\max\{x,y\}^{p-1}\leq p\,\eta\,(x+\eta)^{p-1}$. If $\eta\leq \epsilon x/(2p)$, then $(x+\eta)^{p-1}\leq (1+\epsilon/(2p))^{p-1}x^{p-1}\leq e^{1/2}x^{p-1}$, and therefore $|x^p-y^p|\leq p\cdot \frac{\epsilon x}{2p}\cdot e^{1/2}x^{p-1}\leq \epsilon x^p$. Otherwise, we have $x<2p\eta/\epsilon$ and $y\leq x+\eta<(2p/\epsilon+1)\eta\leq 3p\eta/\epsilon$, so $|x^p-y^p|\leq \max\{x,y\}^p\leq (3p/\epsilon)^p\eta^p$.
\end{proof}

\begin{theorem}[Quantum coreset for $(k,p)$-clustering via the construction of~\citet{hlw24}]
\label{thm:one_shot_main}
Let $A=\{a_1,\ldots,a_n\}\subset \R^d$, let $1\leq k\leq n$ be an integer, $\epsilon\in (0,1)$, and let $p\geq 1$ be a constant. There exists a quantum algorithm that, with probability at least $0.9$, outputs an $\epsilon$-coreset of $A$ for $(k,p)$-clustering that consists of input points only and has size $k+O_p(k^{\frac{2p+2}{p+2}}\epsilon^{-2}\log^2(k/\epsilon)\log n\cdot\log^4(2dk/\epsilon))=\wt O_p(k^{\frac{2p+2}{p+2}}\epsilon^{-2})$, in time
\begin{align*}
    \wt O_p(\epsilon^{-1}n^{0.5}d\,k^{\frac{2p+3}{p+2}})+\wt O(d\cdot\poly(k)),
\end{align*}
using $\wt O_p(n^{0.5}(k+k^{\frac{p+1}{p+2}}\epsilon^{-1}))$ queries to the points of $A$. The size carries the factor $\log n\cdot\log^4(2dk/\epsilon)$, unlike Theorem B.1 of~\cite{hlw24}, whose size depends on neither $n$ nor $d$.
\end{theorem}

\begin{proof}
Throughout the proof, we run each randomized step with failure probability at most $0.01$, and we rescale $\epsilon$ by the constant $2^{O(p)}$ required by Lemma~\ref{lem:one_shot_robust}. Every stage runs on the input points in $\R^d$ without dimension reduction, and the stages below charge $O(kd)$ time for each access to a point after Stage 1. No assumption on $d$ is needed, since the budgets of Lemma~\ref{lem:one_shot_robust} hold for every $d$.

\emph{Stage 1: a constant-factor discrete solution.} Run Lemma~\ref{lem:discrete_constant} to obtain $A^\star\subseteq A$ with at most $k$ points and $\cost(A, A^\star)\leq 2^{O(p)}\cdot \min_{x\in (\R^d)^k}\cost(A, x)$, in $\wt O(n^{0.5}kd)+d\cdot \poly(k)$ time and $\wt O(n^{0.5}k)$ queries.

\emph{Stage 2: decomposition and statistics.} Given $A^\star$, the cluster $P_i$ of a point and its value $\dist^p(a, A^\star)$ can be computed in $O(kd)$ time. We obtain all cell statistics of this stage from Corollary~\ref{cor:cell_mass} which estimates all cells of a partition at once but requires values of bounded range, and we ensure the bounded range by truncation.

\emph{(a) Truncation.} We estimate $\Lambda:={\rm cost}(A, A^\star)$ within a factor $2$ by Lemma~\ref{lem:q_sum}, which gives $\wh\Lambda\in [\Lambda/2, 2\Lambda]$ in $\wt O(n^{0.5})$ queries and $\wt O(n^{0.5}kd)$ time, and we then set the floor $F:=\epsilon^{p+1}\wh\Lambda/(2(3p)^pn)\leq \epsilon^{p+1}\Lambda/((3p)^pn)$. Let $T:=\{a\in A:\dist^p(a, A^\star)\leq F\}$ be the truncated points, $T_i:=T\cap P_i$, $A':=A\setminus T$, and $P_i':=P_i\setminus T$. If $\wh\Lambda=0$, then $\Lambda=0$ and $T=A$, and hence $A^\star$ with the estimated cluster masses of Lemma~\ref{lem:multidim_count} is an $\epsilon$-coreset by Lemma~\ref{lem:center_replacement} with $\|a-a^\star\|_2=0$. Thus, we assume $\wh\Lambda>0$ below. Every $a\in A'$ has $\dist^p(a, A^\star)\in (F, \Lambda]\subseteq [F, RF]$ with $R:=4(3p)^pn/\epsilon^{p+1}$, and therefore $\log R=O_p(\log(n/\epsilon))$. The oracle that maps $a\in T$ to a null cell and reports the value $0$ for it can be computed in $O(kd)$ time, and the points of $T$ will be represented by their centers.

\emph{(b) Cluster statistics.} We apply Corollary~\ref{cor:cell_mass} with the $k$ clusters $P_i'$ as cells, in $\wt O_p(\sqrt{nk}\cdot kd)=\wt O_p(n^{0.5}k^{1.5}d)$ time. This estimates all $|P_i'|$ and ${\rm cost}(P_i', A^\star)$ within $1\pm 1/4$, and since $(5/4)/(3/4)<2$, we obtain estimates $\wh\Delta_i\in [\Delta_i'/2, 2\Delta_i']$ of the averages $\Delta_i':={\rm cost}(P_i', A^\star)/|P_i'|$. If $P_i'$ is empty, then the cluster has no rings and $a_i^\star$ receives the weight $|T_i|$ in (d). The estimates $\wh\Delta_i$ fix the rings of~\citet{hlw24} on the clusters $P_i'$, as in their Definition A.2, and the ring index of a point can be computed in $O(kd)$ time. Their definition has $O_p(\log(k/\epsilon))$ main rings per cluster plus inner and outer rings, and there are at most $O(\log R)$ outer rings per cluster since every value lies in $[F,\Lambda]$. Thus, there are $M=\wt O_p(k)$ nonempty rings in total.

\emph{(c) Ring statistics.} Applying Corollary~\ref{cor:cell_mass} again with the $M$ nonempty rings as cells, we obtain all ring masses and ring costs within $1\pm 1/4$ in $\wt O_p(\sqrt{nM}\cdot kd)=\wt O_p(n^{0.5}k^{1.5}d)$ time. This fixes the groups and the group-cost estimates $\wh c_G$, which are sums of ring costs and are within $1\pm 1/4$.

\emph{(d) Center weights.} Each center $a_i^\star$ must receive the weight $|T_i|+|P_i'\setminus \mathcal{G}|$ within additive error $\epsilon|P_i|$. To estimate these weights, we run Lemma~\ref{lem:multidim_count} with relative accuracy $\epsilon/2^{O(p)}$ on the cells $\{T_i\}_i\cup\{\text{rings}\}$, in $\wt O_p(\sqrt{nM/\epsilon}\cdot kd)=\wt O_p(\epsilon^{-0.5}n^{0.5}k^{1.5}d)$ time.

\emph{Stage 3: one-shot sampling.} Sample every point $a\in A'$ independently with probability $q_a=\min\{1, \Gamma_G\cdot \dist^p(a, A^\star)/\wh c_G\}$, where $G$ is the group of $a$'s ring, and $q_a=0$ for points of $T$ and for points outside all groups, with the budgets $\Gamma_G$ that Lemma~\ref{lem:one_shot_robust} gives for $A'$ and the integer $\max\{2,n\}\geq\max\{2,|A'|\}$, which is also at least $k$ by hypothesis. Each probability is computable in $O(kd)$ time after Stage 2, and $\sum_a q_a\leq 2\sum_G \Gamma_G=\wt O_p(k^{\frac{2p+2}{p+2}}\epsilon^{-2})=:s$ over the $O_p(\log^2(k\epsilon^{-1}))$ groups, so Lemma~\ref{lem:q_sampling_1_d} performs the sampling in $\wt O(\sqrt{ns}\cdot kd)=\wt O_p(\epsilon^{-1}n^{0.5}k^{\frac{2p+3}{p+2}}d)$ time.

\emph{Correctness.} Let $S$ be the set of sampled points with weights $1/q_a$, and let each center $a_i^\star$ receive the estimated weight $\wh w_i$ of $|T_i|+|P_i'\setminus\mathcal{G}|$. We fix $C\in (\R^d)^k$ and split the error into three parts.

First, the sampler runs on $A'$ with the center set $A^\star$ and the clusters $P_i'$, which are built from factor-$2$ estimates. Its Bernoulli weights are the reciprocals of the probabilities used, and it uses the exact center weights $|P_i'\setminus\mathcal{G}|$. The estimation error of these weights is $\epsilon|P_i|$ rather than the $\epsilon|P_i'|$ the lemma tolerates, and we charge it separately in the third part. Since $A^\star$ need not be approximately optimal for $A'$, we use the relative form of Lemma~\ref{lem:one_shot_robust}. With probability at least $0.99$, we have for all $C$,
\begin{multline*}
    \Big|\sum_{s\in S}w_s\dist^p(s, C)+\sum_i |P_i'\setminus\mathcal{G}|\cdot\dist^p(a_i^\star, C)-{\rm cost}(A', C)\Big| \\
    \leq O(\epsilon)\,({\rm cost}(A', C)+{\rm cost}(A', A^\star))\leq O(\epsilon)\,({\rm cost}(A, C)+\Lambda).
\end{multline*}
Second, we represent the truncated points by their centers. Applying Lemma~\ref{lem:center_replacement} to each $a\in T_i$ with $a^\star=a_i^\star$ and $\|a-a_i^\star\|_2^p=\dist^p(a, A^\star)\leq F$, we have
\begin{multline*}
    \Big|\sum_i |T_i|\cdot \dist^p(a_i^\star, C)-{\rm cost}(T, C)\Big|\leq \sum_{a\in T}\big(\epsilon\,\dist^p(a, C)+(3p/\epsilon)^p\dist^p(a, A^\star)\big) \\
    \leq \epsilon\,{\rm cost}(A, C)+(3p/\epsilon)^p\,nF\leq \epsilon\,{\rm cost}(A, C)+\epsilon\Lambda .
\end{multline*}
Third, the center weights are estimated within additive error $\epsilon|P_i|$, which by the relaxed triangle inequality averaged over $P_i$ as in Step 3 of the proof of Lemma~\ref{lem:one_shot_robust} changes the coreset cost by at most $\sum_i\epsilon|P_i|\dist^p(a_i^\star, C)\leq 2^p\epsilon\,({\rm cost}(A, C)+\Lambda)$. Summing the three parts, the weighted set $S\cup A^\star$ has cost within $O_p(\epsilon)\,({\rm cost}(A, C)+\Lambda)$ of ${\rm cost}(A, C)={\rm cost}(A', C)+{\rm cost}(T, C)$ for all $C$. Moreover, by Stage 1, $\Lambda={\rm cost}(A, A^\star)\leq 2^{O(p)}\min_x{\rm cost}(A, x)\leq 2^{O(p)}{\rm cost}(A, C)$. Rescaling $\epsilon$ by $2^{O(p)}$ gives an $\epsilon$-coreset. The at most seven randomized steps, each failing with probability at most $0.01$, are Stage 1, the estimate $\wh\Lambda$, the two applications of Corollary~\ref{cor:cell_mass}, the counts of (d), the guarantee of Lemma~\ref{lem:one_shot_robust} and the sampler of Lemma~\ref{lem:q_sampling_1_d}, so by a union bound, the total failure probability is at most $0.07<0.1$. The size of the coreset is $k+2^{O(p)}\sum_G\Gamma_G=k+O_p(k^{\frac{2p+2}{p+2}}\epsilon^{-2}\log^2(k/\epsilon)\log n\cdot\log^4(2dk/\epsilon))$, since the truncation only removes points from the sampled set and the budgets of Lemma~\ref{lem:one_shot_robust} hold for every $d$ without any preprocessing of the input. Here the factor $\log^2(k/\epsilon)$ accounts for the $O_p(\log^2(k/\epsilon))$ groups, and $\log n\cdot\log^4(2dk/\epsilon)$ is the factor of the main-group budgets, which dominate the outer-group budgets. Theorem B.1 of~\citet{hlw24} has $\log(k/\epsilon)\log^4\epsilon^{-1}$ in place of this factor. Finally, every element of the coreset is an input point.

\emph{Running time and queries.} Stage 1 costs $\wt O(n^{0.5}kd)+\wt O(d\cdot\poly(k))$ time and $\wt O(n^{0.5}k)$ queries, Stage 2 costs $\wt O_p(n^{0.5}k^{1.5}d+\epsilon^{-0.5}n^{0.5}k^{1.5}d)$ time and $\wt O_p(\epsilon^{-0.5}n^{0.5}k^{0.5})$ queries, and Stage 3 costs $\wt O_p(\epsilon^{-1}n^{0.5}k^{\frac{2p+3}{p+2}}d)$ time and $\wt O_p(\epsilon^{-1}n^{0.5}k^{\frac{p+1}{p+2}})$ queries. Since $k\leq k^{1.5}\leq k^{\frac{2p+3}{p+2}}$ for $p\geq 1$, the total time is $\wt O_p(\epsilon^{-1}n^{0.5}d\,k^{\frac{2p+3}{p+2}})+\wt O(d\cdot\poly(k))$ and the total number of queries is $\wt O_p(n^{0.5}(k+\epsilon^{-1}k^{\frac{p+1}{p+2}}))$.
\end{proof}

\subsection{Quantum Algorithm for Data Selection}

As an application, we study the \emph{data selection pipeline} in machine learning, where the goal is to select a weighted subset of data points for training or fine-tuning the model while preserving desirable properties. In this model, data are given as $d$-dimensional embeddings, and a loss function $\ell:\R^d\rightarrow \R_{\geq 0}$ is used to grade the quality of the embedding. $\ell$ can be expensive to evaluate, such as a deep neural network. \cite{ach+24} provides a principled way for data selection using the coreset of $(k, p)$-clustering, under some mild assumptions on $\ell$.

\begin{assumption}
\label{assumption:z_lambda_well_behaved}
Let $\Lambda=(\Lambda_1,\ldots,\Lambda_k)\in \R^k_{\geq 0}, x\in (\R^d)^k$ and let $E\subseteq \R^d$ be a set of embeddings, we say the loss function $\ell$ is $(p, \Lambda)$-well-behaved with respect to $E$ and $x$ if, for every $x_j\in x$ and every $e\in C_j$, where $C_j=\{e\in E: \arg\min_{x_i\in x}\|x_i-e\|_2^p=x_j \}$,
\begin{align*}
    |\ell(e)-\ell(x_j)|\leq & ~ \Lambda_j\|e-x_j\|_2^p.
\end{align*}
\end{assumption}

Define the weighted cost as $\cost^\Lambda(a_i, x)=\Lambda_{j(i)}\cost(a_i, x)$ where we use $j(i)$ to denote the index of the cluster that $a_i$ is assigned to, and similarly $\cost^\Lambda(A, x)=\sum_{i=1}^n \cost^\Lambda(a_i, x)=\sum_{i=1}^k \Lambda_i \sum_{a_j\in C_i} \|a_j-x_i\|_2^p$. \cite{ach+24} essentially proves that under Assumption~\ref{assumption:z_lambda_well_behaved}, one could perform weighted sampling according to $\cost^\Lambda(a_i, x)$. In addition, the expensive loss function only needs to be evaluated on the centers. For convenience, we state an approximate $k$-centers algorithm below.

\begin{lemma}[\cite{mp04}]
\label{lem:approx_centers}
Let $A=\{a_1,\ldots,a_n\}\subset \R^d$ and $X=(\R^d)^k$, let $\delta\in (0,1)$. Then, there exists an algorithm that computes a solution $x'\in X$ such that
\begin{align*}
    {\rm cost}(A, x') \leq & ~ 2^{O(p)}\cdot \min_{x\in X} {\rm cost}(A, x),
\end{align*}
holds with probability at least $1-\delta$. Moreover, $x'$ can be computed in time
\begin{align*}
    O(ndk+nd\log(n/\delta)+k^2\log^2 n+\log^2(1/\delta)\log^2 n) = & ~ \wt O(ndk).
\end{align*}
\end{lemma}

We now state a quantum implementation of the adaptive sampling due to~\cite{ach+24}. 
\begin{algorithm}[!ht]
\caption{Quantum one-round adaptive sampling for data selection.}
\label{alg:q_data_selection}
\begin{algorithmic}[1]
\Procedure{QDataSelection}{$A\in \R^{n\times d}, x\in (\R^d)^k, \Lambda\in \R^k_{\geq 0}, \ell:\R^d\rightarrow \R_{\geq 0}, \epsilon\in (0,1)$}
\State $s\gets 300\epsilon^{-2},\epsilon'\gets 0.01,\delta\gets 0.001$
\State Let $\tau:A\rightarrow x$ be such that $\tau(a_i)=\arg\min_{x_j\in x} \|a_i-x_j\|_2^p$, and let $j(i)$ denote the index of $\tau(a_i)$ in $x$
\State $M\gets \max_{i\in [n]}\Lambda_{j(i)}\|a_i-\tau(a_i)\|_2^p$ via Lemma~\ref{lem:dh_min} with failure probability $\delta$
\State Generate $\wt \cost^{\Lambda}(A, x)=(1\pm\epsilon')\cost^\Lambda(A, x)$ via Lemma~\ref{lem:q_sum} with the bound $M$, accuracy $\epsilon'$ and failure probability $\delta$
\State Let $C_j=\{a\in A: \tau(a)=x_j \}$ and $n_j=|C_j|$
\State Generate $\wt n_1,\ldots,\wt n_k$ via Lemma~\ref{lem:multidim_count} using $\tau$ with accuracy $\epsilon'$ and failure probability $\delta$
\State Compute $\ell(x_1),\ldots,\ell(x_k)$
\State ${\rm sum}\gets \sum_{j=1}^k \wt n_j\cdot \ell(x_j)$
\State Implement an oracle for each $a_i\in A$ as follows:
\State \hspace{4mm} $\wh \ell(a_i)\gets \ell(\tau(a_i)), v(a_i)\gets \Lambda_{j(i)}\cdot \|a_i-\tau(a_i) \|_2^p$
\State \hspace{4mm} $q_{i}\gets \frac{\wh \ell(a_i)+v(a_i)}{\wt \cost^\Lambda(A, x)+{\rm sum}}$ \Comment{$q_i\gets 0$ if the denominator is $0$}
\State \hspace{4mm} $p_i\gets \min\{1, s\cdot q_i\}$
\State $D'\gets \textsc{QSample}(p, 0.001)$, and $w_i\gets 1/p_i$ for $i\in D'$, $w_i\gets 0$ for $i\notin D'$
\State \Return $(D', w)$
\EndProcedure
\end{algorithmic}
\end{algorithm}

We then prove Algorithm~\ref{alg:q_data_selection} implements the data selection procedure in sublinear time. 

\begin{theorem}
Let $\epsilon\in (0,1), p\geq 1, \Lambda\in \R^k_{\geq 0}$, $A\in (\R^d)^n$ and $\ell$ be a loss function that is $(p,\Lambda)$-well-behaved with respect to $A$ and a clustering $x\in (\R^d)^k$. Then, there exists a quantum algorithm (Algorithm~\ref{alg:q_data_selection}) that outputs a weight vector $w\in \R^n_{\geq 0}$ such that, with probability at least 0.99, $\|w\|_0=O(\epsilon^{-2})$ and
\begin{align*}
    |\sum_{i=1}^n \ell(a_i)-\sum_{i=1}^n w_i \ell(a_i)| \leq & ~ \epsilon\cdot (\sum_{i=1}^n \ell(a_i)+2\cost^\Lambda(A, x)).
\end{align*}
Moreover, the algorithm makes at most $k$ queries to the loss function $\ell$, $\wt O(n^{0.5}(\epsilon^{-1}+k^{0.5}))$ queries to the points in $A$, and uses an additional $\wt O(n^{0.5}kd(\epsilon^{-1}+k^{0.5}))$ time.
\end{theorem}

\begin{proof}
Let $u_i:=\wh\ell(a_i)+v(a_i)$ and $U:=\sum_{i=1}^n u_i=\sum_{j=1}^k n_j\ell(x_j)+\cost^\Lambda(A, x)$. Since $\tau(a_i)$ is the nearest center of $a_i$, Assumption~\ref{assumption:z_lambda_well_behaved} gives $\ell(a_i)\leq u_i$ and $\wh\ell(a_i)\leq\ell(a_i)+v(a_i)$, and therefore $L:=\sum_{i=1}^n\ell(a_i)\leq U\leq L+2\cost^\Lambda(A, x)$. Suppose that Lemmas~\ref{lem:dh_min},~\ref{lem:q_sum}, and~\ref{lem:multidim_count} succeed, where each is run with failure probability $\delta=0.001$. Then the denominator $\wt D:=\wt\cost^\Lambda(A, x)+{\rm sum}$ satisfies $\wt D=(1\pm\epsilon')U$, and hence $q_i=u_i/\wt D$ and $\sum_i p_i\leq s\sum_i q_i\leq s/(1-\epsilon')\leq 1.02s$. Algorithm~\ref{alg:q_data_selection} is the one-round adaptive sampling of Theorem 2 of~\cite{ach+24}, with the exact statistics replaced by these estimates and with independent inclusions, and we verify its guarantee directly. The estimator $Z:=\sum_iw_i\ell(a_i)=\sum_i\xi_i\ell(a_i)/p_i$, where the $\xi_i\sim{\rm Ber}(p_i)$ are independent, satisfies $\E Z=L$, and an index with $u_i=0$ has $\ell(a_i)=0$ and contributes to neither side. Moreover,
\begin{align*}
    \var Z= & ~ \sum_{i:\,p_i<1}\ell(a_i)^2\,\frac{1-p_i}{p_i}\leq\sum_{i}\frac{\ell(a_i)^2}{s\,q_i}=\frac{\wt D}{s}\sum_i\frac{\ell(a_i)^2}{u_i}\\
    \leq & ~ \frac{\wt D\,L}{s}\leq\frac{1.01\,(L+2\cost^\Lambda(A, x))^2}{s},
\end{align*}
where we use $\ell(a_i)\leq u_i$ and $\wt D\leq 1.01\,U$. Then Chebyshev's inequality with $s=300\epsilon^{-2}$ gives $|Z-L|\leq\epsilon(L+2\cost^\Lambda(A, x))$ except with probability $1.01/300<0.0034$. The size $\|w\|_0=|D'|$ is a sum of independent Bernoulli variables with mean at most $1.02s$, so by the Chernoff bound, $|D'|\leq 2.04s=612\epsilon^{-2}$ except with probability $e^{-1.02s/3}<10^{-40}$. Finally, a union bound over the three estimates, the sampler of Lemma~\ref{lem:q_sampling_1_d} with failure probability $0.001$, the Chebyshev step, and the size bound gives failure probability at most $0.008$.

To analyze the runtime, note that the oracle $\tau$ can be queried in $O(kd)$ time, so $M$ and $\wt \cost^\Lambda(A, x)$ can be computed in $\wt O(\sqrt nkd)$ time and $\wt O(\sqrt n)$ queries by Lemmas~\ref{lem:dh_min} and~\ref{lem:q_sum}. $\wt n_1,\ldots,\wt n_k$ can be estimated in $\wt O(\sqrt nk^{1.5}d)$ time and $\wt O(\sqrt{nk})$ queries. Finally, each sampling probability can be computed in $O(kd)$ time, so the time for the final sampling is $\wt O(\sqrt{ns}\cdot kd)=\wt O(\epsilon^{-1}n^{0.5}kd)$ with $\wt O(\epsilon^{-1}n^{0.5})$ queries. Computing the weights of the $O(\epsilon^{-2})$ sampled indices takes $O(\epsilon^{-2}kd)$ additional time, which is dominated since $\epsilon^{-1}\leq n^{0.5}$ by our assumption that $n\gg\poly(k,1/\epsilon)$. Thus, the overall runtime is
\begin{align*}
    & \wt O(n^{0.5}kd(\epsilon^{-1}+k^{0.5})). \qedhere
\end{align*}
\end{proof}
Note that computing the weights classically would take $O(ndk)$ time, so ours is the first to achieve this goal in time sublinear in $n$. To compute a set of approximate $k$-centers, one could either use the bicriteria centers of Lemma~\ref{lem:bicriteria} as a proxy, or first compute an $\epsilon$-coreset using Theorem~\ref{thm:one_shot_main} and then apply Lemma~\ref{lem:approx_centers} to the coreset.
\section{Lower Bound}
\label{sec:lb}
In this section, we prove a quantum query lower bound for the generic sensitivity sampling problem, showing that the $\wt O(\sqrt{ns})$ query complexity of Theorem~\ref{thm:general} is optimal up to $\poly(\epsilon^{-1}, \log n)$ and VC-dimension factors. Theorem~\ref{thm:lb_generic} uses the block construction of Theorem 8.1 of~\cite{ag23} (arXiv v3), which shows that finding a constant-factor spectral approximation of $A^\top A$ requires $\Omega(\sqrt{nd})$ quantum row queries. For the dependence on $\epsilon$, Theorems~\ref{thm:lb_eps} and~\ref{thm:lb_eps_S} have a precedent in~\cite{xclj23}. Their Theorem 5.1 (restated as their Theorem D.5) proves that multidimensional counting with $m$ cells, the problem of Lemma~\ref{lem:multidim_count}, requires $\Omega(\sqrt{nm}\,\epsilon^{-1/2})$ queries, by a reduction from the counting problems of~\cite{nw99} on blocks, and the coreset part of their Theorem 5.2 (their Theorem D.7) derives from it the bound $\Omega(\sqrt{nk}\,\epsilon^{-1/2})$ for $\epsilon$-coresets for $(k,p)$-clustering. The leverage-score lower bound quoted in Section~\ref{sec:tech:lb}, Lemma 8.2 of~\cite{ag23}, is credited there to Theorem 2 of~\cite{aw22}, and its hypotheses $n=O(d^2)$ and $\epsilon=\Omega(\sqrt{d/n})$ force $S=d=\Omega(\epsilon^{-2})$.

\begin{theorem}[Query lower bound for generic sensitivity sampling]\label{thm:lb_generic}
Let $n, S$ be positive integers with $S\leq n$, and let $\epsilon\in (0, 1/2)$. There exists a family of instances, consisting of a point set $A=\{a_1,\ldots,a_n\}\subset \R^2$, a universe $X$ with $|X|=S$, and a cost function ${\rm cost}: \R^2\times X\rightarrow \R_{\geq 0}$, such that the total sensitivity of every instance is at most $S$, the dimension of the system (Definition~\ref{def:dim}) is $O(1)$, and any quantum algorithm that outputs an $\epsilon$-coreset of $A$ with probability at least $2/3$ requires $\Omega(\sqrt{nS})$ quantum queries to the points of $A$, provided the output coreset has size $o(\sqrt{nS})$.
\end{theorem}

\begin{proof}
Without loss of generality let $S$ divide $n$, and partition $[n]$ into $S$ blocks ${\cal B}_1,\ldots,{\cal B}_S$, each of size $n/S$. The instance is specified by hidden bits $z\in \{0,1\}^n$: the $i$-th point is $a_i=(j(i), z_i)\in \R^2$, where $j(i)$ denotes the block containing $i$, so that one query to a point reveals its bit. Let $X=\{x_1,\ldots,x_S\}$ and define, for a point $a=(j, z)$,
\begin{align*}
    {\rm cost}(a, x_l) = z\cdot {\bf 1}[j=l],
\end{align*}
so that ${\rm cost}(A, x_l)$ is the number of \emph{marked} points (points with $z_i=1$) in block $l$. The sensitivity of a marked point in block $l$ equals the inverse of the number of marked points in block $l$, and unmarked points have sensitivity $0$. Hence, the total sensitivity is the number of nonempty blocks, which is at most $S$. Moreover, for a subset $M\subseteq [n]$ and a weight vector $v$, every range ${\rm range}_{M,v}(x_l, r)$ in the sense of Definition~\ref{def:dim} is $M$ with the marked points $i\in M\cap {\cal B}_l$ of weighted cost $v_i>r$ removed. All other points of $M$ have cost $0$ under $x_l$, and hence they always belong to the range. Therefore, $M$ has at most $|M|+1\leq |M|^2$ distinct ranges, and the system has $O(1)$ dimension.

Now let $B$ be a valid $\epsilon$-coreset of $A$ with $\epsilon<1/2$. For every block $l$ with at least one marked point, ${\rm cost}(A, x_l)\geq 1$ and thus $\sum_{b\in B} w_b\cdot {\rm cost}(b, x_l)\geq (1-\epsilon)>0$, and since only marked points of block $l$ contribute to this sum, $B$ must contain a marked point of block $l$. Conversely, if block $l$ contains no marked point, then neither does $B\cap {\cal B}_l$. Therefore, reading the bits of the coreset points (at most $|B|$ additional queries) determines the entire vector $({\rm OR}(z_{{\cal B}_1}),\ldots, {\rm OR}(z_{{\cal B}_S}))$, where ${\rm OR}(z_{{\cal B}_l})$ denotes the disjunction of the bits in block $l$. Computing all $S$ disjunctions simultaneously, over independent instances of size $n/S$ each, requires $\Omega(S\sqrt{n/S})=\Omega(\sqrt{nS})$ bounded-error quantum queries by the strong direct product theorem for ${\rm OR}$~\citep{ksw07}. Hence, if the algorithm makes $Q$ queries and outputs a coreset of size $|B|$, then $Q+|B|=\Omega(\sqrt{nS})$, and the claim follows from $|B|=o(\sqrt{nS})$.
\end{proof}

\begin{remark}
As shown in Section~\ref{sec:tech:lb}, the coreset produced by Theorem~\ref{thm:general} for this instance family has size $s=\wt O(\epsilon^{-2}S)$, which is $o(\sqrt{nS})$ whenever $S=\wt o(\epsilon^{4} n)$, and hence Theorem~\ref{thm:lb_generic} matches the $\wt O(\sqrt{ns})$ query complexity of our framework up to $\poly(\epsilon^{-1}, \log n)$ factors. We address the dependence on $\epsilon$ for this family in Theorems~\ref{thm:lb_eps} and~\ref{thm:lb_eps_S} below.
\end{remark}

\begin{theorem}[$\epsilon$-dependence of the query lower bound, $S=1$]\label{thm:lb_eps}
Let $n$ be a positive integer and $\epsilon\in (0, 1/2)$ with $\lfloor 1/\epsilon\rfloor\leq n/2$. There is a family of instances with a point set $A=\{a_1,\ldots,a_n\}\subset \R^2$, a universe $X$ with $|X|=1$ and a cost function as in Theorem~\ref{thm:lb_generic} (total sensitivity at most $1$, dimension $O(1)$), such that any quantum algorithm that, on every instance of the family, outputs with probability at least $2/3$ an $\epsilon$-coreset of $A$ of size $o(\sqrt{n/\epsilon})$ together with its weight vector makes $\Omega(\sqrt{n/\epsilon})$ quantum queries to the points of $A$ on some instance. The bound is tight for this family: an $\epsilon$-coreset of size $1$ can be output with $O(\sqrt{n/\epsilon})$ queries.
\end{theorem}

\begin{proof}
We take $S=1$ in the family of Theorem~\ref{thm:lb_generic}, that is, $a_i=(1, z_i)$ for hidden bits $z\in \{0,1\}^n$, $X=\{x_1\}$, and ${\rm cost}(a, x_1)=z$ for $a=(1, z)$. Thus, ${\rm cost}(A, x_1)=t:=|z|$ is the number of marked points, and, as in Theorem~\ref{thm:lb_generic}, the total sensitivity is at most $1$ and the dimension is $O(1)$.

\emph{Lower bound.} Let $(B, w)$ be the weighted $\epsilon$-coreset output by the algorithm, where the weights are part of the output since $B$ alone carries only membership information. Then $\sum_{b\in B} w_b\cdot {\rm cost}(b, x_1)=(1\pm\epsilon)\cdot t$, and after reading the bits of the points of $B$, which takes at most $|B|$ further queries, the left-hand side is known. Therefore, an algorithm that outputs an $\epsilon$-coreset with $Q$ queries yields, with $Q+|B|$ queries, an estimate $\wt t$ with $|\wt t-t|\leq \epsilon t$. In other words, it solves the $2\epsilon$-approximate relative counting problem of~\cite{nw99} for $z$, and this problem is defined there with the strict inequality $|\wt t-t|<2\epsilon t$. By Theorem 1.13 of~\cite{nw99}, an algorithm that solves the $\epsilon'$-approximate relative counting problem on all inputs with at least one marked point makes, on inputs with $t$ ones,
\begin{align*}
    \Omega\Big(\sqrt{\frac{n}{\lceil \epsilon'(t+1)\rceil}}+\frac{\sqrt{t(n-t)}}{\lceil \epsilon'(t+1)\rceil}\Big)
\end{align*}
queries. Note that for $t=0$ the strict inequality cannot be met, and it is not required. The family consists of all $z\in\{0,1\}^n$, so the algorithm must succeed on every input. We evaluate the bound with $\epsilon'=2\epsilon$ on inputs with $t$ marked points, where $t+1=\lfloor 1/\epsilon\rfloor$, and for these inputs $\lceil 2\epsilon(t+1)\rceil\leq 2$. Moreover, $t\geq 1$ as $\epsilon<1/2$, and $t\leq n/2$ as $\lfloor 1/\epsilon\rfloor\leq n/2$, so the bound is $\Omega(\sqrt{t(n-t)})=\Omega(\sqrt{tn})=\Omega(\sqrt{n/\epsilon})$. Therefore, $Q+|B|=\Omega(\sqrt{n/\epsilon})$, and $Q=\Omega(\sqrt{n/\epsilon})$ whenever $|B|=o(\sqrt{n/\epsilon})$.

\emph{Upper bound.} We compute an estimate $\wt t$ with $|\wt t-t|\leq\epsilon t$ in two stages. First, the relative counter of Theorem 15 of~\cite{bhmt02} with accuracy $1/2$ returns $t_r=(1\pm 1/2)t$ with probability at least $2/3$ in $O(\sqrt{n/t})=O(\sqrt n)$ expected queries, and $t_r=0$ with certainty when $t=0$. If $t_r\leq 3/\epsilon$, then $t\leq 2t_r\leq 6/\epsilon$, and exact counting by Theorem 17 of~\cite{bhmt02} returns $t$ with probability at least $2/3$ in $O(\sqrt{(t+1)n})=O(\sqrt{n/\epsilon})$ expected queries. Otherwise, $t\geq 2t_r/3>2/\epsilon$, and we run the relative counter with accuracy $\epsilon$, which returns $\wt t=(1\pm\epsilon)t$ with probability at least $2/3$ in $O(\epsilon^{-1}\sqrt{n/t})=O(\sqrt{n/\epsilon})$ expected queries. If $\wt t=0$, we output the empty coreset. Otherwise, we find one marked point $b$ by the Grover search of Theorem 3 of~\cite{bhmt02} in $O(\sqrt{n/t})=O(\sqrt n)$ expected queries, and output $B=\{b\}$ with weight $w_b=\wt t$. Then we have $\sum_{b\in B}w_b\cdot {\rm cost}(b, x_1)=\wt t=(1\pm\epsilon)\cdot {\rm cost}(A, x_1)$, so $B$ is an $\epsilon$-coreset of size $1$. Each stage is repeated $O(1)$ times to make its error probability at most $1/20$, taking the median of the count estimates and checking a candidate point with one query. By Markov's inequality, each expected query bound becomes a worst-case bound by stopping the stage at $20$ times the bound. This uses $O(\sqrt{n/\epsilon})$ queries in total, and a union bound over the stages gives success probability at least $2/3$.
\end{proof}

\begin{theorem}[$\epsilon$-dependence of the query lower bound, general $S$]\label{thm:lb_eps_S}
Let $S\geq 2$ be an integer (the case $S=1$ is Theorem~\ref{thm:lb_eps}, and nothing below uses $S\geq 2$), $\epsilon\in (0, 1/8]$ and $n$ a multiple of $S$ with $\lfloor 1/(4\epsilon)\rfloor\leq n/(2S)$. On the block family of Theorem~\ref{thm:lb_generic} (total sensitivity at most $S$, dimension $O(1)$), any quantum algorithm that, on every instance, outputs with probability at least $2/3$ an $\epsilon$-coreset of size $o(\sqrt{nS/\epsilon})$ together with its weight vector makes $\Omega(\sqrt{nS/\epsilon})$ quantum queries to the points of $A$ on some instance. The bound is tight up to a $\log S$ factor: $O(\sqrt{nS/\epsilon}\log S)$ queries suffice on this family.
\end{theorem}

\begin{proof}
Let $N:=n/S$ be the block size and $t:=\lfloor 1/(4\epsilon)\rfloor$, so we have $2\leq t\leq N/2$, $\epsilon t\leq 1/4$, and $t\geq 1/(8\epsilon)$. Note that $\epsilon\leq 1/8$ gives $1/(4\epsilon)\geq 2$, hence $\lfloor 1/(4\epsilon)\rfloor\geq 1/(8\epsilon)$. Let $f:=f_{t,t-1}$ be the partial Boolean function of~\cite{nw99} on $\{0,1\}^N$. It is defined on inputs with exactly $t$ or exactly $t-1$ ones by $f(x)=1$ if and only if $|x|=t$. We apply Corollary 1.2 of~\cite{nw99} with $\Delta_\ell=1$ and $m=t-1$, which is the element of $\{t-1,t\}$ farthest from $N/2$ since $t\leq N/2$. The corollary shows that every quantum algorithm computing $f$ with error at most a constant below $1/2$ makes $\Omega(\sqrt{N}+\sqrt{(t-1)(N-t+1)})=\Omega(\sqrt{tN})=\Omega(\sqrt{N/\epsilon})$ queries, as $t-1\geq t/2$, $N-t+1>N/2$, and $t\geq 1/(8\epsilon)$. We write $Q_{1/4}(f)$ for the $1/4$-error query complexity, which is of the same order.

\emph{Reduction.} Consider an instance of the block family in which block $l$ has $t_l\in \{t-1, t\}$ marked points, and let $(B, w)$ be a weighted $\epsilon$-coreset output by the algorithm. As in the proof of Theorem~\ref{thm:lb_eps}, we read the bits of the points of $B$ with at most $|B|$ queries. This yields, for every block $l$, a number $\wt t_l=\sum_{b\in B\cap {\cal B}_l}w_b\,z_b$ with $(1-\epsilon)t_l\leq \wt t_l\leq (1+\epsilon)t_l$. Note that $(1+\epsilon)(t-1)<(1-\epsilon)t$ if and only if $2\epsilon t<1+\epsilon$, and the latter holds as $\epsilon t\leq 1/4$. Therefore, whenever the coreset is valid, the rule ``output $1$ if $\wt t_l\geq (1-\epsilon)t$'' decides $f(z_{{\cal B}_l})$ correctly for every block simultaneously. Consequently, the algorithm, followed by $|B|$ further queries, computes $f^{(S)}(z_{{\cal B}_1},\ldots,z_{{\cal B}_S})=(f(z_{{\cal B}_1}),\ldots,f(z_{{\cal B}_S}))$ with success probability at least $2/3$ on every input of the product domain, using $Q+|B|$ queries.

\emph{Direct product.} We write $Q_\eta(g)$ for the minimum number of queries of a quantum algorithm computing $g$ with error at most $\eta$ on every input of its domain. Theorem 1.1 of~\cite{lr13} states that for any partial function $f$, any integer $k\geq 1$, and any $\delta\in [2/3, 1]$, we have $Q_{1-\delta^k/2}(f^{(k)})\geq \frac{k\ln(3\delta/2)}{8000}\,Q_{1/4}(f)$. We apply the theorem with $k=S$ and $\delta=1$ to the algorithm above, which computes $f^{(S)}$ with error at most $1/3\leq 1/2$. Therefore, we have $Q+|B|\geq Q_{1/2}(f^{(S)})\geq \frac{S\ln(3/2)}{8000}\,Q_{1/4}(f)=\Omega(S\sqrt{N/\epsilon})=\Omega(\sqrt{nS/\epsilon})$, and $Q=\Omega(\sqrt{nS/\epsilon})$ whenever $|B|=o(\sqrt{nS/\epsilon})$.

\emph{Upper bound.} We run the $S=1$ algorithm of Theorem~\ref{thm:lb_eps} on each block, amplified to error $1/(3S)$ by $O(\log S)$ repetitions, and output the union of the $S$ single-point coresets. As in that proof, the amplification takes the median of the count estimates and one check query per candidate point. This uses $O(\sqrt{nS/\epsilon}\log S)$ queries in total, and a union bound gives success probability at least $2/3$.
\end{proof}

\section{Simulating Random Oracle Access}
\label{sec:QRAM}

In this section, we discuss how to simulate query access to a uniformly random string without impairing the runtime of our algorithm. We observe that this is necessary, as our classical oracle is inherently randomized. Converting it into a quantum circuit will produce a superposition of estimates, incurring additional approximation errors. We resolve this issue by ``derandomizing'' the algorithm, giving it query access to a uniformly random string. Specifically, suppose an algorithm has runtime $q$, then we need to attach a random string of length at most $q$. As observed in~\cite{ag23}, if we only care about query complexity, then we could explicitly sample these bitstrings and use QRAM to access them. In addition, if we want to achieve a sublinear runtime, then explicitly drawing these bitstrings would be too slow. On the other hand,~\cite{bbc+01} shows that any $(k/2)$-query quantum algorithm cannot distinguish a uniformly random string from a $k$-wise independent one, hence it is enough to use a $k$-wise independent hash function to simulate query access to uniformly random strings. The following result is due to~\cite{aw22} and elaborated in~\cite{ag23}:
\begin{lemma}[Claim 1 in~\cite{aw22}, restated in Lemma 3.12 of~\cite{ag23}]
\label{lem:derandomize}
Consider any quantum algorithm with runtime $q$ that uses a uniformly random string, then we can simulate this algorithm with a quantum algorithm
with runtime $\wt O(q)$ without random string, using an additional QRAM of $\wt O(q)$ bits.
\end{lemma}
Hence, to implement our algorithm in QRAM, we apply Lemma~\ref{lem:derandomize} to derandomize the procedure. This allows us to use more standard approaches to deal with deterministic routines: we convert them to quantum oracles using quantum arithmetic gates and Toffoli gates, which is common in the quantum algorithm literature. Each oracle in our applications reads its classical data, such as sketches, factors, and centers, from QRAM, adds up one term at a time, finds minima by binary search, and uncomputes all intermediate values, at a polylogarithmic overhead in time and queries. Hence, besides a register holding the queried row, column, or fiber, as in~\cite{ag23,xclj23}, it uses $O(\log n)$ qubits, or polylogarithmically many if it calls the oracle of Lemma~\ref{lem:q_ann}, provided each kernel evaluation also uses $O(\log n)$ qubits. The exceptions are the oracle of Algorithm~\ref{alg:quantum_nystrom}, which holds $\wt O(k/\epsilon)$ kernel values in a register, and the multidimensional counting and summation of~\cite{xclj23} in Lemmas~\ref{lem:multidim_count} and~\ref{lem:multidim_sum}, whose numbers of qubits we do not bound.

\section{Query Complexity of Our Algorithms}
\label{sec:query}

In this section, we summarize the complexity of our algorithms in terms of \emph{query complexity}, which is independent of the QRAM model. As noted in~\cite{aw22,ag23}, we can remove the QRAM assumption at a price in \emph{time complexity}. In particular, access to a QRAM of $m$ bits can be simulated in the circuit model at a multiplicative cost of $O(m)$ per access, see the discussion in Section 2 of~\cite{ag23}. Our algorithms use a QRAM of $m=\wt\Omega(Q)$ bits, where $Q\geq n^{0.5}$ is the query count in Table~\ref{tab:query_complexity}. The $k$-wise independent string of Section~\ref{sec:QRAM} must fool $Q$ queries, so $k\geq 2Q$~\citep{bbc+01}, and hence the string alone occupies $\Omega(Q)$ bits. Moreover, each of the $Q$ queries is accompanied by at least one access to the string, and therefore this simulation yields a time bound of at least $\wt\Omega(Q\cdot m)=\wt \Omega(Q^2)\geq \wt \Omega(n)$. It is worth noting that this bound is a statement about the simulation, not about the problems. Whether sublinear time without QRAM is achievable for these problems is open. The query complexities below are independent of QRAM.

\begin{table}[!ht]
    \caption{Query complexity of our algorithms. Throughout, we assume $n, d\gg\poly(k,1/\epsilon)$. In all clustering and $\ell_p$ regression rows, $\wt O(\cdot)$ is $\wt O_p(\cdot)$. The $(k,1\leq p<2)$-subspace row assumes the flatness precondition of Theorem~\ref{thm:quantum_lp}. We specify the types of query to measure the complexity for various problems. For most of the problems, the input is $A\in \R^{n\times d}$ that consists of $n$ points in $d$-dimensional Euclidean space, hence the row query corresponds to querying the points. For low-rank approximation, the input is $A\in \R^{d\times n}$ and the queries are column queries, except that the factored output of Theorem~\ref{thm:quantum_lra} also queries rows. For PSD and kernel low-rank approximation, the query corresponds to evaluating the kernel function over two points, where in the PSD low-rank approximation, this is equivalent to querying the entries of $A$. Finally, for tensor low-rank approximation, we allow queries to the rows, columns and tubes of $A$, which would require at most a constant factor more bits to store in the QRAM.}
    \label{tab:query_complexity}
    \centering
    \tablebox{%
    \ifdefined\isarxiv
    \begin{tabular}{|>{\raggedright\arraybackslash}p{5.0cm}|>{\raggedright\arraybackslash}p{2.6cm}|>{\raggedright\arraybackslash}p{2.7cm}|l|}
    \else
    \begin{tabular}{|l|l|l|l|}
    \fi
    \hline
        & Reference & Query Type & Complexity \\ \hline
      $k$-Median Clustering    & Theorem~\ref{thm:one_shot_main} & Rows & $n^{0.5}(k+\epsilon^{-1}k^{2/3})$ \\ \hline
      $k$-Means Clustering & Theorem~\ref{thm:one_shot_main} & Rows & $n^{0.5}(k+\epsilon^{-1}k^{3/4})$ \\ \hline
      $(k,p)$-Clustering & Theorem~\ref{thm:one_shot_main} & Rows & $n^{0.5}(k+\epsilon^{-1}k^{\frac{p+1}{p+2}})$ \\ \hline
      $\ell_{p\leq 2}$ Regression & Theorems~\ref{thm:lewis} and~\ref{thm:ls} & Rows & $\epsilon^{-1}n^{0.5}d^{0.5}$\\ \hline
      $\ell_{p>2}$ Regression & Theorem~\ref{thm:lewis} & Rows & $n^{0.5}d^{p/4}(\epsilon^{-1}+d^{0.5})$\\ \hline
      $(k,1\leq p< 2)$-Subspace Approx. & Theorem~\ref{thm:quantum_lp} & Rows & $\epsilon^{-1}n^{1-p/4}k^{p/4}$ \\ \hline
      $(k,p\geq 2)$-Subspace Approx. & Theorem~\ref{thm:quantum_lp} & Rows & $\epsilon^{-2/p}n^{1-1/p}k^{0.5}$ \\ \hline
      Low-Rank (column coreset) & Corollary~\ref{thm:quantum_pcp} & Columns & $\epsilon^{-1}n^{0.5}k^{0.5}$ \\ \hline
      Low-Rank (factored output) & Theorem~\ref{thm:quantum_lra} & Columns+Rows & $\epsilon^{-2}n^{0.5}k^{0.5}+\epsilon^{-1.5}d^{0.5}k^{0.5}$ \\ \hline
      PSD Low-Rank & Theorem~\ref{thm:quantum_psd}& Entries & $\epsilon^{-1.25}n^{0.75}k^{1.25}$ \\ \hline
      Kernel Low-Rank & Theorem~\ref{thm:quantum_psd}& Kernel Evals & $\epsilon^{-1.25}n^{0.75}k^{1.25}$\\ \hline
      Tensor Low-Rank: Rank-$k$ & Theorem~\ref{thm:quantum_fpt_two}& Rows, Columns and Tubes& $\epsilon^{-2}nk$ \\ \hline
      Tensor Low-Rank: Bicriteria (rank $k^2$) & Theorem~\ref{thm:quantum_bicriteria_two} & Rows, Columns and Tubes & $\epsilon^{-1}nk^{0.5}+\epsilon^{-3}n^{0.5}k^2$ \\ \hline
    \end{tabular}}
\end{table}

\ifdefined\isarxiv
\else
\newpage
\section*{NeurIPS Paper Checklist}

\begin{enumerate}

\item {\bf Claims}
    \item[] Question: Do the main claims made in the abstract and introduction accurately reflect the paper's contributions and scope?
    \item[] Answer: \answerYes{} 
    \item[] Justification: This paper is theoretical. The claimed bounds are summarized in Tables~\ref{tab:coreset} and~\ref{tab:optimization}, Section~\ref{sec:tech} gives the proof overview, and full proofs are in Appendices~\ref{sec:general}--\ref{sec:lb}.

\item {\bf Limitations}
    \item[] Question: Does the paper discuss the limitations of the work performed by the authors?
    \item[] Answer: \answerYes{} 
    \item[] Justification: We clearly discuss the limitations and three open problems in Section~\ref{sec:conclusion}.

\item {\bf Theory assumptions and proofs}
    \item[] Question: For each theoretical result, does the paper provide the full set of assumptions and a complete (and correct) proof?
    \item[] Answer: \answerYes{} 
    \item[] Justification: We provide the full set of assumptions and proofs. An overview can be found in Section~\ref{sec:tech}, and all details are in the appendix.

    \item {\bf Experimental result reproducibility}
    \item[] Question: Does the paper fully disclose all the information needed to reproduce the main experimental results of the paper to the extent that it affects the main claims and/or conclusions of the paper (regardless of whether the code and data are provided or not)?
    \item[] Answer: \answerNA{} 
    \item[] Justification: This paper is theoretical and has no experiments.

\item {\bf Open access to data and code}
    \item[] Question: Does the paper provide open access to the data and code, with sufficient instructions to faithfully reproduce the main experimental results, as described in supplemental material?
    \item[] Answer: \answerNA{} 
    \item[] Justification: This paper is theoretical and has no data and code.

\item {\bf Experimental setting/details}
    \item[] Question: Does the paper specify all the training and test details (e.g., data splits, hyperparameters, how they were chosen, type of optimizer) necessary to understand the results?
    \item[] Answer: \answerNA{} 
    \item[] Justification: This paper is theoretical and has no experiments.

\item {\bf Experiment statistical significance}
    \item[] Question: Does the paper report error bars suitably and correctly defined or other appropriate information about the statistical significance of the experiments?
    \item[] Answer: \answerNA{} 
    \item[] Justification: This paper is theoretical and has no experiments.

\item {\bf Experiments compute resources}
    \item[] Question: For each experiment, does the paper provide sufficient information on the computer resources (type of compute workers, memory, time of execution) needed to reproduce the experiments?
    \item[] Answer: \answerNA{} 
    \item[] Justification: This paper is theoretical and has no experiments.
    
\item {\bf Code of ethics}
    \item[] Question: Does the research conducted in the paper conform, in every respect, with the NeurIPS Code of Ethics \url{https://neurips.cc/public/EthicsGuidelines}?
    \item[] Answer: \answerYes{} 
    \item[] Justification: We strictly follow the Code of ethics of NeurIPS.

\item {\bf Broader impacts}
    \item[] Question: Does the paper discuss both potential positive societal impacts and negative societal impacts of the work performed?
    \item[] Answer: \answerNA{} 
    \item[] Justification: Our work is theoretical, and the quantum hardware it relies on still requires many further developments to realize. Moreover, it is also foundational research and is not tied to any applications or deployments. Hence, we foresee no negative societal impact, and also don't feel compelled to discuss it in the body of the paper.
    
\item {\bf Safeguards}
    \item[] Question: Does the paper describe safeguards that have been put in place for responsible release of data or models that have a high risk for misuse (e.g., pre-trained language models, image generators, or scraped datasets)?
    \item[] Answer: \answerNA{} 
    \item[] Justification: This paper is theoretical and has no experiments.

\item {\bf Licenses for existing assets}
    \item[] Question: Are the creators or original owners of assets (e.g., code, data, models), used in the paper, properly credited and are the license and terms of use explicitly mentioned and properly respected?
    \item[] Answer: \answerNA{} 
    \item[] Justification: This paper is theoretical and has no experiments, hence does not use any existing assets.

\item {\bf New assets}
    \item[] Question: Are new assets introduced in the paper well documented and is the documentation provided alongside the assets?
    \item[] Answer: \answerNA{} 
    \item[] Justification: This paper is theoretical and has no experiments and does not create any new assets.

\item {\bf Crowdsourcing and research with human subjects}
    \item[] Question: For crowdsourcing experiments and research with human subjects, does the paper include the full text of instructions given to participants and screenshots, if applicable, as well as details about compensation (if any)? 
    \item[] Answer: \answerNA{} 
    \item[] Justification: This paper is theoretical and has no experiments.

\item {\bf Institutional review board (IRB) approvals or equivalent for research with human subjects}
    \item[] Question: Does the paper describe potential risks incurred by study participants, whether such risks were disclosed to the subjects, and whether Institutional Review Board (IRB) approvals (or an equivalent approval/review based on the requirements of your country or institution) were obtained?
    \item[] Answer: \answerNA{} 
    \item[] Justification: This paper is theoretical and has no experiments.

\item {\bf Declaration of LLM usage}
    \item[] Question: Does the paper describe the usage of LLMs if it is an important, original, or non-standard component of the core methods in this research? Note that if the LLM is used only for writing, editing, or formatting purposes and does \emph{not} impact the core methodology, scientific rigor, or originality of the research, declaration is not required.
    \item[] Answer: \answerYes{} 
    \item[] Justification: LLMs were used as research assistants that helped establish several of the tighter results in this paper, including the lower bounds, the $\ell_p$ regression results, and the tighter bounds for $(k,p)$-clustering and $(k,p)$-subspace approximation. The authors verified all proofs and take full responsibility for the content, and the acknowledgments state how LLMs were used.

\end{enumerate}
\fi

\end{document}